\newcommand{\columncode}{2} 
\ifcase\columncode
  \errmessage{You have set the columncode to an undefined value  =\columncode. Exit.}
\or
  \documentclass[twoside,preprint,12pt,longtitle]{elsarticle}
  
\or
  \documentclass[twoside,fleqn,3p,twocolumn,times,longtitle]{elsarticle}
  
\else
  \errmessage{You have set the columncode to an undefined value = \columncode. Exit.}
\fi

\usepackage{subfigure}
\usepackage{graphicx}
\usepackage{ifthen}

\usepackage{url}

\input babarsym  

\newcommand{\comment}[1]{}

\newcommand{\sectiondir}{}
\newcommand{\secname}{}

\def\ie{{\em i.e.,}\xspace}
\def\eg{{\em e.g.,}\xspace}

\def\to{\rightarrow}

\def\svt{\mbox{SVT}\xspace}
\def\dch{\mbox{DCH}\xspace}

\def\dirc{\mbox{DIRC}\xspace}
\def\emc{\mbox{EMC}\xspace}

\def\rpc{\mbox{RPC}\xspace}

\def\svtrad{\mbox{SVTRAD}\xspace}

\def\HER{\mbox{HER}\xspace}
\def\LER{\mbox{LER}\xspace}

\def\Bmeson{$B$ meson}
\def\Bmesons{$B$ mesons}

\def\ifmath#1{\relax\ifmmode#1\else$#1$\fi}
\def\BB    {\ifmath{B\overline{B}{}}}

\def\hz   {\ensuremath{\mbox{\,Hz}}\xspace}
\def\khz  {\ensuremath{\mbox{\,kHz}}\xspace}
\def\mhz  {\ensuremath{\mbox{\,MHz}}\xspace}
\def\ghz  {\ensuremath{\mbox{\,GHz}}\xspace}

\def\mv   {\ensuremath{\mbox{\,mV}}\xspace}
\def\volt   {\ensuremath{\mbox{\,V}}\xspace}
\def\kv   {\ensuremath{\mbox{\,kV}}\xspace}

\def\kw   {\ensuremath{\mbox{\,kW}}\xspace}

\def\ohm    {\ensuremath{\mbox{\,$\Omega$}}\xspace}
\def\Mohm   {\ensuremath{\mbox{\,M$\Omega$}}\xspace}

\def\kohm       {\ensuremath{\mbox{\,k$\Omega$}}\xspace}

\def\Amp        {\ensuremath{\mbox{\,A}}\xspace}

\def\muA        {\ensuremath{\mbox{\,$\mu$A}}\xspace}
\def\nA         {\ensuremath{\mbox{\,nA}}\xspace}
\def\Coul       {\ensuremath{\mbox{\,C}}\xspace}
\def\fC         {\ensuremath{\mbox{\,fC}}\xspace}
\def\pC         {\ensuremath{\mbox{\,pC}}\xspace}
\def\pf         {\ensuremath{\mbox{\,pF}}\xspace}
\def\nf         {\ensuremath{\mbox{\,nF}}\xspace}

\def\mbar       {\ensuremath{\mbox{\,mbar}}\xspace}

\def\liter{\ensuremath{\,\ell}\xspace}

\def\tesla{\ensuremath{\mbox{\,T}}\xspace}

\def\gbyte      {\ensuremath{{\rm \,GByte}}\xspace}
\def\gbytes     {\gbyte}
\def\gbyteps    {\ensuremath{{\rm \,GByte/s}}\xspace}
\def\gbytesps   {\gbyteps}

\def\tbyte      {\ensuremath{{\rm \,TByte}}\xspace}
\def\tbytes     {\tbyte}

\def\pbyte      {\ensuremath{{\rm \,PByte}}\xspace}
\def\pbytes     {\pbyte}

\def\mbitps     {\ensuremath{{\rm \,Mbit/s}}\xspace}

\def\gbitps     {\ensuremath{{\rm \,Gbit/s}}\xspace}

\def\mrad       {\ensuremath{\mbox{\,mrad}}}  

\def\Rad        {\ensuremath{{\rm \,Rad}}\xspace}   
\def\MRad       {\ensuremath{{\rm \,MRad}}\xspace}  
\def\kRad       {\ensuremath{{\rm \,kRad}}\xspace}  
\def\mRad       {\ensuremath{{\rm \,mRad}}\xspace}

\def\pt         {\mbox{$p_{\rm t}$}}

\newcommand{\gae}{\lower 2pt \hbox{$\, \buildrel {\scriptstyle >}\over {\scriptstyle \sim}\,$}}
\newcommand{\lae}{\lower 2pt \hbox{$\, \buildrel {\scriptstyle <}\over {\scriptstyle \sim}\,$}}

\def\superb {Super\kern-0.04em\emph{B}\xspace}

\def\xrootd     {\mbox{\tt Xrootd}\xspace}

\newcommand{\BaBarAcronym}[3]{ \noindent {\bf #1}: {\it #2}, #3. \\}
\newcommand{\BaBarAcronymShort}[2]{ \noindent {\bf #1}: #2. \\}

\def\LOneAccept   {\emph{L1~Accept}\xspace}

\usepackage{lineno}
\usepackage{ textcomp }

\begin{document}

\begin{frontmatter}

\title{
{\small
\begin{flushleft}
\babar-PUB-13/009\\
SLAC-PUB-15456\\
arXiv:1305.3560 [physics.ins-det]\\
NIM{\bf A 729} (2013), 615-701\\
\end{flushleft}
}
\vspace*{1.5cm}
%
%
The \babar\ Detector: Upgrades, Operation and Performance
}

\cortext[cor1]{Principal corresponding author}
\cortext[cor2]{Corresponding author}

\author[LAPP]{B.~Aubert}
\author[LAPP]{R.~Barate}
\author[LAPP]{D.~Boutigny}
\author[LAPP]{F.~Couderc}
\author[LAPP]{P.~del~Amo~Sanchez}
\author[LAPP]{J.-M.~Gaillard}
\author[LAPP]{A.~Hicheur}
\author[LAPP]{Y.~Karyotakis}
\author[LAPP]{J.~P.~Lees}
\author[LAPP]{V.~Poireau}
\author[LAPP]{X.~Prudent}
\author[LAPP]{P.~Robbe}
\author[LAPP]{V.~Tisserand}
\author[LAPP]{A.~Zghiche}
\address[LAPP]{Laboratoire d'Annecy-le-Vieux de Physique des Particules (LAPP), Universit\'e de Savoie, CNRS/IN2P3, F-74941 Annecy-le-Vieux, France}

\author[Barcelona]{E.~Grauges}
\author[Barcelona]{J.~Garra~Tico}
\address[Barcelona]{Universitat de Barcelona, Facultat de Fisica, Departament ECM, E-08028 Barcelona, Spain}

\author[BariINFN,BariUniv]{L.~Lopez}
\author[BariINFN,BariUniv]{M.~Martinelli} 
\author[BariINFN,BariUniv]{A.~Palano}
\author[BariINFN,BariUniv]{M.~Pappagallo}
\author[BariINFN,BariUniv]{A.~Pompili}
\address[BariINFN]{INFN Sezione di Bari, I-70126 Bari, Italy}
\address[BariUniv]{Dipartmento di Fisica, Universit\`a di Bari, I-70126 Bari, Italy}

\author[IHEP]{G.~P.~Chen}
\author[IHEP]{J.~C.~Chen}
\author[IHEP]{N.~D.~Qi}
\author[IHEP]{G.~Rong}
\author[IHEP]{P.~Wang}
\author[IHEP]{Y.~S.~Zhu}
\address[IHEP]{Institute of High Energy Physics, Beijing 100039, China}

\author[Bergen]{G.~Eigen}
\author[Bergen]{B.~Stugu}
\author[Bergen]{L.~Sun}
\address[Bergen]{University of Bergen, Institute of Physics, N-5007 Bergen, Norway}

\author[LBNL]{G.~S.~Abrams}
\author[LBNL]{M.~Battaglia}
\author[LBNL]{A.~W.~Borgland}
\author[LBNL]{A.~B.~Breon}
\author[LBNL]{D.~N.~Brown} 
\author[LBNL]{J.~Button-Shafer}
\author[LBNL]{R.~N.~Cahn}
\author[LBNL]{E.~Charles}
\author[LBNL]{A.~R.~Clark}
\author[LBNL]{C.~T.~Day}
\author[LBNL]{M.~Furman} 
\author[LBNL]{M.~S.~Gill}
\author[LBNL]{Y.~Groysman}
\author[LBNL]{R.~G.~Jacobsen}
\author[LBNL]{R.~W.~Kadel}
\author[LBNL]{J.~A.~Kadyk}
\author[LBNL]{L.~T.~Kerth}
\author[LBNL]{Yu.~G.~Kolomensky}
\author[LBNL]{J.~F.~Kral}
\author[LBNL]{G.~Kukartsev}
\author[LBNL]{C.~LeClerc}
\author[LBNL]{M.~E.~Levi}
\author[LBNL]{G.~Lynch}
\author[LBNL]{A.~M.~Merchant}
\author[LBNL]{L.~M.~Mir}
\author[LBNL]{P.~J.~Oddone}
\author[LBNL]{T.~J.~Orimoto}
\author[LBNL]{I.~L.~Osipenkov}
\author[LBNL]{M.~Pripstein}
\author[LBNL]{N.~A.~Roe}
\author[LBNL]{A.~Romosan}
\author[LBNL]{M.~T.~Ronan\fnref{Deceased}}
\fntext[Deceased]{Deceased}
\author[LBNL]{V.~G.~Shelkov}
\author[LBNL]{A.~Suzuki}
\author[LBNL]{K.~Tackmann}
\author[LBNL]{T.~Tanabe}
\author[LBNL]{W.~A.~Wenzel}
\author[LBNL]{M.~Zisman} 
\address[LBNL]{Lawrence Berkeley National Laboratory and University of California, Berkeley, California 94720, USA}

\author[Birmingham]{M.~Barrett}
\author[Birmingham]{P.~G.~Bright-Thomas}
\author[Birmingham]{K.~E.~Ford}
\author[Birmingham]{T.~J.~Harrison}
\author[Birmingham]{A.~J.~Hart}
\author[Birmingham]{C.~M.~Hawkes}
\author[Birmingham]{D.~J.~Knowles}
\author[Birmingham]{S.~E.~Morgan}
\author[Birmingham]{S.~W.~O'Neale\fnref{Deceased}} 
\author[Birmingham]{R.~C.~Penny}
\author[Birmingham]{D.~Smith} 
\author[Birmingham]{N.~Soni}
\author[Birmingham]{A.~T.~Watson}
\author[Birmingham]{N.~K.~Watson}
\address[Birmingham]{University of Birmingham, Birmingham, B15 2TT, United Kingdom}

% ...
\fntext[ccin2p3]{Staff member of the Centre de Calcul IN2P3, Lyon, France}
\author[LAL]{T.~C.~Petersen}
\author[LAL]{S.~Plaszczynski}
\author[LAL]{S.~Pruvot}
\author[LAL]{P.~Roudeau}
\author[LAL]{M.~H.~Schune}
\author[LAL]{J.~Serrano}
\author[LAL,RomaINFN,RomaUniv]{V.~Sordini}
\author[LAL]{A.~Stocchi}
\author[LAL]{V.~Tocut} 
\author[LAL]{S.~Trincaz-Duvoid}
\author[LAL]{L.~L.~Wang}
\author[LAL]{G.~Wormser}
\address[LAL]{Laboratoire de l'Acc\'el\'erateur Lin\'eaire, IN2P3/CNRS et Universit\'e Paris-Sud 11, Centre Scientifique d'Orsay, F-91898 Orsay Cedex, France}

% ...

\begin{abstract}
The \babar\ detector operated successfully at the \pep2\ asymmetric \epem\ collider at the SLAC National Accelerator Laboratory from 1999 to 2008.  This report covers upgrades, operation, and performance of the collider and the detector systems, as well as the trigger, online and offline computing, and aspects of event reconstruction since the beginning of data taking. 

\end{abstract}

\begin{keyword}
\babar\ detector upgrade \sep \babar\ operational experience \sep \pep2\ storage ring operation \sep beam monitoring
\end{keyword}


\end{frontmatter}

\tableofcontents

\cleardoublepage


\renewcommand{\secname}{intro_}
\renewcommand{\sectiondir}{Sec01_Intro}
\section{Introduction}
\label{sec:intro_introduction}

\subsection{Overview}
The \babar\ detector~\cite{Aubert:2001tu} operated at the \pep2\ asymmetric \epem\ collider~\cite{:1994mp,Seeman:2002xa,Seeman:2008zz}
at the SLAC National Accelerator Laboratory from 1999 to 2008.
The experiment~\cite{Boutigny:1995ib} was optimized for detailed studies of \CP-violating asymmetries in the decay of $B$ mesons,
but it was well suited for a large variety of other studies~\cite{Harrison:1998yr}, for instance, precision measurements of decays of bottom and charm mesons and $\tau$ leptons, and searches for rare processes, including many not expected in the framework
of the Standard Model of electroweak interactions.

The \pep2\ collider operated in the center-of-mass (c.m.) energy range of
9.99~\gev (just below the \TwoS\ resonance) to 11.2~\gev, mostly at 10.58 \gev, corresponding to the mass of the \FourS\ resonance.
This resonance decays exclusively to \BzBzb\ and $B^+B^-$ pairs and
thus provides an ideal laboratory for the study of $B$ mesons.  At the
\FourS\ resonance, the electron beam of 9.0~\gev\ collided head-on with the
positron beam of 3.1~\gev\ resulting in a Lorentz boost to the
\FourS\ resonance of $\beta\gamma =0.56$. This boost made it possible to
reconstruct the decay vertices of the $B$ and $\Bbar$ mesons, to determine
their relative decay times, and  to measure the time
dependence of their decay rates, a feature that was critical for the
observation of \CP-violation in \Bz - \Bzb\ mixing.

To reach the desired sensitivity for the most interesting
analyses, datasets of order $10^8$ to $10^9$ $B$ mesons were needed.
For the peak cross section at the \FourS\ of 1.1~\nb, this required
an integrated luminosity of the order 500~\invfb, that is, many years
of reliable and highly efficient operation of the detector, and stable
operation of the \pep2\ storage rings at luminosities exceeding the design of
$3 \times 10^{33} \mathrm{cm}^{-2} \mathrm{s}^{-1}$.

The \pep2\ storage rings gradually increased their performance and towards the end of the first year of data-taking
routinely delivered beams close to design luminosity.  In the following years, a series of upgrades was implemented
to reach a maximum instantaneous luminosity of four times the design and to exceed the design integrated luminosity
per day by a factor of seven~\cite{Seeman:2008zz}.
Among these upgrades, one of the most important was {\it trickle} injection~\cite{Kozanecki:2004wi}, \ie\ continuous injections into both beams, replacing the traditional method of replenishing the beam currents
every 40-50 minutes after they had dropped to about  30-50\% of the maximum.

From the start of \babar\ operation, the goal was to operate the detector at the highest level of efficiency to maximize the data rate and data quality. Once it became obvious that \pep2\ was capable of exceeding its design luminosity, continuous improvements to the hardware, electronics, trigger, data acquisition system (DAQ), and online and offline computing were required. Moreover, the instrumentation to assess, monitor, and control backgrounds and other environmental conditions, and to handle ever-increasing trigger rates had to be enhanced.  These enhancements served the routine operation at higher data rates, and also provided the information needed to understand operational limitations of the detector and software and to subsequently design the necessary upgrades in a timely manner.  To attain such large, high quality datasets and maximize the scientific output, the accelerator, detector, and analyses had to perform coherently in {\it factory} mode, with unprecedented operational efficiency and stability. This factory-like  operation required that experimenters paid very close attention to what were often considered routine monitoring and quality assurance tasks. 
As a result, \babar\ logged more than 96\% of the total delivered
luminosity, of which 1.1\% were discarded during reconstruction
because of hardware problems that could impact the physics analyses.

This review emphasizes the \babar\ detector upgrades, operation, and
performance as well as the development of the online and offline
computing and event reconstruction over a period of almost ten years since the start of data taking in 1999. Following this brief introduction, an
overview of the design of the principal components of the detector, the trigger, the
DAQ, and the online computing and control system is provided.
A brief description of the \pep2\ collider and the interaction region 
is followed by a description of its gradual evolution and upgrades, as well as the performance  
and monitoring of the collider operation, and the
closely related \babar\ background suppression and monitoring.
The following section covers the upgrades to the online computing
and DAQ systems, the trigger, the front-end electronics, and also
the replacement of the muon detectors in the barrel and forward
regions. Next, the operational experience with all detector systems is 
described in detail. The last sections cover selected topics related
to the event reconstruction, and provide an overview of the offline computing, including the provision for long-term access to data and analysis software.

\subsection{Detector System Requirements}
\label{sec:system-requirements}

The need for full reconstruction of $B$-meson decays (which have an average multiplicity
of 5.5 charged particles and an equal number of photons),
and, in many analyses, the additional requirement to tag the flavor of the second $B$ or to fully reconstruct its decay,
place stringent requirements on the detector:

\begin{itemize}

 \item large and uniform acceptance down to small
 polar angles relative to the boost direction;

 \item excellent reconstruction efficiency for charged particles down
 to a momentum of 40~\mevc\ and for photons to an energy of 30~\mev;

 \item excellent momentum resolution to separate small signals from
 relatively large backgrounds;

\item very good energy and angular resolutions for the detection of
photons from \piz\ and $\eta^0$ decays, and from radiative decays
in the full energy range, from 30~\mev to 4~\gev;

 \item efficient reconstruction of secondary vertices;

 \item efficient electron and muon identification, with low
 misidentification probabilities for hadrons;

 \item efficient and accurate identification of hadrons over a wide
 range of momenta for $B$ flavor-tagging (mostly for momenta below 1~\gevc ) and
 the reconstruction of exclusive decays (up to a momentum of 4~\gevc );

\item detector components that can tolerate significant radiation
  doses and operate reliably under high-background conditions;

 \item a flexible, redundant, and selective trigger system,
highly efficient for all kinds of signal events;

 \item low-noise electronics and a reliable, high bandwidth DAQ
and control system;

 \item detailed monitoring and automated calibrations; stable power supplies,
plus control of the environmental conditions to ensure continuous and stable operation;

 \item an online computing system and network that can control,
 process, and store the expected high volume of data;

\item  reconstruction software able to fully exploit the capabilities
  of the detector hardware;

\item a detector simulation of sufficient fidelity to support the
  detailed understanding of detector response appropriate for the
  high-statistics data samples; and

\item an offline computing system scaled to the data 
   flow  arising from {\it factory} operation, and capable of
   supporting a wide variety of highly sophisticated analyses.
\end{itemize}

\subsection{Detector Design and Layout}
\label{sec:detector-design}

The \babar\ detector was designed and built by a large international
team of scientists and engineers.  Details of its original design were
documented in the Technical Design Report~\cite{Boutigny:1995ib}, while
the construction and initial performance of the detector are described
in a later publication~\cite{Aubert:2001tu}.

\begin{figure*}[htb!]
\begin{center}
\includegraphics*[width=16cm]{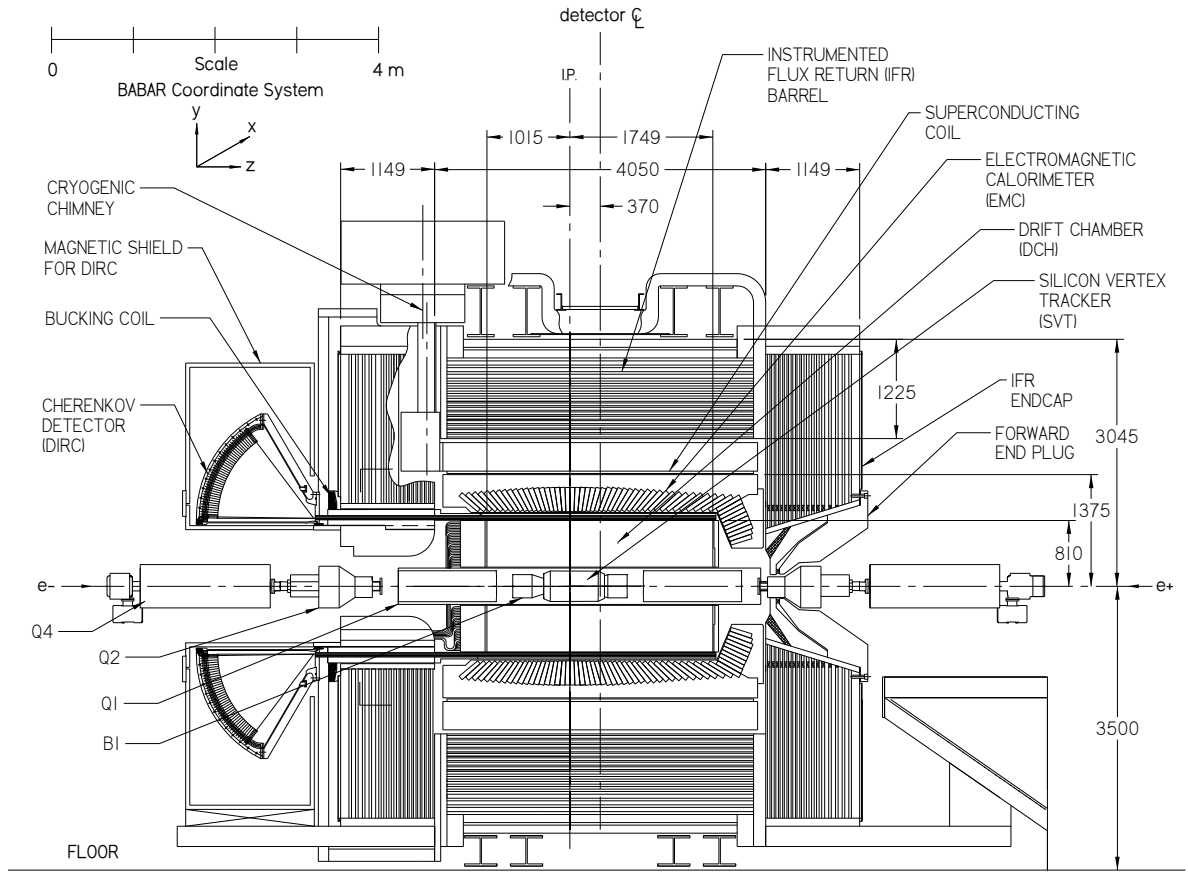}
\caption{\babar\ detector longitudinal section~\cite{Aubert:2001tu}.}
\label{fig:detectorElevation}
\end{center}
\end{figure*}

Figure~\ref{fig:detectorElevation} shows a longitudinal
section through the detector center
with the principal dimensions. To
maximize the geometric acceptance for the boosted \FourS\ decays, the
whole detector was offset from the interaction point by 0.37\m\ in the direction of the high-energy electron beam.

The inner detector consisted of a silicon vertex tracker, a drift
chamber, a ring-imaging Cherenkov detector, and an electromagnetic
calorimeter. These detector systems were surrounded by a
superconducting solenoid which provided a field of 1.5\tesla. The
steel flux return was instrumented for muon and neutral hadron
detection.  The polar angle coverage extended to 350\mrad\ in the
forward direction and 400\mrad\ in the backward direction, defined
relative to the direction of the high-energy beam.  As indicated in
Figure~\ref{fig:detectorElevation},
the right-handed coordinate system was anchored on the main tracking
system, the drift chamber, with the $z$-axis coinciding with its
principal axis. This axis was offset relative to the direction of the $e^-$ beam by 20\mrad\ in the horizontal plane.
The positive $y$-axis pointed upward and the positive
$x$-axis pointed away from the center of the \pep2\ storage rings. For reference,
the detector was located on the eastern section of the storage rings,
with the electron beam entering from the north.

The forward and backward acceptance of the tracking system was constrained
by components of \pep2, a pair of dipole magnets (B1) followed by a
pair of quadrupole magnets (Q1). The vertex detector and these magnets
were placed inside a tube (4.5\m\ long and 0.434\m\ inner
diameter) that was supported from the detector at the backward end
and by a beam-line support at the forward end.
The central section of this support tube was fabricated from a carbon-fiber composite with a thickness of 0.79\% of a radiation length.

The detector was of compact design, its transverse dimension being
constrained by the 3.5\m\ elevation of the beam above the floor.  The
solenoid radius was chosen by balancing the physics requirements and
performance of the drift chamber and calorimeter against the total
detector cost.

Since the average momentum of charged particles produced in $B$-meson
decays is less than 1\gevc, the precision of the measured track
parameters was heavily influenced by multiple Coulomb scattering.
Similarly, the detection efficiency and energy resolution of 
low-energy photons were severely affected by material in front of the
calorimeter.  Thus, special care was taken to keep material in
the active volume of the detector to a minimum. At normal incidence,
a particle would transverse 4\% of a radiation length prior to
entering the drift chamber and
another 26\% to reach the calorimeter.  

\subsection{Detector Components}
\label{sec:detector-components}

The charged particle tracking system was made of two components, the
silicon vertex tracker (SVT) and the drift chamber (DCH).  Pulse height 
information from the SVT and DCH was also used to measure ionization
loss for charged particle identification (PID).
The SVT was designed to measure positions and angles of charged
particles just outside the beam pipe.
It was composed of five
layers of double-sided silicon strip detectors that were assembled from
modules with readout at each end.  The inner three layers primarily provided
position and angle information for the measurement of the vertex
position. They were mounted as close to the water-cooled beryllium beam
pipe as practical, thus minimizing the impact of multiple scattering
in the beam pipe on the extrapolation of tracks to the vertex.  The outer two
layers were at much larger radii, providing the coordinate and angle
measurements needed for linking SVT and DCH tracks. 

The principal purpose of the DCH was the momentum measurement for
charged particles. It also supplied information for the charged
particle trigger and \dedx\ measurements for particle
identification. The DCH was of compact design, with 40 layers
of small, hexagonal cells. Longitudinal information was
derived from wires placed at small angles to the principal axis. By
choosing Al field wires and a helium-based gas mixture, the multiple
scattering inside the DCH was kept to a minimum. The readout
electronics were mounted on the backward endplate of the chamber,
minimizing the amount of material in front of the forward calorimeter.

The DIRC, the detector of internally reflected Cherenkov light, was a
novel device providing separation of pions and kaons up to the
kinematic limit of 4.5\gevc.  Cherenkov light was
produced in 4.9-m long bars of synthetic fused silica of rectangular
cross section (1.7\cm\ $\times$ 3.5\cm), and transported by total
internal reflection, preserving the angle of emission, to a large array of
photomultiplier tubes (PMT).  This array formed the backward wall of 
the toroidal, water-filled {\it standoff box} (SOB) 
which was located beyond the backward end of the
magnet. Images of the Cherenkov rings were reconstructed from the
position of the PMT and time of arrival of the signals~\cite{Adam:2004fq}.

The electromagnetic calorimeter (EMC) was designed to detect
electromagnetic showers with excellent energy and angular resolution
over the energy range from 20\mev\ to 4\gev. This coverage
allowed the detection of low energy $\pi^0$s and $\eta^0$s from $B$ decays
and high
energy photons and electrons from electromagnetic, weak, and radiative
processes. The EMC was a finely segmented array of 
thallium-doped cesium iodide [CsI(Tl)] crystals of
projective geometry.
To obtain the desired resolution,
the amount of material in front of and between the crystals was held to a
minimum. The individual crystals were read out by pairs of silicon PIN
diodes.

The instrumented flux return (IFR) was designed to identify muons and
to detect neutral hadrons.  For this purpose, the steel of the magnet flux return
in the barrel and the two end doors were segmented into layers,
increasing in thickness from 2\cm\ on the inside to 10\cm\ at the
outside. Between these steel absorbers, gaseous detectors were placed.
Initially, single gap resistive plate chambers (RPC) were inserted
to detect streamers from ionizing particles via external capacitive readout strips.
There were 19 layers of RPCs in the barrel sextants and 18 layers in the endcaps.
Starting in 2004, the RPCs in the barrel section were replaced by limited streamer tubes (LST)~\cite{Bauer:1986pc}
and six of the gaps were filled with brass plates to increase the total absorber thickness. Beginning in 2001, the original RPCs in the forward endcap were replaced by RPCs
of improved design and performance. The absorber thickness was also increased~\cite{Anulli:2004zu}.

\subsection {Electronics, Trigger, Data Acquisition and Computing}
\label{sec:ETDAQOL}

The electronics, trigger, DAQ and online computing
were a collection of tightly coupled hardware and software
systems.  These systems were designed to maximize the performance, 
reliability  and maintainability, while minimizing dead time,
complexity, and cost.

\subsubsection{Electronics}
\label{sec:electronics}

All detector systems shared a common electronics architecture.
Event data from the detector flowed through the front-end electronics
(FEE), while monitoring
and control signals were handled by a separate, parallel readout system.  All
FEE systems were located outside the detector, to minimize dead space within
the detector volume. They were directly mounted on the detector to optimize
performance and to minimize the cable plant,
thereby avoiding noise pickup and ground loops in long signal cables.
All detector systems utilized standard interfaces to the data
acquisition electronics and software.

The FEE channel consisted of an amplifier, a digitizer, a trigger
latency buffer for storing data during the Level~1 (L1)
trigger processing, and
a de-randomizing event buffer for storing data between the \LOneAccept\
and the subsequent transfer to the DAQ system.
Custom integrated circuits had been developed to
perform the signal processing.  The digital L1 latency buffers
functioned as fixed-length data pipelines managed by common protocol
signals generated by the fast control and timing system (FCTS).  All
de-randomizing event buffers functioned as FIFOs capable of storing a
fixed number of events.  During normal operation, analog signal
processing, digitization, and data readout occurred continuously and
simultaneously.

\subsubsection{Trigger}
\label{sec:trigger}

The function of the trigger system was to identify signatures of \B\
decays and other interesting events on which to initiate the detector
read-out and, upon further processing, to select a subset of these
events for permanent storage.
Correspondingly, the trigger was built in two subsequent
stages, the second conditional upon the first, with the first stage
being \emph{clock driven}, the second \emph{event driven}.

The L1 hardware trigger was implemented in custom-built
electronics consisting of dedicated processor boards that received
input data continuously
from the drift chamber, electromagnetic calorimeter, and instrumented
flux return.
These boards contained firmware to reconstruct trigger
\emph{primitives}, such as short and long drift chamber tracks and
minimum ionizing, medium and high energy showers in the calorimeter,
which could be counted, matched and combined into triggers via a fully
configurable logic.
Typical L1 Accept rates were about 2\khz at a latency of 11\mus.

The Level~3 (L3) trigger was implemented in software and ran on a small
farm of commercial processors.
It was the first stage of the DAQ system to see complete events, on
which it executed a variety of algorithms based primarily on the
partial reconstruction of DCH and EMC data.
The L3 trigger reduced the event rate to a level that was manageable for offline processing and storage.

The trigger architecture left room to accommodate an intermediate
Level~2, in case future background conditions demanded higher rejection
rates.

\subsubsection{Data Acquisition and Online Systems}
\label{sec:daq-online}

The DAQ and online computing systems were responsible for the transfer of event data from the detector FEE to mass storage with minimum dead time.
These systems also interfaced with the trigger to enable
calibrations and testing.  Other parts of these systems provided 
the control and monitoring of the detector and supporting facilities.
The emphasis was on efficient operation, close monitoring and easy diagnosis of problems online to assure the high quality of the recorded data.  To achieve these goals, the system had to be adaptable to changing conditions and had to be monitored and constantly improved and innovated.

\subsubsection{Reconstruction and Offline Computing}

Recorded data were reconstructed with entirely new, object-oriented,
software developed specifically for \babar, and which - compared
to previous experiments - included advances in the sophistication 
of its algorithms, and
corresponding increases in their computational requirements.  
The software handled reconstruction of
charged particles and vertex finding, the reconstruction of neutral
particles, and the identification of charged leptons and hadrons. It
also handled common analysis tasks such as the generation of lists of
particles originating from the decays of short-lived particles.

By the conclusion of data taking, the offline computing system 
had developed into a complex of six large computer centers, 
at the host laboratory and at national computing facilities in 
\babar\ member countries.
Production computing and user analysis were distributed across these
sites, which hosted calibration, reconstruction, selection of subsets
of data useful for specific analyses, and supported individual analyses 
by a large community of users.  Event simulation, based on a
\geant~\cite{Agostinelli:2002hh}
framework, was run at these computing centers, as well as a
network of smaller laboratory and university sites.

Following the final shutdown in 2008, a new computing system, isolated from the main SLAC computers, was developed to preserve the data and standard analysis and simulation software and support future analyses of the large \babar\ data sample. This long term data access (LTDA) system relies on virtualization technologies and uses distributed computing and storage methods. 

\subsection{Detector Operation}
\label{sec:det-operations}
The \babar\ detector was designed and built by a large international collaboration and its commissioning and operation was shared among the scientists and engineers.  To a large degree, groups who contributed to the construction of a certain detector system also supported its operation, maintenance, and upgrades. 
The overall responsibility for the detector was assigned to the technical coordinator, who was supported by system managers~-- one or two per detector system, for the trigger, the online and data acquisition systems, and the central operation.  The system managers' responsibility extended from the detector components and infrastructure to the online monitoring, the feature extraction of the signals, the calibration and alignment, to the offline reconstruction.  
System managers also oversaw repairs and upgrades.

During the nine years prior to the final shutdown in the spring of 2008, the \pep2\ storage rings operated over long periods.  In total, there were seven such periods, referred to as {\it Runs}, separated by  shutdowns of several months to allow for extended maintenance and upgrades of the collider and the detector. 
During Runs 1--6, data were recorded at the \FourS\ resonance and 40\mev below, while during Run 7 the energy of \pep2\ was changed to record data at two lower-mass resonances, the \ThreeS\ and \TwoS, and to perform a scan at c.m. energies above the \FourS\ resonance, up to a maximum of 11.2 \gev.

For a given Run, detector conditions were kept stable. Access to the detector was kept to a minimum, which meant that, for components that were difficult to reach (like the front-end electronics), their power distribution and cooling systems, their reliability and specific diagnostics were of great importance. The main power supplies, read-out electronics and trigger, 
gas and coolant distributions were accessible during beam operation. 
Short shutdowns of a day or two were arranged with advanced notice for routine detector and \pep2\ maintenance and repairs, and whenever the operation was impaired or the data quality compromised.  Emergency shutdowns were rare.
  
Major maintenance activities, significant improvements and upgrades to the detector and associated electronics and software were executed during the long shutdowns between Runs.  
The principal shutdown activities are summarized in Table~\ref{tab:shutdowns}; more details are presented in the following sections.  

\begin{table}[htbp!]
\caption{ \label{tab:shutdowns} 
Summary of the \babar\ shutdowns, specifying
the dates and upgrades activities.
}
   \centering
   \begin{tabular}{cl}
     \hline
      Shutdown   & Upgrades Activities \cr
     \hline
    11/00 - 01/01 & Add beam line shielding     \cr
                  & Replace 12 FW endcap RPCs   \cr
    07/02 - 11/02 & Replace remaining FW endcap RPCs \cr 
                  & Insert 5 brass absorbers in FW endcap \cr
                  & Upgrade the online computer farm \cr
    07/03 - 08/03 & Install $z$-dependent track trigger \cr 
    08/04 - 03/05 & Install large \pep2\ shield wall \cr
                  & Install LSTs in top/bottom sextants \cr
                  & Insert brass absorbers in 2 sextants \cr 
    10/05 - 11/05 & Upgrade DCH FEE with new FPGAs \cr
    09/06 - 12/06 & Install LSTs in 4 remaining sextants \cr
                  & Insert brass absorbers in 4 sextants \cr
    10/07 - 11/07 & Regular maintenance only \cr
                  \hline
   \end{tabular}
\end{table}


\renewcommand{\secname}{mdi_}
\renewcommand{\sectiondir}{Sec02_PEP-MDI}
\section{\pep2 Operation and Interface to \babar }
\label{sec:mdi:pep2-operation}

\subsection{Overview of \pep2 }
\label{sec:mdi:pep2-layout}  

\pep2\ was an \epem\ storage ring system designed to operate at a
c.m. energy of 10.58\gev, corresponding to the mass
of the \FourS\ resonance.  
The collision parameters of these energy-asymmetric
storage rings are presented in Table~\ref{tab:tabpep2},
both for the original design and for the best performance, which were all 
achieved in the last two years of operation.
\pep2\ surpassed all its design parameters. In particular, the instantaneous luminosity
exceeded design by a factor of four, and the integrated luminosity per day by a factor of seven.
For a  description of the design and operational experience of
\pep2\ we refer to the Conceptional Design Report~\cite{:1994mp}, conference proceedings~\cite{Seeman:2002xa,Seeman:2008zz}, and references therein.
An overview of the various improvements and upgrades is presented below.

\begin{table}[htbp!]
\caption{\pep2\ beam parameters. Values are given both for the design and for the best performance reached in 2006-2008.
HER and LER refer to the high energy (\en) and low energy (\ep) ring, respectively.
$\beta_x^*$  and $\beta_y^*$ refer to the horizontal and vertical $\beta$ functions at the collision point, and
$\xi_y$ refers to the vertical beam-beam parameter limit.
}
\label{tab:tabpep2}
\small
\begin{tabular}{lcc}
\\
\hline					
Parameters 			&Design 	& Best	\\
\hline		
Energy HER/LER~(\gev)	&9.0/3.1	&9.0/3.1 	\\
Current HER/LER~(\Amp) 	&0.75/2.15 	&2.07/3.21 	\\
Number of bunches~   		&1658       &1732		\\
$\beta_y^*$ (\mm)		&15-25	&9-10		\\
$\beta_x^*$ (\cm)		&50	      & 30-35	\\
Emittance $\epsilon_y$ (nm)   & 2   & 1.4   \\
Emittance $\epsilon_x$ (nm)   & 49  & 30-60     \\
Bunch Length (\mm)		& 15  & 10-12	\\
$\xi_y$	            & 0.03      & 0.05 - 0.065 \\
Luminosity~($10^{33}\cms$) &3 	&12 \\
Int. Luminosity~(\invpb/day) & 135   	&911		\\
\hline
\end{tabular}
\normalsize
\end{table}

\begin{figure}[htb!]  
  \begin{center} 
  \includegraphics[width=7.5cm]{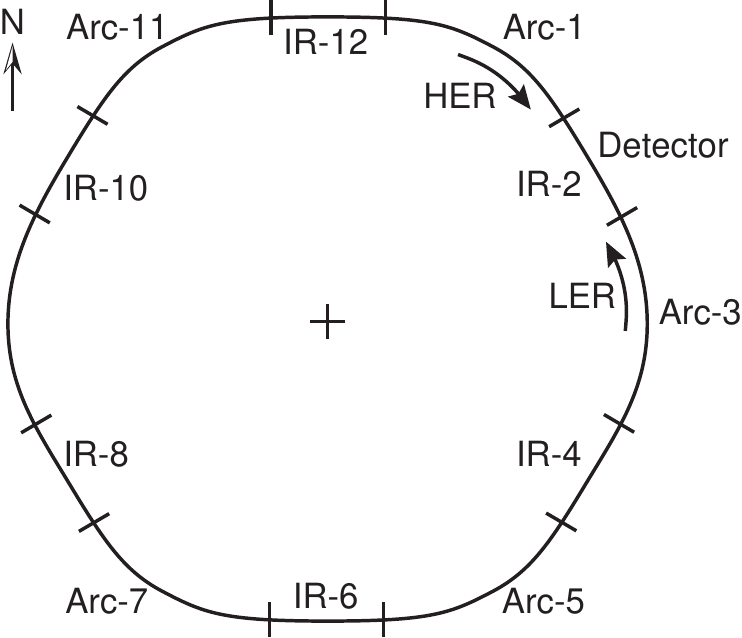}  
    \caption{Layout of the \pep2\ tunnel with six straight sections (IR) and six 
arc sections (Arc), labeled clockwise with even numbers for the IRs and odd numbers for the Arcs.
The \babar\ detector was located at IR-2. }    
    \label{fig:PEPII_Layout}  
  \end{center}  
 \end{figure}

\begin{figure}[htb!]  
  \begin{center} 
  \includegraphics[width=7.5cm]{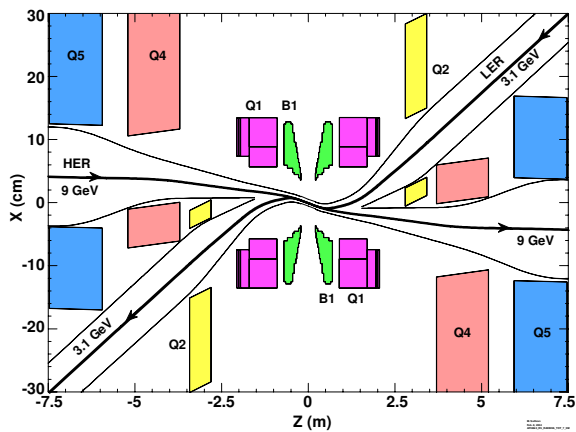}  
    \caption{Plan view of the \pep2\ interaction region layout with dipole and quadrupole magnets.  The scale of the vertical axis is strongly enlarged. The black lines indicate the orbits (thick line) and the physical apertures (thin lines on either side of the orbits) of the two beams.}  
    \label{fig:PEPII_IR}  
  \end{center}  
 \end{figure}

The high beam currents and the large number of closely-spaced bunches,  
all required to produce the high luminosity of \pep2, tightly coupled  
detector design, interaction region layout, machine operation,  
and remediation of machine-induced background.
Figure~\ref{fig:PEPII_Layout} shows the layout of the PEP tunnel with six straight sections (IR) connected by six arc sections (Arc).   The circumference is 2,200 m.  
Since the two rings had the same circumference,  
each bunch in one ring collided with only one bunch in the other ring. 
The bunches collided head-on and were separated in the horizontal plane  
by a pair of dipole magnets (B1), located at $\pm 21$\cm on each side of the collision point (IP),  
each followed by a series of offset quadrupoles.  A schematic of the \pep2\ interaction region is shown in Figure~\ref{fig:PEPII_IR}.  
The tapered B1 dipoles and the Q1 quadrupoles were samarium-cobalt permanent magnets located inside the field of the \babar\ solenoid. The Q2, Q4, and Q5 quadrupoles, located outside or in the fringe field of  
the solenoid, were standard iron magnets. The collision axis was offset  
from the $z$-axis of the \babar\ detector by  20\mrad\ in the  
horizontal plane \cite{Sullivan:1997ex} to minimize the perturbation of the  
beams by the detector solenoid field.  
 
The beams were enclosed by a water-cooled beam pipe of
27.9\mm\ outer radius, composed of two layers of beryllium (0.83\mm
and 0.53\mm\ thick) with a 1.48\mm\ water channel between them.  To attenuate photons from synchrotron radiation, the inner surface of the pipe was
coated with a 4\mum\ thin layer of gold. In addition, the beam pipe
was wrapped with 150\mum\ of tantalum foil on either side of the IP,
beyond $z=+10.1$\cm\ and $z=-7.9$\cm.
The total thickness of the central beam pipe section at normal
incidence corresponded to 1.06\% of a radiation length.

The beam pipe, the permanent magnets, and the \svt were assembled,
aligned, and enclosed in a 4.5\m-long support tube, which spanned the
IP.  The central section of this tube was fabricated from a carbon-fiber
epoxy composite with a thickness of 0.79\% of a radiation length.

\subsection{\pep2 Evolution and Upgrades}

\subsubsection{\pep2\ Instrumentation}
\label{pep2-instrumentayion}

The \pep2\ storage rings were equipped with instrumentation and electronics designed to monitor the beam orbits, measure beam profiles, optimize the lattice parameters, facilitate efficient injection and thereby achieve  efficient operation and high luminosity.  Fast feedback systems were critical to maintain the beams in collision.
The very high bunch frequency and high currents required novel devices and controls, specifically fast beam position monitors and bunch-by-bunch current monitors.  In the following, a brief description of three of the important beam monitoring devices is given, followed by a description of the beam collimation system.

\paragraph{Beam Profile Monitors}
The transverse and longitudinal bunch profiles were measured by synchrotron-light monitors (SLM) in each ring using visible to near ultraviolet light (600 to 200 nm). The transverse beam sizes were determined by direct imaging~\cite{Fisher:1998nv} and by interferometry~\cite{Fisher:2002hd}.  A streak camera~\cite{Fisher:2005rs} recorded longitudinal profiles. 
HER measurements were made at the high-dispersion point in the center of Arc-7. 
The LER SLM was installed in IR-2, 30 m downstream of the IP. 
For each ring, a water-cooled mirror directed the light, incident at 3 degrees to grazing, through a fused-silica window to a nearby optical system to record the image with a CCD video camera. In addition, a beam splitter directed the light to an interferometer, designed to measure the small vertical beam size.  A synchrotron x-ray monitor~\cite{Fisher:2004ne} was added later at the downstream end of Arc-7 in the LER.

Calculations to translate beam sizes measured with synchrotron light in the arc sections to those at the IP~\cite{Kozanecki:2008zz} relied on a one-turn matrix formalism~\cite{Cai:2003ef} and matrices taken from machine characterization measurements obtained for a single beam at low bunch current.  This translation of the single-beam measurements at low currents to collisions at high currents accounted for lattice changes and beam-beam interactions. 

\paragraph{Beam Position Monitors}
Beam position monitors (BPMs) were required to determine the transverse positions of the beams over an extremely wide range of beam intensities:  from a multi-bunch beam in the full ring with bunches of up to $8 \times 10^{10}$ electrons or positrons, with 4.2-ns spacing, averaged over 1024 turns with a resolution of 1 $\mu$m, to a single injected bunch of $5 \times 10^8$ particles, measured for a single turn with a resolution of better than 
1 mm. Each BPM consisted of four 15-mm diameter pick-up buttons
matched to 50 $\Omega$ and designed to contribute low impedance to the ring. There were 318 BPMs in the LER and 293 BPMs in the HER,
located at the quadrupoles in the rings, and rotated by $45 \deg$ relative to the horizontal plane to avoid synchrotron radiation.

\paragraph{Beam Current Monitors}
In order to measure the total current in each ring with a 5 mA resolution and for a maximum current of 5~A, commercial DC current transformers (DCCT)~\cite{dcct} were installed.   The high resolution of the DCCTs was driven by two requirements: 1) to monitor a 1~A beam with a 3-hour lifetime, {\it i.e.,} a charge loss of  93~$\mu$A/s, and 2) to record the injection of $5 \times 10^8$  particles per pulse, {\it i.e.,} 
$11~\mu$A, into individual bunches.  

In addition, each ring was equipped with a bunch-current monitor (BCM), which performed measurements of the charge for each of the 3,492 RF buckets in each beam at a rate of 1 Hz. In each beam, every other RF bucket was filled, except for a gap of 300 to 400 ns to provide time for the abort kicker magnet to turn on. 
As the vacuum in the rings improved and with an upgrade to the abort kickers, this gap was reduced to 50 ns. This improved the stability of the beams by diminishing the current transient induced by the gap.

The maximum charge per bunch was $8 \times 10^{10}$ particles, resulting in a total beam current of 3\,A. To optimize the beam-beam tune shift and maximize the luminosity, it was desirable to operate each ring with equal population in all of its bunches, and, therefore, it was necessary to precisely monitor the charge of each bunch. The BCM system was developed for top-off injection (injecting beam into coasting bunches) and  became essential for the implementation of trickle injection (injection of a few pulses per second into different bunches to maintain a constant current). The Bunch-Injection Controller (BIC) recorded bunch currents and DCCT readings for both rings at 1~Hz. It normalized the bunch currents to the total ring current and derived the individual bunch lifetimes. The BIC determined the sequence of injection for the master pattern generator, which reset the SLAC linear accelerator timing pulse-by-pulse to add charge in any of the RF buckets in one of the two storage rings.

\paragraph{Beam Collimation}
Though not normally referred to as beam instrumentation, the \pep2\ collimators were important tools to reduce background in the \babar\ detector caused by off-energy beam particles and transverse beam tails.  Collimators were installed at selected locations around the rings.
For the energy collimation, they were placed at locations where the dispersion function dominated the $\beta$ functions, and for $x$ and $y$ collimation, they were positioned at places with large horizontal or vertical $\beta$ functions, respectively. The collimator openings were typically set at 6 to 8 times the size ($\sigma$) of the circulating beams. The energy collimator was placed in Arc-1 
for the HER and in IR-10 for the LER. Four transverse (horizontal and vertical) collimators were installed in IR-12 for the HER and in IR-4 for the LER. 

In each transverse plane, two collimators were used, one as primary and the other as secondary collimator located at about $90^{\circ}$ in betatron phase after the primary. The HER energy collimation consisted of a single, primary collimator. The collimating devices were fixed, single-sided copper jaws protruding to a distance of about 10 mm from the nominal beam orbit. The degree of collimation was controlled  by closed-orbit bumps setting the distance of the jaw to the beam orbit. Beam sizes ($1 \sigma$) were roughly 1~mm horizontally, 0.2~mm vertically at location of the collimators. The jaw settings were large enough to maintain a beam lifetime of about 300~min in the HER and 50 min in the LER. Long ramps mitigated the impedance of the jaws and the generation of Higher Order Modes (HOM). A movable horizontal collimator was installed near the interaction region in the HER,  about 10 m upbeam of the  detector, to absorb off-energy particles generated in the upbeam straight section. Beam-loss monitors and thermocouples monitored the effect of the collimation
and provided protection from overheating.

\subsubsection{Gradual Enhancement of Performance}

The PEP-II performance gradually improved throughout the life of the machine~\cite{Seeman:2002xa,Seeman:2008zz}. Initially, improvements in the peak luminosity were achieved by gradually increasing the number of bunches in each ring. As the beam currents increased, the walls of the beam vacuum pipes in each ring were {\it scrubbed}, resulting in improved dynamic vacuum pressures and lower beam-related backgrounds in the detector. The increase in beam currents was made possible by slowly raising
(and eventually doubling) the total RF power and gap voltage available to both rings, by improving the design of the vacuum chambers, and by the addition of HOM absorbers. The bunch-by-bunch feedback systems (both transverse and longitudinal) were also upgraded as the beam currents were increased. 

The \pep2\ components were designed to function above the design luminosity and currents. 
But as the currents increased over the years, a number of problems arose.  Wake fields induced in the vacuum chambers resulted in HOM generation that led to local heating 
and required improved designs of vacuum components (especially for the high power expansion bellows), and additional cooling. Several RF stations were added;  in total there were 15 stations to handle the maximum beam currents of 3.2~A in the LER and 2.1~A in the HER.

The \pep2\ vacuum system was designed to be upgradable to meet the requirements for beam currents of up to 3~A.  Specifically, it had to sustain the large photo-desorption gas load caused by the bombardment of the chamber walls by intense synchrotron radiation. In its final configuration, 
the HER vacuum pumping system consisted of lumped ion pumps installed every 16~m along the arc and straight sections, a total of 198 distributed ion pumps in the arc dipole magnets, and a few non-evaporable getter (NEG) pumps in the straight sections. The LER vacuum was maintained by lumped ion pumps and 198 Ti sublimation pumps (TSP) in the arc sections. At full currents, the average pressure was about 3~nTorr in the HER and about 2~nTorr in the LER. In the two half straight sections upstream of the \babar\ detector, additional pumps (lumped ion pumps, NEG pumps, and TSPs) were used to reduce beam-gas backgrounds in the detector.  As a result, the average pressure in these sections was lower by a factor of ten, about 0.2~nTorr in the LER (30~m upstream of the IP) and 0.3 nTorr in the HER (50~m upstream of the IP).

The high beam currents were made possible by large RF systems consisting of 1.2~MW klystrons operating at 476~MHz and high power cavities with HOM absorbing loads. Each cavity had three loads with a capability of 10~kW. At peak currents, a HER cavity received on average 285 kW and a LER cavity 372~kW. The average klystron power was 1.01~MW. This overhead of about 20\% in power was needed to enable stable RF feedback systems. 
The longitudinal and transverse bunch-by-bunch feedback systems were used to measure and to control beam instabilities. The coupled bunch instabilities were damped up to the highest current, but the HER growth rates were somewhat anomalous and stronger than predicted~\cite{Wienands:2006iz}. The kickers 
for the longitudinal and transverse systems were upgraded, and both systems turned out to be highly robust.

\begin{figure}[htb!]  
  \begin{center}  
  \includegraphics[width=7.0cm]{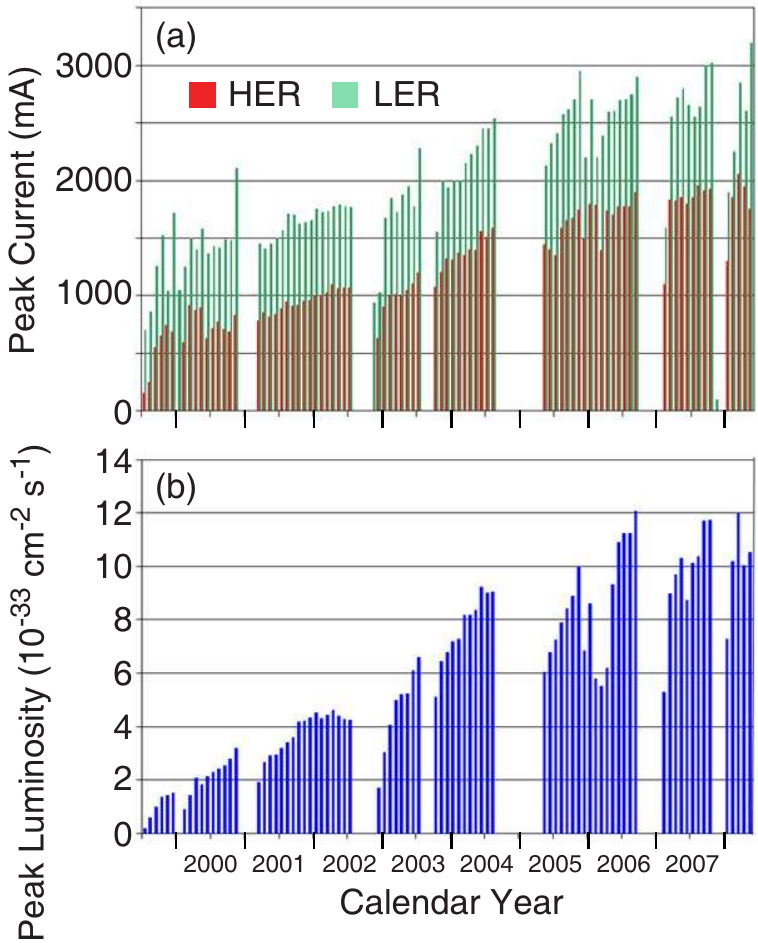}  
    \caption{
\pep2: (a) Maximum beam currents and (b) maximum instantaneous luminosity
achieved each month. 
}  
    \label{fig:Luminosities_Month}  
  \end{center}  
 \end{figure}

Figure~\ref{fig:Luminosities_Month} shows the peak beam currents and the instantaneous peak luminosity per month, demonstrating the continuous improvements in the performance 
of the \pep2\ storage rings.  

Part of the luminosity enhancement in 2004 can be attributed to improved control of the lattice functions in the LER, which reduced the equilibrium emittance and improved betatron matching at the IP.   Further improvements were achieved by continuous adjustments to the beam tunes as well as corrections and upgrades to components
~\cite{Wienands:2006iz,Novokhatski:2004fj,Novokhatski:2007zzb}.
Beam simulations  were continuously improved to understand the complex beam properties, for instance,  the $x-y$  coupling in the LER, the beam-beam interaction, and tune-shift limits.

As predicted, the impact of parasitic collisions away from the IP was small, because the beam trajectories were separated by the B1 dipoles. At full currents, parasitic crossings reduced the luminosity by a few percent.

As the current in both the LER and the HER increased, different types of beam instabilities were 
observed. Various methods were developed to enable stable collisions up to currents of 2~A for electrons and 3~A for positrons.

\begin{figure}[htb!]  
  \begin{center}  
  \includegraphics[width=7.0cm]{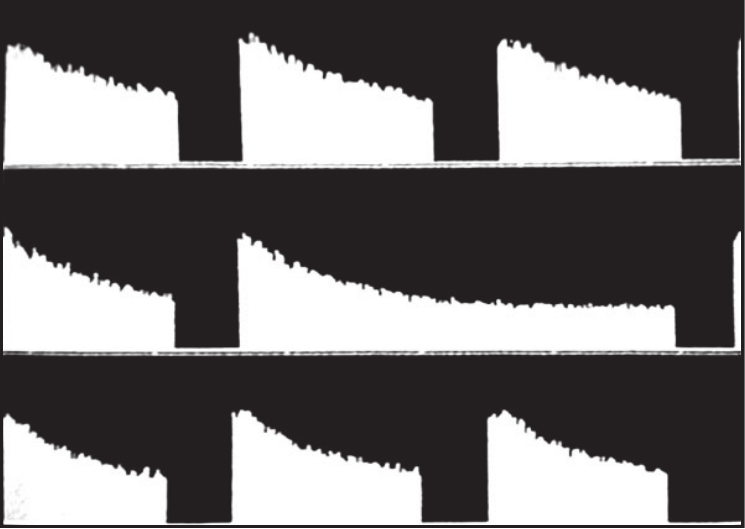}  
    \caption{
\pep2: Oscilloscope record of bunch trains: The online luminosity per bunch (vertical axis) for  830 stored bunches (horizontal axis).  The bunches are shown on three consecutive traces from top left to bottom right.  There are eight bunch trains, separated by gaps that are much larger than the 8.4~ns separation between the consecutive bunches. 
}  
    \label{fig:PEPII_ECI}  
  \end{center}  
 \end{figure}

One effect of the higher beam currents was the multi-bunch transverse instability arising from interactions of the positron bunches with a low energy electron cloud created by synchrotron radiation photons interacting with the walls of the vacuum chambers. This so-called electron-cloud instability was not 
unexpected~\cite{Zimmermann:2001ax}.
A spectacular example of how the electron cloud accumulation in the LER affected the luminosity is shown in Figure \ref{fig:PEPII_ECI}. This effect was first observed, when the number of bunches in each ring was increased from 415 to 830, thus decreasing the bunch spacing from 16.8~ns down to 8.4~ns.
A large decrease in the instantaneous luminosity was observed. This was the result of the build-up of the electron cloud during a train of closely spaced bunches. Whenever a large gap occurs, the electron cloud disperses and at the beginning of the next bunch train the positron bunches are restored to their normal size and the luminosity recovers.
The problem was mitigated by wrapping the LER beam pipe with solenoid windings (1.8 km in all) to generate a 30~G field and to confine the very low-energy electrons close to the walls.
 
The effect was largest in the stainless steel vacuum chambers in the straight sections, 
and much smaller in the arc sections where the ante-chambers were Titanium nitride (TiN)-coated, which suppressed the electron cloud formation.

Beam-beam effects can lead to higher luminosity, but they also introduce beam instabilities. 
The PEP-II tunes were chosen to optimize the beam-beam performance (0.508 in the horizontal plane and 0.574 in the vertical one). The proximity 
of the horizontal working point to the half-integer induced a beneficial dynamic-$\beta$ effect that increased the instantaneous luminosity  more than 50\%. This was   at the cost of a large betatron mismatch (also known as $\beta$-beat) in the entirity of the rings, which required considerable effort to be brought back to
within tolerance.

Computer simulation of the luminosity contours on the 
$x-y$ tune plane supported the empirically chosen tune values. Operation away from optimal tunes often led to increased beam instability and aborts caused by short beam lifetimes and high detector backgrounds~\cite{Sullivan:2006jz}. 
The measured luminosity varied linearly with the product of the bunch currents at low currents, but more slowly at higher currents. In this regime, beam-beam forces caused the bunches to increase in size, typically in the vertical plane. Since the beam-beam force acts as a focusing lens, the resulting tune shifts tend to limit the allowed tune values. The resulting maximum vertical beam-beam parameters $\xi_y$ in the two rings were 0.05 to 0.065. At the highest currents of colliding beams, the HER current was limited by the LER lifetime and the LER current by HER-generated IR backgrounds in the detector~\cite{Sullivan:2006jz}.

Whenever conditions indicated that the accelerator or detector might be at risk, beam aborts were initiated. 
For every abort, signals and conditions were recorded, and subsequently categorized and analyzed offline. At the end of \pep2 operation, there were on average eight beam aborts per day, about three due to RF problems, three caused by unstable beams, and two by high detector backgrounds.

\subsubsection{Trickle Injection}
Electrons were generated in a thermionic gun, whereas positrons were collected from secondary particles  produced in a target that was struck by a 20\gev electron beam.  Both beams were bunched, injected into the SLAC linear accelerator, and accelerated to 1.2 \gev , before their transverse emittances were reduced in separate damping rings.  Individual bunches of electrons and positrons were then re-injected into the linear accelerator, accelerated  to 9.0 and 3.1 \gev,  respectively, and then extracted for injection into the \pep2\ rings.  To reduce the oscillation of the injection bunches,  a pair of injection kickers in the HER and LER moved the stored beam close to the injection septum magnet.  The typical number of particles in the injected bunch was in the range of 3 to 8 $\times 10^9$.

Initially \pep2\ operated on a 40-50 minute fill cycle.  At the end of each fill, when the currents had dropped by about 30-50\%,
data taking was interrupted, certain detector voltages were ramped down to avoid damage by large backgrounds during the injection, and beams were injected to the maximum sustainable currents.
It took about three minutes to replenish the beam to full currents, and a few more minutes 
to raise the detector voltages to their nominal operational values and to restart the DAQ.  In this top-up-and-coast operation, the average luminosity was about 70-75\% of the peak luminosity.
After an interruption of beam operation, either due to a beam loss or for other reasons, it took  approximately 10-15 minutes to fill the two beams under favorable conditions.

A substantial increase in luminosity was achieved in 2004, when the beam currents were 
maintained constant by injecting continuously at rate of 1-5\hz, while \babar\ recorded data~\cite{Kozanecki:2004wi}, thereby keeping the luminosity close to maximum and avoiding the ramping of detector voltages.
The primary challenge for this {\it trickle injection} was to deliver sufficient current to replenish the stored beams while avoiding excessive exposure of the \babar\ detector to radiation.

In the HER, the vertical injection combined with large $\beta$ functions outside of the detector necessitated better control of the incoming bunches.  In the LER, the maximum vertical $\beta$ 
function occurred closer to the interaction point resulting in smaller $\beta_y$ at the IP, and, thereby, lower sensitivity to the position and angle of the injected bunches.  
Systematic improvements to the electron beam in the Linac (injection energy, and RF phase relative to the \pep2\ rings, betatron and longitudinal tails), careful matching of the injection orbit with the closed orbit of the stored beams, and 
general stabilization of the trajectories via feedback control were critical to this mode of operation.   

First tests of trickle injection showed that the newly injected bunches 
resulted in very high rates of background events
that contained many hits in almost all detectors and saturated the \babar\ DAQ bandwidth.
These effects were studied by analyzing the additional triggers and dead-time associated with the injected bunches, separately for electrons and positrons.
The \babar\ FCTS provided a time-stamped record of each trigger 
input and recorded whether the trigger was accepted or rejected.
For each injected bunch, an {\it injection marker} that was
locked to the time of injection was recorded. Because the FCTS was locked to the PEP-II timing system, the
trigger time-stamps could be related to the number of electron or positron bunch revolutions and phase
since their injection. In this way, recorded events could be analyzed 
as a function of the delay relative to the last injected bunch. This phase was measured by a 119~MHz clock, such that 873 clock ticks corresponded to a complete revolution.

\begin{figure}[htb!]
\begin{center}
\vspace{1.4cm}
\includegraphics[width=7.0cm]{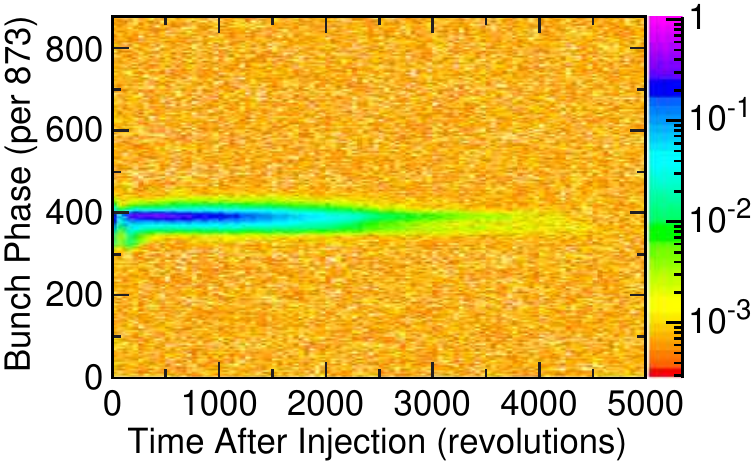}
\caption{\pep2 trickle injection background: EMC single cluster trigger rate per 420~ns (color coded as shown on the right) as a function of the bunch phase (873 clock ticks per revolution) for the first 5000 revolutions ($36.7 \ms$) after the injection into the $\LER$.  Note that 420~ns correspond to 50 clock ticks.}
\label{fig:trick-inj-2d}
\end{center}
\end{figure}

Figure~\ref{fig:trick-inj-2d} shows the distribution of EMC  trigger delays versus  the time of injection into the LER, specified in terms of the number of revolutions and phase since injection.
For several hundred revolutions, an increase of the trigger rate by about two orders of magnitude is observed at a phase coincident with the beam injection.

\begin{figure}[htb!]
\begin{center}
\vspace{0.4cm}
\includegraphics[width=7.0cm]{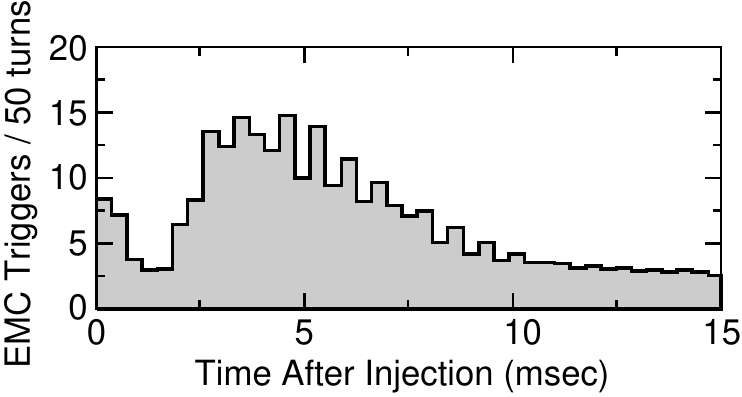}
\caption{\pep2 trickle injection:  EMC single cluster trigger rate during the first 15 ms after $\LER$ injection.  The rates are not corrected 
for contributions from the stored beams, visible at delays exceeding 12 ms.
}
\label{fig:trick-inj}
\end{center}
\end{figure}

The trigger records were analyzed online. The number of  EMC single cluster triggers and DCH single track triggers in phase with the injected bunch were recorded, after correction for out-of-phase triggers.
This information was transmitted to \pep2\ operators as a continuous feedback for injection tuning.
As an example, the EMC single cluster trigger counts after LER injection are shown in 
Figure~\ref{fig:trick-inj}.
The enhancement at 1-2 ms is related to the conditions at injection,
{\it i.e.}, the angle and position of the bunch relative to the stored beam.
The broad enhancement from 2 to 12~ms is due to longitudinal oscillations 
caused by energy and phase mismatch between the injected bunch and the stored beam.

Offline studies of the trigger records revealed that dead-time due to injection backgrounds could be mitigated by inhibiting the DAQ for triggers generated within
the first several hundred revolutions of the injected bunch.
For this purpose, a special gating module was built to veto triggers in phase with the injected bunch.
 
The impact of LER trickle injection backgrounds on quantities relevant to the \babar\ 
physics analyses was quickly investigated in a dedicated offline study.
Quantities including particle efficiencies and resolutions were evaluated in several time intervals related to the  bunches injected in the LER.
It was concluded that the benefit of the increase in the integrated luminosity outweighed the small degradation in detector performance restricted to short time intervals after injection.  Similar studies were later performed on data recorded during a HER trickle injection test.

\begin{figure}[htb!]	 
\begin{center}		  
\includegraphics[width=0.45\textwidth]{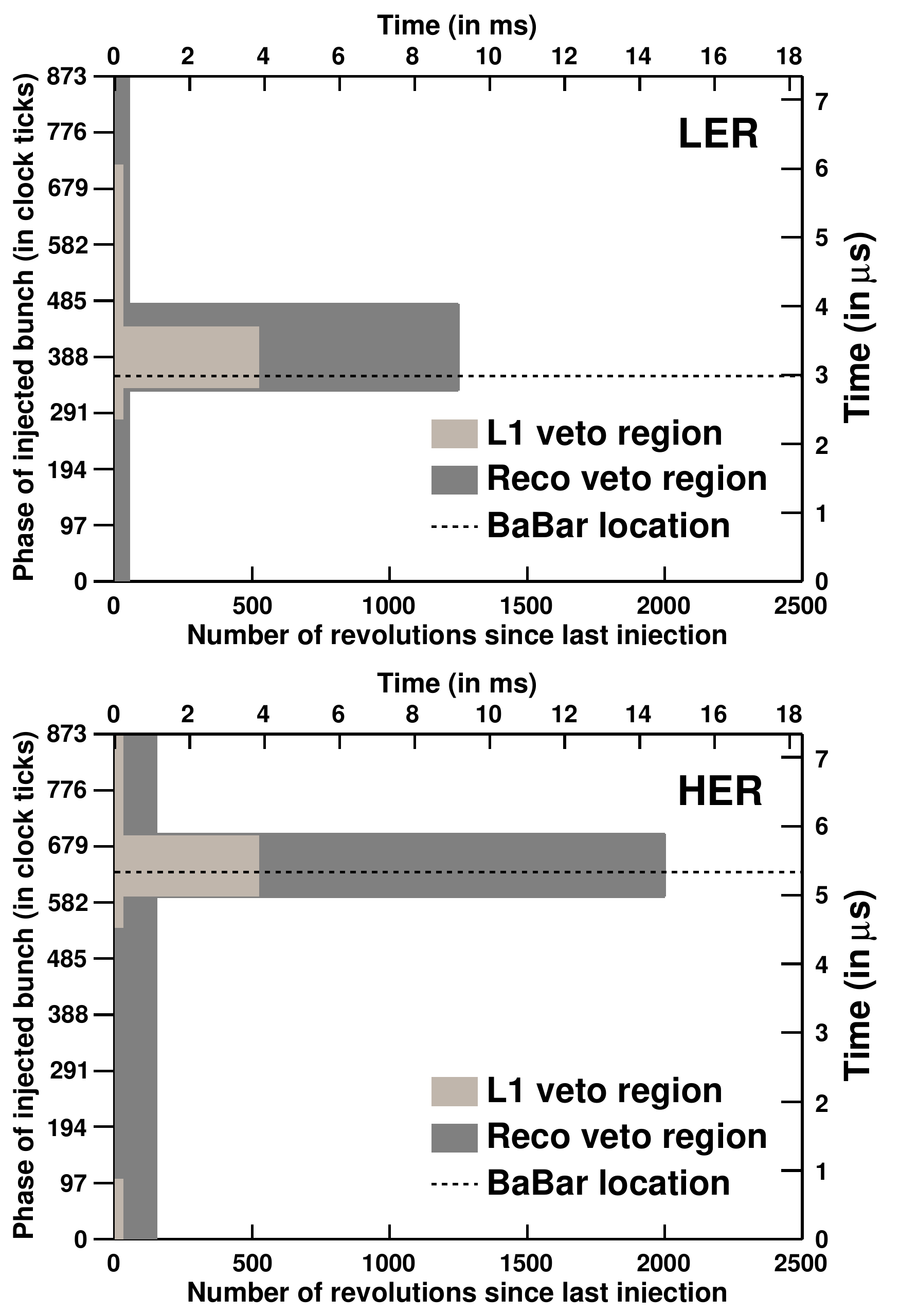}	 
\caption{\pep2 trickle injection: Illustration of the timing vetoes imposed by the L1 trigger and event reconstruction to reduce the backgrounds generated by trickle injection, separately for the LER and HER.}   
\label{fig:TIvetos}	  
\end{center}		   
\end{figure}

Offline filters of injection backgrounds extended the vetoes beyond the
hardware trigger inhibits.  
Table~\ref{tab:tkl_windows} and Figure~\ref{fig:TIvetos} summarize the typical online and offline 
vetoes and their effective loss of luminosity.  The timing vetoes are expressed 
in terms of two two-dimensional restrictions of
the injected bunch phase versus the number of bunch revolutions.  
Specifically, these include the online trigger vetoes for each beam rejected 
events occurring in 440 of the 873 clock ticks of each of the first 30 revolutions 
of the bunch and in 106 of the 873 clock ticks of each of the next 495 revolutions.  
These vetoes added up to $0.55\ms$ of dead-time for each injection.  The LER veto 
was extended offline by $1.33\ms$ for each LER injection.  Thus, for a typical 
LER injection rate of 5\hz, the luminosity loss was 0.94\%.  The HER veto was extended
offline by $2.23\ms$  for each HER injection; a wider window was chosen
to cover the effects of electron source intensity jitters.  For a typical 
HER injection rate of 1.5\hz, the luminosity loss was 0.42\%.   
By design, the HER and LER filter windows did not overlap.  
Thus, the total luminosity dead-time was the sum of  contributions from both rings.
These veto settings remained unchanged for the remainder of the data taking.

\begin{table*}[hbt]
\begin{center}
\caption{\pep2\ Trickle injection veto settings, online and offline (in units of revolutions and clock ticks), and typical injection rates. The dead-time per injection for the LER and HER and the impact on the analyzed data also are listed.}
\label{tab:tkl_windows}
\begin{tabular}{lcc} \hline
                  & LER & HER \\ \hline
L1 Inhibit Window &
  \begin{tabular}{r}
        30 rev $\times$ 440/873 \\
     + 525 rev $\times$ 106/873 \\
       0.55 $\ms$/injection \\
  \end{tabular} &
  \begin{tabular}{r}
        30 rev $\times$ 440/873 \\
     + 525 rev $\times$ 106/873 \\
       0.55 $\ms$/injection \\
  \end{tabular} \\ \hline
Reconstruction Filter &
  \begin{tabular}{r}
        50 rev $\times$ 873/873 \\
    + 1250 rev $\times$ 150/873 \\
       1.88 $\ms$/injection \\
  \end{tabular} &
  \begin{tabular}{r}
       150 rev $\times$ 873/873 \\
    + 2000 rev $\times$ 108/873 \\
       2.78 $\ms$/injection \\
  \end{tabular} \\ \hline
Average Injection Rate & 5\hz & 1.5\hz \\
Average Luminosity Loss & 0.94\% & 0.42\% \\ \hline
\end{tabular}
\end{center}
\end{table*}

\begin{figure}[htb!]  
  \begin{center}  
  \includegraphics[width=7.0cm]{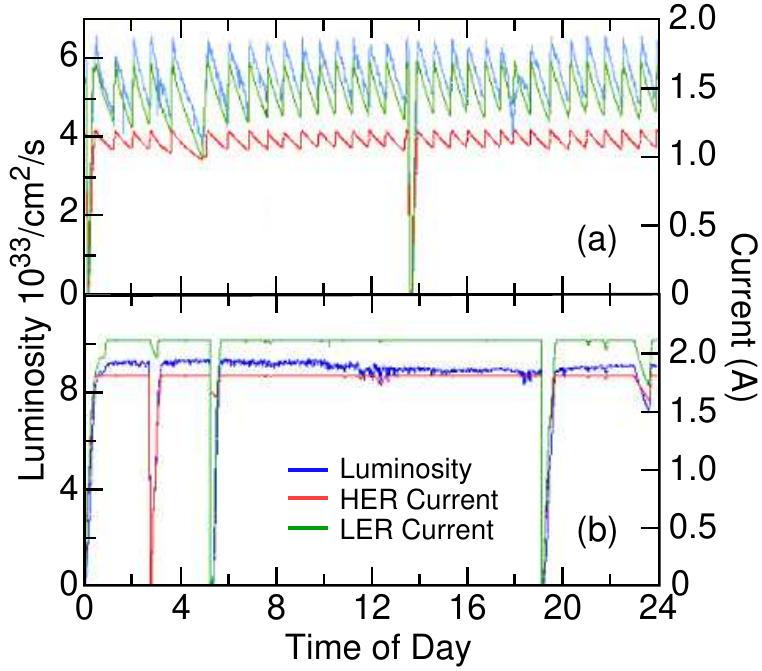}  
    \caption{
\pep2: Luminosity and beam currents for a 24-hour period (a) before and 
      (b) after the implementation of trickle injection.  
}  
    \label{fig:PEPII-ops}  
  \end{center}  
 \end{figure}

After successful tests, trickle-charge injection became the default mode of \pep2\ operation. 
Figure~\ref{fig:PEPII-ops} compares the 24-hour history of the beam currents and average luminosity before and after the implementation of continuous injection. 
In the fill-coast mode, current and luminosity decreased during the time interval between fills, while they remained almost constant with trickle injection. 
Trickle injection improved the average luminosity by about 40\% immediately after implementation in 2004.  
Three major factors contributed to this enhanced performance: 1) the accelerator operated at the peak luminosity at all times, 2) the constant beam currents (and luminosity) allowed for tuning of the beams to a significantly higher level of performance and stability, and 3) the interruption of the \babar\ data taking during injection was very significantly reduced.

The observed increase in the integrated luminosity 
was higher than expected from the increase in availability 
of the beams. The continuous injection and almost constant beam currents and luminosity 
led to more stable operating conditions for \pep2\ as well as for \babar. 
Key to the commissioning of this mode of operation were the diagnostics provided by \babar, the stabilization of the injected beams, the 
optimization of the parameters of the injected bunches and the move of the stored orbit 
closer to the septum magnet.
The \babar\ diagnostics were incorporated into the standard online monitoring of 
the trigger and data acquisition.  This information  was also
available to the \pep2\ operators for beam tuning and diagnostics.
Offline comparisons of the reconstruction efficiency and the mass resolution of benchmark exclusive final states revealed no significant degradation for events which occurred close in time to the bunch injection into the LER or HER.

\label{sec:mdi:trickle}

\subsection{\pep2 Operation and Monitoring}

\subsubsection{Beam Energies}
\label{sec:mdi:pep2-energies}

The energies of the two circulating beams were calculated based on total magnetic bending strength (including effects of off-axis quadrupoles, steering magnets, and wigglers) and the average deviations of the RF frequencies from their central values.
The systematic uncertainty on the calculated c.m. energy was typically $5-10\mev$; the relative setting of each beam was reproducible and stable to about 1\mev.

Changes in the setting of the c.m. energy of the rings were implemented by adjusting the energy of the HER and keeping the LER energy unchanged. 
This method was chosen because the final-focus 
permanent magnet QD1 made energy changes difficult in the \LER.
Furthermore, the permanent dipole magnets B1 near the collision point had less impact on the HER orbits than the LER orbits.
The same procedure was followed for operation outside to the \FourS\ region, for instance, at the \ThreeS\ and \TwoS\ resonances and at energies above the \FourS~\cite{Sullivan:2009zzc}.

\begin{figure}[htb!]  
  \begin{center}  
  \includegraphics[width=7.5cm]{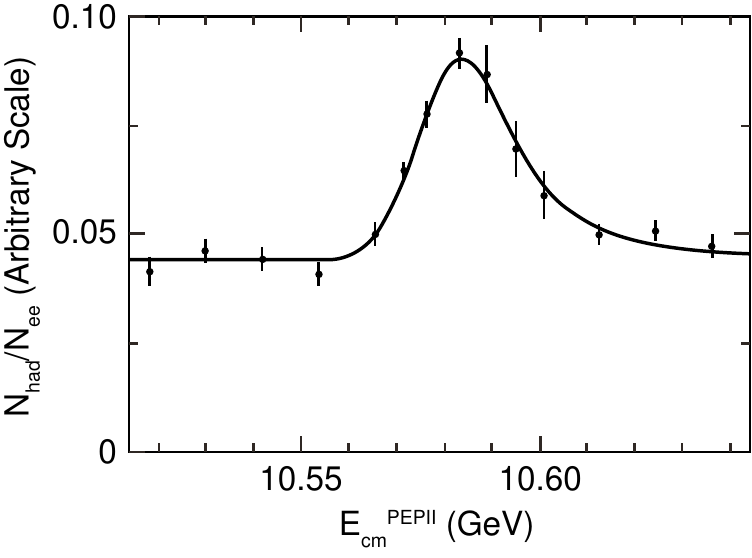}  
    \caption{
\pep2 c.m. energy calibration at the \FourS\ resonance: the ratio of the observed rates for hadronic to Bhabha events as a function of the \pep2\ c.m. energy. The curve represents the fit to the data which was used to determine the energy setting of \pep2. 
}  
    \label{fig:Y4S_scan}  
  \end{center}  
 \end{figure}

Most of the data were recorded on or near the peak of the \FourS\ resonance,
which has a mass of $10.579 \pm 0.001 \gev$ and a full width of $20.5 \pm 2.5 \mev$~\cite{Beringer:1900zz}.
About 10\% of the data were taken below the threshold for \BB\ production, \ie\ 40 \mev below the \FourS\ resonance. 
To set the c.m. energy of the colliding beams to the peak of the \FourS\ resonance, an absolute energy calibration was necessary.  For this purpose, 
scans were performed regularly, during which data were recorded at a number of discrete energies below and above the expected mass of the resonance.  At each energy,
the ratio of multi-hadron events to \epem\ pairs was determined. Multi-hadron events were required to have at least four charged tracks with a restriction on the second 
Fox-Wolfram moment~\cite{Fox:1978vw}, $R_2<0.2$, eliminating jet-like event topologies.  
Bhabha events were identified as two oppositely charged electrons, emitted 
back-to-back in the c.m. system.  
Figure~\ref{fig:Y4S_scan} shows the very first scan of this kind performed at \pep2, recorded for a total integrated luminosity of $3 \invpb$. 
The data were fit to a Breit-Wigner resonance, convoluted with the beam energy spread, and modified to take into account radiative events of the type 
$\epem \to \FourS \gamma$.  Free parameters of the fit were the mass and width of the resonance.   
The value of the peak energy was $10.5828 \pm 0.0007\gev$, deviating from the well-measured \FourS\ mass by 3.4 \mev.  This offset corresponds to a setting error of the 
\pep2\ energy  of $3 \times 10^{-4}$.  

To ensure that the data were recorded at the peak of the $\FourS$ resonance,
the observed ratio of \BB -enriched hadronic events to lepton pairs was monitored online.
Near the peak, a 2.5\% change in this ratio corresponds
to a 2\mev\ change in c.m. energy, a value comparable to the precision with which the
energy of \pep2\ could be maintained.

The best monitor and absolute calibration of the c.m. energy was derived from the measured c.m. momentum of fully-reconstructed hadronic $B$ decays, combined with the known $B$-meson mass.
An absolute uncertainty on the c.m. energy of  1.1\mev was obtained for an integrated luminosity of 1\invfb.
This error was limited by the  uncertainty in the $B$ mass and the detector resolution.

Although the \pep2 collider was designed as a fixed energy machine
and mostly operated on or near the \FourS\ resonance, smaller data samples were recorded at c.m. energies ranging from 10 \gev to 11.2 \gev. During the last few months of operation, data were recorded at the $\ThreeS$ and the $\TwoS$ resonances. Their masses and total widths are $10.3552 \pm 0.0005 \gev$ and $20.32 \pm 1.85 \kev$ and  
$10.0233 \pm 0.0003 \gev$ and $31.98 \pm 2.63\kev$, respectively~\cite{Beringer:1900zz}.  
For this purpose, the HER energy was lowered by 378 \mev to reach the $\ThreeS$ and then again by 546 \mev to reach the $\TwoS$. 
Operating the HER at an energy of almost 1\gev below the \FourS\ to reach the \TwoS\ resonance proved to be very challenging, since the HER beam orbit was very close to the physical aperture of the beam pipe at the entrance and exit of the detector.
As a result, \babar\ backgrounds became very sensitive to small changes in the HER orbit through the interaction region. A compromise was made
to optimize the achievable luminosity with acceptable backgrounds.  

\begin{figure}[htb]
\begin{center}
\includegraphics[width=\columnwidth]{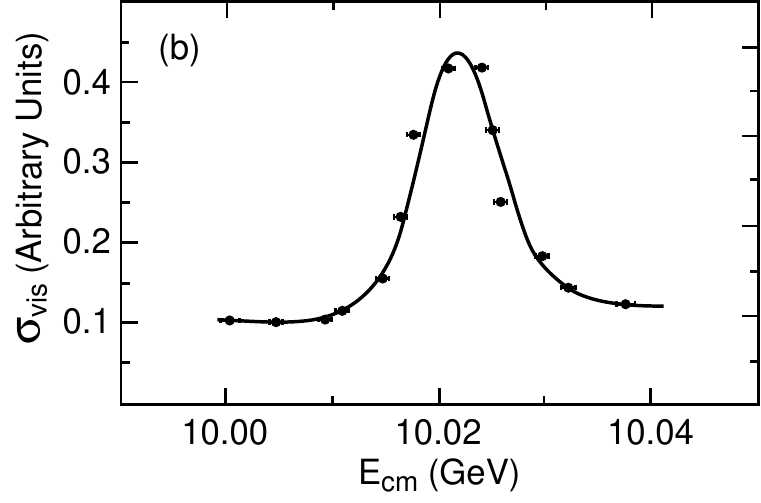}
\includegraphics[width=\columnwidth]{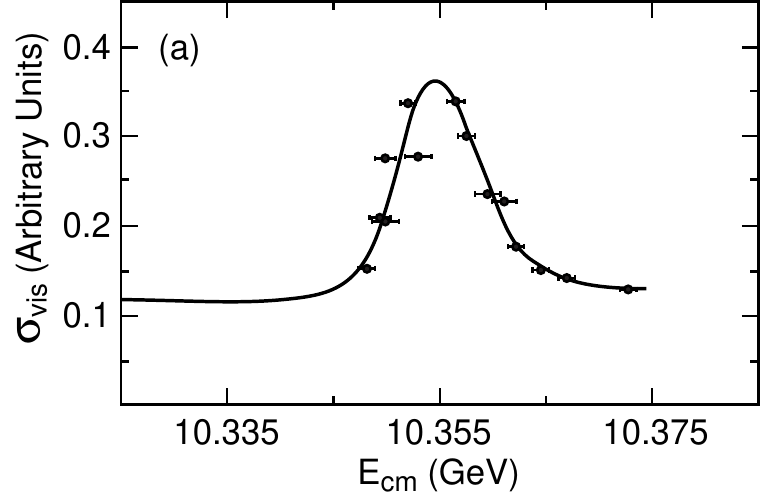}
\caption{
\pep2  c.m. energy calibration at (a) the $\TwoS$ and (b) the $\ThreeS$ resonance. The data show the observed hadronic cross section.  The curve represents the fit to the data which was used to determine the energy setting of \pep2.  
}
\label{fig:resonance_scans}
\end{center}
\end{figure}

Figure~\ref{fig:resonance_scans} shows the energy scans for the $\ThreeS$ and $\TwoS$ resonances.  Since these states are 
extremely narrow, the observed widths (full width at half maximum of $8\mev$) reflect the energy spread of the \pep2\ rings.
These narrow widths made it challenging and necessary to monitor the hadronic cross section online, so as to keep the beam energies set for the maximum signal rate.

During the last few weeks of \babar\ operation, the c.m. energy was 
increased to record data from the $\FourS$ 
resonance up to the maximum possible energy of 11.2\gev.  
The energy was raised in steps of
5\mev\ and data were recorded for 50 minutes at each step. 
The luminosity was roughly constant at $1 \times10^{34} \cm^{-2}s^{-1}$
up to about 11\gev. At this point, the synchrotron radiation
of the HER beam current of 1.7~A caused excessive heating in the arc vacuum chamber. 
To operate at higher energies, the HER beam current was lowered and as a result the luminosity decreased.

\subsubsection{Bunch Sizes, Positions, and Angles}
\label{sec:mdi:pep2-bunches}

To maximize the overlap of the colliding beams at the collision point and thereby the luminosity,
the beam positions and trajectory angles were continuously adjusted by a feedback system.
This system moved the HER beam position and its vertical angle to coincide
with the LER beam position and angle at the IP.
Initially, the feedback operated at an update rate of $\simeq 0.1$\hz.
It was later upgraded and its rate was raised to $\simeq 1$\hz~\cite{Fisher:2007zzb,Gierman:2006zz}.

  Measuring the beam parameters at the IP with high precision was 
  an  essential 
  and challenging task. The specific luminosity, defined as the luminosity 
  normalized to the number of stored bunches and to the product of the bunch 
  currents in each beam, is proportional to the inverse product of 
  the overlapping beam sizes at the IP.  
  Two shared BPMs were used to bring the beams close to each other, and
  a scan was performed and the observed deflection was used to determine 
  the central beam position.   
  The transverse sizes of the two beams were measured using synchrotron-light 
  profile monitors in the two rings.
  The fast, high-rate signal from a  small-angle photon detector 
  for radiative Bhabha scattering (see Section~\ref{subsec:pep2-luminosity})
   was used 
  to maximize the luminosity. 

  The very high event rate, combined with the high precision of the \svt and the advances
of data acquisition and online processing capability of the \babar\ detector,
made it possible to reconstruct in real time the spatial and transverse momentum distributions
of  $e^+ e^-$ and $\mu^+ \mu^-$ final states, which reflect the phase space distribution of the colliding electron and positron beams~\cite{Kozanecki:2008zz}.
This provided feedback to the \pep2\ operators on the time scale of ten minutes (using typically 1000 selected $e^+ e^- \to e^+ e^- , \mu^+ \mu^- $ events per minute) on the position, orientation, and longitudinal and transverse size and shape of the luminous region, and, thereby, the vertical $\beta$ function and the angular divergence.
  
  Offline analysis of the three-dimensional luminosity distribution and of the 
  angular distribution of muon pairs enabled the reconstruction of the history, under 
  normal high-luminosity conditions and on timescales from less than 30 
  minutes to a couple of weeks, of many
  phase-space parameters. The use of complementary techniques, with very 
  different sensitivities to the beam parameters, provided redundancy and 
  consistency checks. Several of the observables accessible to these luminous
  region analyses were either not directly measurable by conventional 
  techniques, or could only be determined at very low current or in single
  bunch mode, greatly complicating the interpretation of accelerator performance 
  during physics running.
 
\begin{figure}[htb]
\begin{center}
\includegraphics[width=0.46\textwidth]{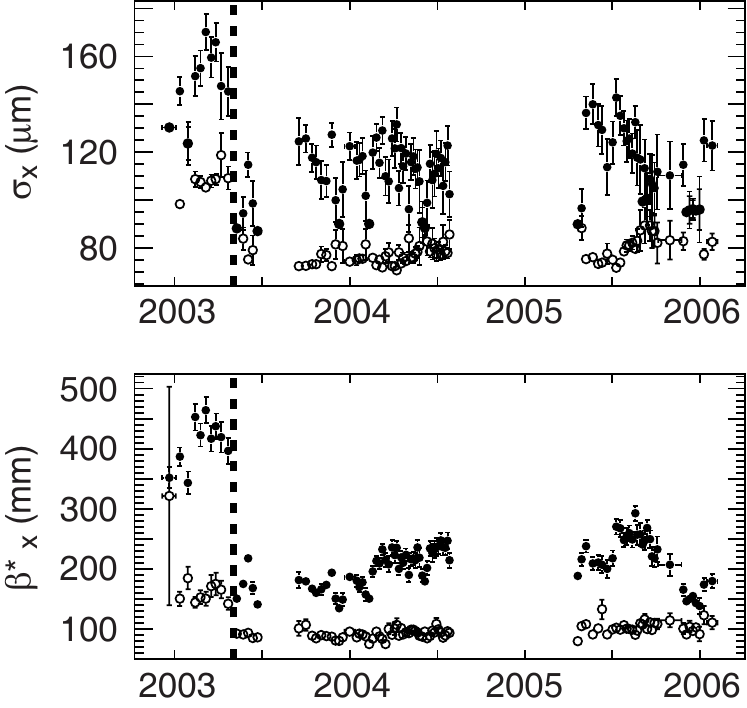}
\caption{\pep2: History of horizontal IP spot sizes (top) and $\beta*$
functions (bottom) in the \LER (open circles) and the \HER
(black dots), extracted from luminous-region observables
measured by \babar. The dotted line indicates the time
when both $x$ tunes were moved close to the half-integer,
resulting in a sizable luminosity improvement.~\cite{Kozanecki:2008zz}}
\label{fig:kozanecki}
\end{center}
\end{figure}

  Combined analyses of all these measurements provided constraints on the 
  transverse and longitudinal emittances, the horizontal and vertical beam 
  spot sizes, and the $\beta$  functions and bunch lengths of both beams at the 
IP. Figure~\ref{fig:kozanecki} illustrates the results from these efforts
in terms of the horizontal beam size and $\beta$ function.

\subsection{Luminosity Measurements}
\label{sec:pep2-luminosity}

\subsubsection{\pep2\ Peak and Integrated Luminosities}
\label{subsec:pep2-luminosity}

The instantaneous luminosity was monitored by a signal originating from a zero-angle photon detector,
located at 9~m from the IP on the upbeam side of the \HER beam.
This detector recorded photons from radiative Bhabha events at a rate exceeding 100\mhz for luminosities of $10^{34} \cms$.

At this high rate, the luminosity could be monitored on a 
bunch-by-bunch basis, a feature that was used to optimize the machine performance.
The online luminosity measurement 
was calibrated 
on a regular basis every two weeks, using more accurate offline \babar\ measurements.

\begin{figure}[htbp!]
\begin{center}
\includegraphics[width=0.5\textwidth]{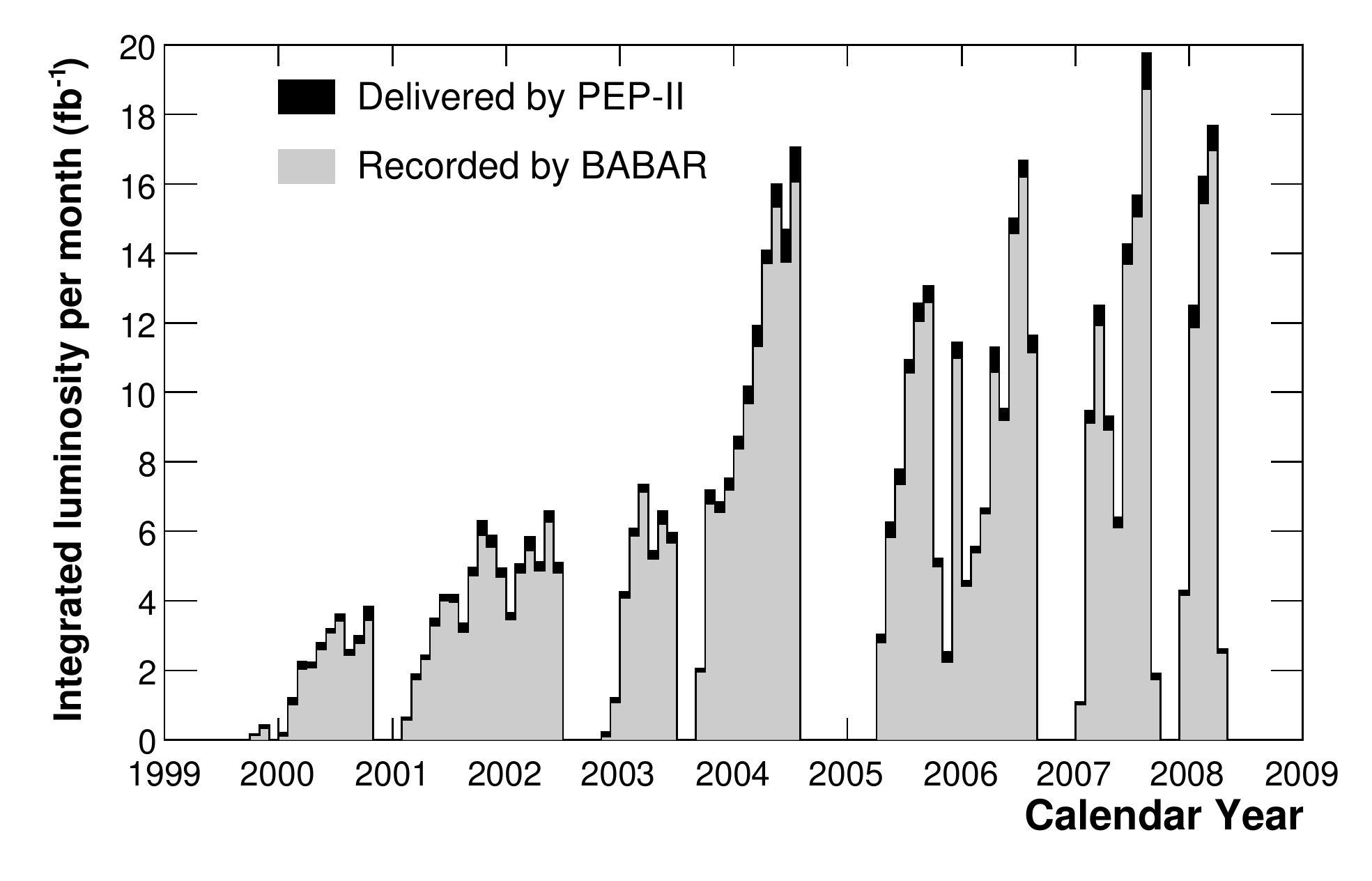}
\end{center}
\vskip -0.5cm
\caption{Integrated luminosity per month, differentiating
the luminosity delivered by \pep2\ from that recorded by \babar.}
\label{fig:finalLuminosities_perMonth}
\end{figure}

\begin{figure}[htbp!]
\begin{center}
\includegraphics[width=0.5\textwidth]{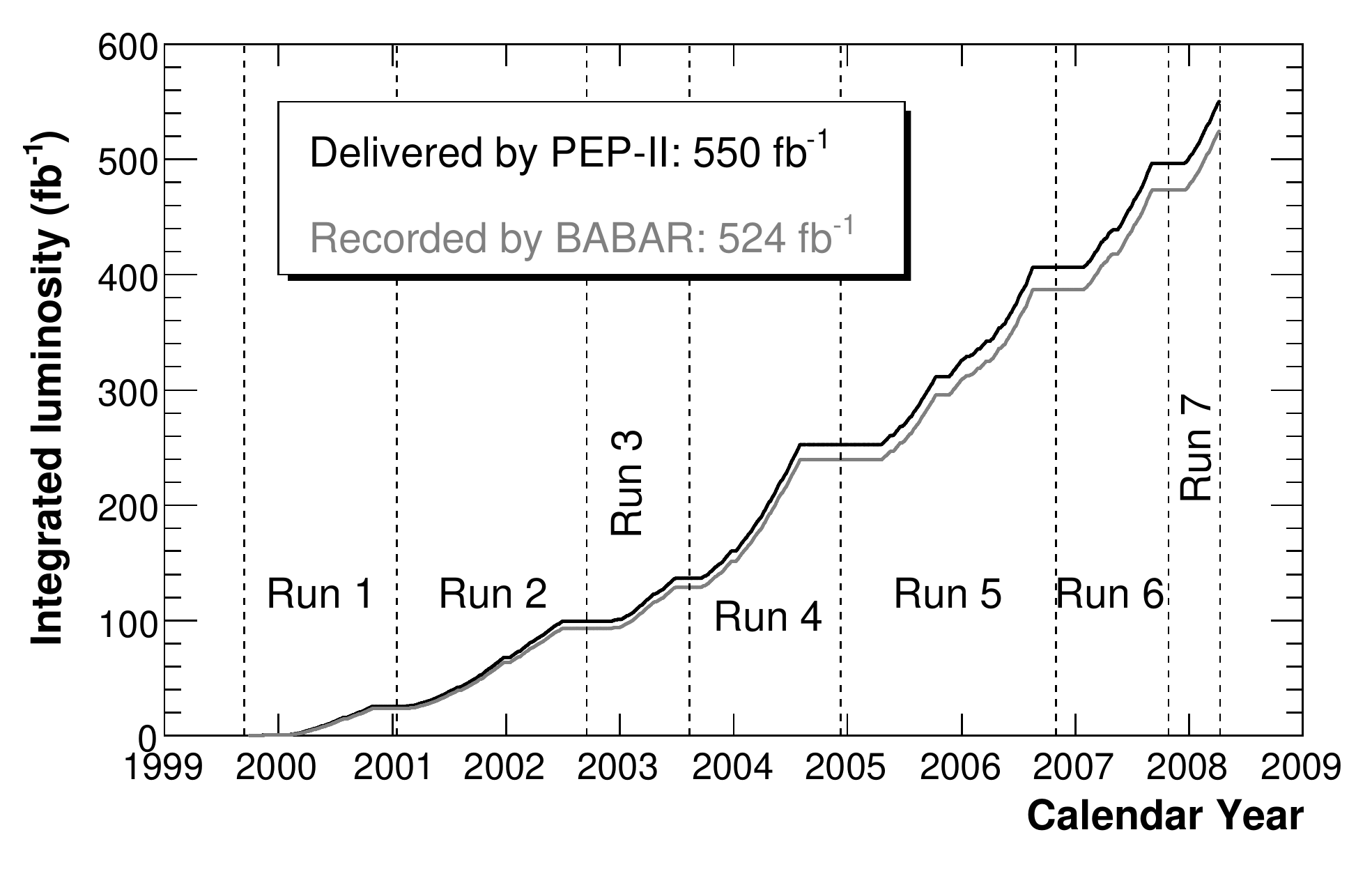}
\end{center}
\vskip -0.5cm
\caption{\pep2: Evolution of the integrated luminosity, differentiating
the luminosity delivered by \pep2\ from that recorded by \babar.}
\label{fig:finalLuminosities}
\end{figure}

Figure~\ref{fig:finalLuminosities_perMonth} summarizes the integrated luminosity per month, and Figure~\ref{fig:finalLuminosities} shows the integrated luminosity over the nine years of operation.   Of the 550\invfb delivered by
\pep2, 95.3\%  were recorded by \babar, of which only 1.1\% were  discarded due to some hardware problem that might affect the data analysis.  

\subsubsection{Precision Measurement of the Integrated Luminosity}

The most precise measurement of the integrated luminosity~\cite{lees:2013rw} was based on large samples of  Bhabha 
($\epem \to \epem (\gamma)$) and dimuon ($\epem \to \mu^+ \mu^- (\gamma)$)  events. The luminosity for these two final states was determined separately  using the relation,  
\begin{equation}
\mathcal{L} = (N_{\rm cand} - N_{\rm bkg})/\sigma_{\rm vis}, 
\end{equation}
where $N_{\rm cand}$ is the number of selected events, $N_{\rm bkg}$ is the number of background events satisfying the selection criteria, and 
$\sigma_{\rm vis}$ is the visible cross section for the selected process. 

Bhabha events were identified online and a  polar-angle-dependent 
fraction was recorded. More stringent criteria were applied offline, 
\begin{itemize}
\item  $|\cos\theta^*|<0.70$ for one track and 
$|\cos\theta^*|<0.65$ for the other, 
where $\theta^*$ is the c.m. polar angle. 

\item $x_1>0.75$ and $x_2>0.50$, where $x_i \equiv 2p^*_i/\sqrt s$ refers to the normalized momentum, and
$p^*_1$ and $p^*_2$ are the higher and lower c.m. momenta of the two tracks.

\item $\alpha < 30^\circ $, where the acolinearity $\alpha$ is
defined as $180^\circ$ minus the c.m. angle between 
the two tracks.
\item $E_\mathrm{EMC}/p>0.7$ (ratio of the calorimeter energy to the track momentum)
for at least one track, and $E_\mathrm{EMC}/p>0.4$ for the second track, provided
that it could also be associated with an EMC shower.
\end{itemize}

For dimuons, the following requirements were imposed: 
\begin{itemize}
\item  $|\cos\theta^*|<0.70$ for one track and 
       $|\cos\theta^*|<0.65$ for the other; 
\item  $x_1>0.85$ and $x_2>0.75$; 
\item  acolinearity  $\alpha < 20^\circ$; 
\item  $E_\mathrm{EMC}<0.5~\gev$ or  $E_\mathrm{EMC}<1~\gev$  for
        one or two clusters in the EMC.
\end {itemize}

At the \FourS\ resonance, backgrounds to the Bhabha sample were negligible. 
At the \TwoS\ and \ThreeS\ resonances, backgrounds from $\Upsilon \to \epem X$ decays contributed  $(1.4\pm0.1)$\% and $(0.9\pm0.1)$\%, respectively. 

At the \FourS, $\epem \to \tau^+ \tau^-$ events contributed 0.08\% and Bhabha events 0.02\% to the \mumu\ sample.
At the \TwoS\ and \ThreeS, the $\Upsilon \to \mu^+\mu^-$ decay was a
sufficiently high background in the dimuon sample that dimuon events were not used to determine the luminosity.   

The visible cross sections were calculated from Monte Carlo (MC) simulation,  
corrected for small differences with data. 
BHWIDE \cite{Jadach:1995nk} was used to generate the Bhabha  events, 
while BABAYAGA \cite{Balossini:2006wc} was used to estimate systematic errors. 
The KK generator \cite{Jadach:1999vf} was used for dimuon events. 
Corrections of 0.3\% and (0.14--0.27)\% were 
applied to the trigger and track-finding efficiencies, respectively.
 
Uncertainties in the theoretical calculations of the cross sections were estimated to be 0.21\% for Bhabha scattering and 0.44\% for dimuons.  Experimental uncertainties were 0.13\% from track finding, 0.10\% from trigger requirements, and 0.4--0.7\% from small differences in kinematic distributions  of data and simulation. Variations of the data with time added 0.05--0.17\% to the systematic uncertainties. The results for \epem\ and \mumu\ pairs were combined, taking into account correlations in their systematic uncertainties. 

Table~\ref{table:datatakingperiods} summarizes this measurement of the integrated luminosity for the six Runs recording data at the \FourS\ 
resulting in $(424.17 \pm 0.04 \pm 1.82) \invfb$ and for smaller samples recorded during Run~7 at the \ThreeS\ and \TwoS\ resonances of
$(27.96 \pm 0.03\pm 0.21)\invfb$ and $(13.60 \pm 0.02\pm 0.11)\invfb$, respectively.
The systematic error of  0.4\% of the \FourS\ on-peak data is dominated by the data-simulation comparison for Bhabha scattering and the uncertainty of the theoretical calculations for dimuons. On average, about 9\% of the data were recorded 
off resonance, primarily for studies and subtraction of non-\BB\ background.  The precise knowledge of the ratio $\lum_{\rm on}/\lum_{\rm off}$ is important for this subtraction.

\begin{table*}[htbp!]
\caption{ \label{table:datatakingperiods} 
Summary of information for the seven \babar\ Runs, showing the $\Upsilon$ resonance,
the run period, the integrated luminosity for data recorded on ($\lum_{\rm on}$)
and just below ($\lum_{\rm off}$) the $\Upsilon$ resonance, and their ratio.
In each entry, the first uncertainty is statistical and the second is systematic.
For the scan at energies above the \FourS\ the luminosity information is less accurate
since the energy was changed typically every 50 minutes. 
}
   \centering
   \begin{tabular}{cccccc}
     \hline
     Res. & Run & Mon./Year & {$\lum_{\rm on}$~(\invfb)} & {$\lum_{\rm off}$~(\invfb)} & {$\lum_{\rm on}/\lum_{\rm off}$} \cr
     \hline
     $\FourS$    & 1 & 10/99 - 10/00 &  $20.38 \pm 0.01 \pm 0.10$   & $2.565 \pm 0.002 \pm 0.012$   &  $7.947 \pm 0.006 \pm  0.012$ \cr
             & 2 & 02/01 - 06/02 &  $61.37 \pm 0.01 \pm 0.28$   & $6.875 \pm 0.004 \pm 0.032$   &  $8.926 \pm 0.005 \pm  0.012$\cr
             & 3 & 12/02 - 06/03 &  $32.30 \pm 0.01 \pm 0.14$   & $2.445 \pm 0.002 \pm 0.012$   &  $13.211 \pm 0.013 \pm 0.028$\cr
             & 4 & 09/03 - 07/04 &  $99.59 \pm 0.02 \pm 0.44$   & $10.017 \pm 0.006 \pm 0.046$  &  $9.942 \pm 0.006 \pm 0.014$\cr
             & 5 & 04/05 - 08/06 &  $132.34 \pm 0.02 \pm 0.61$  & $14.277 \pm  0.007 \pm 0.068$ & $9.269 \pm 0.005 \pm 0.014$ \cr
             & 6 & 01/07 - 09/07 &  $78.32 \pm 0.02 \pm 0.36$   & $7.752 \pm 0.005 \pm 0.037$   & $10.103 \pm  0.007 \pm 0.015$ \cr
             & Total &            & $424.18 \pm 0.04 \pm 1.82$   & $43.92 \pm 0.01 \pm 0.19$     & $9.658 \pm 0.003 \pm 0.006$ \cr
     \hline
     $\ThreeS$    & 7 & 12/07 - 02/08 & $27.96 \pm 0.03 \pm 0.21$  & $2.623 \pm 0.008 \pm 0.021$  & $10.66 \pm 0.03 \pm 0.04$ \cr
     \hline
     $\TwoS$    & 7 & 02/08 - 03/08 & $13.60 \pm 0.02 \pm 0.11$  & $1.419 \pm 0.006 \pm 0.013$  & $9.58 \pm 0.04 \pm 0.05$ \cr
     \hline
     $>\FourS$  & Scan & 02/08 - 03/08 & 3.9 & n/a & n/a \cr
     \hline
   \end{tabular}
\end{table*}

\subsubsection{\BB\ Event Counting}
\label{sec:B-counting}

Since the event rate for \BB\ production at the \FourS\ resonance depends critically on the PEP-II c.m. energy setting and beam energy spread,
it cannot be used to derive the total number of produced
\BB\ pairs from the total luminosity. Therefore, an alternate procedure was
developed.  

The full \babar\ data sample recorded at the \FourS\ resonance  contains
$N_{\BB} = \left( 471.0 \pm 2.8\right) \times 10^6$
\BB\ pairs~\cite{McGregor:2008ek}.
This number was determined by using the off-peak data to
subtract the number of hadronic events produced in continuum processes
from the total number of on-peak events.  Specifically,
\begin{equation}
N_{\BB} = (N_{\rm had} - N_{\mu\mu} \cdot R_{\rm off} \cdot \kappa)/\epsilon_{\BB}, 
\end{equation}
with
\begin{equation}
\kappa \equiv
\frac{\epsilon^\prime_{\mu\mu} \cdot \sigma^\prime_{\mu\mu}}{\epsilon_{\mu\mu} \cdot \sigma_{\mu\mu}}
\cdot
\frac{\sum_i \epsilon_{X_i} \cdot \sigma_{X_i}}
{\sum_i \epsilon^\prime_{X_i} \cdot \sigma^\prime_{X_i}}
\end{equation}

where
\begin{itemize}
\item $N_{\rm had}$ is the number of events satisfying the selection of multi-hadron events in the on-peak data;

\item $N_{\mu\mu}$ is the number of events satisfying muon-pair
selection criteria in the on-peak data;

\item $R_{\rm off}$ is the ratio of selected hadronic  to muon-pair events 
in the off-peak (continuum) data;

\item $\kappa$ 
%
corrects for the differences in continuum production cross sections ($\sigma$) and
efficiencies for satisfying the selection criteria ($\epsilon$)
between on- and off-peak c.m. energies. $\kappa$ has a value close to 1. Off-peak quantities are denoted by a prime and the subscript $\mu\mu$ refers to
muon pair events. The multi-hadron cross sections and efficiencies for continuum
processes, primarily $\epem \to \qqbar$,
are denoted by the subscript $X_i$. The muon pair
and $\qqbar$ cross sections vary as $1/E^2_{cm}$
(0.7\% increase between on- and off-peak).

\item $\epsilon_{\BB} = 0.940$ is the efficiency for produced \BB\ pairs to
satisfy the hadronic event selection, determined under the
assumption that
$\mathcal{B}(\FourS\to\BpBm) = \mathcal{B} (\FourS\to \BzBzb) = 0.5$.
Contributions of non-$B\bar{B}$ decays of the $\FourS$ are small. The uncertainties in the relative production of $\BpBm$ to $\BzBzb$ are treated as
systematic uncertainties.

\end{itemize}

Hadronic events were selected by requirements on the number of charged tracks
($\ge 3$), the total measured energy, the event shape, the location  of
the event vertex, and the momentum of the highest momentum
track. Muon pairs were identified by the invariant mass
of the two tracks, the angle between them, and the  energy
deposited in the EMC.  In a small
fraction of events, backgrounds in the EMC (such as out-of-time \epem pairs) caused a timing offset,
and the EMC clusters could not be associated with the tracks. 
In this case, at least one of the
tracks was required to be identified as a muon in the IFR.

The selection criteria were chosen to be insensitive to beam
backgrounds, while retaining high efficiency for \BB\ and
$\mumu$ events. Simulated events were produced with detector
conditions corresponding to each month of data collection, and
were mixed with randomly-triggered beam crossings from the same
time period, thereby accurately reflecting the varying
detector and background conditions.

The residual time variations of the efficiencies
resulted in an uncertainty in $\kappa$ and a corresponding
0.27\% systematic error on $N_{\BB}$. The other significant
contributions to the overall  uncertainty of 0.6\% on $N_{\BB}$ include
0.36\% from uncertainties in low-multiplicity \BB\ decays, and
0.40\% from the uncertainty in the total energy requirement.

Most $\babar$ publications  have used $N_{\BB}$ from an earlier analysis~\cite{Aubert:2002hc}
with a quoted uncertainty of 1.1\%. The new analysis 
benefited from overall improvements in the charged
particle reconstruction efficiency, and modifications to the
selection criteria to reduce sensitivity to background conditions. 
The central values of $N_{\BB}$ 
from the earlier analysis are consistent with the new ones within the systematic error of the new analysis.

The numbers of hadronic events and
muon pairs were determined for each run, and analysts could then obtain 
\BB\ counts and luminosity values for any subset of the full dataset.

The numbers of produced  $\ThreeS$ and $\TwoS$ events  in data samples 
collected at these resonances were determined  in a similar way. 
The off-peak continuum scaling was performed using $\epem \to \gamma \gamma$
events, to avoid the non-negligible rate of $\Upsilon \to\mumu$ decays.
The hadronic selection criteria were  modified for these analyses.
The data samples contain
$\left( 121.3 \pm 1.2\right)\times 10^6$ $\ThreeS$ and
$\left( 98.3 \pm 0.9\right)\times 10^6$ $\TwoS$  decays, respectively.
The primary contributions to the systematic errors are
uncertainties on the efficiency of the total energy selection (0.6\%)
and the requirement on the number of tracks (0.4\%), plus the 0.5\% impact of the 
uncertainty on the $\Upsilon \to \ell^+ \ell^-$ branching fractions.

\subsection{\babar\ Background Protection and Monitoring}
\label{sec:back-prot}
\subsubsection{Beam Background Sources}
\label{sec:mdi:background-sources}

The primary sources of steady-state accelerator backgrounds were synchrotron radiation in the vicinity of the interaction region, interactions
between the beams and the residual gas in either ring, radiative Bhabha scattering generating electromagnetic showers and also leading to electroproduction of hadrons, resulting in low energy neutrons.  

Synchrotron radiation from the IR quadrupole doublets
and the B1 dipoles generated up to 150~\kw\ of power and was potentially a severe
background. The beam orbits, apertures and masking schemes had been designed such that
almost all of these photons passed through the detector region and were channeled to a distant 
absorber. The
remainder were forced to undergo multiple scatters, before  eventually entering the \babar\ acceptance.
The resulting synchrotron radiation background was dominated by x-rays generated in the HER final focus
quadrupoles and scattered from the tip of a tungsten mask.

Bremsstrahlung and Coulomb scattering off residual gas molecules resulted in losses of beam particles.
The intrinsic rate of these processes is proportional to
the product of the beam current and the residual pressure (which itself increases with
current). Their relative importance, as well as the resulting spatial distribution and
absolute rate of lost particles impinging the vacuum pipe in the vicinity of the detector,
depended on the beam optical functions, the limiting apertures, and on the residual-pressure
profile around the entire rings. The separation dipoles bent energy-degraded particles from
the two beams in opposite directions and consequently most \babar\ detector systems exhibited
occupancy peaks in the horizontal plane, \ie\ near $\phi=0^{\degrees}$ (East side of the rings) for the LER and
near $\phi=180^{\degrees}$ (West side of the rings) for the HER. After several months of pumping, the ring-averaged
dynamic pressure reached typical values of 2.5 nTorr/A in the HER and 1 nTorr/A in the LER.

\begin{figure}[htb!]  
  \begin{center}  
    \includegraphics[width=7.5cm]{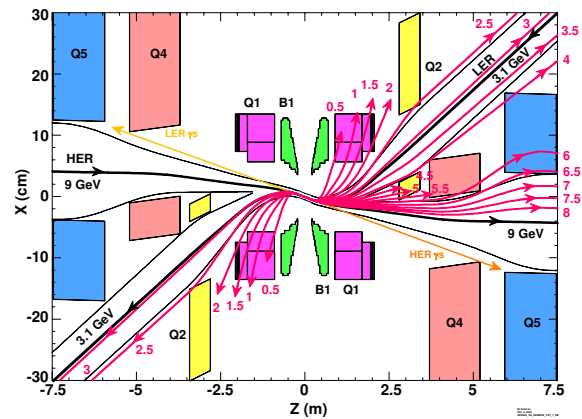}  
    \caption{Plan view of the \pep2\ interaction region with trajectories of 
photons and  electrons and positrons from radiative Bhabha scattering. Their energies in units of \gev\ are indicated.
The scale of the vertical axis is strongly enlarged.}  
    \label{fig:PEPII_Bhabhas}  
  \end{center}  
 \end{figure}

Radiative Bhabha scattering resulted in energy-degraded electrons and positrons hitting aperture
limitations within a few meters of the IP and spraying the \babar\ detector
with electromagnetic showers.
This background was proportional to the instantaneous luminosity.
Electrons and positrons interacted with the copper or steel of the masks and vacuum pipe flanges  and beam-line elements
and produced unstable resonances which emitted neutrons with a typical energy of 1-2~$\mev$.  This source of background, which primarily affected the DIRC, had not been anticipated. Its effect was greatly enhanced by the B1 bending magnets and  the off-axis passage of the LER through the quadrupoles Q1.
Figure~\ref{fig:PEPII_Bhabhas} shows the trajectories of electrons and positrons of different energies from radiative Bhabha pairs and illustrates where they might interact with different beam-line components.  

In addition to these steady-state background sources, 
there were other background sources that fluctuated widely and led to very large instantaneous rates, thereby disrupting stable operation.
Among them were backgrounds generated by
\begin{itemize}
\item
beam losses during injection;
\item
non-Gaussian beam tails from beam-beam interactions (possibly in
conjunction with electron-cloud-induced blow-up of the positron beam) that were highly sensitive
to adjustments in collimator settings and ring tunes;
\item
intense bursts of radiation in the HER only, varying in duration from a few \ms to several
minutes, attributed to micron-sized dust particles, most likely originating from the non-evaporable
getter pumps and from distributed ion pumps, attributed to pump saturation; and
\item
HOM power penetrating the RF screens inside the LER vacuum chamber resulted in heating of the NEG vacuum pumps and 
subsequently produced large, sustained (extending over several minutes) outgassing in the beam line 10-80~m upbeam of the detector. Consequently,
some of the pumps in this region and also two movable collimators were removed.
\end{itemize}

\subsubsection{Survey of Beam Background}
\label{sec:mdi:background-Survey}

During the first year of operation, it became evident that 
beam-related backgrounds were higher than anticipated and their origins were poorly understood.  In particular, it was not known that radiative Bhabha scattering and the neutrons they produced were the dominant source of background in the DIRC.  An initial survey was performed with a Geiger counter to locate areas of high radiation along the beam line and decide on the placement of permanent monitors and shielding. 

A large number of radiation monitors were installed along the beam pipe and throughout \babar\ to survey and record beam-generated backgrounds. More than 40 CsI crystals with PIN diode readout were installed along the beam pipe to measure the absorbed dose from ionizing radiation. They were complemented by several quartz radiators connected to fibers with PMT readout to sense electromagnetic showers.
Three small wire chambers filled with BF$_3$ which were sensitive to thermal neutrons and insensitive to electromagnetic radiation were installed, one on each Q4 quadrupole, and a third far forward outside of \babar. 

Special studies and measurements of background rates versus beam currents were carried out with both single beams and colliding beams. 
The observed neutron rate could be expressed in terms of the HER and LER currents 
($\rm{I}_\mathrm{HER}$ and $\rm{I}_\mathrm{LER}$) and the luminosity $L$,
\begin{equation}
  {\rm Rate \; kHz)} \approx 1.0 \, {\rm I_{\LER}} + 0.6 \, {\rm I_{\HER}} + 1.74 \, {\rm L};
\end{equation}
here the currents are measured in Ampere, and the luminosity 
in units of $10^{33}\cms$.
The neutron rate near the beam on both sides of the IP showed a strong dependence on luminosity, similar to rates measured in various \babar\ detector components. 

Based on high counting rates, several hot spots were located along the beam line indicating high neutron activity, typically near beam pipe flanges. Assuming that there was one dominant neutron source and that the neutrons were emitted uniformly, the source location could be determined based on the rates in two counters. To do this assessment, the absolute efficiency  of the BF$_3$ neutron counters was
calibrated  with a $^{238}{\rm Pu}$ neutron source; it was close to 2-3\%,
not atypical for these types of detectors. 
The estimated location was consistent with expectations based on other surveys of radioactivity and with ray tracing of electrons and positrons from radiative Bhabha events.  Based on the location and the observed rate, an estimate of the expected neutron rates was derived at various positions within \babar.  The results are presented in Table~\ref{tab:neutron}.  The predicted rate range reaches levels of kHz~$/\cm^2$.

\begin{table}[htbp!]
\caption{Prediction of neutron fluxes into the \babar\ detector for a luminosity of $7 \times 10^{33} \cms$, assuming spherical emission from a point source located along the beam line on the forward side of \babar.}
\label{tab:neutron}
\small
\begin{tabular}{lcc}
\\
\hline					
Detector    & Distance to     & Rate  \\
Component   & Source ($\cm$) & (kHz/$\cm^2$)	\\
\hline		
\emc ~~  Inner wall, forward       &  $ 90$   &  16   \\
\svt ~~  Central section           &  $130$   &  7.5  \\
\rpc ~~  Inner radius, forward     &  $140$   &  6.5  \\
\dch ~~  Backward end              &  $250$   &  2.0  \\
\dirc ~~ Entry to SOB             &  $450$    &  0.62 \\
\hline
\end{tabular}
\normalsize
\end{table}

\subsubsection{Detector Shielding}
\label{sec:mdi:beam_shielding}

Beam backgrounds impact the detector systems in a variety of ways.
As charged particles, photons and neutrons interact
with the detector components, they increase occupancy, may degrade the response or lead to HV break-down. Integrated
over time, they may damage the active detector components 
and the electronics.
The best way to reduce their impact is to shield the detector and
electronic components by inserting absorbers near the beam.
As the luminosity increased throughout the years, several beam
background shields were installed along the beam line.

As mentioned in the previous section, it was discovered shortly after the start of data taking 
that the DIRC was impacted by high rates of beam-generated background, primarily
from electrons and positrons from radiative Bhabha scattering producing
electromagnetic showers and neutrons of a few\mev.
The estimated neutron flux at the entrance to the DIRC SOB was about $6 \times 10^2$\hz/$\cm^2$.
These neutrons either scattered elastically off hydrogen in the water or
efficiently thermalized in about 1 m of water and were
captured on hydrogen, producing $2.2 \mev$ photons.
The thermalization typically took $1\mus$ and the subsequent capture
occurred within $200\ns$. The Compton photons scattered in the water and the
resulting electrons produced Cherenkov light.

The high rate of particles resulted in activation of the beam pipe, masks, and other IR components. 
Shielding, installed at the hottest spot, immediately improved the
DIRC background rate by about a factor of three.  
To allow future access, this shield was made in two parts:
a horse-shoe shape piece (158\cm long and 10\cm thick, 5\cm each of lead and steel) that could slide along the beam pipe,  complemented by several fixed layers
of lead (1.5\cm) and steel (5\cm), also 158\cm long. The shielding was augmented 
by bags filled with lead shot underneath the beam elements. The overall improvements in the background rates were consistent with simulations of radiative Bhabha scattering.

The outermost forward endcap layers of the IFR were exposed to backgrounds 
generated in the LER.  Beam losses in upstream 
collimators generated electromagnetic showers causing extremely high rates 
in the outermost layer (400~kHz), often tripping the HV supply and rendering the layer
 inoperable.  The next outermost 
layer was beneath 10 \cm of steel and had lower backgrounds (80~kHz). 
Small lead shielding walls were placed downstream of collimators and quadrupole 
magnets in the \pep2\  beam line up to 40 m from \babar. Due to the variability 
of the background rate, it was difficult to quote a quantitative improvement 
from these installations. However, these shielding efforts and improvements 
in \pep2\ beam steering resulted in fewer high background periods. 

\begin{figure}[!htb]
\centering
\includegraphics[width=7.5cm]{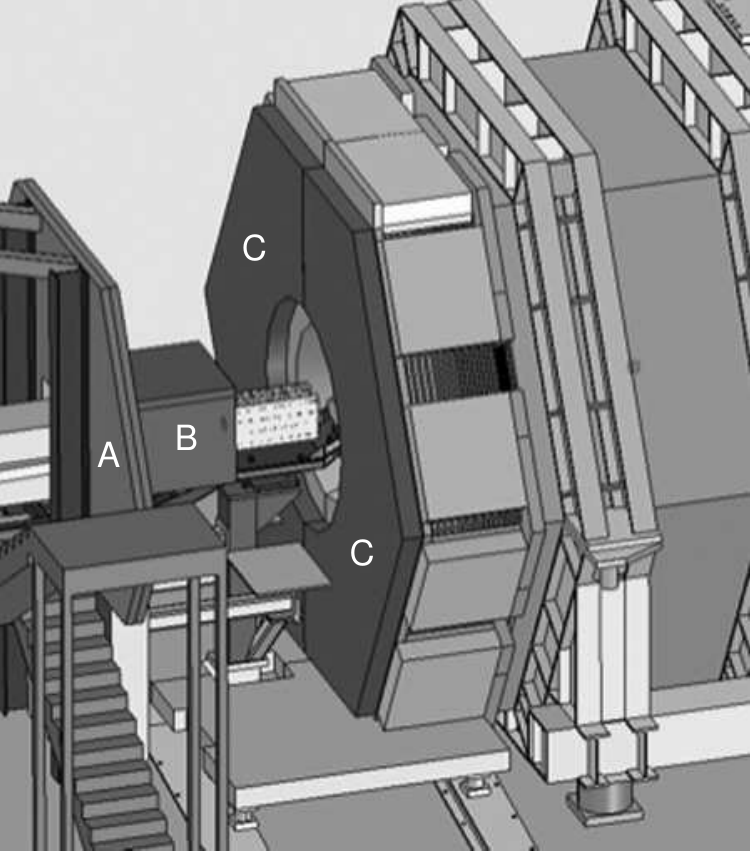}
\caption{IFR forward endcap shielding:  (A) wall covering the tunnel exit and (B) extension around the beam line components built to protect 
the outer RPC layers of the upgraded endcap (C) from beam-generated background.}
\label{RPC_8821A22}
\end{figure}

After the 2002 endcap upgrade, the outermost IFR layers (15,16), were protected by 
2.5\cm of steel and layers (13,14) were protected by an additional 10\cm of steel.
However, layers 15 and 16 were not used in data analyses due to high occupancy. In late 2004 
a large (4~m radius) 20 cm thick steel wall, shown in Figure~\ref{RPC_8821A22},
was placed between \babar ~and the \pep2 tunnel.  

\begin{figure}[!htb]
\centering
\includegraphics[width=7.5cm]{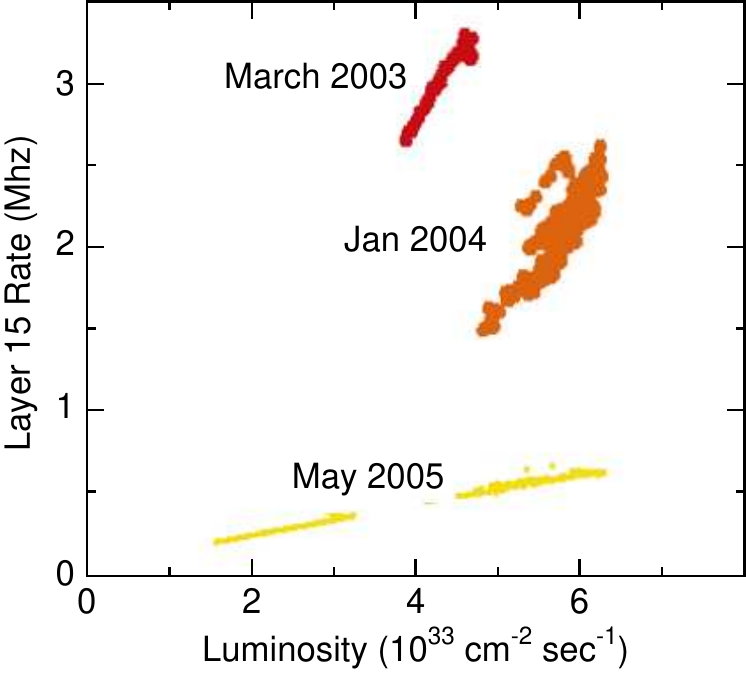}
\caption{RPC forward endcap background rates in layer 15 as a function of luminosity. Three representative time periods are shown: March 2003 (after the endcap upgrade), January 2004 (during implementation of trickle 
injection), and May 2005 (after the installation of the shielding).
}
\label{RPC_8821A21}
\end{figure}

Backgrounds were significantly reduced by both the introduction of the wall and 
the implementation of continuous injection of the positron beam (Section 2.2.2).
Figure~\ref{RPC_8821A21} shows how the noise 
rate of outer layer 15 varied with luminosity after the endcap upgrade, after the start of trickle
injection, and after the installation of the shielding wall. Not only were singles rates reduced,
but the correlation with increased luminosity was reduced.
All of the remaining \babar ~data  were collected with this shielding configuration. 
All RPC layers were active and 
occupancies in the outer layers were less than 5\%. A smaller 10\cm\ steel 
extension  was added in May 2007
around the Q5 magnet between the shielding wall and \babar\ to 
complete the shielding efforts.

\subsubsection{Active Detector Protection Systems}

In order to protect the detector from high radiation dose due to large background
bursts during injection or steady data taking, several active protection systems were installed.

\paragraph{\svtrad System}
\label{sec:SVTRAD}

The \svt radiation monitoring (\svtrad) system was designed to monitor the instantaneous
and the total integrated radiation doses received by the SVT, and, thereby, to  reduce the potential for
permanent damage of the silicon detectors and the associated electronics.
The original system, installed in 1999,  consisted of 12 Hamamatsu S3590-08 silicon PIN diodes ($1\cm \times 1\cm \times 300\mum$).  At an operating voltage of 500~V, their sensitivity was measured to be 200 pC/mRad.

These PIN diodes were mounted at a radial distance of $3\cm$ from the beam axis on both sides of the IP at $z=+12\cm$ and $z=-8\cm$~\cite{Meyer:2000tu},
monitoring the dose at beam level and $3\cm$ above and below, on the inside and outside
of the storage rings. The sensor currents were sampled at a rate of 10\khz, and the radiation-induced doses
were derived taking into account the diode leakage currents and temperature variations.
At luminosities of $1.2 \times 10^{34} \cms$, the typical measured dose rate 
during stable beam operation was 50 mRad/s in the mid-plane and 10 mRad/s elsewhere at the same distance from the beam.

While the PIN diodes accumulated 2-3\MRad\ of absorbed dose, their leakage currents increased from an initial level of 1\nA to more than 2\muA.
These  currents were very large compared to typical signals from beam-induced background of 10\nA or less. Furthermore, the leakage
current changed with temperature by approximately 10\% per degree $C$, requiring very accurate tracking of temperature changes
and making accurate rate estimates increasingly difficult.

To improve the radiation monitoring, two polycrystalline
chemical vapor deposition (pCVD) diamond sensors of $1\cm \times 1\cm \times 500\mum$ were
installed \cite{Bruinsma:2004iu,Bruinsma:2007zz} in 2002.  
At an operating voltage of 500\,V, the sensitivity of the diamond sensors was measured to be 100 pC/mRad.

The diamond sensors were placed in the mid-plane at the backward end of the SVT, where the highest radiation levels were expected. Due to the limited space, the diamond sensors
were located slightly further away (15\cm) from the IP than the PIN diodes,
and at a radial distance of 5\cm
from the beam. The leakage current in the pCVD diamond sensors
remained in the range of 1-2\nA with no sensitivity to
temperature variations, even after a total absorbed dose of close to
2\MRad. For the last five years of the detector operation, the two diamond sensors
served as the primary radiation monitors.

The SVTRAD system was also used as an active protection system.
Whenever the SVTRAD current exceeded a predefined limit for a specific time interval, the system initiated an abort of the circulating \pep2\ beams.
Initially, the abort limits were set at very
low level to ensure the safety of the SVT, but this also led to very frequent beam aborts. 
As experience was gained, luminosity rose and the SVT was found to be
more radiation-tolerant than expected, the abort limits were relaxed
by an order of magnitude.

\begin{figure}[t]
\begin{center}
\includegraphics[width=7.0cm]{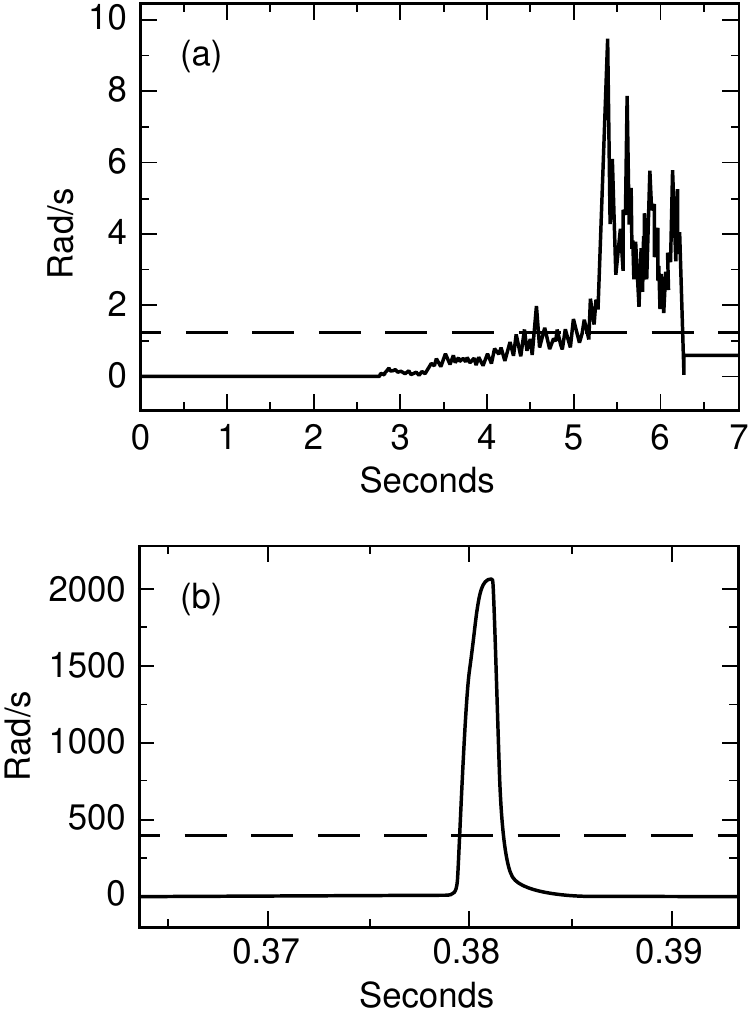}
\caption{SVTRAD: Radiation levels for two types beam aborts:
(a) integrated limit of 5\Rad, and (b) instantaneous burst exceeding 400\Rad/s ~\cite{Bruinsma:2007zz}.
}
 \label{fig:svtrad_aborts}
\end{center}
\end{figure}

The following describes the typical limits set during the last few years of
operation. To limit excess exposure to radiation,
the integrated rate was monitored over periods of 1 and 10 minutes.
If the dose exceeded 100\mRad/s for more than 1 minute, an alarm was sent to the operators.
If conditions persisted for 10 minutes, a beam abort was initiated~-- though this could be delayed
by the operator if it was expected that radiation levels would go back to normal within a few extra
minutes.
In addition, two criteria for beam aborts were implemented to respond to faster radiation bursts:
(a) the rate exceeded 1.25\Rad/s and the time-integrated dose was larger than 5 Rad or (b) the instantaneous rate exceeded 400\Rad/s.
Figure~\ref{fig:svtrad_aborts} shows typical radiation levels during these two types of beam aborts.
During the year 2000 Run with a peak luminosity of $2-3 \times
10^{33}\cms$, there were an average of 10 SVTRAD-induced beam aborts per  day; the average daily dose accumulated in the most sensitive area was about 1.2~kRad.
During the last running period of \babar\ with the
peak luminosity exceeding $1.2 \times 10^{34}\cms$, the SVTRAD abort rate was about 3.5 beam aborts per day
while the average accumulated dose reached almost 4\kRad\ per day.

The \svtrad system, particularly the highly reliable pCVD diamond
detectors, provided valuable feedback to the \pep2\ operation team on
background conditions near the IP, thereby assisting in the tuning of
the storage rings and the reduction of beam backgrounds~\cite{Bruinsma:2007zz}.
With the introduction of trickle injection, the \svtrad system was
extended to also monitor the radiation doses associated with each
injected pulse of electrons and positrons during trickle injection at
a rate of a few \hz. The total currents were measured in a 10 ms time interval immediately following the injection pulse and compared to those measured during a
second 10 ms interval that started 10 ms after injection. The difference of these two measurements translated to a
dose of typically 0.02\mRad\ and 0.01\mRad\ per injected bunch of electrons and positrons, respectively. 

\begin{figure}[htbp!]
\begin{center}
\includegraphics[width=7.0cm]{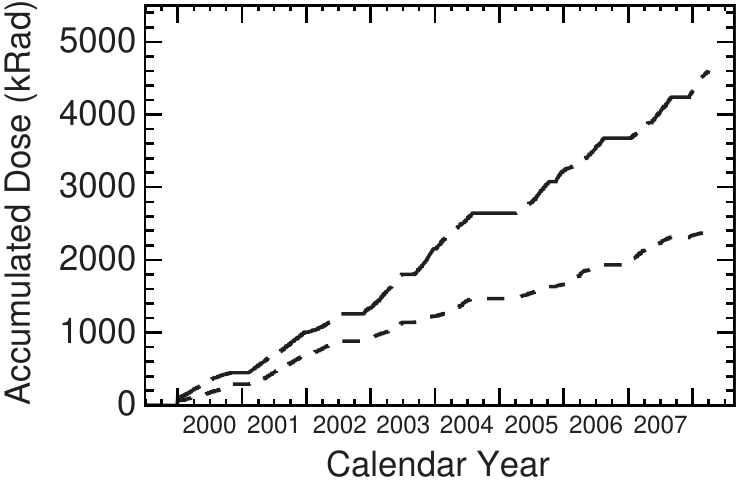}
\caption{SVTRAD: The integrated dose rate recorded by sensors placed in the mid-plane at azimuth angles of $\phi=0\degrees$ (dotted line) and  $\phi=180\degrees$ (dashed line) .
\label{fig:svtrad}}
\end{center}
\end{figure}

During the nine years of operation the highest integrated dose the SVT received 
exceeded 4\MRad, still below the 5\MRad\ limit set for the SVT.
Figure~\ref{fig:svtrad} shows the accumulated dose in the horizontal plane over the nine years of \babar\ operation. The higher dose at $\phi=180\degrees$ is primarily due to HER backgrounds, while the LER backgrounds dominate at $\phi=0\degrees$.

\paragraph{\dch Active Protection System}
\label{sec:DCHprotection}

For the \dch, the principal concern  related to surges in beam-generated
background was potential damage to the wires and to the analog and
digital electronics that were mounted on the backward end plate.  The
over-current limiting system of the CAEN HV supply provided the necessary
protection.  The original power supply boards, CAEN model A944, imposed a
current limit of 40~$\mu$A per channel (there were 55 channels total). It
soon became clear that this would not be sufficient as the luminosity and
backgrounds increased.  The power supplies were replaced by CAEN A833 boards, which
allowed for up to 200~$\mu$A per channel, although they were operated with a
software limit of 60~$\mu$A. Typical operating currents were 8--10~$\mu$A
per channel at the start of the project, and 15--20~$\mu$A at the end.

The system was operated in two different configurations. Prior to the
implementation of trickle injection, an overcurrent ($> 40\muA$ on any of
the 55 individual HV channels) caused the entire chamber to ramp down from
the ${\sim} 1930$~V operating point (see Section~\ref{sub:opsdch}) to
1200~V. The voltage would automatically ramp up when the SVTRAD system
indicated that the backgrounds had subsided to an acceptable level. The
detector trigger was inhibited during this process, invoking a minimum of
30~s dead-time per incident. For instance, there were a few trips per day in the months
immediately prior to the start of trickle injection.

Once trickle injection became the default mode of operation, short-duration
current surges (a few seconds or less) were observed on individual HV
channels, changing from one channel to another at intervals of several tens
of seconds.  These high background bursts could last for minutes or
hours, depending upon the performance of the accelerator; the protection
system above resulted in a high fraction of dead-time.

To reduce this dead-time, while still protecting the drift chamber from
potential damage, the HV system was configured so that in case of a current
surge, the voltage of an individual channel would be lowered to keep the
current below the limit of $40\muA$ (increased to $60\muA$ mid-way through
Run 5).  As a result, the affected small region of the chamber had
temporarily a slightly lower gain, potentially affecting efficiency in
regions of very high backgrounds.  The trigger was no longer inhibited.
During Run 6, there were typically 10-20 such occurrences per day, in total
extending over just ${\sim} 100$~seconds.

\subsubsection{Background Monitoring}
\label{sec:mdi:background-monitoring}

Beam induced backgrounds were carefully monitored during data taking since they affected the trigger rates and dead-time, the data quality (resolution and efficiency degradation), and the integrity and lifetime of the subdetectors.

\begin{figure*}[htbp!]
\begin{center}
\includegraphics[angle=-90,width=0.9\textwidth]{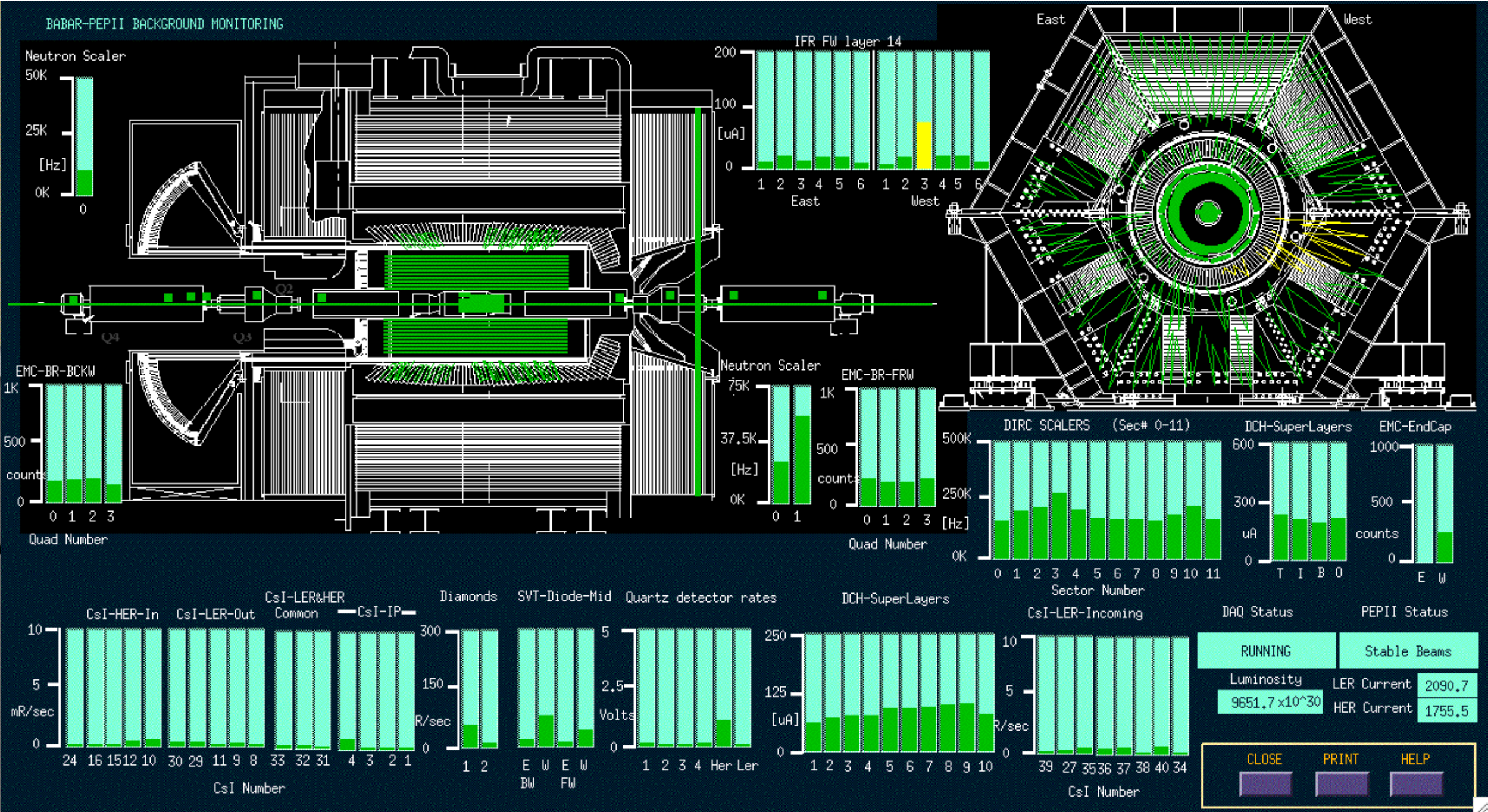}
\includegraphics[angle=-90,width=0.9\textwidth]{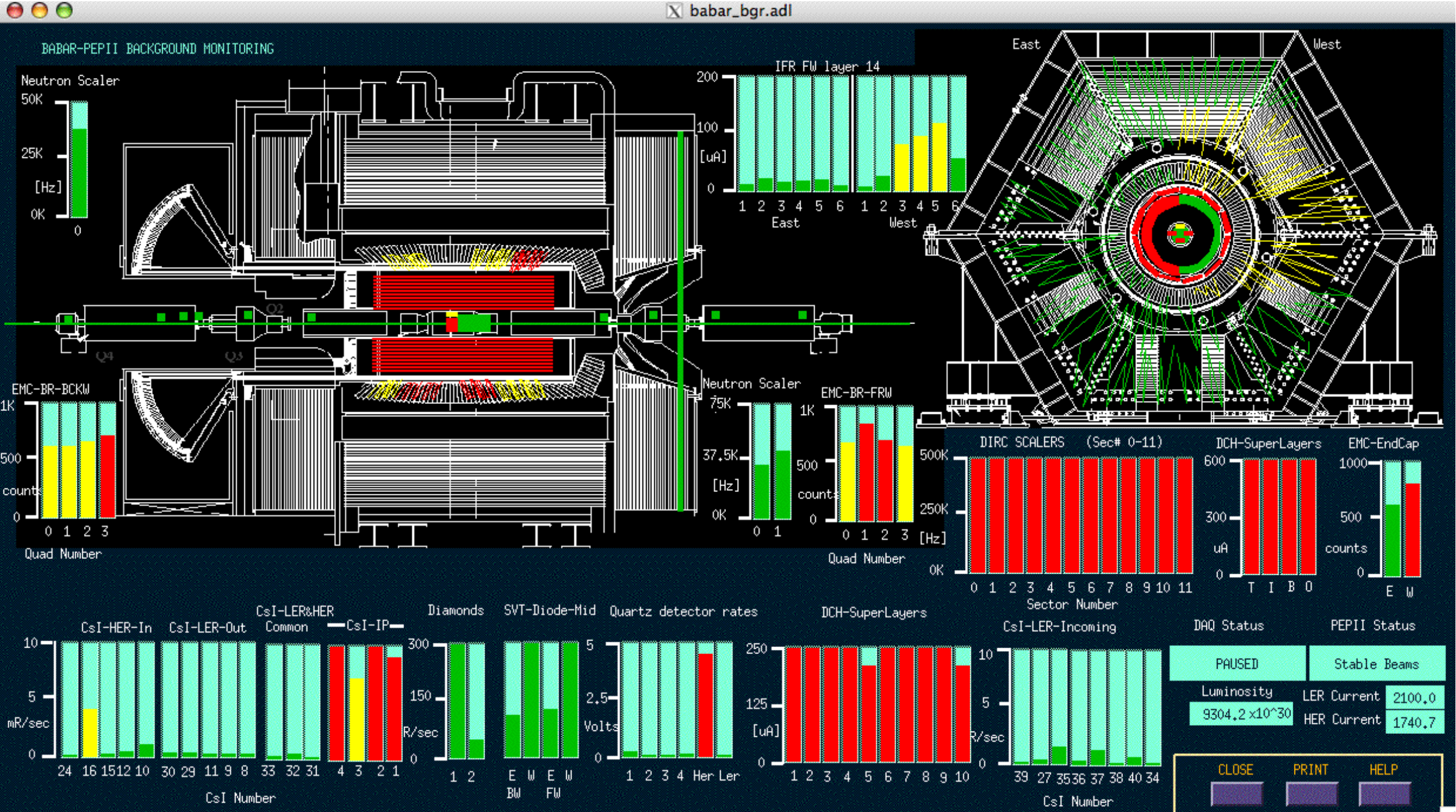}
\end{center}
\caption{
\babar\ background display for normal (top) and very high (bottom) background conditions. The background meters are color-coded: green for normal, yellow for elevated, and red for critical background levels.
}
\label{fig:bkgdisplay}
\end{figure*}

The readings of the various background detectors were available online on a large display on which their location was overlaid on the \babar\ mechanical drawing, thus assisting the operators to identify the location of high backgrounds in real time.  Examples of this display are presented in Figure~\ref{fig:bkgdisplay}, one
for a situation with normal background conditions and a second with a very high background level. 
At the bottom, the readings of selected CsI(Tl) detectors indicate the doses in mRad/s
at various locations, along the HER and LER beam lines and at the IP; the two diamond detectors near the SVT measure the dose in Rad/s, and the quartz detector near the IP readout measure the pulse height in volts. 
The BF$_3$ neutron counters indicate typical rates of 5-10 kHz.  For the DCH and the IFR endcaps, the currents are presented for various detector layers.  The singles rates in the 12 DIRC sectors and the quadrants of the EMC forward and backward are also presented.
The color coding is designed to draw the operator's attention to unusually high radiation and detector backgrounds. 

This instantaneous information was complemented with scrolling histories of background levels in various detectors and allowed the operator to correlate incidents of high background. It should be noted that these displays were updated once per second to indicate chronic background conditions; short background bursts could be missed.

\paragraph{DIRC Scalers}
High background levels in the \dirc\ PMTs were of concern, since they could accelerate ageing and also saturate the readout, thereby affecting the quality of the signal reconstruction due to missing signal hits and large numbers of background hits. To monitor the rate in real-time, twelve PMTs, one per sector, had their outputs split: one  
connected to the regular DAQ system and the other to a NIM scaler. 
The rates were integrated over 0.5~sec every 4~sec; they were available to \babar\ and \pep2\ operators, and also stored for offline analysis. 
The dependence of the scaler rate, averaged over the 12 sectors, on the beam currents and the luminosity could be expressed as,
\begin{equation}
  {\rm Rate \; (kHz)} \approx 3.9 \, {\rm I_{\LER}} + 12.9 \, {\rm I_{\HER}} + 14.9 \, {\rm L}.
\end{equation}
The currents are in units of Ampere, and the luminosity in units of $10^{33}\cms)$.

During high-luminosity operation, the normal PMT rates were a few hundred \khz, but peaks close to 1\mhz were observed occasionally. 
Except for these very short bursts, the backgrounds in the DIRC were not high enough to be a serious concern for the \dirc\ operation. On the other hand, the \dirc\ scalers were sensitive to sources of beam backgrounds that were difficult to diagnose for the accelerator experts, and therefore they were very useful in tuning the \pep2\ beams for low background conditions.   

\paragraph{\dch\ Current Monitors}

\begin{figure}[htbp]
\includegraphics[width=0.9\columnwidth]{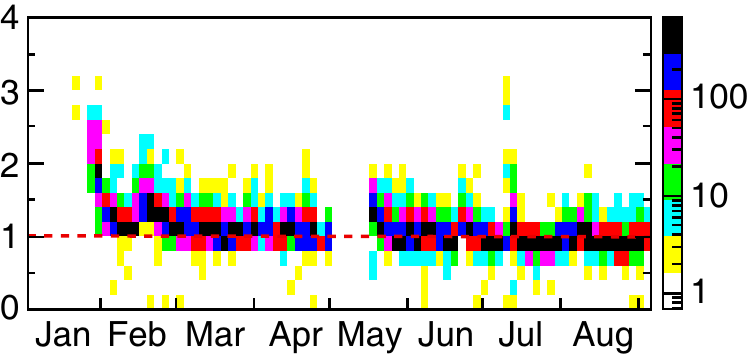}
\caption{\dch Background monitoring:
Scatterplot of the ratio of the measured total current to that predicted from \pep2\ beam currents and luminosity, sampled during Run 6 (2007). The left scale shows the ratio, the right scale the color code for the number of entries per sample.}
\label{fig:dch_current}
\end{figure}

The currents of the CAEN HV supplies also served as background monitor for both \babar\ and \pep2. After trickle injection was introduced,
the CAEN A833 boards were modified to provide in real time an analog sum of the currents of each of the ten superlayers to the \pep2 control room for beam tuning.

\begin{figure}[htbp]
\includegraphics[width=0.9\columnwidth]{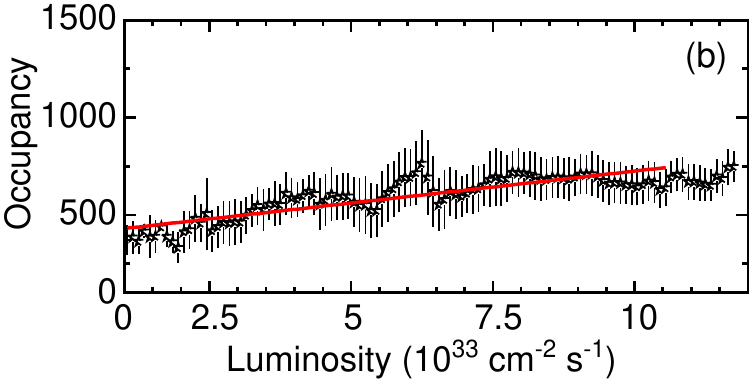}
\caption{\dch Background monitoring: The occupancy, defined as the
number of drift chamber hits in a triggered event, as a function of luminosity in
the second part of Run~5 and Run~6. The error bars indicate the RMS of the distribution.}
\label{fig:dch_occupancy}
\end{figure}

In addition, \dch\ backgrounds were extensively monitored offline, specifically the currents, the number of HV trips or incidents of overcurrents, and the occupancy.   
Figure~\ref{fig:dch_current} shows for Run 6 the observed total current \dch\ relative to its predicted value based on the actual \pep2\ currents and luminosity.  This was an effective way of monitoring over longer periods the excess backgrounds caused by poor beam conditions and tuning or poor vacuum conditions.  
At the highest luminosity, there were on average 750 hits per triggered
event in the \dch, corresponding to 10.5\% occupancy, as illustrated in Figure~\ref{fig:dch_occupancy}. The dependence of the occupancy on the luminosity 
can be approximated by a linear function,  
$N= 397 + 32.8 \times L$, where $N$ is the number of hits in the chamber,
and $L$ refers to the luminosity measured in units of $10^{33}\cms$.     
The tracking efficiency was affected whenever the occupancy exceeded $15 {\sim} 20$\%.

\paragraph{EMC RadFET Monitors}
\begin{figure}[ht!]
  \begin{center}
	  \includegraphics[width=0.5\textwidth]{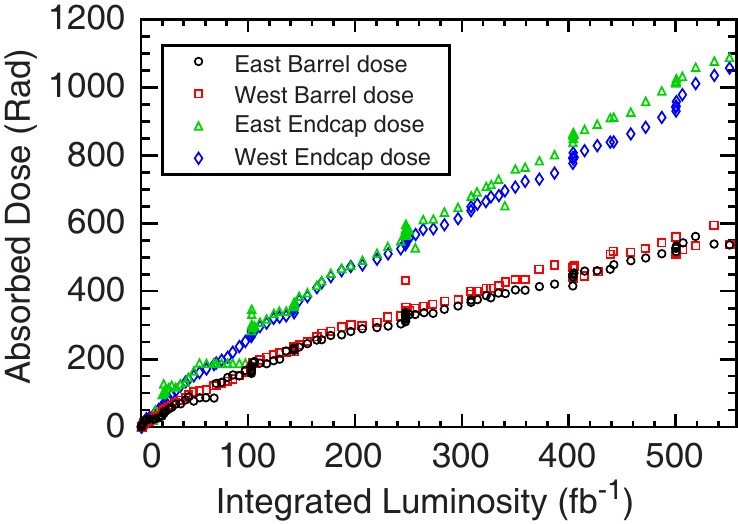}
    \caption{EMC: Absorbed radiation dose as a function of the delivered luminosity.}
    \label{fig:emc_radFET_dose}
  \end{center}
\end{figure}

Accelerator-generated backgrounds can result in large depositions of radiation in the EMC.  This radiation creates color centers in the crystals which lead to changes in light transmission.

In total, 115 radiation-sensitive field-effect transistors (RADFETs) were installed throughout the EMC to record the absorbed dose.
There were 55 RADFETs in the barrel, arranged in eight strips in $\phi$, each supporting
seven RadFETs (except for one strip with six),
and 60 RADFETs in the endcap, arranged in twenty strips in $\phi$, each supporting three RADFETs.
All RADFETs were placed in front of the crystals.
A more detailed description of the RADFETs and associated hardware can be found in Ref.~\cite{Camanzi:2002nk}.

The doses absorbed in each section of the EMC are shown as a function of delivered luminosity in Figure~\ref{fig:emc_radFET_dose}, indicating that, as expected, rates were lower in the barrel than in the endcap. 
The absorbed dose is not quite a linear function of luminosity.
The principal reason is the introduction of trickle injection (see Section~\ref{sec:mdi:trickle} for details). 
Prior to the introduction of trickle injection, close to half of the dose was absorbed during injection or tuning of the beams, when no data were recorded. 
Thus, while initially the dose rate was
$4-12\phantom{i}/\mathrm{Rad\phantom{i}fb^{-1}}$, it decreased to
$1-2\phantom{i}/\mathrm{Rad\phantom{i}fb^{-1}}$ after trickle injection was adopted as the normal mode of operation.
Earlier estimates for the absorbed dose  were on the order of $0.6 \phantom{i} \mathrm{Rad \phantom{i} day^{-1}}$.
The absorbed dose in the endcap over all runs was ${\sim} 1.1 \phantom{i} \kRad$.
Dividing this number by the number of days with operating beams (2289 days) results in an average daily dose of $0.48 \phantom{i} \Rad$.

The absorbed dose at the end of the experiment for the different parts of the \emc\ is shown in Table \ref{tab:emc_radFET_dose}.  As expected, the doses in the endcap were larger than in the barrel, decreasing with radial distance from the beam.
The largest contribution to the uncertainty in the measured doses originates from the calibration  of the RADFETs and their temperature dependence.
The total light yield decreased measurably 
at exposures of 100\Rad\ and light yield uniformity over the length of the crystals became an issue at exposures exceeding 1\kRad.

The projected maximum integrated dose for the EMC was estimated to be 
$10 \phantom{i}\mathrm{kRad}$ over the lifetime of \babar,
based on an expected integrated luminosity of $1~\mathrm{ab^{-1}}$.
For the duration of the experiment the overall dose was well below this limit. 

\begin{table}[!htb]
\caption{EMC: Integrated radiation dose absorbed over the lifetime of the \pep2\ B Factory, as measured by the RADFET system.
}
\label{tab:emc_radFET_dose}
\centering
\begin{tabular}{lc}
\\
\hline
Domain              & Absorbed Dose (\Rad) \\
\hline
East barrel         &  $ 540 \pm  35 $ \\
West barrel         &  $ 591 \pm  39 $ \\
East endcap inner   &  $1541 \pm 120 $ \\
East endcap center  &  $1056 \pm  87 $ \\
East endcap outer   &  $ 731 \pm  54 $ \\
West endcap inner   &  $1511 \pm 116 $ \\
West endcap center  &  $ 946 \pm  74 $ \\
West endcap outer   &  $ 675 \pm  58 $ \\
\hline
\end{tabular}
\end{table}


\renewcommand{\secname}{upgrades_}
\renewcommand{\sectiondir}{Sec03_Upgrades}
\section{\babar\ Detector Upgrades}

\subsection{Overview}

To handle the increase in both signal and background rates, the detector 
underwent a series of upgrades and operational changes to achieve and maintain the highest data collection efficiency.

The only detector system which required major upgrades was the IFR.  
To cope with the high background rates 
in the forward endcap, the original RPCs were replaced with chambers of
improved design. In the barrel section, the RPCs were replaced
by LSTs in 2004 and 2006. In the barrel and endcap the absorber thickness was 
increased to improve pion rejection.  

Extensive upgrades to the front-end electronics, trigger, and
data acquisition system were implemented to handle the increased event
rates and occupancy, and to maintain the nearly dead-time-free
operation originally envisioned. Improvements in automation and error
diagnosis and recovery, driven by detailed monitoring and rigorous
evaluation of the lessons learned during operation,
allowed \babar\ to maintain a long-term average of better than 96\%\
operational efficiency.

\subsection{Online System Upgrades}
\label{online-upgrades}

\subsubsection{Evolution of Requirements}

Of all the changes in the \pep2\ operation, the advent of trickle
injection (see Section~\ref{sec:mdi:trickle}) had the greatest impact
on the online system.  It resulted in a fundamental change in the
detector operation, requiring the online system to support continuous
data acquisition over periods of a day or more, and necessitating the
development of new features in the trigger and data acquisition to
suppress backgrounds generated by freshly injected beam pulses.

Another important contribution to operation at increased data rates
was the meticulous focus on operational efficiency, driven by the goal
of {\it factory operation}, requiring continuous monitoring and
attention to operational details. The original design of the trigger
and online system had already aimed at minimizing the data acquisition
dead-time. However, to improve the overall data taking efficiency,
this was now complemented by system enhancements and refined
operational procedures, that focused on maximizing detector and online
system uptime.
 
As the experiment made the transition from commissioning to stable
operation, it became desirable to reduce the staffing requirements for
detector operation.  This was achieved through extensive automation of
the online control and monitoring functions, and simplification of
the user interfaces.

During the lifetime of \babar, the online computing system made the
transition from a vendor-specific CPU and operating system environment
to the use of commodity systems. This was necessary to deliver
significant performance upgrades at a modest cost, meeting the demands
of luminosity and background increases and providing a substantially
larger operational headroom.

\subsubsection{Overall Architecture}

\begin{figure*}[htbp!]
\begin{center}
\includegraphics[width=15cm]{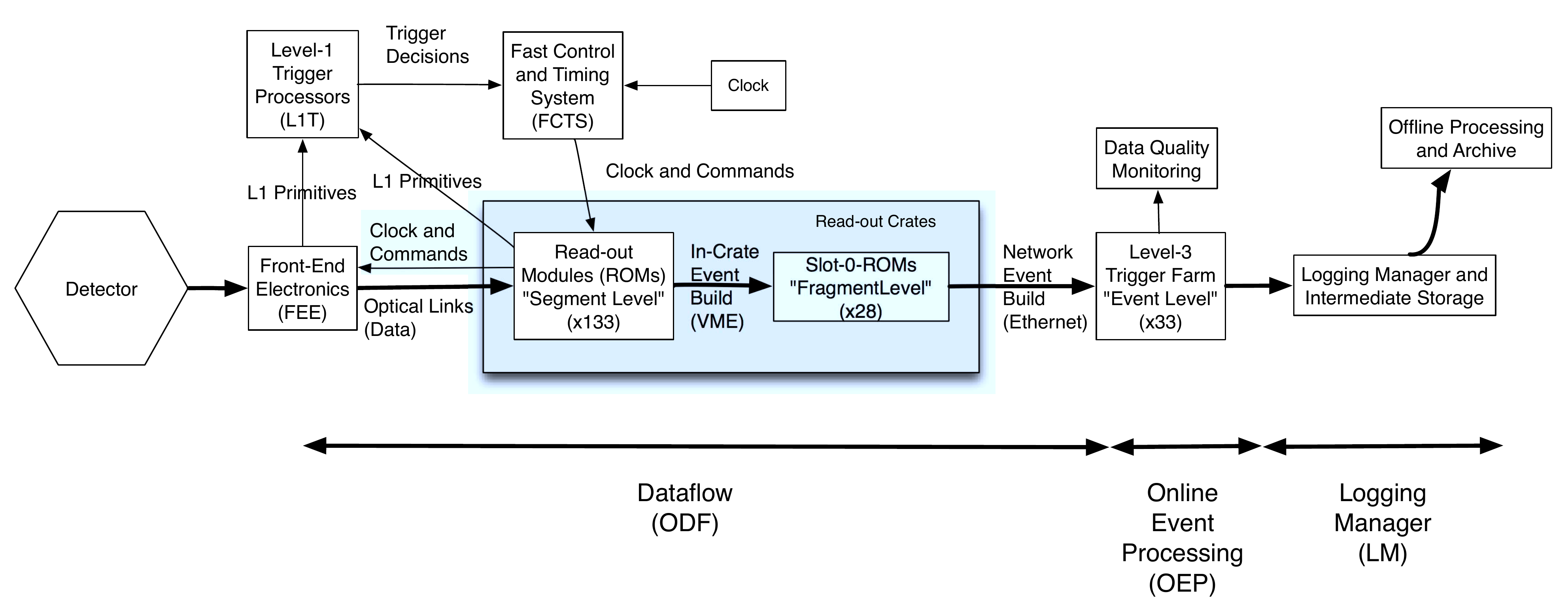}
\caption{Schematic diagram of the dataflow in the data acquisition chain.}
\label{fig:online_dataflow}
\end{center}
\end{figure*}

Throughout \babar\ data taking, the high-level design of the online
system was retained as described previously~\cite{Aubert:2001tu}. 
The data acquisition chain from the common front-end electronics
(FEE), through the embedded processors in the readout modules (ROMs),
the network event builder, the Level~3 trigger, and the event data
logging system is depicted in Figure~\ref{fig:online_dataflow}. Its
major components were:

\paragraph{Online Dataflow (ODF)} Responsible for communication with,
and control of, the detector system front-end electronics, and the
acquisition and building of event data from them; includes the Fast
Control and Timing System (FCTS), responsible for distributing
triggers and the detector timebase.

\paragraph{Online Event Processing (OEP)} Responsible for processing
of complete events, including Level~3 (software) triggering, data
quality monitoring, and the final stages of online calibrations.

\paragraph{Logging Manager (LM)} Responsible for receiving from OEP
events selected by the L3 trigger and for writing them to disk
for use as input to offline computing.

\vspace{1.5ex}   

The online system also included systems to control the data
acquisition and the detector, to monitor data quality and to
orchestrate online calibrations. The major components were:

\paragraph{Detector Control (ODC)} Responsible for the control and
monitoring of environmental conditions of the detector systems and the
handshake with the accelerator control system.

\paragraph{Run Control (ORC)} Responsible for tying together all the other 
components, and for sequencing their operation, interlocking them
as appropriate, and providing a \emph{graphical user interface} (GUI)
for operator control.

\paragraph{Online Databases (ODB)} Comprised the databases for the
storage of detector configurations, conditions, and calibration
constants.

\paragraph{Support Infrastructure} Comprised the servers, consoles, 
local networking, system administration tools, electronic logbook, etc.

\subsubsection{Upgrades and Improvements}

A comparison of parameters of the \babar\ online system in
2001~\cite{Aubert:2001tu} and at the end of data taking in 2008 is
shown in Table~\ref{tab:online-upgrades}.

\begin{table*}
\caption{\label{tab:online-upgrades} Comparison of the online
  system in 2001 and at the end of data taking in 2008}
\centering
\begin{tabular}{lll}
\hline
System Parameter & 2001 & 2008 \cr
\hline
Number of readout crates & 23 ($+$ 1 master crate) & 27 ($+$ 1 master crate) \cr
Number of ROMs & 158 & 158 \cr
Network & 100~MBit/s Ethernet & 1GBit/s Ethernet \cr
Farm Nodes running Level~3 & 32 300~MHz Sun UltraSparc II & 33 Dual-CPU/Dual-Core 2.2~GHz AMD Opteron\cr
L1-Accept rate capability & 2\khz & 5.5\khz \cr
Logging capability & 300~Hz & 5.5\khz (for special tests) \cr
\hline

\end{tabular}
\end{table*}

\paragraph{Online Dataflow}

The primary responsibility of the dataflow system was the acquisition
and delivery of assembled event data from the detector front-end
electronics (FEE) to the L3 processing farm.  The critical measure of
performance of the dataflow system was the accumulated dead-time.  As
expected luminosity upgrades projected higher trigger rates and larger
event sizes, the performance of the dataflow system had to be
extrapolated and improvements were developed to maintain the \babar\
goal of near dead-time-free operation.  Upgrades were first directed
at maximizing the performance of the existing hardware.  Hardware
upgrades were undertaken only after these efforts had been exhausted.

A new component of the dataflow software was implemented to collect
monitoring and performance data throughout the data
path~\cite{Perazzo:2002aj}.  A performance model of the data transport
was created, and information was collected at each stage of transfer
within the VME read-out modules (ROMs) and crates.  Combined with
projected trigger rates and event data sizes, the model was used to
formulate an upgrade plan for the dataflow system.  The model
accounted for transfer times on these major links in the data path:

\begin{itemize}
\item the serial link from the detector front-end electronics to the
  ROM personality card buffers;
\item the i960 bus DMA from the personality card processors to the ROM
  host processor memory;
\item the VME bus DMA from the slave ROMs to the VME master
  ROM in each crate (slot-0); and
\item the network transfer from the VME master ROM to the
  OEP farm network switch.
\end{itemize}

The model estimated the computation time in each host processor as
well as memory bus and peripheral bus utilization. The result was a
tabulation of maximum transfer rates for each stage,
allowing the identification of bottlenecks. 

The FEEs for each of the detector systems had been implemented with
four event buffers to hold event data prior to transfer via the serial
link to the ROM personality card buffers. Fluctuations in the
instantaneous L1 trigger rate could fill all four front-end buffers
before the first was emptied via the serial link.  The dead-time from
bottlenecks in the FEE serial link transfer was estimated with a model
that included these fluctuations. In all subsequent readout stages
enough buffering was available so that dead-time was only created when
the {\it average} trigger rate exceeded the capacity of a link or
processing stage.

The first upgrade effort was directed at the ROM VME master modules
(see Figure~\ref{fig:online_dataflow}) which collected data from the
slave ({\it segment level}) ROMs and transmitted the packaged events
to the event processing farm ({\it event level}).  The projected
performance of the VME master ROMs was impacted by three factors.
First, the host processors spent a large fraction of time in the
VxWorks operating system network stack.  Second, the distribution of
detector system input data was unbalanced among the VME crates,
requiring some of the master ROMs to handle a much larger share of
data than others.  Third, the 100\mbitps\ network interfaces of the
VME master ROMs to the event building switch were projected to become
overloaded at higher future data rates.

The upgrade plan first addressed the network stack performance.  A
significant fraction of the processing time was devoted to overhead in
the transmission of datagrams, independent of the event size.  Thus,
the processing time could be reduced from 168\mus\ to 83\mus\ per
accepted event by packing two events into a single datagram prior to
transmission.  This {\it batching} of events into datagrams required
negligible additional processing in the OEP event level hosts to
unpack the individual events.

The unequal distribution of the data load among VME crates was reduced
by adding more crates and simplified VME master ROMs (with slightly
faster PowerPC processors) designed exclusively for the fragment-level
event building task. The number of readout crates was increased from
24 to 28, providing a greater number of fragment level event builders
for the DCH, DIRC, and SVT readout.

Commercial Gigabit Ethernet interface cards in a PMC form factor were
purchased to replace the on-board 100\mbitps\ interfaces as network
outlets in the VME master ROMs, eliminating the network link
throughput limitations. A custom zero-copy UDP network stack was
written specifically for use with the new interfaces, reducing the VME
master ROM CPU use to 73\mus\ per event.

The upgrade simulation model also projected a limit on the data
transmission within the segment level readout of several detector
systems.  The DCH readout processing, which extracted track hit
information from the raw data, and the oversubscribed serial link from
the DCH FEE, were the most significant bottlenecks.  A major hardware
upgrade was performed to address both issues (see
Section~\ref{sec:electronics:dchupgrades})

The readout software for several other detector systems was modified
or rewritten to enhance the processing speed. In the most significant
software upgrade, the time to perform the EMC digital filter
processing was reduced from 250\mus\ to 175\mus\ per event.
Implementation of the hardware sparsification scheme, anticipated in
the original design, further reduced the average processing time to
less than 140\mus.  Nevertheless, digital filtering remained the key
bottleneck, setting the ultimate trigger rate limit of 7\khz.
Upgrades to the SVT, DIRC, and trigger readout software are described
below, along with the hardware upgrades.

\paragraph{Trickle Injection Support}
One of the most significant upgrades to PEP-II was the trickle
injection described in Section~\ref{sec:mdi:trickle}. This required
modifications to the online system in two areas: changes to process
control and operational procedures to enable near-continuous
data-taking, and the development of special background filtering and
data quality analysis tools.

Prior to the implementation of trickle injection, there was a regular
cycle of a few minutes of beam injection, with detector voltages at
safe levels and data acquisition stopped, followed by 40-60 minutes of
data-taking. Operational procedures were focused on maintaining safe
interlocks against raising voltages or taking data during injection,
and on efficient transitions between the two states. To allow trickle
injection, some of the interlocks (such as the hardware to prevent
injection while the detector high voltages were ramped up) were
modified and additional monitoring and protection software was added
to ensure safe detector operation during injection.

With trickle injection, data taking periods typically continued for
periods ranging from a few hours up to one day, motivating many
changes in operational strategy and requiring improvements to run and
detector control to minimize time spent in error recovery,
calibration, and other non-data-taking states.

A major new requirement was the handling of the backgrounds associated
with trickle injection.  It was observed that the injection of current
into a single circulating bunch produced elevated backgrounds
associated with that bunch, lasting for 5--10\ms.  Because of the
finite time resolution of the L1 trigger, accepted events generated by
these backgrounds extended for 10--15\%\ of a full revolution.  This
effect is illustrated in Figure~\ref{fig:trick-inj-2d}.

The transit of perturbed bunches through the detector was found to
produce events where large fractions of the detector channels had
signals above thresholds. The readout of these very large events
produced unacceptable data acquisition dead-time. It was therefore
necessary to inhibit the L1-Accept signals during the passage of the
perturbed bunches. Programmable hardware was added to the FCTS to
enable this inhibit for a specified number of revolutions following an
injected pulse, and only for the affected fraction of a revolution
with respect to the phase of the pulse.

In addition, events in a larger time window around the passage of the
perturbed bunches were rejected in software as part of the L3
trigger, because they were found to contain background levels that
compromised their usability for physics.

The hardware and software inhibits were facilitated by a design
decision in the original FCTS to synchronize the timebase for the
event time-stamps to the PEP-II timing fiducial.

The combination of hardware and software event rejection produced an
effective dead-time of 1--2\% under the typical conditions of trickle
injection, operating at about 5\hz per beam.

\paragraph{Raw Data Format and Schema Evolution}

All \babar\ raw data were recorded in a format called XTC, for {\it
  eXtended Tagged Container}.  An XTC file is a sequence of partially
self-describing 32-bit-oriented binary data records. Each record
corresponds to a state transition in the data acquisition system, an
acquired event resulting from a L1-accept in the FCTS,
or a simple time-stamped message sent through the data acquisition
system (a marker used to record trickle injection pulses). Each record
consists of a hierarchical collection of data from the ROMs, strongly
typed to facilitate interpretation in a C++ memory model.

The ODF system permitted the transmission of either the raw front-end
data or the feature-extracted data; in normal operation only the
latter was used.

Over the lifetime of the experiment, the feature extraction output
from several detector systems was modified, sometimes significantly,
initially to correct problems, and subsequently to upgrade or improve
the feature extraction algorithms (see
Section~\ref{sec:electronics:electronics} for details).

The XTC data access APIs were designed to permit a simple schema
evolution. Detector systems were able to provide simple plug-in code
snippets that converted older feature extracted data formats to newer
ones, allowing the most recent code releases to continue to read all
of the raw data recorded during the entire operational history of the
experiment.  This proved essential in permitting uniform reprocessing
of all acquired data. 

\paragraph{Online Farm Upgrades}

The computer farm used for event building and L3 triggering at the
start of \babar\ operation was a cluster of 32 Sun Ultra 5
workstations, with single 333\mhz\ UltraSPARC-IIi CPUs. While the
capacity of this cluster proved sufficient for early operation, as
PEP-II began to exceed the design luminosity, it became clear that the
L3 processes would soon require additional CPU power and memory.

As an initial step, the original software limitation of the event
builder to 32 target nodes was removed and the cluster was expanded to
60 nodes by making use of machines originally intended for data
quality monitoring.

By 2001, a further upgrade became necessary. The evolution of trends
in commodity hardware and software now strongly suggested the use of
Intel x86 processors and a Linux operating system.

At the start of \babar\ operation, the offline software was in the
process of being ported to Linux (early software development had taken
place on a variety of commercial Unix variants), while the central
offline and online production systems at SLAC were all based on Sun
Microsystems SPARC processors and the Solaris operating system.

Upgrading the online farm to Intel x86 processors and Linux required the
extension of this porting activity to the online software. This
introduced an additional set of C++ and system library issues (\eg\
relating to the then-limited support of POSIX standards in Linux,
especially for inter-process communication) as the online software
made use of numerous system interfaces and language features not
otherwise used in \babar.

The original event builder, as well as the software framework for the
L3 trigger, relied on the compatibility of the byte ordering within
machine words of the PowerPC processors used in the ROMs and the
UltraSPARC processors used in the online farm. With the introduction
of Intel x86 processors the byte ordering in the ROMs and the online
farm was no longer the same. Since the raw detector data were written
in a mix of native data types of varying size, a content-dependent
byte reordering became necessary to make the data usable in the x86
CPUs.

From the beginning, the data analysis environment provided by the
online system for access to XTC data had a layer of abstraction in the
data interface, originally motivated by the expected need to support
schema evolution in the raw data format.

This interface readily provided a place to interpose a
byte-rearrangement layer in the software, leaving it to the detector
systems to provide short code snippets that implemented the specific
type-dependent transformations required by their data formats.  As a
result, no Level~3 trigger or offline reconstruction code needed to be
changed to accommodate the new processors.  The byte ordering
transformations were performed on an in-memory copy of the raw data.

The format of the XTC files was kept in PowerPC byte ordering
during the whole lifetime of the experiment.  This avoided having to
introduce byte reordering support at the ROM level.

The event builder and farm upgrade were designed and developed in
2001; initial tests were conducted in the spring of 2001. The new farm
was installed during a four-month maintenance period in the
summer of 2002, and regular data taking with the new system began in
the fall.  The new farm consisted of 50 machines with 1.4\ghz\ dual
Intel Pentium III CPUs and optical (Intel e1000-based) Gigabit network
cards for the event builder~\cite{Jacobsen:2003kw}.

To support Gigabit Ethernet in the event builder, the central event
building switch had to be replaced. The original device, a Cisco
Catalyst 5500 with three 1\gbitps\ backplane buses, would not have been
able to support the faster event building network. In fact, it had
already caused packet loss in the 100\mbitps\ event building network
due to its limited input and output buffers and the very high
instantaneous rates in the event builder. The new event building
switch was a Cisco Catalyst 6500-series device with a 16\gbitps\
backplane bus and sufficient per-port buffers to handle the
ever-increasing event building dataflow.

In 2006, the L3 trigger farm was replaced again. The timing of the
upgrade was driven by a typical 4-year equipment life cycle.  The
number and configuration of the new machines (45 servers, each
equipped with two 2.2\ghz\ dual-core AMD Opteron 275 CPUs), was based
on the anticipation of increased luminosity and increased backgrounds,
and therefore a significant increase in event size. At the same time,
the event building switch was replaced with a Cisco Catalyst
6500-720-series device, featuring multiple 20\gbitps\ full-duplex
links from the line cards.
The network
switch upgrade was performed partly in anticipation of the
significantly larger event sizes, and partly because of concerns about
long-term vendor support.

\paragraph{Node Group Management}
The original online system design recognized the need for coordinated
operation of groups of processes distributed over several nodes,
carrying out a common task. The two original distributed applications
envisioned for this were the L3 trigger and the online data
quality monitoring. Originally, it was planned to create a distributed
process manager to supervise such distributed tasks in a
general way.  This system was not implemented in time for the start of
operation, and the management of 30-60 L3
processes was carried out with rudimentary tools based on Unix shell
scripts invoking remote shells.  Detailed data quality monitoring was
based on a single process, to avoid replicating this awkward
system. 

A distributed process manager was finally designed and implemented in
2002. For the worker nodes a process-management daemon was developed,
intended to be started at system boot time.  Whenever an instance of a
specific worker process (\eg\ a L3 trigger process) was required, the
daemon could be commanded to start it, using a predefined command
line, and to monitor the lifetime and termination status (normal,
non-zero return code, or signal) of the spawned process. A central
manager process for each type of worker coordinated the individual
daemons, maintained a summary status for the ensemble of worker
processes, and provided basic error detection and recovery mechanisms.
In particular, the system permitted a configurable number of workers
to terminate unexpectedly without raising an overall error, providing
support for simple fault-tolerance.

The system also provided tools for organizing the log output (merged
standard output and standard error I/O) resulting from the operation
of these distributed worker groups.

\paragraph{Data Quality Monitoring Upgrades}
The L3 processes carried out a small number of data quality
monitoring tasks on each event, but this was limited to analyses that could
be performed within the tightly controlled L3 event processing time budget.

During the first two years of operation~\cite{Aubert:2001tu},
additional time-consuming data quality monitoring tasks were performed
by a single process connected via TCP/IP to one of the L3
processes. In order to limit the impact on the L3 latency, only a very
limited fraction of the recorded events were sampled.

The introduction of the node group management tools allowed the
data quality monitoring to be expanded to a pool of processes paired
with L3 processes.  Together with the use of the much faster x86
processors in the upgraded online farm, the rate of events subjected
to the detailed data quality analysis was able to be increased by a
factor of $O(50)$ compared to the single process described above.
Histogram collection from the monitoring processes was
performed by a distributed histogramming system~\cite{Metzler:1998qq}.

The ease of use of the node group management tool subsequently enabled
the introduction of a second set of specialized monitoring processes
which performed full track reconstruction on di-lepton events,
permitting the reconstruction of the luminous region with high
precision, as described in Section~\ref{sec:mdi:pep2-bunches}.

\paragraph{Logging Manager Upgrade}

Events selected by the L3 trigger were logged to sequential files in
the XTC data format.

In the original online system design~\cite{Aubert:2001tu}, data
logging was the responsibility of a single process, the Logging
Manager (LM), on a dedicated server, receiving events over TCP sockets
from each of the L3 nodes.  The LM recorded these events on disk in
the order it received them, interleaved across all the input nodes.

While this design proved capable of handling the average rates seen in
the early years of operation, it turned out to be an important
serialization point and an unanticipated source of dead-time on
occasion. The design of the LM used one thread per L3 node to receive
events, and a further, single thread to write them to disk, with an
in-memory queue for derandomization.

When the writing thread was momentarily unable to keep up with the
output rate, the queue could rapidly grow to fill its designated
memory on the logging server, at which point the process would no
longer read from its input sockets. This could fill the kernel memory
for the sockets, and eventually lead to backpressure to the
originating L3 processes.  Once the L3 processes were blocked,
backpressure was asserted through the event builder to the data
acquisition system, leading to true dead-time.

Scheduling and I/O priority issues in Solaris and in the Veritas file
systems used on the logging server were found to prevent the rapid
clearing of the backlog even after backpressure had been asserted,
leading to occasional minute-long interruptions of the data
acquisition.

When performance tuning of the logging application and the logging
server operating system and even major hardware upgrades of the server
and its RAID disk arrays failed to resolve these problems, we
concluded that the existence of this serialization point in the system
was itself the core of the problem, and undertook a redesign that
avoided this altogether~\cite{Hamilton:2004tz}.

In the new design, introduced in 2004, each L3 node was responsible
for logging its selected events locally. To facilitate this, the Linux
x86 farm nodes were equipped with additional local disks, sufficient
to store approximately a day's worth of data distributed across the
farm. A centrally-located merge controller selected the completed data
runs (corresponding to 40-60 minutes of data taking) that were
available to be merged and assigned the merge for each single run to
one of a small pool of merge servers. The merge server, in turn,
queried each of the farm nodes for the availability of the
corresponding per-node raw data files and then requested that the
files be transmitted.

The merge servers were able to produce a time-ordered combined file by
maintaining a list of the next event available from each node and
simply repeating the process of determining and copying the earliest
one from any node to its output.

Because of the large amount of local storage available, the existence
of multiple merge servers, and the network capacity, the system could
tolerate large fluctuations in latency. In practice, it was able to
merge the files of a data run within 10 minutes of its completion.

This design removed the possibility of backpressure from a single
point of failure and enabled extremely high peak logging rates for
periods of tens of minutes or more. The latter capability was used to
permit the logging of data at the full L1 trigger rate, from time to
time, for calibration and diagnostic purposes. The ability to log data
at $O(5\khz)$ was used on several occasions. At the narrow $\Upsilon$
resonances, data were logged continuously at rates of 900-1000~Hz.

\paragraph{Online Detector Control Upgrades}

The detector control system was implemented using
EPICS~\cite{Online:EPICS}, a toolkit for building control systems.
The core components of EPICS are a driver model for device
abstractions, a network communications protocol and a distributed
system for acquiring and processing sensor data and controlling
devices (the EPICS database). In the EPICS model, all device
interaction and data processing take place on I/O Controllers (IOCs).
Simple data processing can be performed in the database. More
complicated calculations can be implemented using an add-on C-like
state machine language. EPICS also provides tools for constructing
GUIs and for managing and processing thresholds and alarms.

Over the nine years of data taking, the EPICS toolkit evolved
significantly. In the earlier versions, the IOC code used the
proprietary vxWorks API for accessing system functions. IOCs could
therefore only run on vxWorks while Unix/Linux systems were limited to
running client applications. Moreover, the Unix-side APIs for
providing and accessing EPICS data were complicated and restrictive.
Later versions of EPICS allowed native IOCs to run on other operating
systems and also introduced the capability to run software IOCs on
Unix/Linux.

For ambient data archiving, the initial \babar\ design implemented a
custom system that stored the time-series data in an Objectivity
database ({\it ambient database}). This system turned out to have
severe limitations in the number of channels it could handle and in
the per-channel logging frequencies. In addition, there were severe
reliability problems caused both by the input side interface to EPICS
and by the performance and availability of the Objectivity database.
Database down time or lock problems would immediately cause loss of
ambient data. These issues were addressed by a complete redesign of
the archiving system.  The input side was changed to use a simpler and
more modern EPICS APIs and the process of acquiring data was decoupled
from the database by logging to an intermediate local disk buffer that
would be flushed to the database in regular intervals.

In the initial, somewhat naive detector control system network
configuration, all IOCs were directly connected to the main control
room network which also hosted operator consoles and development
machines and allowed unrestricted access from EPICS clients
anywhere on the SLAC network. This caused intermittent performance
problems on the IOCs whenever a large number of clients connected to the
system. The problems were aggravated by the introduction of
automated network security scans at SLAC which caused IOCs to randomly
disconnect or sometimes to stop working completely. These issues, and
general concerns about control system and network security, led us to
move the IOCs to a separate, isolated network, which only allowed
direct access for a small number of clients critical to the operation
(such as the operator consoles). In the new design, all other access
to the control system was mediated by EPICS application-level
gateways. These EPICS gateways provided performance isolation
through caching and protected the sensitive IOCs from direct
uncontrolled access~\cite{Kotturi:2003wg}.

The addition of more channels to monitor and the processing
requirements for system automation started to approach and sometimes
exceeded the capabilities of the original IOCs (Motorola
MVME177 single board computers with 66\mhz\ 68060 CPUs). In
anticipation of further growth, upgrades of the system to a faster IOC
hardware platform were explored. Since the licensing
terms and cost for newer versions of vxWorks had become much less
favorable, the use of RTEMS~\cite{Online:RTEMS}, a free and open
source real-time operating system, on the MVME5500 (PowerPC) based
single-board computer platform was explored.

The new CPUs solved the performance issues, however, RTEMS turned out
to be too immature, especially in the areas of networking and network
file system support, and it lacked memory protection, so that
application errors would often cause system crashes. Running Linux on
the MVME5500 boards seemed to be a more promising option. The {\it
  Embedded Linux}\cite{Online:EmbeddedLinux} distribution was finally
chosen, because a board support package for the MVME5500 was
available, and EPICS and the device drivers were ported to this
platform. Because none of the applications had hard real-time
requirements, true real-time support from the operating system was not
needed. This enabled us to implement IOC functionality simply by
running a software IOC process that could be restarted without the
need to reboot the operating system.  During the final three years of
\babar\ data taking, a number of critical IOCs were replaced with
these Embedded Linux/MVME5500 systems.

The combination of all measures described above significantly
increased the stability and reliability of the detector control
system.

In addition to controlling and monitoring the \babar\ detector and
environment, the detector control system also provided the
communication channel between \babar\ and the PEP-II accelerator
control system (which was based on EPICS). A primary bi-directional
connection using a dedicated network link between the IOC controlling
the PEP-II injection and the IOC controlling the overall detector
status was used to implement the handshake and interlock for
injection, and to exchange limited amounts of data between the two
control systems.  For example, beam spot information calculated by the
\babar\ online monitoring system was transmitted to the accelerator
control system.  A secondary read-only path (via EPICS gateways over
more general parts of the SLAC network) provided the \babar\ detector
control system with access to a broad variety of diagnostic
information (such as temperatures or vacuum pressures in PEP-II). Some
of this information was archived in the \babar\ ambient database in
order to be able to study correlations with ambient data generated by
\babar.

\paragraph{Online Run Control Upgrades}
\label{sec:onl-runcon}

The run control system coordinated all components of the online system
to accomplish the routine tasks of data taking, calibration and
monitoring. It provided the primary user interface for operators to
control these functions. The system relied on a model of the online
components as simple state machines. Components were represented by
{\it proxy} processes which performed the translation from the detailed
behavior to the state machine model. Implementation of the system
relied on the SMI~\cite{Franek:2006jw,Online:SMI} and
DIM~\cite{Gaspar1996257,Online:DIM} libraries, which were migrated
from Fortran to C++ implementations.

At the beginning of \babar\ operation, the use of the run control
system required a fairly detailed understanding of the architecture
and functioning of the online system and provided very limited
capabilities for error detection and recovery, other than a complete
restart. This necessitated extensive training of \babar\ operators as
well as frequent on-call access to experts.

Based both on feedback from operators during normal operation and
comments during operator training sessions, run control developers
improved the user interfaces and the internal functioning of the
system, including additional automation.  In particular, the startup
and shutdown of the online system as a whole were greatly simplified
and accelerated.  Automated responses to commonly encountered error
conditions were programmed into the state machine model, reducing the
need for human intervention.  By the third year of taking data, this
process had significantly reduced the need for specialized training,
and led to the reduction in the number of operators on shift from
three to two.

Most of the original user interface was generated automatically from
the state machine model and it frequently confused the operators by
exposing them to too much detail about the online system. As part of
the ongoing improvement process, a new user interface was developed
that contained the logical model for the user interaction and
communicated with the state machine manager through the appropriate
APIs.  This allowed the user interaction to be decoupled from the
internal workings of the run control system, while also implementing
automated error detection and recovery, and adding more user-friendly
interface components, such as GUIs to partition the system or to
execute automated system-wide calibrations.

The detector systems had calibration procedures that required the
collection of dedicated data, generally involving the injection of
test signals and requiring beam-off conditions.  Early in the life of
the experiment, these procedures were controlled by unique scripts
developed for each detector system and only loosely tied into the run
control system.  With the pressure to improve operational efficiency,
common features of all these calibrations were identified and were
brought into a more structured framework under the run control system,
improving the ability to detect and recover from error conditions.
Data acquisition was cleanly separated from the subsequent analysis of
the calibration events and determination of calibration constants,
allowing the analysis to overlap with other operations and further
improving efficiency.

\paragraph{Database Improvements}
\label{sec:onl-dbimprove}
The configuration database was used to maintain hierarchical
associations of system-specific configuration data, identified with a
single numeric {\it configuration key}.  This key was distributed to
all online components and  could be used to retrieve the required
configuration information.

At the start of \babar\ operation, the originally envisioned user
interface for creating and managing multiple configurations had not
been developed. The management tools that were provided exposed
implementation details of the database and were found to be very
difficult to use. Most configuration updates could not be performed by 
detector system experts, but required the intervention of the
database developers.

As part of the online system upgrades, the user interface was
redesigned and reimplemented, including both graphical and command
line tools.  The ability to script configuration updates was improved.
This had become mandatory to support rapid operational changes and
avoid down time.  Instead of exposing raw database objects, new {\it
  alias tree} tools allowed users to give symbolic names to
sub-configurations at every level of the hierarchy. This ability to
use logical names made it much easier for detector system experts to
manage their own configuration data. More detailed descriptions of the
configuration database and its improvements can be found in
~\cite{Bartoldus:2003xn,Salnikov:2006ck}.

When the Objectivity-based event database was replaced by a ROOT-based
object event store (discussed in Section~\ref{sec:offline:evolution}),
it was initially decided to retain the Objectivity implementations of
the non-event databases (configuration, ambient, and conditions).
Unfortunately, many of the same concerns that motivated the event
store migration also applied to these databases, \ie\ vendor platform
support, poor lock management, and ongoing maintenance cost.

In 2006, the collaboration therefore decided to phase out Objectivity
completely. Each of the non-event databases had an important role in
the online system, so this migration had a much more profound effect
on the online system than the earlier event store migration, and had to be
carefully prepared to avoid interference with normal operation.

The new ambient database implementation was based entirely on ROOT
files. It was introduced with minimal disruption because of the
separation of data collection and archiving that had already been
introduced, as described above. The new implementation offered much
better performance; it also required much less storage because of data
compression.

Both the configuration and conditions databases were migrated from
Objectivity to hybrid object-relational implementations. 
Because of complexity, size, and access requirements, different
implementations were chosen: For the configuration database,
serialized ROOT objects were directly stored as Binary
Large Objects (BLOB) in a MySQL database. For the significantly larger
and more complex conditions database, serialized ROOT objects were
stored in ROOT files and referenced by a relational database. To
provide simple large-scale read access to the conditions database, a
read-only ROOT-only implementation was developed.

\paragraph{User Interfaces and Electronic Logbook Improvements}

Over the lifetime of the experiment, major improvements and upgrades
of the user interfaces for the operators were implemented. The
original Sun Ultra-5 workstations with two or three CRT screens per
machine were replaced with personal computer systems running Linux,
each one driving two or three flat screens. XFCE~\cite{Online:XFCE}
was chosen as desktop environment and window manager because of its
speed and configurability.  It was modified to better support multiple
screens and functions to take screen shots of full screens or screen
areas and automatically insert them into the electronic logbook were
added.

The web-based electronic logbook played a central role in data taking
and detector operation. It was implemented in Perl/CGI and used an
Oracle database backend. From the beginning, the logbook interface
combined a free-form log entry tool with a front-end to the \babar\
bookkeeping database.  The logbook originally required operators to
manually enter information about data runs, such as the run type (\eg\
colliding beams, cosmic data taking, and diagnostic data runs) and
other per-run meta-data such as the reason why the data run ended and
detector and global data quality assessments. Over time meta-data
fields were added and entry of run information was automated where
possible, \eg\ if the system detected a beam loss, the end of run
reason would automatically be set to {\it beam loss}. The logbook also
provided separate sections for detailed information relating to each
detector system.  Most of the \babar\ online luminosity accounting was
recorded in the electronic logbook.  Delivered and recorded
luminosities were added to the run entries and shift meta-data, and
totals and efficiencies were automatically calculated.

\subsection{Trigger Upgrades}
\label{upgrades-TRG}

\subsubsection{Overview}

The two primary requirements for the design of the trigger system were: to maintain
a maximally high data-logging efficiency and to keep dead-time to a
minimum, even during challenging running conditions.
The trigger system was designed as a sequence of two separate stages.
The L1 trigger dealt with incoming detector signals,
recognizing physics, calibration and diagnostics signatures to produce
an event rate acceptable for event building. The subsequent
L3 software trigger reduced that rate to a level
manageable for offline storage and processing.

\pep2 bunch crossings occurred with a spacing of 4.2\,\ns\ --
essentially continuous with respect to the 59.5\,\mhz clock of the
\babar\ electronics.
Thus, the L1 trigger had to process data continuously while
generating \LOneAccept\ signals within a latency of approximately
11\,\mus.
The L1 trigger processors produced clocked outputs to the fast control
system at 30\,\mhz, the time granularity of resultant \LOneAccept\ signals.
The arrival of an \LOneAccept\ from the data acquisition system caused
a portion of the L1 latency buffer for each detector system to be read out, ranging in depth from 500\,\ns for the SVT to 4-16\,\mus for the EMC.
Absolute timing information for the event, \ie\ the association with a
particular beam crossing, was first determined by the L3 trigger
and later refined offline.
The event time from the L3 reconstruction was derived from DCH track segment timing to better than 2\,\ns.
The L1 trigger system was configured to produce an output rate of
approximately 2\,\khz at design luminosity.

The L1 trigger output was based on data from the DCH, EMC, and IFR
triggers, referred to as the DCT, EMT and IFT, respectively.
Each of these three trigger subsystems generated trigger
\emph{primitives}, \eg\ charged particle tracks in the DCH and energy
deposits in the EMC, which contained information about the location
and energy/momentum of particles detected in an event.
The primitives were sent to the global L1 trigger (GLT), which aligned
them in time, counted and matched them, and formed up to 24
specific trigger lines from their combinations, which were then passed on to
the FCTS.
The FCTS masked or pre-scaled the triggers as required, and if an
active trigger remained, an \LOneAccept was issued to initiate the
global event readout and subsequent event build on the L3
processing farm.
L3 algorithms analyzed and selected events of interest,
to be stored for offline processing. All accepted events were
flagged for physics analysis, calibration and diagnostic purposes.
More details of the system design are given in~\cite{Aubert:2001tu}.

Aside from the regular maintenance efforts, various smaller upgrades
were implemented in the L1 systems over the years of running.
The following section describes some of the more significant hardware
upgrades that were implemented.
These upgrades were motivated by the ever-increasing luminosity and
the need for the L1 rate to remain within acceptable limits.
Beam-generated backgrounds dominated the L1 output rate, and they
increased with beam currents. 
In addition, some headroom was needed in the system to allow for
fluctuations in running conditions.
Even under poor beam conditions, dead-time of the trigger system
remained at or below 1\%.

\begin{figure*}[htb!]
\begin{center}
\includegraphics[width=14.0cm]{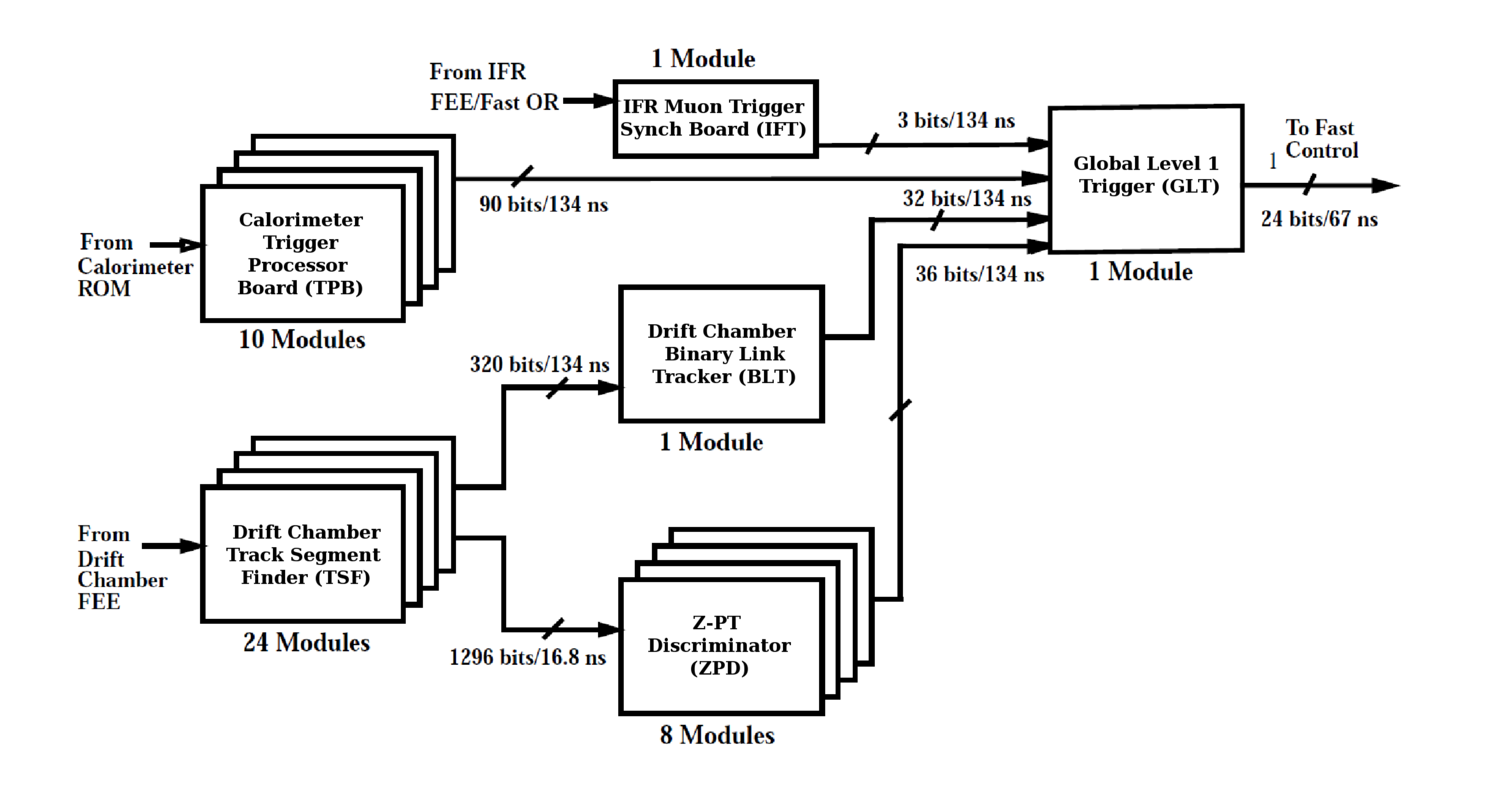}
\caption{Trigger: simplified L1 schematic at the end of the data taking. Indicated on the figure are the transmission rates
between components in terms of total signal bits. }
\label{fig:trg-schem}
\end{center}
\end{figure*}

The L1 trigger schematic at the end of \babar\ running is shown in Figure~\ref{fig:trg-schem}.

No major upgrades to the L3 software were required.
Bug-fixes, adjustments and a series of incremental improvements were
made over the years, but due to the comprehensive and extensible
design of the L3 system, aside from a single upgrade of the
online farm hardware, no significant changes were necessary.

\subsubsection{Drift Chamber Trigger Upgrade}

\begin{figure}[htb!]
\begin{center}
\includegraphics[width=0.5\textwidth]{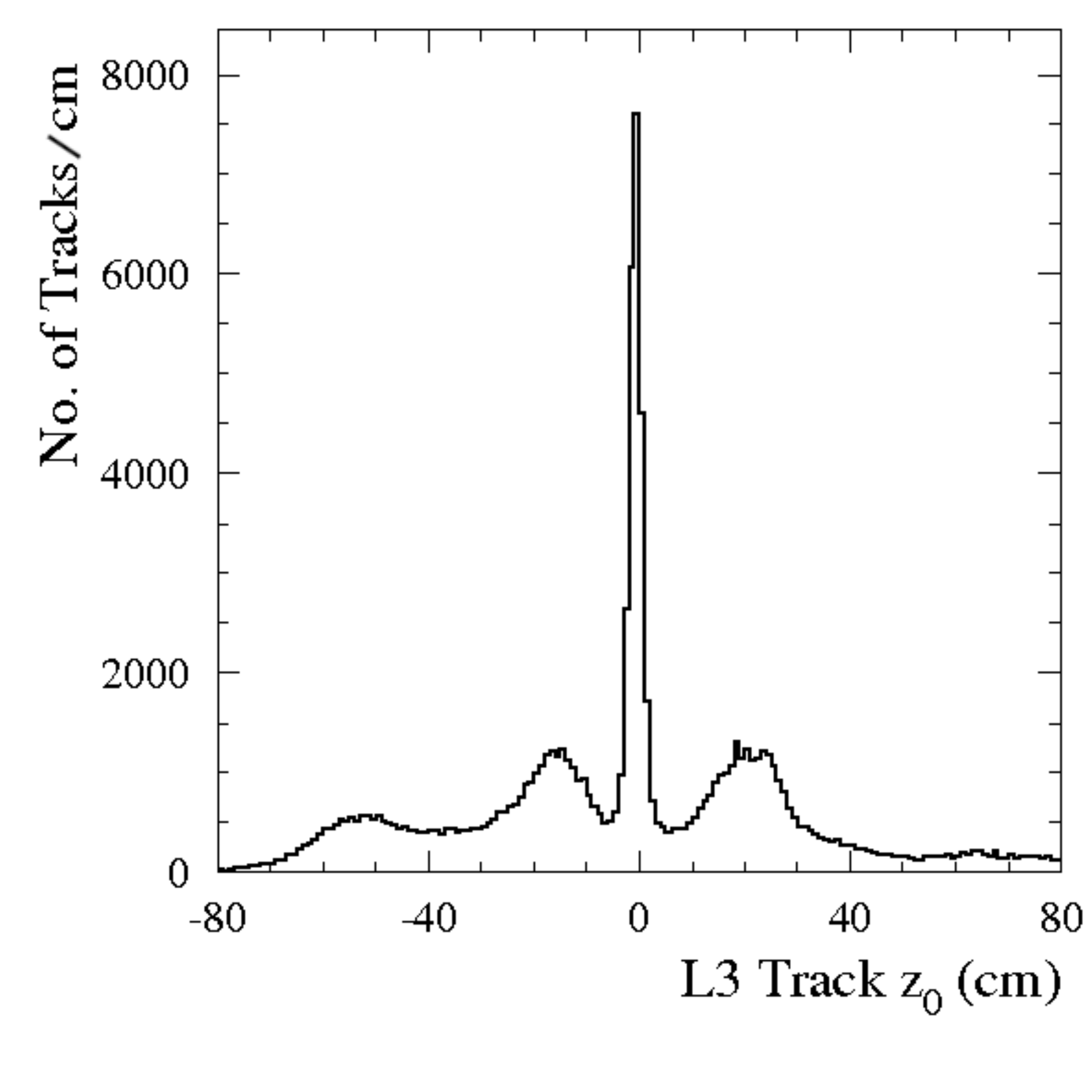}
\caption{Trigger: Longitudinal track impact parameter $z_0$ of L3
  reconstructed tracks passing the L1 trigger.  For each track, $z_0$ is defined as the distance, along the $z$-axis, between the IP and the point of
closest approach to that axis.
}
\label{fig:trg-z0}
\end{center}
\end{figure}

PEP-II exceeded its design luminosity quite early on,
and by Run~5 (2005--2006) was regularly delivering data at three
times the design value.
The L1 trigger system was initially expected to operate at an output rate of
about 1\,\khz, but was soon operating at 3\,\khz\ and above.
If the luminosity had continued to rise as projected, the trigger
dead-time would have increased significantly or the selection criteria 
would have had to be tightened, 
even for modest beam backgrounds.
To benefit from the increase in the rate of physics events, and to
avoid losing data due to dead-time or lower efficiency, the L1 system was upgraded to cope with luminosities of up to ten times the design.
A large fraction of the background was generated by single beam particles
interacting with beam-line elements located approximately 
20\,cm from the interaction point, as illustrated in
Figure~\ref{fig:trg-z0}. 
The longitudinal track impact parameter $z_0$ for tracks
originating from an
\epem\ collisions peaks at zero, and the enhancements 
  near $\pm$20\,\cm are due to gas-scattered beam particles that hit the
  faces of the B1 magnets on both sides of the IP. Beam particles 
  scattering off 
  masks and other beam-line components contribute to higher $z_0$ values.  
A trigger with track $z$-information could significantly
suppress the background from scattered single-beam particles.
Such a trigger would require three-dimensional track finding.  

The original track segment finders (TSFs) processed data from both
axial and stereo layers, but they passed only the axial
segments, \ie the transverse ($r$ and $\phi$) hit information, to the
\pt\ discriminator (PTD).
To enable 3D tracking, all TSF boards had to be redesigned to provide
both axial and stereo information.
This resulted in an increase of the TSF board output rate by about a factor of
four.
To stay within the latency budget, the internal clock speed was raised
from 30\,\mhz to 60\,\mhz.

The original TSFs passed information to both the binary link tracker (BLT)
and the PTD.
The original BLT board could be retained, but the
original PTD modules, which extracted only the transverse momentum \pt,
were replaced by $z$-\pt\ discriminator (ZPD) boards.
The ZPDs received information from the TSF boards for the input
segments of all DCH superlayers, with better precision in azimuth
($\phi$) and more segments per supercell than the original PTDs.
In this respect, the ZPD was a 3D-enhanced version of the PTD, with
the ability to select tracks with a seed track in superlayer seven or ten, and 
their origin along the beam line.
Each ZPD processed TSF data for a $45^{\circ}$ azimuthal wedge of the DCH, taking into account 
hits in adjacent wedges.

The ZPD algorithm was composed of two sections:
a seed track finder and a fitter.
The finder reconstructed tracks using a Hough transformation
\cite{Hough:1959}.
The fitter received up to ten track segments associated with a track
candidate, and fit them to a helix in the $r - \phi$ projection to improve the
measurement of $1/\pt$.
The five track parameters~\cite{Aubert:2001tu} were determined from a linear fit in the $r - z$ projection,
using the difference in $\phi$ between the track and stereo segments.
The trigger primitives were constructed using different sets
of selection criteria on the track parameters; most important was a restriction on
$z_0$ to remove the relevant backgrounds.
The total ZPD latency was about 2\,\mus,  well within the trigger
time constraints.

\begin{figure}[htb!]
\begin{center}
\includegraphics[width=8.0cm]{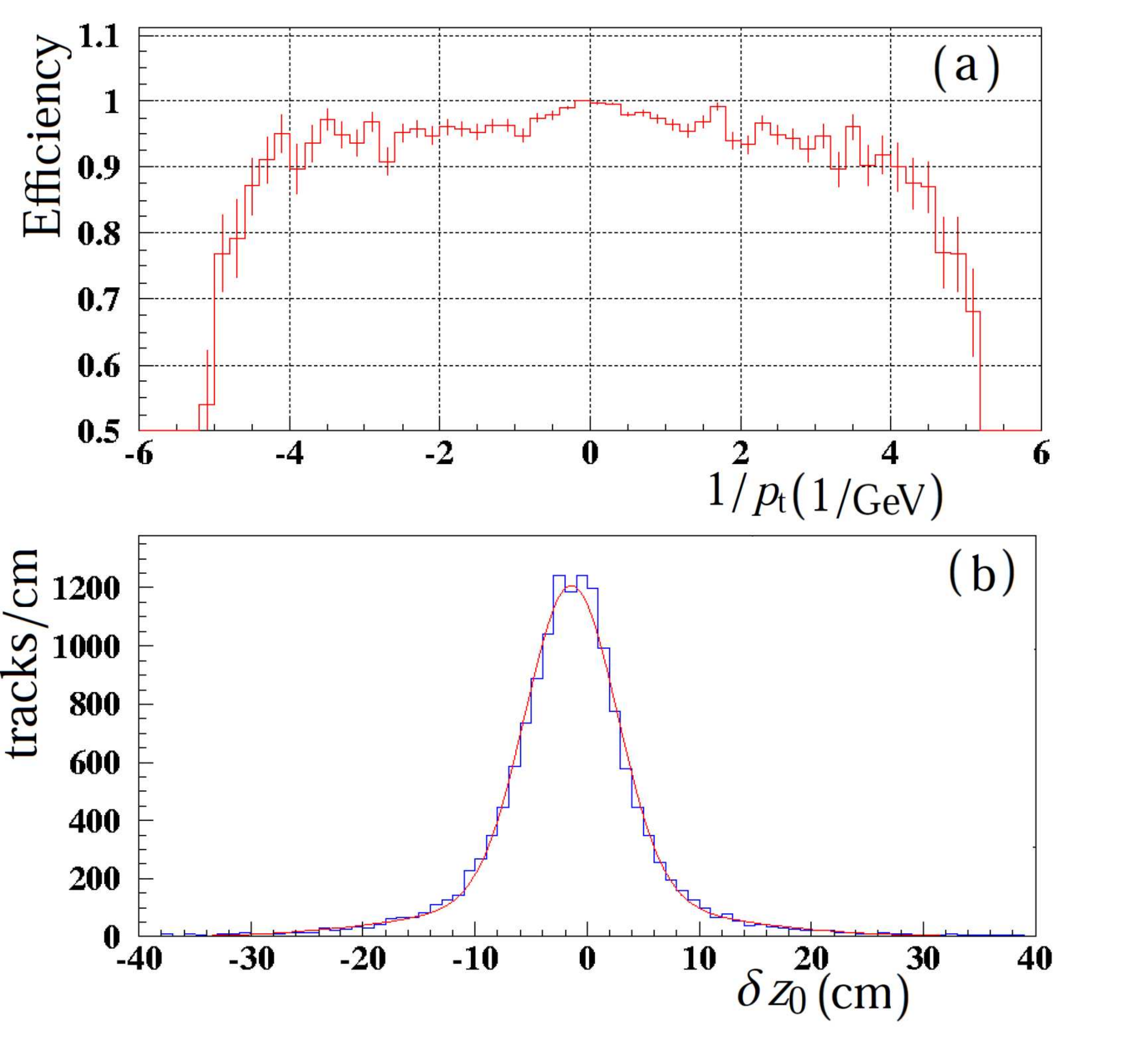}
\caption{Trigger: Performance of the ZPD track trigger: (a) the ZPD
  tracking efficiency as the fraction of tracks reconstructed offline,
  that are matched to tracks reconstructed by the ZPD; (b) the ZPD
  $z_0$ resolution defined as the difference between longitudinal 
  track impact parameter obtained from the ZPD and
  offline track reconstruction. }
\label{fig:trg-zpd}
\end{center}
\end{figure}

The new TSF and ZPD boards were installed in the summer of 2004, near
the end of Run 4.
Initially, the new system operated parasitically, with the use of a
fiber splitter that allowed the two systems to run in parallel.
It was adopted at the beginning of Run~5 in April 2005.
The ZPD tracking efficiency exceeded 97\%\ for
isolated tracks, and 92\%\ for tracks in hadronic events
down to $\pt = 250\mevc$, as shown in Figure~\ref{fig:trg-zpd}.
The $z_0$ resolution was measured to be $4.3\,\cm$, consistent with expectations from MC simulation.

This ZPD trigger, providing $z$ information for individual charged tracks, 
was the first 3D-trigger system of its kind, implemented at a collider.
It had not been anticipated in the original trigger design, but
the flexibility of the system allowed for a very successful adaptation.
The conservative implementation with loose
selection criteria on $z_0$ resulted in a modest
decrease in trigger rate and excellent single track efficiency.
The DCZ system was exploited to its full capacity during Run~7
when the c.m. energy was changed several times, leading to very different data
taking conditions. 
Operation at the \ThreeS\ resonance resulted in very high trigger rates.
To maintain low dead-time, and to take advantage of the new physics
processes at this resonance, significant changes were made to the
trigger configuration (see Section~\ref{trg-run7}) which benefited
greatly from these upgrades to the DCT system.

In summary, despite the luminosity exceeding the design early in the experiment lifetime and larger than expected backgrounds, the flexibility and space capacity in the trigger system
meant that in most cases, only minor upgrades were necessary over
the course of the experiment.
This flexibility also came into play with the DCT upgrade, allowing
the implementation of the hardware and software changes to be
relatively smooth.
An important aspect to all upgrades to the trigger system was the
inclusion of the operating parameters into the MC simulated data.
The trigger conditions could be simulated to provide a
luminosity-weighted sampling of real running conditions.

\subsection{Electronics Upgrades}
\label{sec:electronics:electronics}

\subsubsection{DCH Front-End Electronics}
\label{sec:electronics:dchupgrades}

In addition to the increased currents in the drift chamber
(Section~\ref{sec:DCHprotection}), the increase in the L1
trigger rate that accompanied the higher luminosity and backgrounds 
had to be addressed.
When the L1 trigger rate exceeded 2~kHz, the time required to transfer \dch\
data from the FEE on the endplate to the ROM created dead-time 
that reached several percent.
 
To maintain an open trigger, it was necessary to reduce the volume of data
transferred to the ROM per trigger. 
The modifications to the electronics and firmware were implemented in two stages, 
one that could be completed prior to the start of Run 5,
and the second that was installed mid-way through Run 5.

In the first stage, every second FADC sample was removed from the data stream
before it was transferred to the ROM. The TDC information, which was interspersed
in the waveform, was kept unchanged. 
The missing FADC samples were estimated by interpolation and reinserted in the ROM
prior to the feature extraction, \ie, the signal time and charge, from the 
waveform. 
Extensive tests  verified that this algorithm did not noticeably 
affect the track efficiency or \dedx\ resolution. Unfortunately, during
the implementation of these changes, 
a minor error was introduced into the feature extraction assembly code that corrupted 
one-half of the waveforms that contained two TDC hits, 4.2\% of the 
waveforms in total, which resulted in a  2\% reduction of tracking
efficiency.  Some 30~\invfb (5.6\% of the full dataset) were recorded
before this error was identified and corrected.

In the second stage, the feature extraction was implemented on 
Xilinx Spartan 3 FPGAs on each of the new 48 readout interface boards 
on the \dch\ endplate. This reduced the amount of data transferred per 
waveform from 32 bytes per waveform to approximately 6.5 bytes. 
The new firmware could be uploaded to the FPGA without access
to the electronics. 
As protection against single-event upsets (see Section~\ref{sec:DCHradiation}),
a fall-back image of the firmware was stored in EPROM and 
uploaded at the end of each run, or in the case of errors.

These upgrades to the \dch\ electronics enabled the system to process L1
trigger rates  of up to 5~kHz without dead-time. 
Thanks to this change, \babar\ was able to activate a single-photon
trigger which was needed for dark matter searches at the 
\TwoS\ and \ThreeS\ resonances (see Section~\ref{trg-run7} for details).

\subsubsection{DIRC Front-End Electronics}
\label{sec:electronics:drc_fee}

Up to the \pep2\ design luminosity of $3 \times 10^{33} \cms$,
the dead-time induced by the DIRC data acquisition was well below 1\%.
However, the DIRC time-to-digital converter integrated circuit (IC)
TDC1~\cite{Bailly:1999xe} showed sensitivity to high background rates.
Its dead-time  rose steeply when the average rate per PMT channel exceeded 
200\khz, indicating that its capacity was insufficient
for efficient operation at luminosities above $10^{34} \cms$.

To cope with the increase of the instantaneous luminosity and mitigate the 
effects of higher beam-generated background, the shielding of the SOB
(see Section~\ref{sec:mdi:beam_shielding}) was improved and a new
version of the TDC IC was designed and installed in the fall of 2002.

TDC2~\cite{Adam:2004fq} was fabricated in CMOS 0.6\mum\ triple-metal technology 
by AMS~\cite{AMS}, and met the following specifications:
reduce the dead-time to extend the rate limit per PMT up to 2\mhz;
accept L1 trigger rates up to 10\khz, rather than 2\khz; accept 
successive L1-Accept decisions that are separated
by at least 2.2\mus; and flag overflow of the TDC pipeline 
buffer in real time.

TDC2 had the same footprint and contact layout as TDC1
for easy replacement on the DIRC front-end board (DFB).
Like TDC1, TDC2 was a 16-channel custom integrated circuit
with 0.52\ns\ binning, input buffering, and selective readout of the data
in time for the L1 trigger decision.
It simultaneously handled input and output data.
As in TDC1, the precision time measurement was achieved using 
voltage-controlled delay lines.  Also, the IC was self-calibrating.
Successive signals in a given channel that were more than 33.6\ns\ apart
 were separately digitized.
The selective readout was simplified and more robust.
Though the trigger latency and acceptance windows were programmable,
they were kept fixed, at 11.5\mus\ and 600\ns\ respectively,

\subsubsection{Dead-Time Reduction}

A significant fraction of the DAQ dead-time originated from the feature extraction (FEX), the first stage of data processing in the ROMs, 
with the leading contribution from the SVT. Therefore, in preparation for 
Run 7, the SVT readout time was reduced based on extrapolations of data taking conditions to luminosities of $2 \times 10^{34} \cms$.

Originally set at 2\mus, the SVT readout window had already been shortened to 1\mus before Run~7. Nominally, it was made up of 15~clock ticks, each 66\ns\ apart, with the beam crossing time coinciding with the fifth clock tick, allowing an adequate margin to account for time jitter. In 2007, this window was  reduced further by eliminating the first and the last three clock ticks.  Simulation verified that this change would eliminate accidental hits, without any measurable impact on the SVT hit efficiency or resolution. This prediction was confirmed by a tests with the modified readout configuration prior to adopting the shorter time interval as the default.

\begin{figure}[htb!]
\begin{center}
\includegraphics[width=0.49\textwidth]{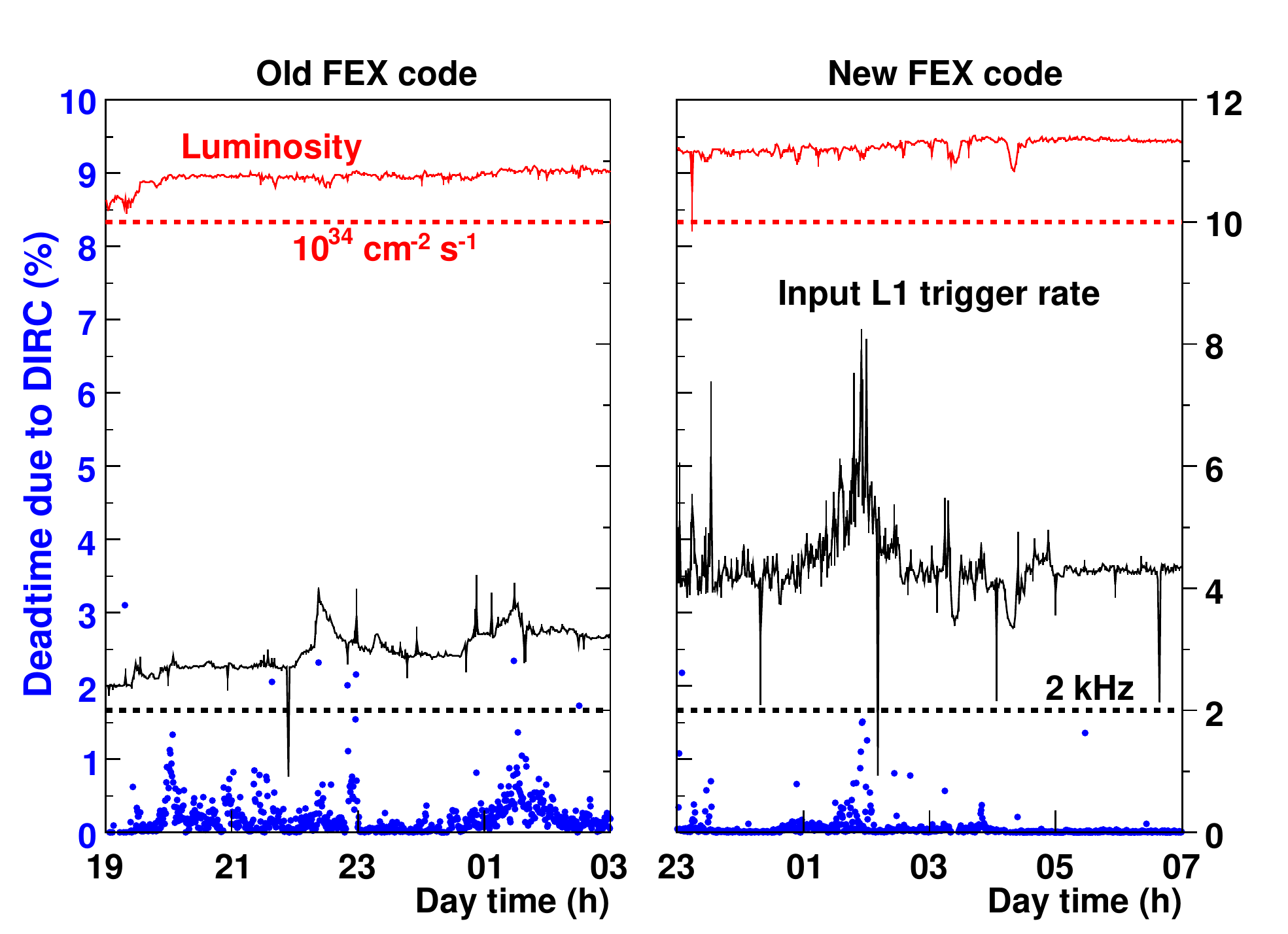}
\end{center}
\vspace{-0.5cm}
\caption{Dead-time reduction by the new DIRC FEX code: 
Comparison over periods of eight hours of  the old
(left) and new (right) FEX code in the DIRC ROMs. 
The $y$-axis on the left shows the DIRC contribution 
to the overall \babar\ dead-time (blue data points, in $\%$).
The $y$-axis on the right gives the scale for the instantaneous luminosity 
(red curves, in unit of $10^{33}\cms$) on the top, and  the 
input L1 trigger rate (black curves, in \khz).}
\label{fig:drc_fex}
\end{figure}

The second slowest readout system was the DIRC with a FEX code that required on average 170\mus\ per channel, with spikes of more than 200\mus\ during periods of high beam background, limiting the L1 rate to about 5\khz. Therefore, the DIRC FEX code 
was rewritten in early 2007. 
The new code was highly optimized, much faster (only 70\mus on average) and more reliable.
The dead-time reduction is illustrated in Figure~\ref{fig:drc_fex}.
Although the input L1 trigger rate was significantly higher, 
the average DIRC dead-time after the implementation of the new code was much smaller.
The few spikes are due to background bursts and are also visible in the trigger rate.

\subsection{Instrumented Flux Return}

\begin{figure}[!htb]
\centering
\includegraphics[width=0.5\textwidth]{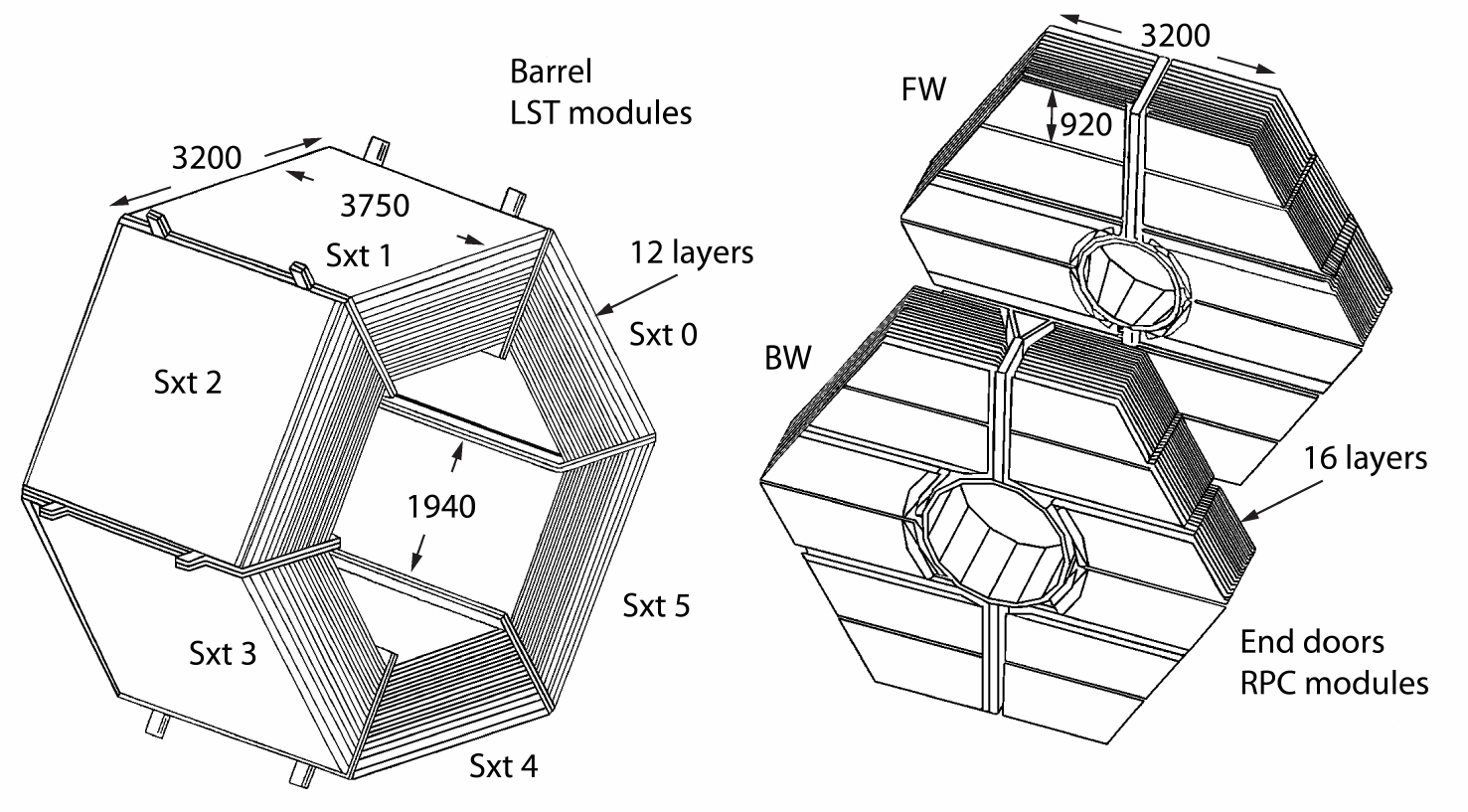}
\caption{Layout of the IFR: barrel sextants; forward (FW) and backward (BW) end doors. The dimensions are given in mm.}
\label{fig:ifr:structure}
\end{figure}

The Instrumented Flux Return (IFR) used the steel flux return of the superconducting solenoid to absorb hadrons and 
identify muons. It was composed of a hexagonal barrel and two end doors covering the forward and backward regions~\cite{Aubert:2001tu}.
Figure \ref{fig:ifr:structure} shows the overall final layout of the IFR.

Originally the finely segmented flux return was instrumented with 806 planar Resistive Plate Chambers (RPCs), 19 layers in the barrel and 18 in the end doors.   Within the first year of data taking, the RPCs showed serious aging problems~\cite{Piccolo:2001zy,Piccolo:2002ku,Anulli:2002kd,Anulli:2003bk}, necessitating 
replacement in the forward end door and the barrel. The original RPCs in the 
backward endcap were retained, given the lower rates and smaller solid angle coverage.

\begin{figure}[!htb]
\centering
\includegraphics[width=0.45\textwidth]{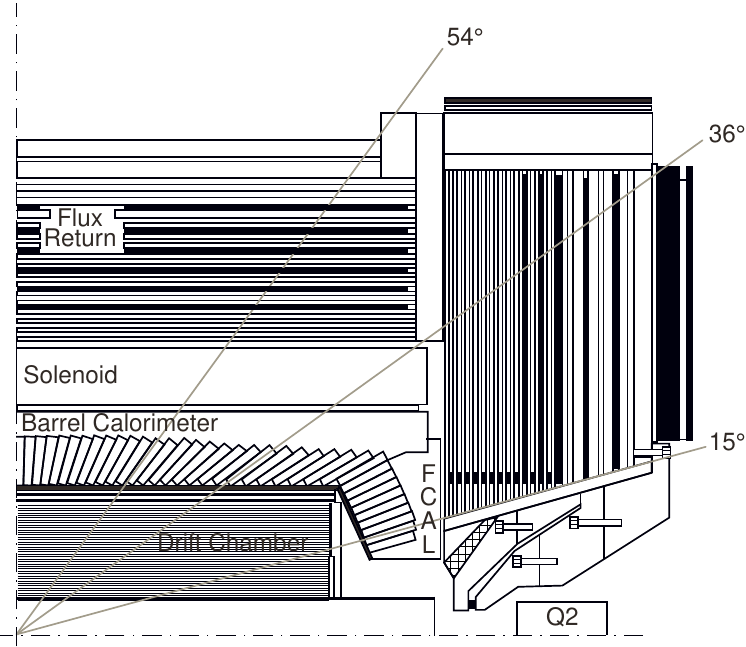}
\caption{Final configuration of the IFR shown as a longitudinal section of the \babar\ detector with 
details of the IFR barrel and forward endcap surrounding the inner detector elements. LST modules fill 
the gaps in the barrel, and RPC planes fill the forward endcap. The added brass and steel absorbers are marked in black.}
\label{A01}
\end{figure}

More than 200 second-generation RPCs were installed in the forward endcap in 
2002~\cite{Anulli:2004zu} to replace the original chambers.
The barrel RPCs were replaced by layers of Limited Streamer Tubes (LSTs)~\cite{LSTproposal} during shutdowns in 2004 and 2006. 
In the final IFR configuration shown in Figure~\ref{A01},
there were 12 layers of LSTs  in the barrel sextants and 16 layers of second generation RPCs in the forward endcap.
In parallel to the chamber replacement, some of the flux return slots were filled with brass absorber plates
instead of chambers, and some external steel plates were added, thereby increasing the total absorber thickness
and improving the pion rejection ability of the muon identification algorithms.

\subsubsection{RPC Infrastructure Upgrades}
\label{sec:upgrades-RPC}

The second-generation RPCs performed much better than the original ones~\cite{Anulli:2005wi}, 
but required several improvements~\cite{Band:2006ig,Ferroni:2009zz} to maintain good efficiency over the remaining 
life time of the experiment.  The humidity of the input gas was increased in 2004 to stabilize the Bakelite resistivity.  
Starting in 2006, the RPCs with the highest rates were converted to avalanche mode operation.
In addition to these major upgrades, numerous improvements to the original infrastructure 
(cooling, gas, high voltage, low voltage, and monitoring) were required to achieve and maintain the 
expected efficiency and reliability.

\paragraph{Detector Cooling}
Data from cosmic rays and first colliding beams showed in 1999 that the temperature 
of the IFR barrel was unstable. After the endcap doors were closed,
it increased by ${\approx} 0.5{^\circ}C$ per day. 
The heat source was traced to the front-end electronics cards (FECs) which were 
directly mounted on the RPCs inside the barrel gaps. While a single FEC dissipated 
only 1.4~W, the power dissipated by more than 1000 FECs in the barrel alone had a large impact.
Since RPC gain and current were strongly temperature dependent, the HV current rose to unsupportable 
levels after a few weeks. The high temperatures and currents also exacerbated the 
observed RPC aging. The following changes and the maintenance of stable, moderate 
temperatures were therefore crucial to continued operation of the RPC system.

Since forced air cooling was ruled out by fire safety concerns and limited access, the decision was made 
to cool the RPCs indirectly by water cooling the steel  absorber structure. Copper pipes were bolted 
onto the ends of the barrel gaps and water at $15{^\circ}C$ was circulated by shallow-well pumps. After a successful prototype test, the entire 
barrel cooling system was installed in early 2000. The system was expanded to 
include the endcaps the following summer.  The completed system of pumps, 
heat exchangers, and cooling loops  maintained the barrel temperature to about $20{^\circ}C$. 

The endcaps FECs were located in external crates (after the first year of 
operation), but the temperature in the outermost layers also depended on the air 
temperature in the experimental hall which could exceed $28{^\circ}C$ during the
summer months.  Insulation and additional cooling were added to the endcap exterior 
to stabilize the temperature to about  $20-24{^\circ}C$.

\paragraph{Front-End Electronics Cards}
The FECs were found to have another undesirable feature. Their low voltage power was supplied by 
thin individual cables of more than 10~m length. Whenever a large background 
burst generated multiple hits in a RPC the resulting large signals in the FEC caused the ground 
voltage to drop locally
by more than 1V. This temporarily reversed-biased transistors in the digital 
communication portion of the device.  The resultant damage to the communication protocol disabled not only
affected FEC, but also the seven other FECs sharing the 
same readout.  Since the barrel RPCs were well shielded 
from the beam lines, the barrel FECs were generally stable with the 
exception of the widespread damage from asynchronous beam losses (see Section 4.5.2).

Originally, one half of the forward endcap FECs were installed inside the gaps between the absorber plates.  During the first year of operation, 
these FECs suffered multiple failures as described above. 
The remaining FECs in the forward endcap and all of the FECs in the backward 
endcap were placed in mini-crates which
provided power and readout to 16 FECs. Presumably 
because of the more robust ground and power connections, FECs in mini-crates were not vulnerable to background bursts.
During the fall of 2000, RPC signal lines were extended and the FECs in the forward endcap 
gaps were moved into mini-crates mounted outside of the steel structure.   
After the FEC relocation, no further background related failures were observed in the endcap FECs.

\paragraph{High Voltage System}
The large currents placed considerable demands on the high voltage supplies
and distribution network. The average current per RPC was six times larger than expected in the initial design.  The high voltage distribution system and connectors failed at a rate of several percents per year. It was found that the HV segmentation was inadequate to deal with constantly changing currents and background conditions. To remedy this situation, the number of HV crates was increased from two to five (in 2000-2002) and higher current CAEN pods of the type A328 enhanced the current capability per crate by a factor of two.

As part of the forward endcap upgrade in the fall of 2002, all HV distribution boxes in the barrel and forward endcap were replaced by boxes with 20 kV connectors and upgraded components. 
The number of HV lines from the power supplies to the detector was 
doubled. HV fan-outs near the power supply allowed easier re-configuration of the HV distribution network.  This upgraded HV system proved to be very reliable.  After this upgrade,  all observed HV problems  were due to HV shorts in the RPC or the cables. Most of these problems were overcome by reversing the HV wiring and changing the HV polarity. 

\paragraph{Gas Supply System} 
Although the gas supply system performed reliably under normal conditions, it was observed that the operation was sensitive to low ambient temperatures.  The gas supply cylinders were located outdoors to reduce fire and asphyxiation hazards. During cold nights, the Freon or isobutane re-condensed in the supply lines. A significant improvement in the reliability of the gas supply was achieved by installing heated insulating jackets
around the Freon and isobutane supply lines and bottles.  

After operation at high gas flow for several years, some of the new endcap RPCs in the outermost layers exposed to high rates showed rate dependent inefficiencies. 
These inefficiencies were attributed to changes in the Bakelite bulk resistivity. Studies at CERN had shown that a rise in Bakelite resistivity could be caused by long-term exposure to dry RPC gas.
\begin{figure}[!htb]
\centering
\includegraphics[width=0.4\textwidth]{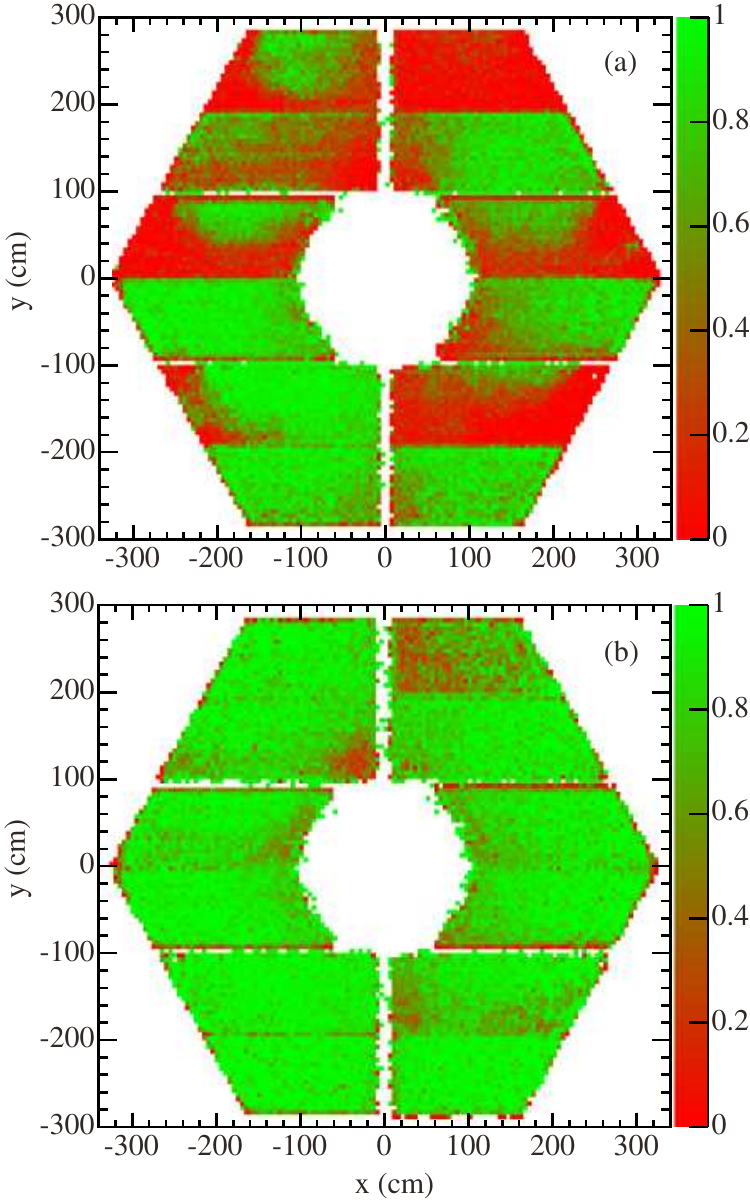}
\caption{RPC: Two-dimensional efficiency of FW endcap layer 16 (a) after 2 years 
of operation with dry gas and (b) after four months of operation with humidified
gas. The light (green) areas indicate high efficiency and the darker (red) areas low efficiency.}
\label{RPCfig1}
\end{figure}

In order to increase and control the relative humidity (RH) 
of the gas mixture in the chambers, the
flow was split so that a fraction of the input gas was diverted to run through a bubbler filled with water. Combining the nearly 100\% RH output from the bubbler with the dry gas, allowed a tuning of the humidity to the appropriate level.  The RPC input gas was adjusted to about 30\% RH.
Figure~\ref{RPCfig1} compares the efficiency of Layer 16 of the FW endcap
before and after the gas humidification.
The efficiencies continued to improve and eventually all areas reached nominal efficiency. 

When some of the forward endcap RPCs were converted to avalanche mode in 2007, a different gas mixture was required.
A new gas supply system, similar to the one designed for the LST chambers
(see Section~\ref{subsubsec:LST_power_utilities}), was built and operated.

\subsubsection{Forward Endcap Upgrade}

\begin{figure}[!htb]
\centering
\includegraphics[width=0.4\textwidth]{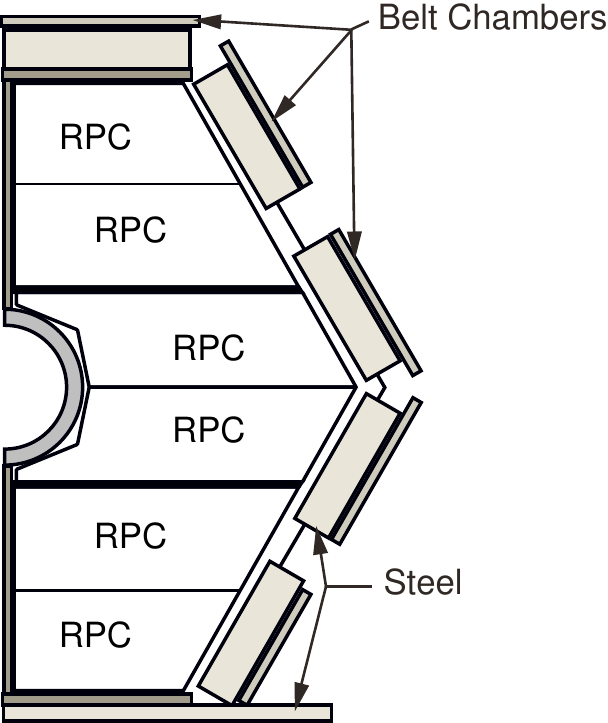}
\caption{End view of one half of the IFR FW Endcap: 
There are three large chambers each assembled from two high-voltage RPC modules. On the periphery, ten smaller RPCs are mounted on top of a pre-existing thick steel absorber outside of the flux return.
}
\label{A02}
\end{figure}

In the fall of 2002, the forward endcap RPCs were replaced to restore efficiency; brass plates and additional steel absorbers were installed to enhance the hadron absorption. There were 12 layers of single gap RPCs separated by steel absorbers of gradually increasing thickness. These were followed by two layers, each containing two single gap RPCs, separated by 10\cm of steel.  Each layer was assembled from three chambers, each of which consisted of two high-voltage modules with interconnected gas lines to form a single gas volume and read-out strips extending 
across the modules. The new IFR configuration is shown in Figure~\ref{A01}. 
To increase the  absorber thickness traversed by tracks exiting the side of the endcap, additional steel absorbers with RPCs attached were installed to cover the outer perimeter of the endcap, as shown in Figure~\ref{A02}.  

The design of the second generation of RPCs was carefully planned and stringent quality control procedures were imposed to eliminate 
many of the problems which had been encountered with the original RPCs. 
Molded gas inlets, more uniform graphite coatings, and thinner linseed oil coatings were implemented. The gas distribution and the HV distribution systems were upgraded.  
Details of the upgrade~\cite{Anulli:2004zu} and gas system~\cite{Foulkes:2004qc} were reported previously.   

\begin{figure}[!htb]
\centering
\includegraphics[width=0.35\textwidth]{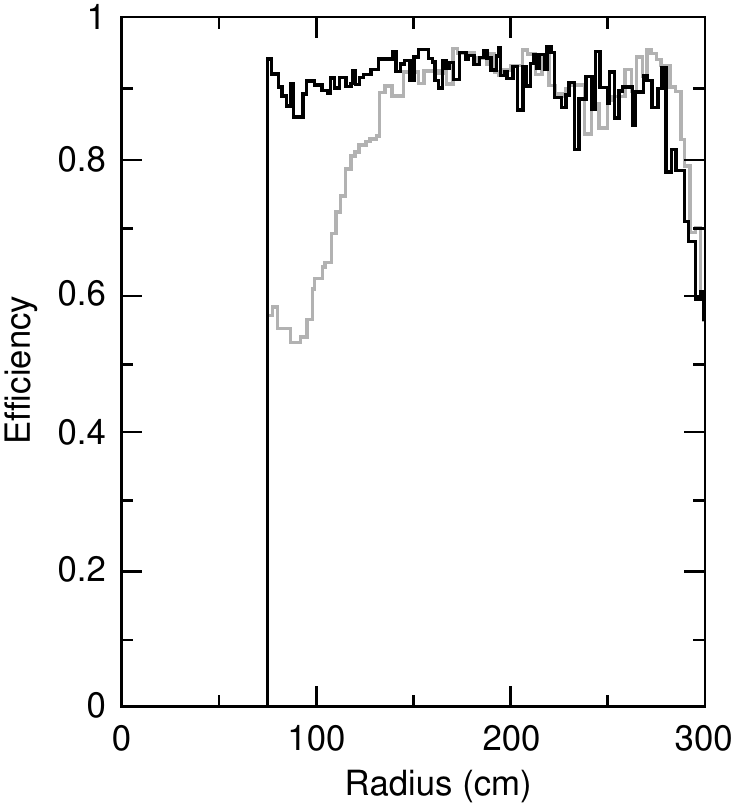}
\caption{FW endcap RPC efficiency (layer 1) as a function of radius before (gray) and after (black) the avalanche mode conversion.}
\label{RPC8821A05}
\end{figure}

\paragraph{Avalanche Mode Conversion of the FW Endcap}
After several years  of operation, some of the new RPCs showed efficiency losses in the high rate region around the beam-line. The loss was first evident in the innermost layers where the background rates were highest.  These high rates were proportional to the 
\pep2\ luminosity and were attributed to electrons from radiative Bhabha scattering striking the vacuum pipe and other beam-line elements inside the detector.
The efficiency was rate-dependent, and the inefficiency at a given luminosity slowly increased with exposure. The efficiency deteriorated to about 55\% in 
some regions as seen in Figure~\ref{RPC8821A05}.
Investigations performed on RPCs after the end of the data taking were inconclusive in establishing the 
cause of the diminished rate capability. 
Small changes in surface and bulk resistivities were observed, but these were within the normal range observed for Bakelite sheets.

The RPC rate capability can be increased by operating in avalanche mode~\cite{Cardarelli1996470},
which reduces the charge per track. For a given rate, the lower current through the gas reduces
the voltage drop across the Bakelite cathodes and anodes. The disadvantage of the avalanche mode
operation is that the induced signal on the RPC strips is reduced to ${\approx} 1/100$ of the streamer mode.
Preamplifiers, mounted directly on the pickup strips, are therefore needed to enlarge the signal
while minimizing electronic noise.

Since the RPCs were already installed, this option was 
unavailable. Preamplifiers could only be connected via a cable, $0.5 - 2.0$ m in length.  Based on studies of various options for avalanche gas mixtures and preamplifier designs, a gas mixture which produced larger than normal avalanche signals and a preamplifier design were chosen (75.5\% Freon 134a, 19.4\% Argon, 4.5\% Isobutane, and 0.6\% $SF_6$)
that reduced the RPC current, improved their efficiency, and operated well for the typical RPC counting rates.
The new electronics and gas were first tested in 2006. The following year, 
the first six layers of the two middle chambers (see Figure~\ref{A02}) were modified. 
The improvement in the RPC efficiency of Layer 1 is shown in Figure~\ref{RPC8821A05}. 

\subsection{IFR Barrel Upgrade}
\label{sec:upgrades-LST}

The continuous deterioration of RPC performance and the potential for further 
reduction in muon identification called for the replacement of the barrel RPCs 
in addition to those in the forward endcap.
This required a significantly larger intervention than any other \babar\ upgrade.
The proposed schedule for the replacement in the barrel demanded that the 
new detectors for the first two barrel sextants needed to be ready for installation by 
the summer of 2004, one and a half year after the decision to 
proceed with this upgrade.
  
In the endcaps, the shapes of the gaps were irregular trapezoids and therefore
RPCs were favored as the most suitable technology for a replacement.  
In the barrel, however, 
the gaps in the flux return were rectangular, of fixed length and variable width, and two additional detector 
technologies were considered as an alternative to upgraded RPCs:
extruded plastic scintillator bars and limited streamer tubes (LSTs).
The new RPCs, though promising, had not established a satisfactory performance record, and 
there were also concerns about the availability of a commercial company to produce 
large quantities in the time frame of one year.  
The plastic scintillator option relied on readout via avalanche photodiodes, 
which were not considered a sufficiently mature technology at the time.
The design and fabrication of LSTs were conceptually well understood. 
LSTs had performed well in more than a dozen HEP experiments and proved  
to be very robust~\cite{Bauer:1986pc,Bari:1988bx,Decamp:1990jra,Benvenuti:1989gf,
Ahmet:1990eg,Abbiendi:1993mi,Akikawa:2003zs}.
Furthermore, an experienced commercial vendor was available to produce the 
tubes on the very tight schedule.  Design studies and prototype 
tests by \babar\ groups had demonstrated the robustness of LSTs at the 
highest expected rates and integrated charge over the projected lifetime 
of the experiment.
Thus, LSTs were chosen to replace the RPCs in the barrel sextants.

The plan was to replace 12 out of 19 layers of RPCs with LSTs, and to fill the remaining gaps with 22 mm thick brass absorbers.  This increased the total absorber thickness and 
compensated for the loss in steel absorber of the outermost gap, layer 19, that could not be instrumented because of geometrical interference.
The barrel upgrade  was completed in two phases: 
two sextants (top and bottom) were replaced in summer 2004, and the remaining four 
in fall 2006.  Details of the upgraded IFR barrel detectors are presented in Table~\ref{tab:ifr-geom}.

\begin{table}[htb!]
\begin{center}
\caption{IFR barrel upgrade:  Overview of the mechanical parameters for the 18 gaps in each of the sextants, LST and brass absorber locations,  
the inner radius of each gap, 
the total number of interaction lengths at normal incidence, 
and the number of LST cells per layer.
}\label{tab:ifr-geom}
\begin{tabular}{ccccc}
\hline
Gap  &   Element  & Radius    &  Interaction         & Number  \\
     &            & (cm)      &  length $(\lambda)$  & of cells \\
\hline
1	&	LST	&	192.5	&	0.00	&	98	\\
2	&	LST	&	198.0	&	0.12	&	100	\\
3	&	LST	&	203.5	&	0.24	&	104	\\
4	&	LST	&	209.0	&	0.36	&	108	\\
5	&	brass &	214.5	&	-	&	-	\\
6	&	LST	&	220.0	&	0.73	&	114	\\
7	&	brass &	225.5	&	-	&	-	\\
8	&	LST	&	231.0	&	1.10	&	120	\\
9	&	brass &	236.5	&	-	&	-	\\
10	&	LST	&	242.0	&	1.47	&	128	\\
11	&	brass &	248.5	&	-	&	-	\\
12	&	LST	&	254.7	&	1.96	&	134	\\
13	&	brass &	260.9	&	-	&	-	\\
14	&	LST	&	267.1	&	2.44	&	142	\\
15	&	brass &	275.3	&	-	&	-	\\
16	&	LST	&	283.5	&	3.17	&	152	\\
17	&	LST	&	291.7	&	3.47	&	156	\\
18	&	LST	&	305.3	&	4.07	&	160	\\
\hline
\end{tabular}
\end{center}
\end{table}

\subsubsection{LST Design and Construction}

\begin{figure*}[!htb]
\centering
\includegraphics[width=0.90\textwidth]{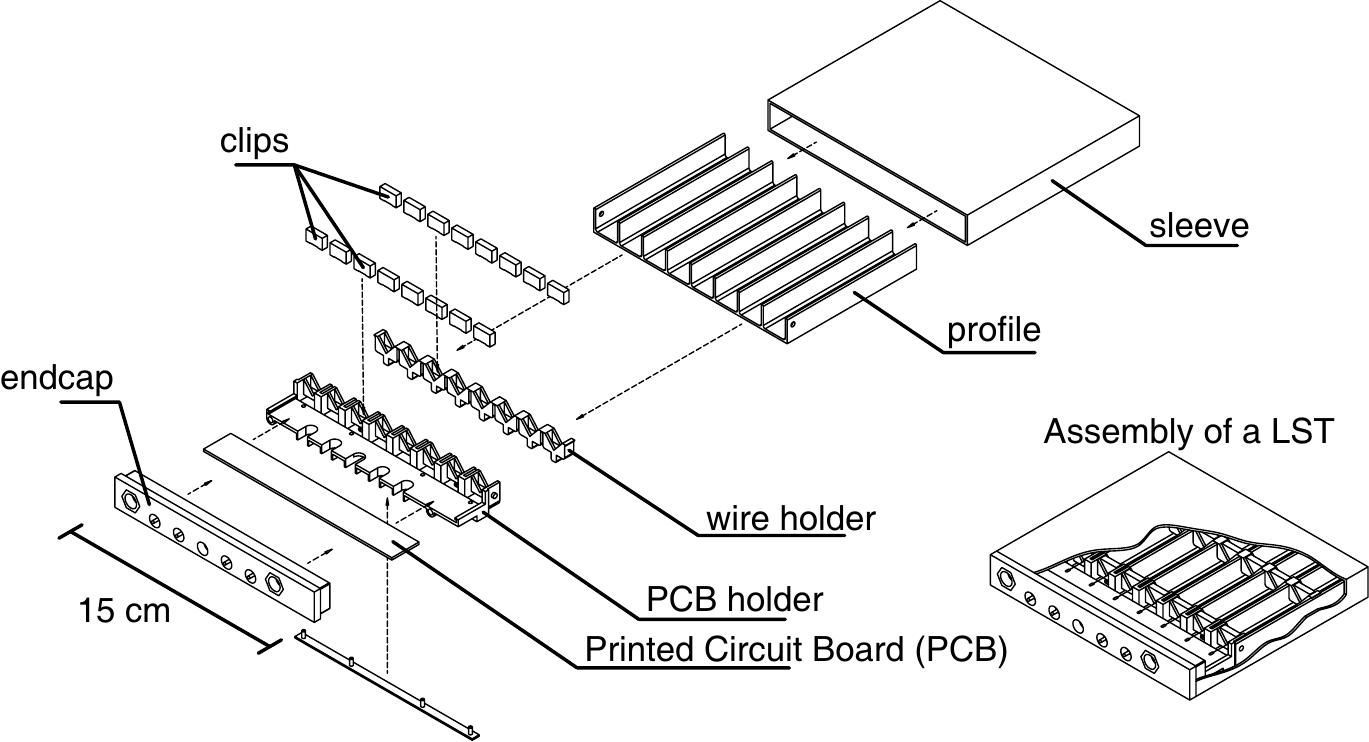}
\caption{LST: Exploded view with all the small components.}
\label{lst}
\end{figure*}

The design of a standard LST tube is shown in 
Figure~\ref{lst}.
A silver-plated anode wire of 100\mum\ diameter is placed at the center of a square
cell~\cite{Battistoni:1978mk}.
Seven or eight such cells are produced as an extruded comb structure made of PVC, with common side walls and open on one side.  The inner walls of the cells form the cathode. They are coated with a resistive layer of graphite, with a typical
surface resistivity of $0.1\Mohm$/$\Box$ to $1\Mohm$/$\Box$.
After the wires are installed, the profiles are inserted in plastic tubes, called sleeves, of matching dimensions for gas containment.

The wires are connected to an external positive HV supply.
A charged particle crossing one of the gas-filled cells ionizes the gas
and, given an appropriate gas mixture and electric field, will initiate an avalanche and streamer discharge around the anode wire.
The signal on the wire is read out and provides one coordinate derived from the wire location.
In the IFR barrel, these wires run parallel to the beam and measure the $\phi$ coordinate of the hit.
A simultaneously induced signal on external strips that are 
perpendicular to the wires provides the $z$-coordinate.
The third coordinate is taken from the radial distance of the wire from the $z$-axis.

More than one year of R\&D preceded the choice of the final LST design.
The R\&D program concentrated on several critical issues:
optimum cell geometry, selection of a safe gas mixture, rate capability,
wire surface quality and uniformity, aging effects, and performance tests 
of the prototypes~\cite{Lu:2004su}.

LST tubes were mechanically fragile, so careful design, assembly, 
and handling were important to prevent failures.
The longevity of the LSTs depended on the design of the cell and anode wire supports, on
attention to detail during construction, on strict quality controls and acceptance tests, and on well-conceived installation procedures.

In previous LST systems, double layers of tubes were installed,  with the cell centers offset by half a cell width.
This arrangement mitigated the inherent loss of efficiency due to cell walls.
Given the limited gap height (${\sim} 22\mm$) available in the IFR barrel,
a single layer structure was chosen, with a larger cell size.
Though efficiency losses from the cell wall are inevitable, these larger cells
reduced the sensitivity to the anode wire placement. Other advantages of the large
cells were reduction in the number of cells, and better overall mechanical robustness.

Water-based graphite paint was selected over the Methyl-Isobutyl-Ketone
(MIBK) paint that had been used in most previous experiments.
The latter had good adhesion to the PVC surfaces, a stable resistivity and was not 
sensitive to humidity. However, its toxicity complicated the application.
The resistivity of water-based paint depended on humidity so that it was essential to thoroughly
clean the PVC surface to achieve a uniform coating that firmly adhered to the PVC.
The water-based paint was not hazardous, which greatly simplified its application.
The humidity in the cell could be maintained at adequately low levels.

For the aging studies, prototype tubes with water-based paint were commercially 
produced~\cite{PHT} 
and tested up to a total accumulated charge of $\sim0.6\Coul/\cma$.
The current remained stable, indicating that no gas gain degradation had taken place.
The total charge accumulated in a barrel LST module during the full \babar\ operation was estimated to be $\sim$0.12\Coul/\cma.  Thus, there was quite a large safety margin.

\begin{figure}[!htb]
\centering
\includegraphics[width=0.45\textwidth]{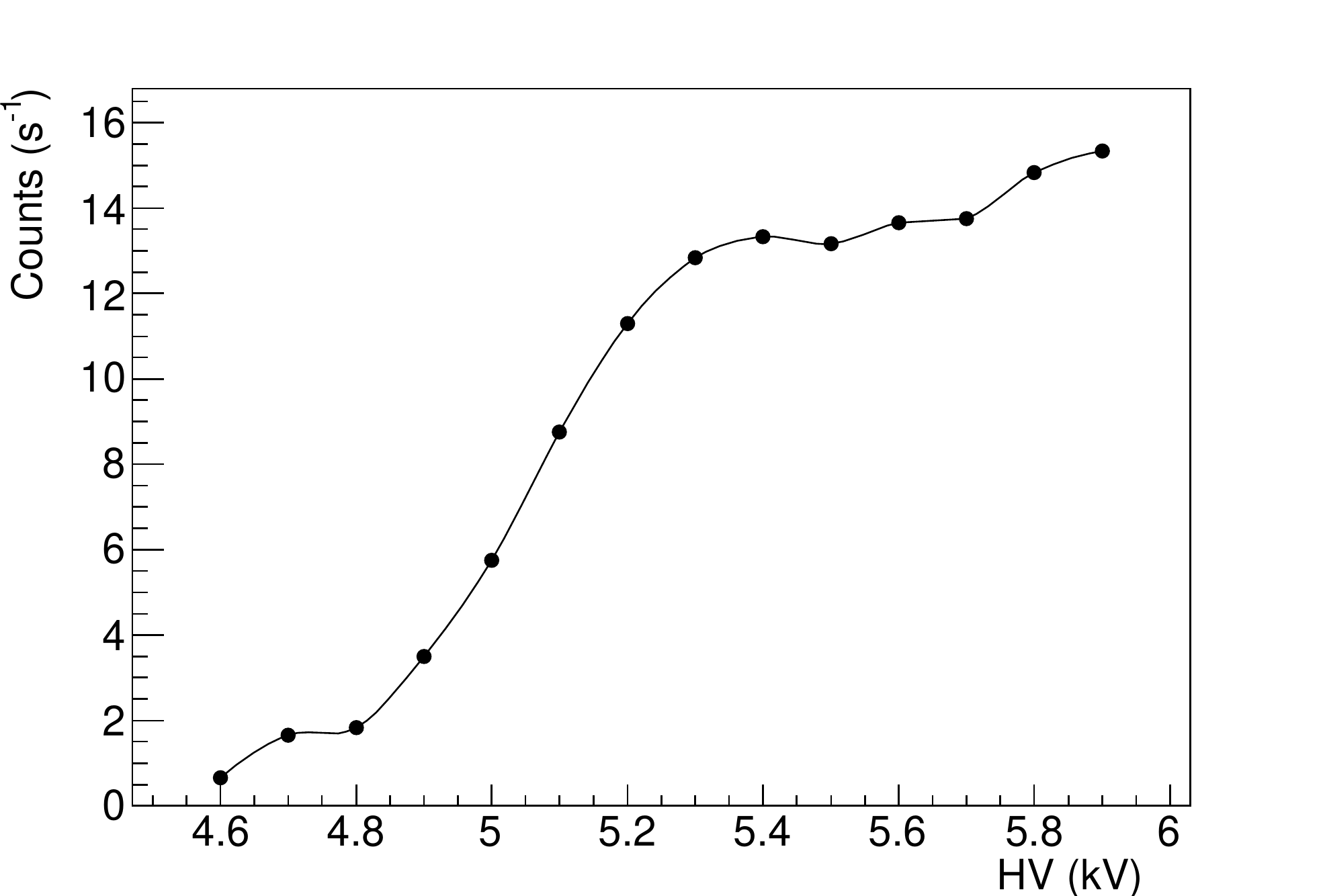}
\caption{LST: Example of single rate plateau for a single module.}
\label{fig:lstupgrade:plateau}
\end{figure}

A gas mixture of 89\% CO$_2$, 3\% argon, and 8\% isobutane was selected.
This  mixture was non-flammable and provided good performance~\cite{Alekseev:1980rd}.
The measured plateau for singles rates was about 500~V wide, as shown in Figure \ref{fig:lstupgrade:plateau}.
The operating point was typically 5500~V.  The tube efficiency was measured to be
about 92\%, with the inefficiency mostly due to the inactive areas of the cell wall.

\begin{figure}[!htb]
\centering
\includegraphics[width=0.45\textwidth]{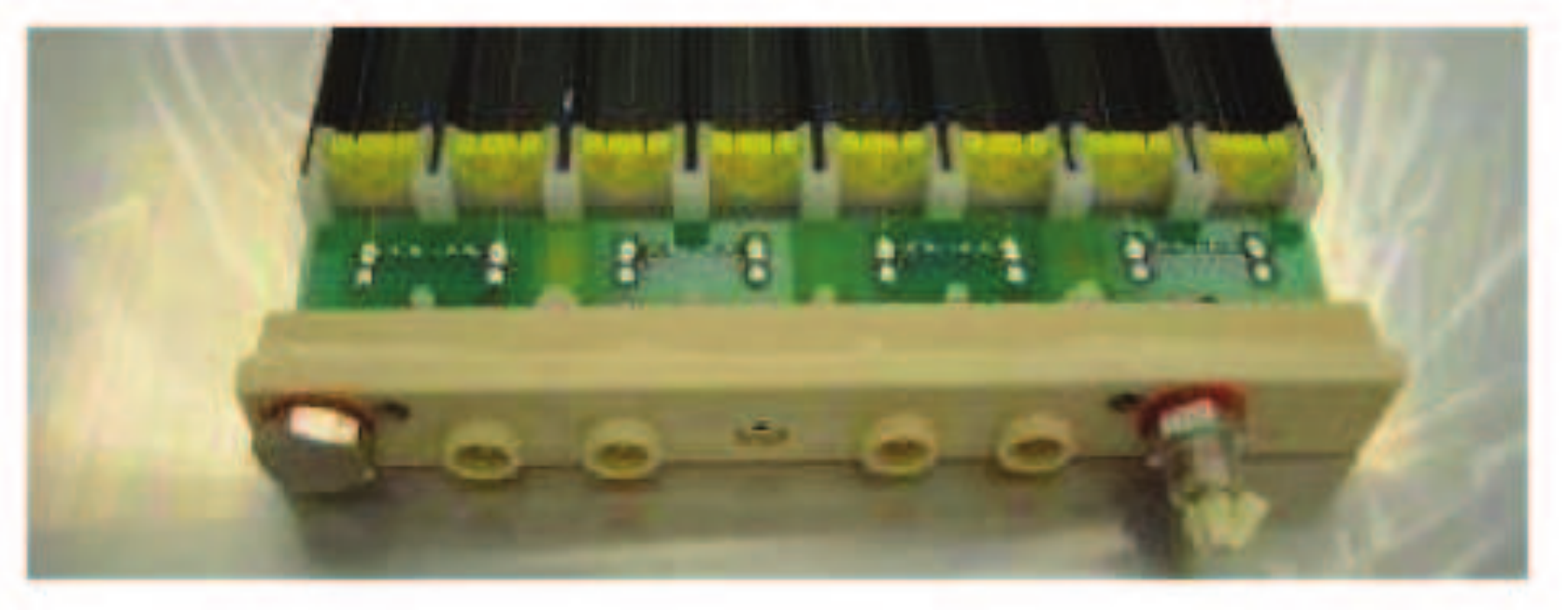}
\caption{LST: Picture of a tube endcap with HV and gas connectors.}
\label{fig:lstupdate:babarlst}
\end{figure}

These design studies and prototype tests led to the large-cell LST tube design, 
very similar to the one shown in Figure~\ref{lst}.
Each tube was composed of seven or eight cells; each cell was 17\mm\ wide, 
15\mm\ high and 3.75~m long.
The anode wire was chosen to be gold-plated tungsten of 100\mum\ diameter.
Six wire supports were equally spaced over the length of a cell to prevent
the wires from sagging and to thereby provide electrostatic stability.
Three sides of the cell were
painted with the water-based graphite paint and kept at ground potential.
The top side of the cell was open to the sleeve without graphite paint.
Both tube ends were equipped with gas connections, while HV connectors 
were attached to one end (see Figure~\ref{fig:lstupdate:babarlst}). 
Two wires were connected to one independently powered HV channel so that
a failing cell could be readily disconnected  
without losing the entire tube.
Tubes were assembled into modules to minimize the dead space, to achieve 
better mechanical stability, and to reduce the number of detector components.
The streamer signals on the anode wires were read out through a coupling 
capacitor 
to measure the azimuthal coordinate, $\phi$.
The $z$ coordinate was determined from the induced signal on strips, installed on the outside of the modules, orthogonal for the wires. These Cu strips made up a large plane covering the entire surface of an LST layer.

\subsubsection{LST Fabrication}

\paragraph{Production and Assembly of LST Tubes}

The LST tubes were manufactured by a company~\cite{PHT} located in Carsoli (Italy). 
Stringent quality control (QC) and testing during fabrication demonstrated that
specifications were being met while also providing continuous feedback to improve
each fabrication step~\cite{Cibinetto:2005ec}, and allowing the LST team to reject poor quality tubes at
an early stage. QC tests included:
\begin{itemize}
\item {Mechanical inspection of the profiles to make sure there were no significant bends, dents, or cracks.}
\item {Visual inspection of the graphite coating to spot major defects immediately after the painting.}
\item {Automated surface resistivity measurements of the graphite paint which imposed tolerances on the average resistivity of 100~k$\Omega$/$\Box$ to 600~k$\Omega$/$\Box$.}
\item {Leak tests to ensure the gas tightness, requiring less than one small bubble per minute from the assembled
tube submerged in 15-20\mbar\ over pressure in water.}
\end{itemize}

Following these mechanical tests during fabrication, each tube was subjected to electrical conditioning to burn off small pieces of debris and dust.  This was followed by a plateau measurement to verify that the working point was close to 5500~V, and a scan
with a 10~MBq $^{90}$Sr source inducing a current of ${\sim} 800\nA$ to spot graphite surface defects.  Prior to the final acceptance, each tube
was kept at its maximum operating voltage (determined by the plateau measurement) for a period of 30 days.
During this test, the current drawn by each tube was monitored and tubes with currents exceeding 200~nA for more than 5\% of the duration of the test were rejected. 
Of the 1500 LST tubes produced, about 8\% were rejected by the QC procedures.
More than half of the rejected tubes failed the long-term test at full voltage.

\paragraph{Module Assembly}

LSTs that had passed the QC procedures were shipped to Ohio State University and Princeton University for assembly into modules.
A module was composed of two or three tubes which were glued along their full length to 1\mm\ thick steel ribs to enhance the mechanical rigidity of the module.  Grounded copper foils were glued to one surface and the signal transfer plane (STP) to the other. The STP, which is described in detail below, connected the signal wires to the forward end of the LST module.

Since the width of the IFR gaps varied, 
efficient coverage required four types of modules,
assembled from different sets of tubes:
\begin{itemize}
\item{two 8-cell full-length tubes (by far the most common modules).}
\item{three 8-cell full-length tubes.}
\item{two 7-cell full-length tubes.}
\item{two 8-cell shorter tubes (for use in the outermost layer).}
\end{itemize}
Before being assembled into a module, each tube had to pass a plateau measurement and source scan test. Once assembled, a module was kept at HV for one week prior to being shipped to SLAC for the final QC and installation~\cite{Menges:2006xk} into \babar.

\paragraph{z-Plane Assembly}
To read out the induced signal from the streamer, a single large plane~\cite{Convery:2005vs} with strips running transverse to the anode wires was required for each LST
layer.  The length of all $z$-planes was $3719\mm$, except for
the outermost layer which was shorter by $370\mm$, due to
mechanical interference. The widths of the 
$z$-planes varied from $1908\mm$ for the innermost layer to $3110\mm$ for the
outermost layer. The thickness of the $z$-planes was precisely controlled
to $1.00 \pm 0.02\mm$ to assure adequate clearance for the insertion of the 
modules into the gaps between the absorbers.  

The $z$-planes were assembled from layers of Mylar, copper foils, and Mylar ribbon cables,
which were vacuum laminated in a single operation.
Each plane was composed of a common external Cu-foil ground plane, separated by a thick
dielectric film from a layer of 96 parallel, conductive readout strips
(each 35\mm\ wide and separated by $2.0\pm0.5\mm$).
A unique component was the custom-made, flat flexible cable that carried signals from the strips to the readout electronics. 
Each of the 16 conductors in these cables was soldered to a different strip and the cables were produced as part of the thermal vacuum lamination process
for the $z$-planes.

The design of the $z$-planes was very robust. Nevertheless, they
were tested to ensure their proper functionality.
In addition, careful monitoring during and after installation ensured
that any damage could be repaired before the $z$-planes became inaccessible.
For monitoring, the capacitance between ground and each readout channel
was measured. The expected value was about ${\sim} 5\nf$,
whereas very small or zero capacitance was a clear indication of a broken connection.
Only a few broken connections were found and almost all of them were
repaired prior to or during installation.

\subsubsection{High Voltage Distribution}
\label{subsubsec:LST_power_utilities}

\paragraph{High Voltage Supplies}

The operating HV settings of the individual LST tubes ranged from 5 to 6\kv
while currents ranged from 15 to 100\nA\ without
beams, and from 50 to 1000\nA\ (depending on the position
within the detector) with colliding beams. Since the occupancy, and
hence the current, changed as function of the radial distance from the beam,
the HV settings had to be independently adjustable, at least on a layer-by-layer basis.
Moreover, a high level of granularity in connections made the LST detector less
sensitive to a single-cell failure, as long as the high granularity was matched by the
HV system and individual HV channels could be disconnected or adjusted separately.
There were 192 tubes per sextant, each  split into four two-cell segments, 
requiring a total of 768 independent HV connections per sextant. 
The current drawn was independently monitored 
for each of the 192 tubes per sextant with a precision better than
10\nA, at a rate of 10\hz\ or more.

A sudden current increase to more than 2000\nA\ was generally a
sign of a self-sustaining discharge in one of the cells.
To protect the detector, a hardware over-current protection stopped the discharge by
automatically lowering the HV. For other high current situations, such as high beam-induced backgrounds, a sophisticated configurable trip logic was
implemented.

While high granularity of the HV distribution and the current
monitoring was necessary, independent settings of the operating HV 
of the individual two-cell segments within a tube were not needed. 
On the other hand, the HV settings had to be adjustable to reduce the
currents and thereby avoid potential damage to the tubes without 
compromising the detection efficiency.

The  high voltage power supplies designed and produced by the Ohio State University~\cite{Benelli:2006pa} 
represented a cost-effective solution that fulfilled all of the stated requirements. The LSTs in each sextant were powered by three power supplies. Each supply provided 80 independent HV outputs with voltages adjustable from 0 and 6\kv .  Each output was connected to a separate current monitor. 
Each of the 80 output lines was split into four lines which were connected to the four two-cell segments in a single LST tube.  Thus, there was one voltage setting and one current monitor for the eight anode wires in each of the LST tubes. 
Even though the four outputs were connected in parallel to the power supply, having four separate connections preserved the ability to disconnect
individual two-wire segments of the LST tubes in case of problems.

Each current monitor module was capable of measuring a full scale 
current of 12\muA\ with a resolution of 1.5\nA. 
The module included  
an overcurrent protection circuit 
using a HV diode.
During normal operation, the output current was small and the HV diode
was forward biased. The forward bias condition held the output voltage
equal to the set voltage plus the HV diode drop. When the
output current rose above the overcurrent protection limit
the diode became reverse biased.

If the output current demand continued to increase, the output voltage
dropped linearly.
This mechanism provided each channel with overcurrent
protection without affecting any other channel on the same supply.
Since the internal HV DC-DC converter could be set remotely,
it was possible to vary the current protection limit to adapt to
changing operating conditions, without hardware modifications.
A software-based monitor was implemented to turn off the HV whenever one or more channels were drawing
high currents for an extended period of time.
\paragraph{HV Boxes and Cables}
The HV was delivered to each LST by a circuit enclosed in a box filled with dielectric material.  For each anode wire, the circuit consisted of a 1\Mohm\ isolation resistors, an impedance matching resistor,
and a 300\pf\ HV coupling capacitor. The circuit connected the wire signals to the signal transfer plane, and
via isolation resistor to HV supply cables. Each of these HV cables had eight independent wires to supply the eight channels for a standard 2-tube
module, or 12 for a 3-tube module. The HV cables had  inline connectors to simplify the installation.

\subsubsection{Gas System}
The gas mixture of (89:3:8 parts) of CO$_2$, argon
and isobutane is non-flammable and has good quenching properties.
A redundant system of mass flow meters monitored the gas component flows and
was able to shut off the system in case the mixture deviated significantly
from the preset standard. This provided protection against
accidental changes to a flammable mixture.

Half of the LST tubes per layer in a given sextant were connected in sequence to one gas line,
which resulted in a total of 144 separate gas circuits.
The gas was supplied to each of these circuits by a fully passive
system.
Thin capillary tubes were used as flow resistors to provide a pressure drop
that effectively balanced the flow in each circuit.
This simple, passive system kept the flow in each circuit balanced at the 10\% level
without the use of adjustable needle valves.
The gas flow in each circuit was monitored at the outlet with digital bubblers~\cite{Little:2003me}.
The total flow rate of the system was 2.5\liter/min,
which corresponded to about one volume change per day. A much lower gas flow  would have resulted in reduced performance, because water permeating the tube walls would have resulted in higher humidity.

\subsubsection{Electronics}
\begin{figure}[!htb]
\centering
\includegraphics[width=0.45\textwidth]{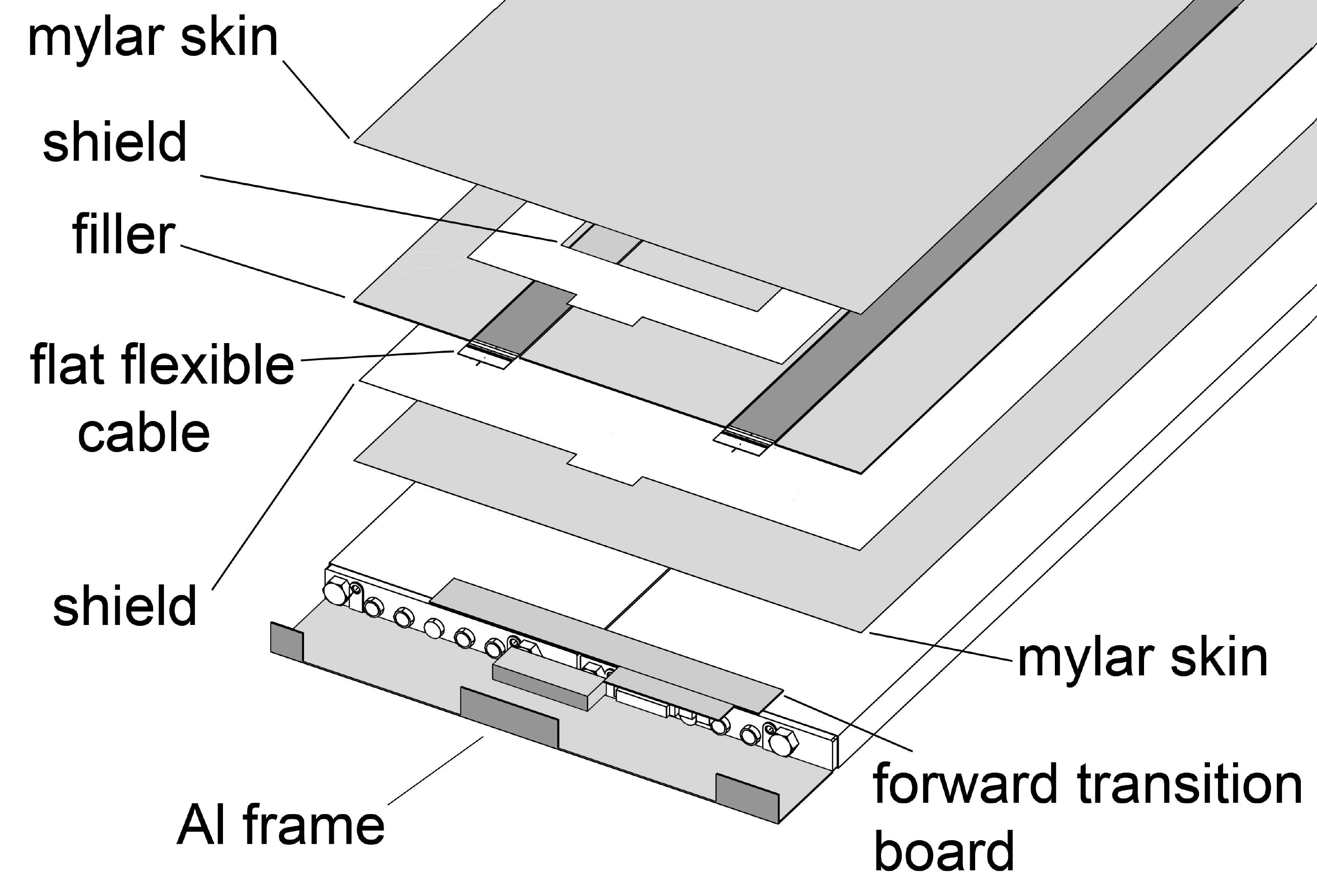}
\caption{LST: A sketch of the signal transfer plane and LST module.}
\label{fig:lstupgrade:readout}
\end{figure}

New front-end electronics cards (LST-FE) were developed for the LST detector. 
They interfaced to the existing IFR-FIFO Board (IFB) and were read out by the standard 
\babar\ DAQ system~\cite{Aubert:2001tu}. The IFB 
represented the sink for the  LST signals.
The main component was a mother board on which four daughter cards could be installed, each with 16 analog input channels. The signals from the wires and the strips had different polarity and shape. This difference was accommodated at the daughter board level. The signals were amplified, discriminated with an adjustable threshold, and
converted to 1-bit digital signals. No other information was retained,
\eg\ signal timing, charge, and other pulse information were not passed on to
the DAQ system. The crates with the front-end electronics were
not mounted on the detector, but installed nearby for easy access.
\paragraph{The Wire Signal Path}
The anode wires of the LST detectors were soldered to printed
circuit boards (PCBs) which were fitted and glued to the two tube endcaps.
The wire-holder PCBs performed both mechanical and
electrical functions.  At one side of the LST, 
the PCBs were soldered to the HV pins inserted in the endcap mould.
The PCBs distributed the HV to the wires through 220\ohm\ resistors.
The PCB insulator was regular FR4, 1.6\mm\ thick. 
To avoid surface discharge, the planar clearance between traces
of different HV circuits was about 3\mm\ per kV of potential difference.
The anode signals were decoupled from the HV potential in the HV connector modules.
Each module contained a PCB supporting four AC-coupling 470\pf\ capacitors and a ``bleed-off''
10\kohm\ resistor to ground.
The ground referenced signals were routed to a 10-pin connector accessible at the back side of the HV connector, through which they entered the signal transfer plane (STP).
The STP (see Figure~\ref{fig:lstupgrade:readout}) was made of a long segment of flexible flat cable (FFC),
thin copper sheets as ground planes and insulating Mylar foils. The FFC was made of 10 laminar conductors  with a 2.5\mm\ spacing between two conducting sheets. On one side the FFC carried
the 10-pin connector mating to the  HV connector.
The other end of the FFC was terminated with pins soldered to a transition board which was glued to the top of the module. Its function was to merge the signals from two or three FFCs into a single PCB-mount header for the signal cable.

\paragraph{The Induced Signal Path}
The top layer of  a $z$-plane was divided into 96 strips (80 strips for the outermost layer) of equal widths and lengths.
The bottom layer was a solid copper foil acting as a ground plane.
Between the two, there were insulating sheets and six FFCs with 16 conductors each, used to carry the strip signals to the backward end.
The free ends of the FFCs extended for about 10~cm beyond the backward
edge and were terminated at plugs mounted in a plastic casing.
The $z$-strip transition boards (ZTBs) were designed to connect the FFCs to the
shielded cables carrying the strip signals to the front-end cards. The ZTBs were also used
to connect the shields of the $z$-strip planes to the shields of the signal cables and to the detector ground (magnet steel).
\paragraph{Grounding and Shielding}
Though the LST tubes produced large charge signals, much care was required 
to avoid noise pick-up from many potential sources.
In particular, single-ended signals needed to be carried on average 
10~m from the detector to the front-end signal processing cards.
A flexible grounding scheme was designed to connect the ground
shields from the STP or the $z$-strip planes to the front-end
cards with or without connection to the magnet steel. 
After repeated measurements of the noise at the output of the front-end
amplifiers for different grounding and shielding configurations, it was decided to attach all shields from the detectors to the front-end cards and to connect all
of them to detector ground.
\paragraph{The LST Front-End Cards}
The LST-FE card provided amplification, comparison to a programmable
threshold, storage and readout processing for 64 LST signals. Its basic functions were equivalent to those performed by four of the FEC cards developed for the RPCs~\cite{Cavallo:1997pf}.
The LST-FE card was built, for the sake of flexibility, as a modular unit, with the following components:
\begin{enumerate}
 \item A motherboard, featuring:
   \begin{itemize}
      \item  DC power regulators and power monitors.
      \item  64 discriminators, connected in four groups of 16, each group with a common, programmable,
                threshold voltage.
      \item  A FPGA that performed the standard data acquisition functions plus programmable
               threshold, programmable channel masking, stand-alone rate counting, and other diagnostic
               tasks. These additional features were handled by 
               a NIOS processor and
               its associated peripherals, implemented in the FPGA.
   \end{itemize}
\item Integration-amplification daughter (IAD) cards,  each featuring 16 dual-stage
          integrating amplifiers.
\end{enumerate}
The LST-FE motherboard was a single width VME-Eurocard module, connected to the custom backplane
through a standard 96~pin DIN41612-C connector.
The core of the motherboard was a Cyclone II FPGA by ALTERA~\cite{ALTERA}.
For the power-up configuration of the FPGA, a dedicated FLASH RAM serial configuration device, which could be reprogrammed in-circuit to reflect changes in the FPGA design, was used. The NIOS processor could also access a bank of locations in this non-volatile RAM to store and retrieve operating parameters such as comparator thresholds and channel
enable masks.
These operating parameters were  maintained even through power cycles, making the board independent of the detector slow controls system.
A suitable voltage regulator provided the supply voltages needed by the FPGA. It included timing networks to provide the proper sequencing during
power cycles.
The main task of the FPGA was to provide storage and readout functions equivalent to those of the FECs designed for the RPC readout. 
This was a specification of the LST-FE front-end, which was supposed to
transparently interface to the existing Barrel DAQ system.
The FPGA inputs from the 64 discriminators on the LST-FE motherboard were grouped by 16 and then routed to four identical processing units, obtained by cascading the following blocks:
\begin{itemize}
  \item Input synchronizing latches, used to make sure that no input pulse was lost.
  \item Pipeline memories which held the sequence (sampled at 10\mhz) of states 
             of the 16 input signals of each block over a time interval corresponding to the L1 trigger latency.
  \item Output shift registers, which were continually refreshed by being parallel-loaded with the
            output of the pipeline memories. When the L1 trigger fired 
with a latency of 11$\pm$0.5\mus, the Shift/Load input changed state; at this point, the output registers started shifting their contents at a frequency of 15\mhz.
  \item The serial streams from the output shift registers were routed to the ECL level translators
            on the motherboard and then, after the conversion, to the backplane connectors. The
            LST-FSD card collected the serial streams
            from eight LST-FE cards via the backplane and routed them to a twisted
            17-pair flat cable connected to one port
            of an IFB.
\end{itemize}

The IAD cards featured 16 dual-stage, bandwidth-limited amplifiers;
one right-angle, 34-pin header mating to the cable connector, accessible from the front panel;
and one board-to-board connector to the motherboard for routing power, ground, test
pulse, and the 16 amplified output signals to the IAD.
The amplifier was designed to function as an
inverting amplifier for LST wire signals
or as a non-inverting amplifier for the positive signals induced on the
$z$-strips external to the modules.
The gains were different in the two cases. They were set at  80\mv/\pC\ for the $z$-strip amplifier
and at 40\mv/\pC\ for the wire amplifier. Another difference was the dual stage resistor-diode
network, protecting against high voltage input transients from possible LST breakdown.
\paragraph{Noise and Threshold}
The noise level was evaluated by measuring singles rates for cosmic data
recorded with two different configurations:
\begin{itemize}
  \item A first test with the high voltage set to about 3000~V,
  much below the operating point,
           changing the thresholds from 100\mv\ to 1000\mv\ to evaluate 
the noise level of  the complete readout
           chain. For threshold settings of 200\mv, no counts were 
           observed for wire 
           or  strip readout.
  \item A second scan with the high voltage set to 5.5 kV, performed to check the noise
            related to the LST operation: the electronic noise was negligible for thresholds greater than 200\mv.
\end{itemize}
In addition, the module efficiencies were measured at different thresholds without showing any variation.
The thresholds were set to 900\mv\ for both the wire and the strip signal discriminator,
to take into account a possible  increase in noise due to the changes in \pep2\ operation.


\renewcommand{\secname}{running_}
\renewcommand{\sectiondir}{Sec04_Operation-Performance}

\section{Detector Operation}
\label{sec:opsperf}

\subsection{Overview}

The operation and maintenance of the detector was the responsibility of the collaboration at large. For periods of about six months, two Run Coordinators, 
usually assisted by a deputy, were in charge of the day-to-day operation. They were assisted by the managers of the central operation system and the trigger, the online coordinator, and their teams of experts.  Two physicists were assigned to eight-hour shift duty in the \babar\ control room, for a few consecutive days  a few times per year. The lead operator was responsible for monitoring the trigger rate, dead-time, beam backgrounds, and the state of the detector components. He/she also maintained contact with the \pep2\ control room,  executed calibrations, and managed the electronic logbook. The second operator was responsible for monitoring the data quality, relying on online monitoring software.
During the first few months of operation, and thereafter, whenever major changes 
were introduced either in \pep2\ operation or \babar\ background monitoring, a \babar\ physicist was assigned as liaison to assist in the \pep2\ control room. This greatly enhanced communication between these two teams, thereby facilitating the integration of new procedures. 

Operating the \babar\ detector required not only contending with routine data taking, but also coping with number of less routine problems. The later included minor device failures, and predicting and coping with backgrounds and the effects of radiation damage. During routine data taking, the detector systems and their operational parameters (such as voltages, temperatures, gas flows, fluid levels, etc.), electronics, and DAQ were continuously monitored and calibrations performed on a daily to semi-weekly cycle, depending on the particular calibration. A fraction of the incoming data were analyzed as it passed through the DAQ to provide monitoring of all aspects of the detector performance. Typical problems that were uncovered from electronic failures of single channels which could be replaced or repaired during the next short down, to subtle, complicated problems that required extended investigation \eg, the increase in the SVT leakage currents  during trickle injection. 

Efforts to ameliorate backgrounds continued throughout the lifetime of the experiment.
Although there had been an extensive program of test beam measurements performed on individual detector components to understand the impact of ionizing radiation, the radiation levels encountered at the high luminosities motivated additional studies of radiation damage on detector samples at high-intensity neutron, photon and electron sources.

The main tools employed in the \babar\ control room for operational tasks were the Experimental Physics and Industrial Control System (EPICS)~\cite{Online:EPICS} and the \babar\ dataflow software. These systems provided extensive diagnostics, status and history charts, and warning/alarm indicators. Electronic calibrations of detector response, which for most systems served as an overall functional check, were performed typically once per day or when a specific failure occurred. 

The tasks of the two operators were simplified by a high degree of automation. For example, during normal data acquisition, the lead operator simply monitored the status displays. Conditions that could potentially endanger personnel or equipment initiated automated procedures to put the detector in a safe state and alerted the shift team.

Even though most shift operators had limited expertise and training, this minimal staffing resulted in an extremely efficient operation. The operation and protocols were documented in a detailed and current operation's manual\cite{CareAndFeeding}. Each detector system provided an expert on call at all times. Whenever the operators encountered a problem they did not understand or could not solve quickly, they called the appropriate expert. These system and online experts had remote access to the detector status and data, and many of them monitored operation regularly, without being assigned to shift duty.  The system experts, usually on duty for six months, were backed up by support teams they could contact for additional help

In addition to the data taking and online operation, extensive cross checks and calibrations were performed offline by scientists. These included checks of the event selection at the trigger level, alignment and calibrations for different detector systems, assessments of data quality, and stability of efficiencies and resolutions based on physics analyses. Isolated failures and temporary changes were recorded and implemented in the MC simulation to improve its accuracy in reproducing the run conditions. Only 1.1\% of the recorded data were discarded
during reconstruction, because of hardware failures that could impact the physics analyses.

\
This excellent performance could only be achieved because the control of the quality of the \babar\ data~-- from the raw data acquired in the control room to the processed events ready for physics analysis~-- was a priority for the collaboration. The first level of data quality checks was performed in real-time during data taking in the Fast Monitoring (FM) framework. An extensive set of histograms, newly collected for each run, was accumulated using data from a fraction (typically 20\%) of the events accepted by the L1 trigger. These live histograms were automatically compared to reference histograms, selected by the data acquisition system depending on the running mode (colliding beams, single beam tests, cosmic rays). Problems could be spotted by the on-duty crew who would rely on written instructions (provided by each detector system) to assess the situation and decide whether it was a known problem or one that should be reported to an on-call expert. After the end of a run, the FM data were automatically processed to produce a set of plots to be checked daily by system experts. They were also saved 
for offline re-analysis. The same software could also be run offline.

After event reconstruction, the quality of the processed data was surveyed by the Data Quality Group (DQG).
The group met on a weekly basis to report on the data processed during the past few days: only runs validated by the DQG were included in the datasets prepared for physics analysis. During the data processing, a set of histograms  based on 10\% of the data was automatically generated and made available to the DQG. It included both raw and reconstructed quantities and allowed DQG experts to identify runs with problems and diagnose their origin. If the processing had failed, the run was processed again, whereas if the raw data were compromised, online experts would check whether parts of the data could be recovered. 
In addition, during the final reprocessing of the full dataset, all runs which had been declared bad were reviewed again to see if they could be repaired. These two operations recovered about 1~$\invfb$ of data.

\subsection{Silicon Vertex Tracker}
\label{sec:opsSVT}

\begin{figure*}[htbp!]
\centering
\includegraphics[width=0.9\textwidth]{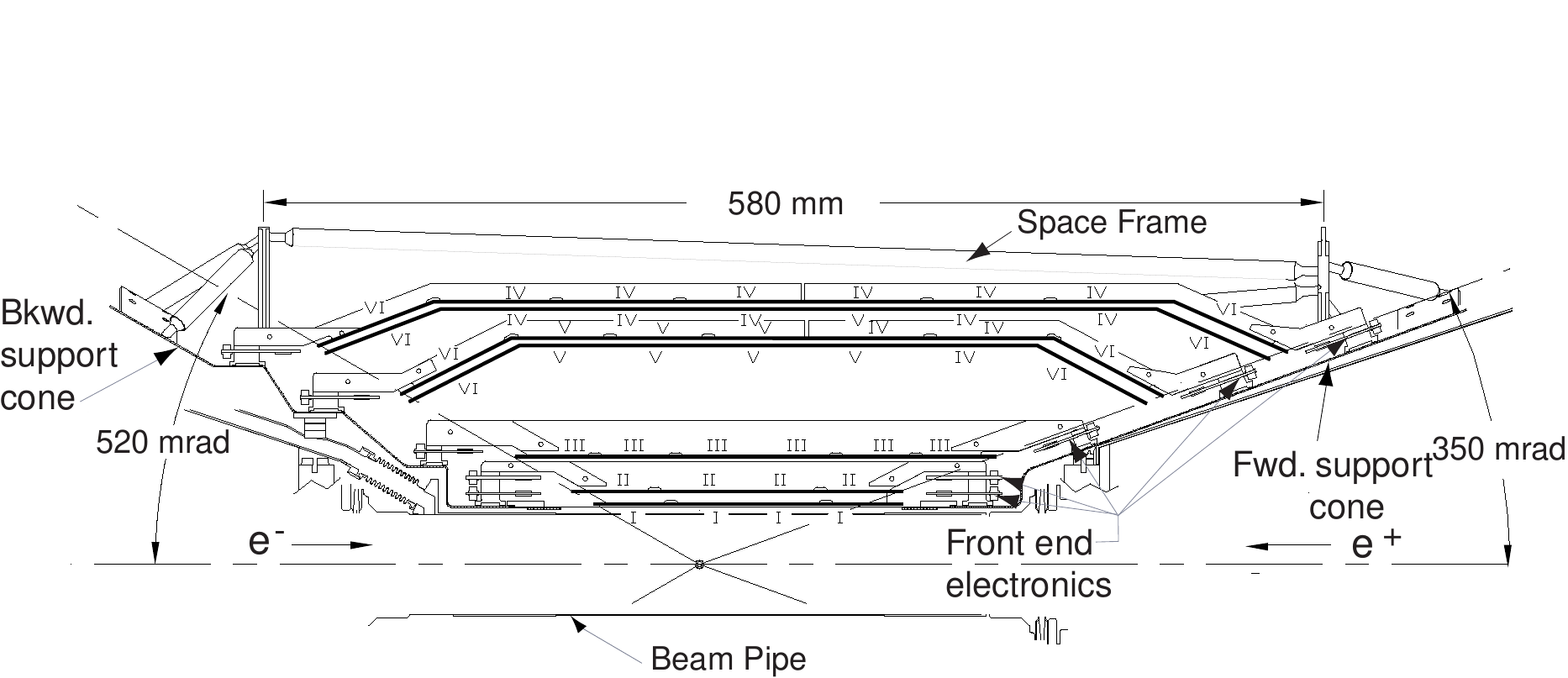}
\caption{Longitudinal section of the SVT. 
The roman numerals label the six types of sensors with different dimensions and strip pitch.}
\label{fig:svt_sideview}
\end{figure*}

The \babar\ Silicon Vertex Tracker (SVT) was designed to provide precise reconstruction of charged particle trajectories and decay vertices near the interaction region. 
The separation of the two $B$ decay vertices and the average decay lengths of  tau lepton and charm decays required vertex resolutions on the order of 100\mum.
This translated into  on-sensor spatial resolution of about 10-15\mum\ for the three inner layers and 40\mum\ for the two outer layers. 
A detailed description of the structure, geometry and features of the SVT was presented previously~\cite{Aubert:2001tu};  a longitudinal section is shown in Figure~\ref{fig:svt_sideview}. 
There are five layers of double-sided silicon strip sensors which are arranged 
in 6, 6, 6, 16, and 18 modules, respectively.  The modules of the inner three
layers are planar, while the modules of layers 4 and 5 are arch-shaped.
The strips on the opposite sides of each sensor are orthogonal to each other:
the \mphi\ measuring strips run parallel and the $z$ measuring strips  
transverse to the beam axis.  

Since the SVT was mounted on the beam pipe inside the support tube,  any access for repairs would take more than one month;  
therefore the requirements on reliability were extremely high. All SVT components inside the support tube were chosen and tested to have a long
mean-time-to-failure, and redundancy and flexibility of the readout were built into the design.

\subsubsection{SVT Performance}
\label{sec:svtperf}

Among the most critical measures of the SVT performance were the single layer hit efficiency and resolution.
The SVT hit efficiency for $\phi$ or $z$ strips
was measured using high energy tracks from dimuon events. The efficiency for each half module was determined as the fraction of tracks with a signal detected close to the intersection of the projected track with the active area of the detector module. Non-functioning
readout sections were excluded, but otherwise no
attempt was made to avoid noisy or problematic readout channels, known pinholes, and other defects. 

\begin{figure}[htbp!]
\centering
\begin{tabular}{c}
\includegraphics[width=0.45\textwidth]{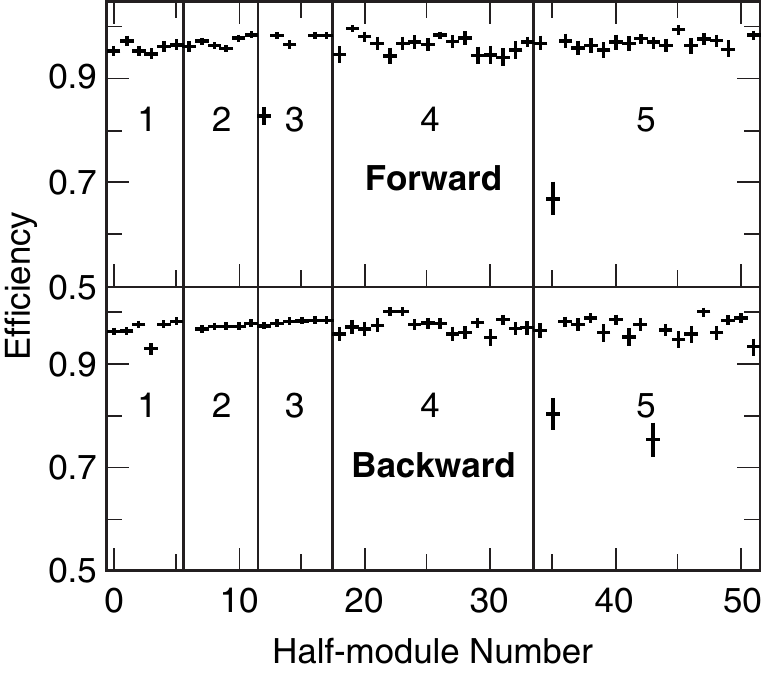} \\
\includegraphics[width=0.45\textwidth]{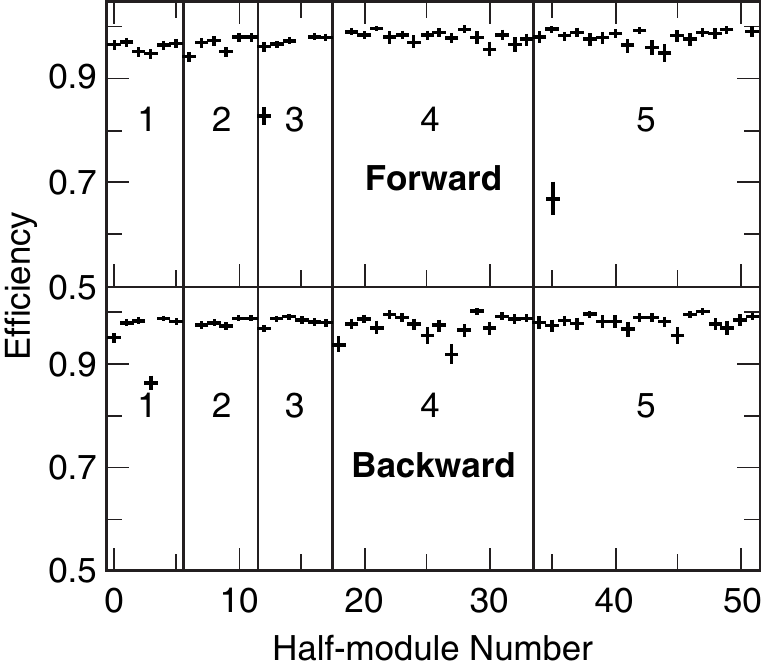}    \\
\end{tabular}

\caption{SVT hit efficiency for five $\phi$ layers (upper plots) and five $z$ layers (lower plots) during normal operation in 2003, separately for forward and backward modules, in each of the detector layers (indicated by numbers 1-5). Each data point represents one SVT half-module. Note the suppressed zero on the vertical axis. 
}
\label{fig:hit_efficiency}
\end{figure}

Figure~\ref{fig:hit_efficiency} shows the SVT hit
efficiency for $\phi$ and $z$ strips for each SVT half-module. The hit efficiencies exceeded 95\% , except for a few layers
with malfunctioning readout or dead
modules (\eg\ layer 3, module 4, forward, $\phi$ side) which are not shown in these plots.

\begin{figure}[htbp!]
\centering
\begin{tabular}{c}
\includegraphics[width=0.45\textwidth]{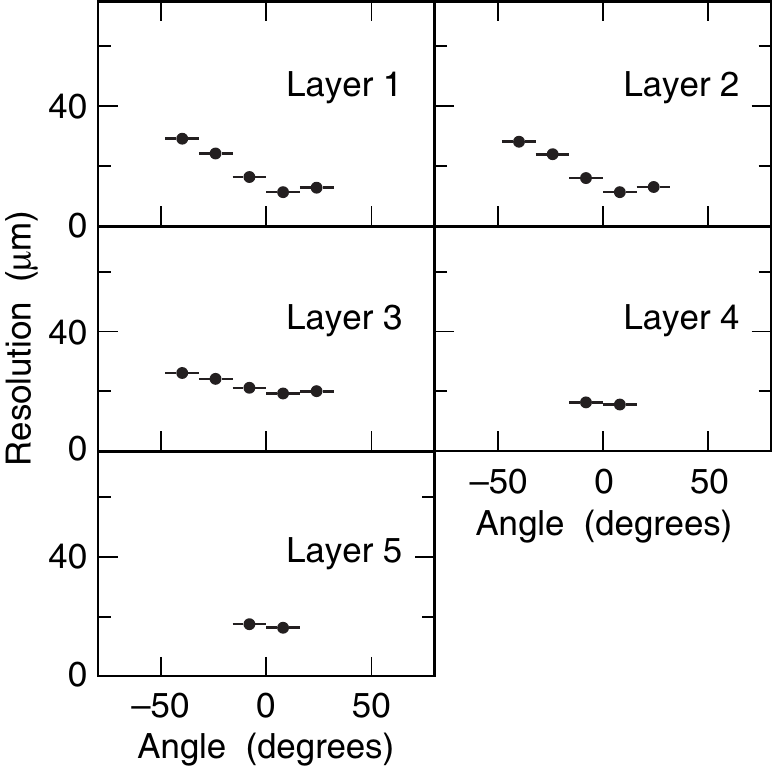} \\
\includegraphics[width=0.45\textwidth]{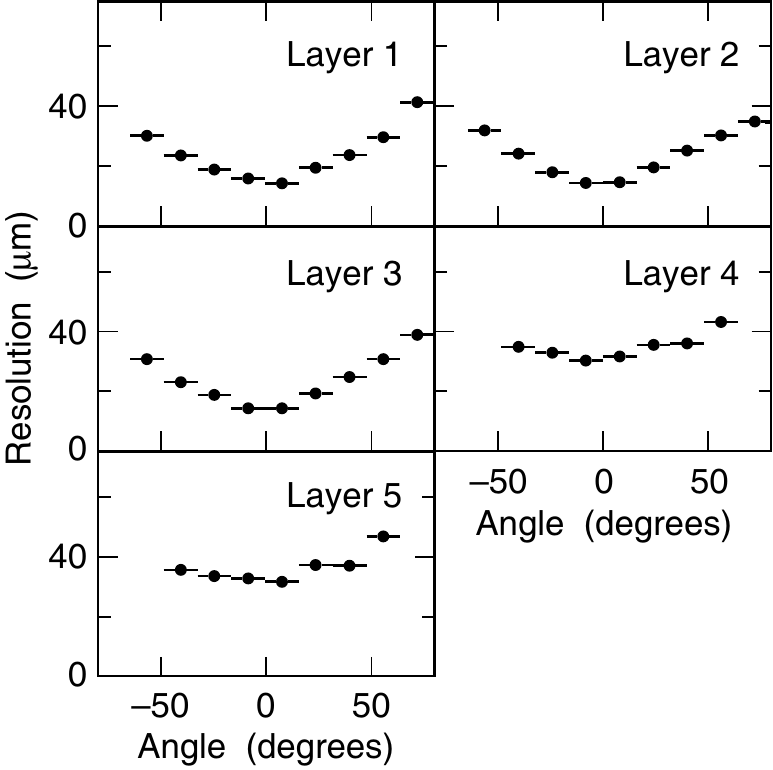} \\
\end{tabular}
\caption{SVT resolution for $\phi$ (upper plots) and $z$ (lower plots) strips as a function of incident angle of the track. $\phi=0^{\circ}$ corresponds to perpendicular incidence. }
\label{fig:hit_resolution}
\end{figure}

Figure~\ref{fig:hit_resolution} shows the $\phi$
and $z$ hit resolutions as a function of the incident angle of the
track with respect to the wafer.  The observed asymmetry relative to normal incidence ($\phi=0$) 
is introduced by  the magnetic field and by 
the tilt of $5^{\circ}$ of the detectors in the inner layers, introduced to create an overlap between adjacent modules.   
The measured resolutions are consistent with the original design goals for perpendicular incidence, $(10-15)~\mum$ for the inner layers, and $40~\mum$ for the two outer layers. 

\subsubsection{SVT Operation and Challenges}
\label{sec:svtops}

During the decade of operation, the SVT running conditions were generally smooth, and, with a few exceptions, most of the problems were not unexpected and dealt with routinely. 
The SVT tracking performance was constantly  monitored and
extrapolated well to operation at higher beam currents and increasing integrated luminosity. 
The primary concerns were detector occupancy and radiation damage.
Increases in occupancy, the fraction of read-out channels with pulse heights exceeding a preset lower limit, were generally instantaneous, usually caused by large beam backgrounds. The occupancy could  be controlled to some degree by adjustment of the pedestals in individual layers. In contrast, radiation damage to the detectors and the front-end electronics was mostly an integrated effect. The main effect was a reduction of the collected charge and an increase in the noise, resulting in lower overall track reconstruction efficiencies. This degradation in performance could not be recovered, except by the replacement of damaged detector module.

During the first few months of data taking it became evident that the actual running conditions were very different from the expected ones.
The rise in luminosity was accompanied by background levels
which were five to ten times higher than those set as the design criteria for the detector.  Including a safety margin,
the maximum integrated dose, the SVT would be able to sustain over its lifetime was estimated to be 2\MRad.  With the much higher than expected radiation exposure,
it became evident early on that this limit was likely to be reached in certain modules of the SVT~\cite{svt-ieee-2002} after a few years of operation, unless special measures were introduced to limit the exposure.

\subsubsection{Radiation Damage to SVT Sensors}
\label{sec:SVT-issues}

There are two different mechanisms of radiation damage to the SVT sensors. The first is caused by instantaneous, intense bursts of
radiation due to beam losses or beam interactions with residual gas and dust particles. Large bursts may induce voltage breakdown of the AC capacitors separating the metal readout strips from the implant, thereby damaging the diodes ({\it p-stop shorts}).
The second mechanism is the accumulation of radiation dose over long periods 
of time which causes damage to the bulk of the highly-depleted silicon and thereby changes the sensor characteristics. Of particular concern were the SVT modules located in the central horizontal plane of the \babar\ detector, {\it i.e.}, the bending plane of the colliding beams, where the radiation exposure was the largest.

\paragraph{P-stop shorts}

On each face of the sensor the readout strips were separated from the bias strips by a thin oxide layer, operating as an AC coupler.  
During normal conditions the potential difference was small.
For large bursts of radiation, the charge
released inside the silicon could induce a discharge, resulting in
a drop of the voltage on the bias line.
The minimum deposited charge per strip required to fully discharge the detector 
was the equivalent of $7 \times 10^{5}$ minimum-ionizing particles (MIPs) at normal incidence~\cite{Bosisio} corresponding to a dose of about 1~Rad in a half module in the innermost layer.  In such incidences,
the bias network restored the voltage with a time-constant of the order $R_{\mathrm{bias}}C_{\mathrm{AC}} \approx $1 ms. During this time, the voltage drop across the AC couplers could amount to a significant fraction (up to 2/3) of the total bias voltage $V_{\mathrm{bias}}$, inducing a breakdown of the oxide layer, especially on the n-side of the sensors which had thinner oxide layers isolating the p-stop blocking strips.

During the construction, all SVT sensors had been tested up to 20~V for AC breakdown.  During the commissioning phase, severe
damage occurred to a SVT module during a beam test designed to evaluate the effect of radiation bursts.  A number of p-stop shorts were produced.
A subsequent study of detectors with a pitch
similar to the inner-layer SVT modules showed that a
few percent of the AC capacitors in each sensor were subject to
breakdown. The expected failure rate, taking into account the number
of sensors in each SVT module, was quite significant. 
Some of the failures were attributed to mechanical damage to the pads during the wire bonding. Repeated exposure to high levels of radiation may also have contributed to oxide damage. Furthermore, there was evidence of
p-stop shorts in modules located above or below the central plane, and also in layer 3, areas of much lower exposure to radiation.

The SVT modules experienced p-stop shorts only when the detectors
were biased. After commissioning in 2001, the SVTRAD system (see Section~\ref{sec:SVTRAD} and Refs.~\cite{Meyer:2000tu,Bruinsma:2004iu,Bruinsma:2007zz}) effectively prevented the beams from delivering dose rates exceeding 1Rad in
less than 1~ms, and no new p-stop shorts occurred.

\paragraph{Detector Bulk Damage}

The second mechanism of damage was the accumulation of absorbed radiation
over time, which damaged the bulk and led to three effects: an increase in leakage current, a change in depletion voltage, and a drop in the charge collection efficiency.
The leakage current caused an increase in noise proportional to the square root of the single channel current. This effect was not expected to result in a large rise of the total noise, even at high doses.
The change in the depletion voltage could affect the operational conditions and the position resolution.

\begin{figure}[htbp!]
\centering
\includegraphics[width=0.45\textwidth]{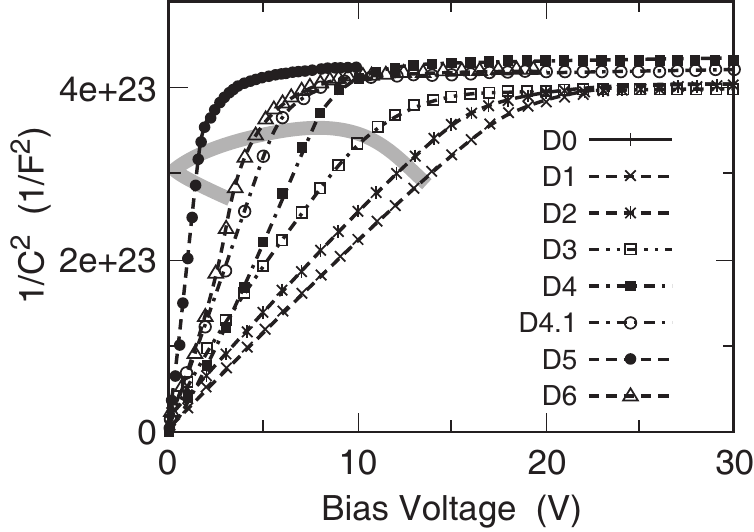} 
\caption{SVT depletion voltage variation under irradiation:  The inverse of the measured capacitance squared ($1/C^{2}$) as a function of the applied bias voltage for eight different
absorbed doses of ionizing radiation D0 to D6. The plateau corresponds to the range in which the detector is fully depleted. The grey band indicates how the radiation damage affects the depletion voltage (knee of the curve). }
\label{fig:cv-curve1}
\end{figure}

\begin{figure}[htbp!]
\centering
\includegraphics[width=0.45\textwidth]{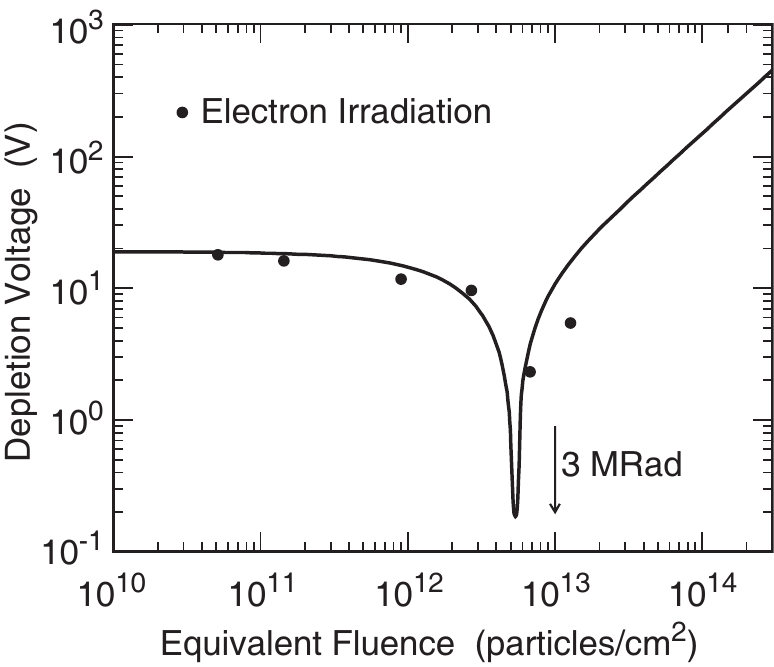}
\caption{SVT depletion voltage variation as a function of the fluence of particles. The minimum corresponds to type-inversion. Points are data, the solid line is the expected behavior calculated according to the NIEL model~\cite{niel}. }
\label{fig:cv-curve2}
\end{figure}

To study bulk damage to the SVT sensors, beam tests were performed at SLAC and at Elettra (in Trieste) with electrons in the energy range
(0.9 - 3)\gev~\cite{calderini:svt-ieee-bulk}. 
Wafer irradiation was carried out up to a total dose of 5\MRad\  using both primary electron beams and  electron beams scattered off a copper target. Detectors were also
irradiated in a non-uniform way to create the high radiation environment of the central plane modules in \babar. The measured increase in leakage
current for the exposed sensors was measured to be about 
2~$\muA$/cm$^2$/\MRad\ at $27~\degrees$C, a result which is fully consistent with the increase detected with colliding beams at \babar.
In test structures produced in the same batches as the SVT sensors,
the measurement of the detector capacitance as a function of the
applied bias voltage showed a shift in the
depletion voltage after each of a series of irradiations 
(see Figure~\ref{fig:cv-curve1}).   This decrease in the
depletion voltages as a function of the equivalent radiation dose
led to a {\it type-inversion} of the SVT detectors at 
about 2-3\MRad (see Figure~\ref{fig:cv-curve2}).
The measurements agree well with the predictions derived from 
estimates of the total non-ionizing energy loss~\cite{niel}.

The measured electrical properties of the sensors (edge currents, inter-strip
capacitance, and insulation between the $p$-strips at low bias voltages)
proved that the detectors can be operated even after
{\it type-inversion}, indicating that this effect would not limit the SVT lifetime as long as the sensor could be depleted. 

A further concern was that the high non-uniformity of the
radiation dose in modules located in the plane of the storage rings could affect the SVT spatial resolution. Since the change in the absorbed dose was located in a very narrow band in the horizontal mid-plane, the effective doping varied dramatically
across the silicon wafer, 
typically over a distance of tens to hundreds of \mum. 
In these regions, the average electric
field would no longer be perpendicular to the wafer and therefore  
the direction of the charge drift and the collection would be affected. 
A simple calculation showed that for a silicon thickness of 300\mum\ 
and a 1 cm-wide transition region, the expected shift in the position of the charge collected on the strips was about 9\mum.   
Thus, the potential effect was small compared to the spatial resolution
of the SVT and therefore of limited concern.

\begin{figure}[htbp!]
\centering
\includegraphics[width=0.45\textwidth]{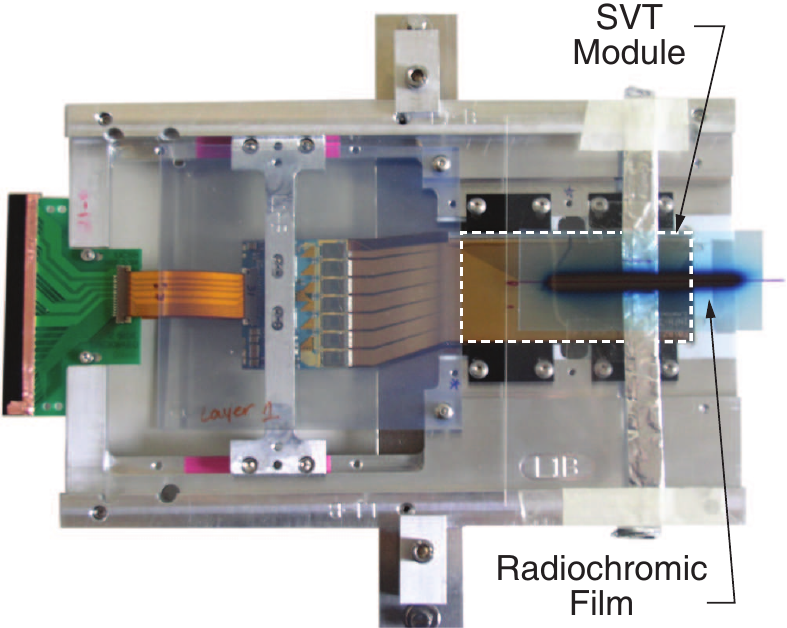}
\caption{SVT layer-1 module (its rectangular contour is marked) after irradiation with an  beam  at Elettra with a radiochromic film placed on the silicon sensors.  The area of the film region exposed to the 
electron beam is dark.}
\label{fig:elettra-module}
\end{figure}

The expected drop in charge collection efficiency (CCE) with radiation damage in the SVT sensors was measured in an electron beam.
An inner-layer SVT module was irradiated in a region
parallel to the $\phi$ strips, far from the front-end electronic circuits
(see Figure~\ref{fig:elettra-module}).
This choice was motivated by the need to reproduce the conditions
of the  \babar\ detector in the bending plane of the beams, 
and to avoid damage to the front-end electronics.
A maximum dose of  9\MRad\ was accumulated in several steps at the
0.9\gev electron beam at Elettra. After each step, 
the module was tested and the CCE was measured with a 1060~nm
light-emitting-diode focused to a spot of 500\mum\ in diameter. The light
penetrated the silicon wafer and generated charge in the bulk.
An identical module which had not been exposed to radiation was used
to correct for any change in the conditions of the experimental setup.
The modules were read out via the standard SVT front-end electronics and the
measured charge generated by a LED was compared with the value 
obtained prior to irradiation.
In the central portion of the irradiated zone, we observed a reduction of the
charge collection efficiency by about $(10 \pm 2)\%$ for a maximum dose of
9\MRad.  This moderate effect was observed for much higher than average doses at \pep2.

\subsubsection{Radiation Damage to Front-End Electronics}
\label{sec:life elec}

Ionizing radiation can cause damage to the readout electronics resulting in a reduction in gain, an increase in noise, or an operational impairment of the circuits.

\begin{figure}[htbp!]
\begin{center}
\includegraphics[width=0.45\textwidth]{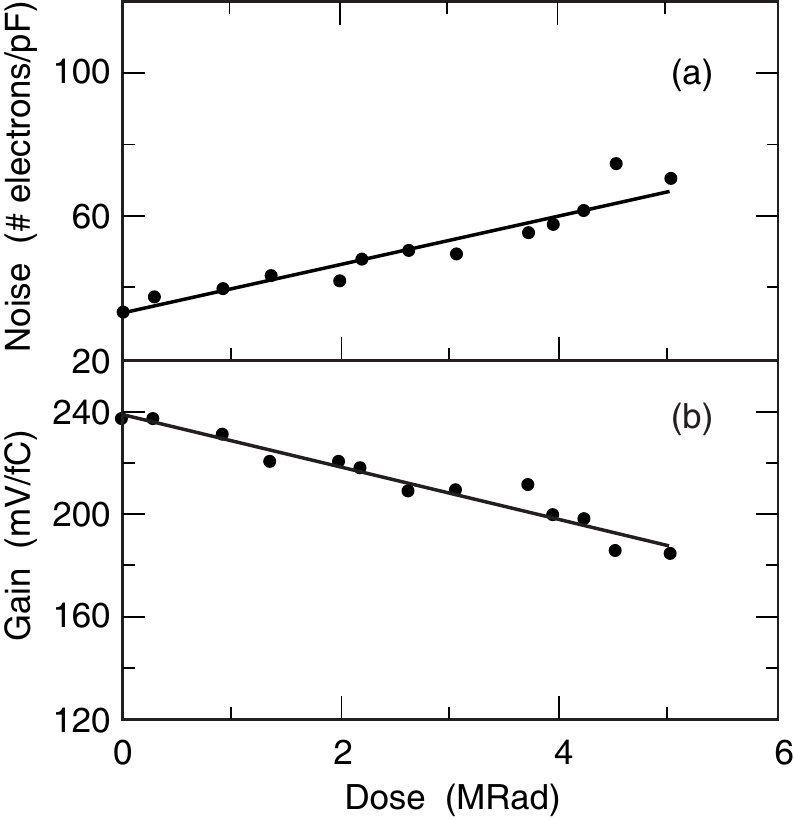} 
\vspace{-0.5cm}
\caption{SVT: Impact on (a) the noise and (b) the gain of the AToM IC as a function of the absorbed dose on an individual channel with an input capacitance of 33\pf .  
}
\label{fig:gain}
\end{center}
\end{figure}

A set of measurements were performed to evaluate the
effects of high radiation doses on the front-end electronics.
The AToM front-end circuits~\cite{svt-ieee-atom} 
and hybrid substrates were exposed to doses of up to 5\MRad
at SLAC and LBNL using $^{60}$Co photon sources.
The Integrated Circuits (IC) were powered during the irradiation and received external clock signals. 
A few identical ones were left unpowered to serve as a control sample.  
After each irradiation cycle the noise and the gain were measured (see Figure~\ref{fig:gain} for results 
for a single channel).

In general, the noise, expressed in equivalent electrons, depends linearly on the capacitive load $C_{\rm load}$, and thus can be parameterized in the form
$N_{\rm ENC}=\alpha + \beta \times C_{\rm load}$.
The two parameters $\alpha$ and $\beta$ were measured to increase as a function of the absorbed dose, by 15\% per \MRad\ and 19\%  per \MRad, respectively. 
The single channel gain decreased linearly by about 3\% per MRad.  These measurements agreed very well with studies based on standard calibrations of the electronics channels of the installed SVT modules.

The ICs which were not powered during the irradiation showed digital
failures starting at a total dose of about 2\MRad, whereas 
ICs powered during the irradiation did not show any digital failures 
up to 5\MRad.  This test was repeated, and this time 
the ICs failed at 5.5\MRad.
A subsequent investigation could not determine whether this failure was caused by radiation damage or by another effect.

\begin{figure}[htbp!]
\centering
\begin{tabular}{l}
\includegraphics[width=0.45\textwidth]{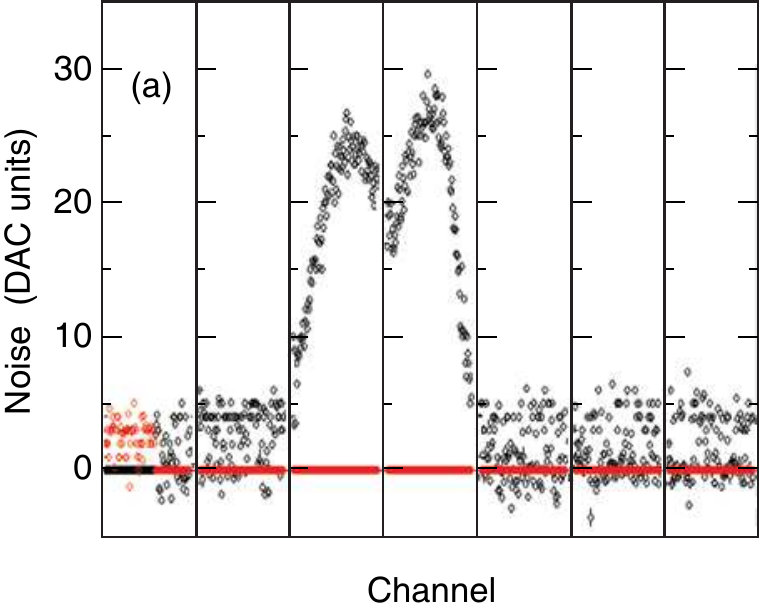} \\
\includegraphics[width=0.45\textwidth]{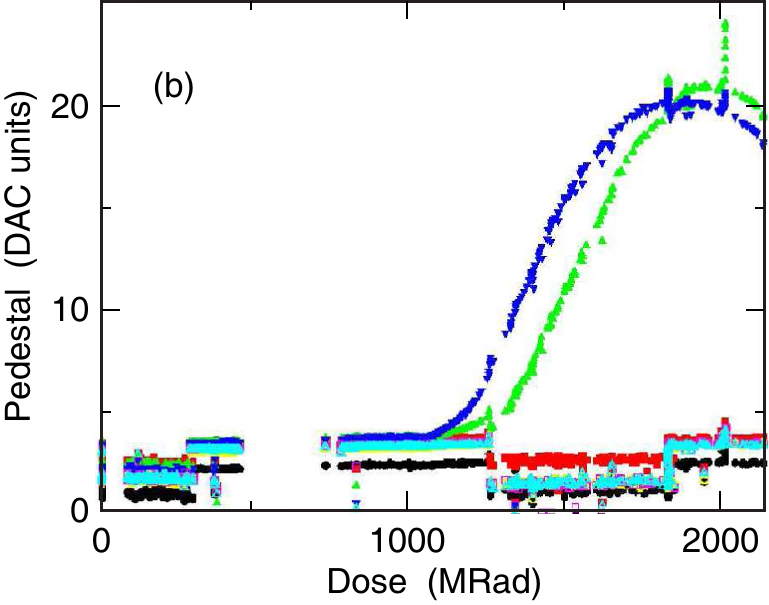}
\end{tabular}
\vspace{-0.5cm}
\caption{SVT pedestal measurements for layer-1 modules: (a) Noise calibration for a module located in the central plane,
(b) variation of pedestals for different ICs as a function of the dose recorded by the nearest PIN-diode.}
\label{fig:calib}
\end{figure}

In general, the results of laboratory studies agreed with experience with the  SVT modules operating in \babar, except for an unexpected effect first observed in 2003.  Some of the ICs in the innermost layer near the central plane showed a large increase in the occupancy with respect to a
reference calibration recorded in December 2002 (see Figure~\ref{fig:calib}).
The effect observed on both $z$ and $\phi$ strips, was comparable
in the forward and backward modules,  and most pronounced for modules
on the inside of the \pep2\ rings, the area most effected by HER background.

Further studies revealed that the effect was not just an increase in noise,
but it appeared to be a shift in the pulse height for a given charge or a change in the IC discriminator.
The effect was first observed in channels close to the central plane and 
was similar for ICs bonded to $\phi$ and $z$ strips, indicating it was
related to changes in the ICs and not to the sensors. 
After an initial increase observed at an absorbed dose of about 1\MRad, the pedestals for the affected channel stabilized and reached a plateau at
about 1.8\MRad, and then decreased at higher doses~(see Figure~\ref{fig:calib}).  
At the plateau, the typical pedestal corresponded to about
half of the dynamical range of the IC (0 to 63 counts, 
with 1 count~=~0.05\fC).

One of the direct consequences of the increase in channel occupancy was
a reduction in efficiency. This can be explained by the single-hit readout which 
only registered the first pulse on a given readout channel. Thus, 
whenever a true signal was preceded by a noise pulse on the same channel, it would be lost.
During track reconstruction, the narrower window imposed on the time-stamp of the hit would often discard the early noise hit, with the result that both the noise and genuine signal were lost.  In some areas, the loss rate reached 50$\%$.

In order to overcome the effect of the pedestal changes, 
the IC electronic calibrations were updated periodically.
Since a common threshold was applied for all 128 channels on an IC, the threshold setting was chosen on the basis of noise calibrations to optimize the average hit efficiency over the 128 channels.
  
Additional studies were performed to monitor the dependence of the observed pedestal shifts versus time and versus absorbed radiation dose, since there was some concern that these shifts had not been observed in irradiation tests with $^{60}$Co sources.  Additional comprehensive
irradiation tests of AToM circuits were carried out at Elettra~\cite{svt-ieee-atom}, to better understand the difference between
radiation damage by high energy electrons and low energy photons.

\subsubsection{Impact of Radiation Damage on Signal-to-Noise}
\label{sec:svt-S-N}

\begin{figure}[htbp!]
\centering
\includegraphics[width=0.45\textwidth]{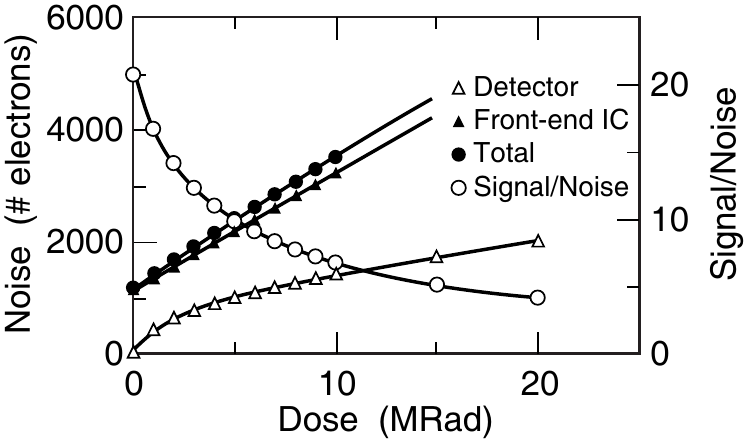}
\vspace{-0.5cm}
\caption{SVT: Expected evolution of the noise level and signal/noise ratio
as a function of the integrated radiation dose.}
\label{fig:svt-sn}
\end{figure}

To develop a better understanding of the SVT performance under exposure to radiation during the \babar\ lifetime, the results of all laboratory and in-situ studies were compiled. 
The signal-to-noise ratio (S/N) for a single readout channel was taken as a
measure of the SVT operability. At the time of installation,
this ratio exceeded 20.  The primary affects that were expected to contribute to the S/N reduction were, an increase in noise due to the leakage current
and electronic noise in the front-end IC, a decrease in the magnitude of the signal due to reduced charge collection efficiency caused by radiation damage in the silicon, and an increase in the noise and a reduction in 
the gain of the AToM circuits.
For 5\MRad\ of integrated dose, the S/N ratio was expected to be reduced to
about 10. The dominant affect was the increase in noise
in the front-end circuit with absorbed radiation
(see Figure~\ref{fig:svt-sn}).

On the basis of these tests and the monitoring of the S/N ratio,
it was concluded in early 2002 that no significant affect on physics analyses was 
expected from the reduction in the detector performance for doses below
 5\MRad. This value was adopted 
as a limit of operability of the SVT.  Based on the projected radiation doses in the \pep2 interaction region, the expected lifetime of the layer-1 SVT modules in the central plane was expected to exceed the remaining period of \babar\ operation. This information was of critical importance, because the replacement of the inner-layer SVT modules with available spares would have required a very long shutdown.
 
\subsubsection{Unexpected Leakage Currents in Layer 4}
\label{sec:leakage}

\begin{figure}[htbp!]
\centering
\includegraphics[width=0.5\textwidth]{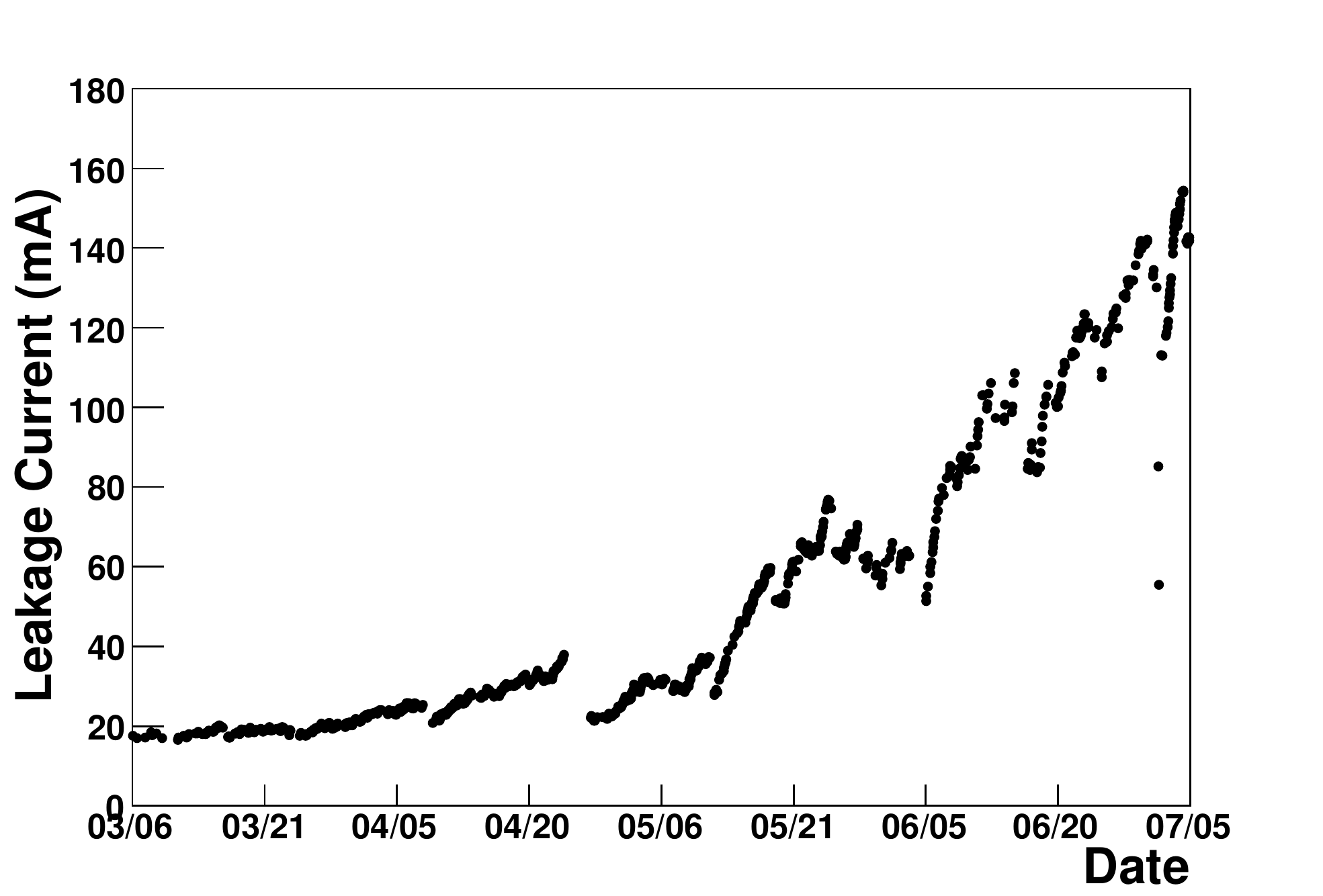}
\caption{SVT: Increase of the leakage current for a layer-4  SVT module 
in the months before and after the implementation of trickle injection to \pep2\ in 2004.}
\label{fig:svt-linc}
\end{figure}

Shortly after the introduction of trickle injection, many SVT modules in layer 4 started to show a very large increase in leakage currents (see Figure~\ref{fig:svt-linc}) and very high occupancies, causing  a significant reduction in the track reconstruction efficiency.  
The effect was dramatic. The leakage current, which had been roughly constant at a few~\muA\ per channel since 1999, rose in some cases by tens of \muA\ per day. 
The bias currents in these modules also increased, probably causing additional noise and an increase in occupancy.  There was no obvious correlation with enhanced levels of radiation in the inner or outer  SVT layers.

This effect was observed in most of the layer-4 modules, although at different levels, while the layer-5 modules, which were located at a  13\mm  larger radius did not show any affect.  In some layer-4 modules, the noise sources appeared to be near the wedge sensors.

It became evident that beam conditions played a critical role: during periods of stable beams or no beam, the current decreased, with and without detector bias voltage applied.
During periods in which the layer-5 bias voltage was off, the leakage currents remained low, with or without beams. 
In other words, the high leakage currents occurred only in the presence of beams with trickle injection, when both outer layers were biased.

\begin{figure}[htbp!]
\centering
\includegraphics[width=0.40\textwidth]{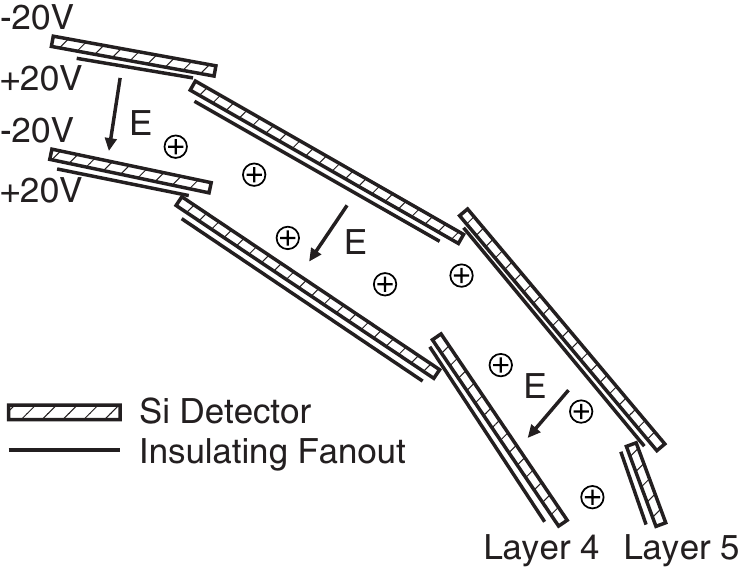}
\caption{SVT: Default bias voltage setting and drift of positive charge in the electric field between the modules of the layer 4 and layer 5 }
\label{fig:svt:layers-sketch}
\end{figure}

The fact that the leakage current increase did not depend on the position
of the modules, the power distribution, cabling nor on the readout configuration, indicated that it was due to a local problem in the SVT external layers.  Further investigation revealed a difference in the configuration of the detector modules in the inner and outer layers.  
For the two outer layers, the $z$ strips were on the $n$-side of the sensors (set to +20~V) and fully covered by a UPILEX fanout~\cite{upilex}, while the $p$-side (set 
to -20~V) with the $\phi$-strips was fully exposed. Thus, positive charge induced during the trickle injection would drift in the electric field between the detector surfaces of layer 4 and layer 5 towards the uncovered $p$-side of layer  (see Figure~\ref{fig:svt:layers-sketch}).  This explained why the current build-up only occurred when the sensor bias voltage in layer 5 was turned on.  

For the three inner layers, no current build-up was observed, because the configuration  was inverted: the $z$-strips were on the $p$-side and the 
$\phi$-strips on the $n$-side, with positive voltage, were exposed.

Microscopically, the charge accumulation on the passivated surface of the sensor could explain the large change in the leakage current.
The field induced on the tip of the $p+$ implants by the accumulated positive charge can create a 
junction breakdown which was responsible for the increase in current.  
Simulations of the field configuration confirmed that the observed effect could be explained by surface charge build-up and subsequent junction breakdown in layer 4.

\begin{figure}[h]
\centering
\includegraphics[width=0.45\textwidth]{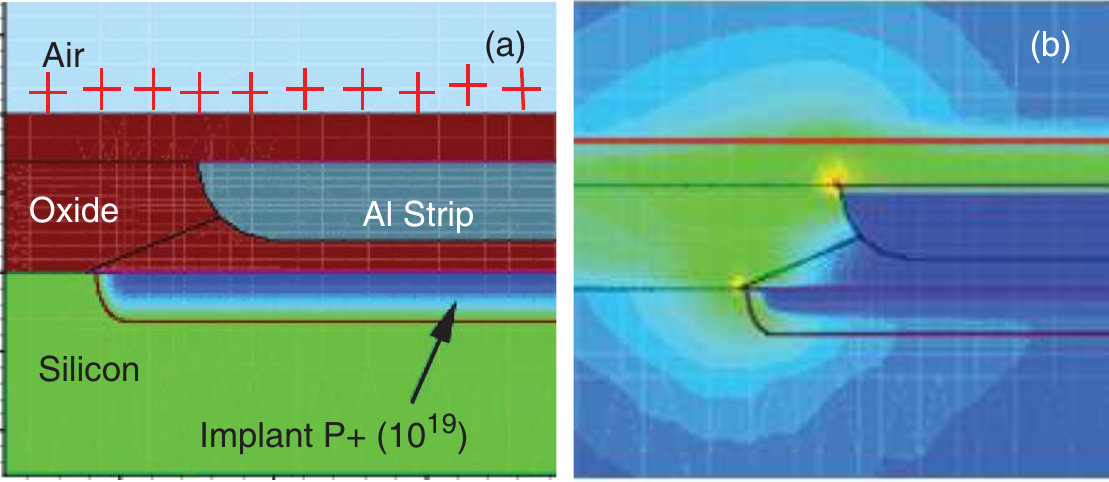}
\label{svt:simu2}
\caption{ 
SVT:  Simulation of the electric field configuration inside the junction region of a silicon detector, in the presence of accumulated positive charge on the external insulating surface: (a) geometrical features of the device in the edge region of a strip implant, and (b) the electric field map, indicating in yellow and orange the high induced fields at the tip of the Al pad and the p+ implant.
} 
\end{figure}

A laboratory test was performed to reproduce the conditions of the SVT operation at \pep2.
A module identical to the ones used in the SVT outer layers was exposed to a strong electric field and charge was generated by an ionizer in the dry-air environment surrounding it. The increase in leakage current observed in the SVT was reproduced in a qualitative way.

To decrease the charge accumulation,  the bias voltage reference was changed.  
The bias voltage supplied to the front-end electronics had an electrical reference ground which was set for every module by the CAEN~\cite{CAEN} power supply, to the common shield potential, through a soft ground. 
To modify the intermodule electric field, the reference voltage of the layer-4 modules was changed so that all its internal potentials were shifted by the same value, typically +30~V. Thus, the field
configuration internal to the module remained unchanged, but the voltage difference across the gap between layer 4 and layer 5 was significantly reduced, 
thereby suppressing the accumulation of the induced charge on layer 4, and
reducing the leakage currents and occupancy in layer 4 to levels observed prior to trickle injection.

In an effort to further reduce the charge deposition, the humidity of the air flow around the SVT  was increased from typically 10\% to 16\%.
This led to a reduction of the accumulated charge,
and after some time the current stabilized at a level reduced by 35\%.

\subsubsection{Summary}
In summary, the SVT performed reliably during the decade of operation.   As the detector closest to the beam, the primary task was to monitor the beam-generated radiation and its effect on the performance, with increasing beam currents and luminosity. 
The increase in noise due to the leakage currents was much smaller than the increase of the electronics noise. 
The observed shifts in the depletion voltage and the bulk damage modestly 
affected the position resolution.  The increase in occupancy and readout dead-time was controlled by frequent updates of the pedestals and a reduction in the readout time window. The surprising impact of trickle injection on the leakage current in the outer two layers was effectively remedied by a shift in the layer-4 bias voltage setting.

\subsection{Drift Chamber}
\label{sub:opsdch}

\subsubsection{DCH Operations}

The reliable operation of the Drift chamber (DCH) required a stable gas mixture at constant temperature,
stable voltages, calibrated electronics, and robust offline
calibration and reconstruction code.  To achieve these goals, routine operation of the DCH included:
\begin{itemize}
\item daily calibrations of gains and pedestals to be used by the feature-extraction algorithm;
\item online monitoring of operational parameters, including
sense wire voltages and currents, electronics voltages, gas flow and
mixture, temperatures and pressures;
\item gain correction for pressure and temperature variations, 
performed during the first pass of the offline event reconstruction.
\end{itemize}

The DCH was conservatively designed and required little maintenance.
Much of the work on the drift chamber was software-related, including alignment, time-to-distance calibration, \dedx\ calibration, pattern recognition, and simulation.

The time-to-distance relations were originally parameterized as a
function of distance of closest approach only, with corrections
applied for azimuth  and dip angles. One relation was derived
per layer, with an additional correction for each sense wire.
This procedure was later replaced
by a fit to the two-dimensional distribution of the distance of closest approach versus azimuth angle, with a correction for dip angle.
This change resulted in an increase in the average number of DCH hits 
associated with low momentum tracks by 15\%. The typical
single-cell hit resolution did not change significantly;
it remained at 100\mum for tracks in the center of the cell.

The \dedx\ performance was substantially improved by increasing the number of parameters of the calibration function and using a 
wider range of control samples of charged leptons and hadrons.
This led to  a significant improvement in the overall performance of particle identification (see Section~\ref{sec:cpid}).

The pattern recognition efficiency was improved by reducing the number of required hits per segment in a superlayer from four to three.

\begin{figure}[htbp!]
\includegraphics[width=\columnwidth]{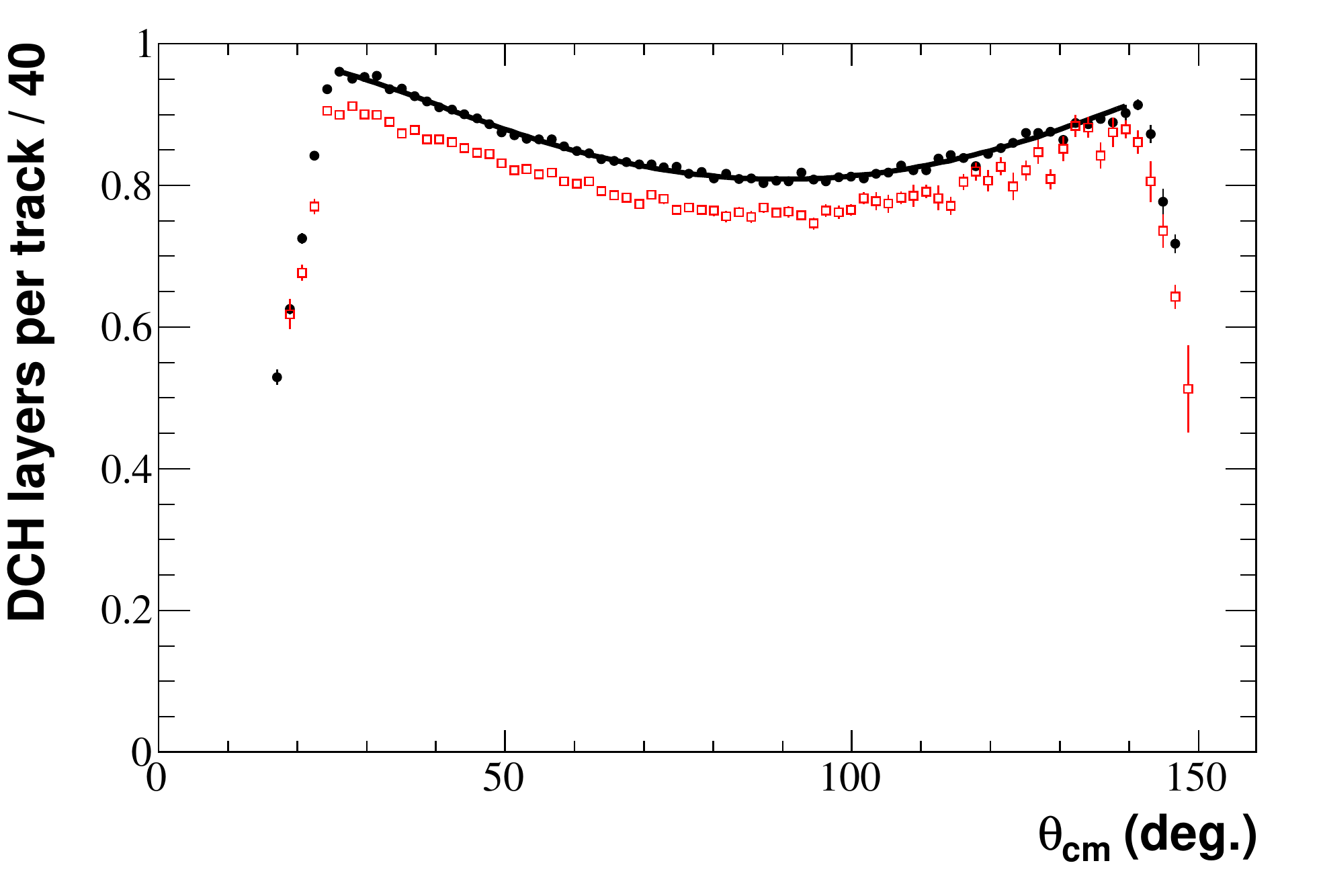}
\caption{DCH data from Run 6:
The mean fraction of DCH layers used in the reconstruction of high-momentum tracks as a function of the polar angle $\theta$ measured relative to the $z$-axis in the 
c.m.~frame, separately for isolated muons (dots) and pions from $D^0$ decays (open squares).
}
\label{fig:layereff}
\end{figure}

Despite the substantial work on calibration and pattern recognition, the probability for a particular layer to
be included in the fit of a high-momentum track at vertical incidence
($\theta_{cm} = 90^\circ$) was 80\% for tracks in dimuon events, and lower for tracks in multi-hadron $B \overline B$ events.
Figure~\ref{fig:layereff} shows data from Run 6.
The hadron efficiency was somewhat higher in earlier 
Runs which had lower beam backgrounds and lower luminosity,
leading to lower pattern recognition losses.

The reduction at $\theta_{cm} < 25^\circ$ and 
$\theta_{cm} >140^\circ$ corresponds to tracks exiting the DCH through the endplate.
The muon data are fitted to a function 
$\epsilon(\theta_{cm}) = \epsilon_0 [1 - (1 - \epsilon_1)^{1/\sin\theta_{cm}}]$,
with $\epsilon_0 = 0.978$ and $\epsilon_1 = 0.827$.

Closer inspection of tracks with missing hits 
revealed that there were hits in the layer, but they were not
close enough to the trajectory to be included in the fit. 
The underlying cause of this inefficiency is not completely understood.
A possible explanation may be that the charge deposited near 
the point of closest
approach to the sense wire is too small to pass
the relatively high discriminator threshold of 2.5 electrons.

\subsubsection{Incidents}
\label{sec:DCHincidents}

There were two notable incidents during the DCH lifetime.  The first~\cite{Aubert:2001tu} occurred shortly after the start of
operation when a voltage of 2000~V was mistakenly applied to guard wires, which normally
operated at 340~V; operational changes were introduced to protect the chamber from further damage.

The second incident occurred during the summer shutdown in 2002, during
the removal of the support tube from within the drift chamber. A bolt head, which extended unexpectedly far beyond the profile of the support tube,
dragged against the DCH Beryllium inner cylinder, causing a 210\mm long
crack. Although the cylinder lost gas tightness, it did not lose structural
integrity. Taking into consideration the potential safety
hazards, the crack was  filled with epoxy. A
50\mum thick aluminum foil was then glued over the region and
electrically connected to the cylinder.  The incident and its repair had no impact on the amount or quality of the data recorded subsequently.

\subsubsection{Radiation Effects}
\label{sec:DCHradiation}

\begin{figure}[htbp!]
\includegraphics[trim = 0mm 0mm 0mm 18mm, clip, width=\columnwidth]{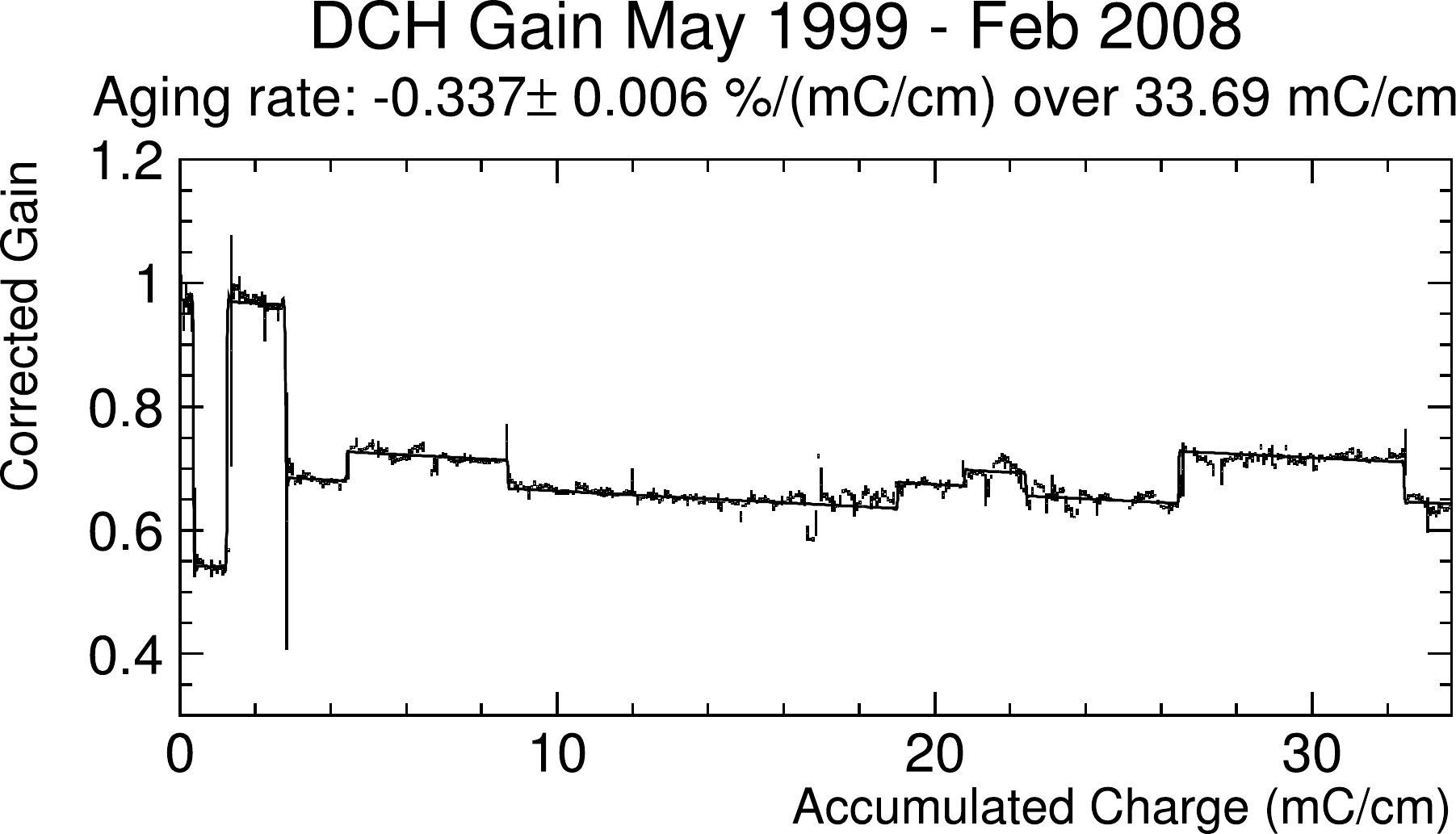}
\caption{Relative DCH
gain (corrected for temperature and pressure variations) as
a function of accumulated charge on the wires. The steps correspond to changes in operating voltage;
the curve is a fit to the reduction in gain, giving $\delta G/G = (0.337 \pm 0.006)\%$ per mC/cm.
}
\label{fig:dchgain}
\end{figure}

A variety of actions were taken to respond to aging of the
DCH. The most obvious manifestation of aging was a decrease 
in signal gain. Figure~\ref{fig:dchgain} shows the
gain relative to its initial value
as a function of the average accumulated charge
per centimeter of sense wire. The steps in the curve come from
changes in the operating voltage: a 10~V change in voltage corresponded to a gain change of 9\%. The large variations in the first
3~mC/cm are due to changes in the HV setting following the guard wire incident.
The DCH voltage was set to 1930~V at the start of Run 2 (${\sim} 3$~mC/cm). At the start of Run 6 (${\sim}
27$~mC/cm), the voltage was increased to 1945~V to accommodate the reduced gain. 
At a fixed voltage, the gain dropped $(0.337 \pm 0.006\%)$ per mC/cm.
A total charge of 34~mC/cm was accumulated over the lifetime of the experiment. 
 
Radiation damage effects on the electronics were small, but not negligible.
Three front-end electronics assemblies, all of them in locations closest
to the beam line, failed and were replaced during Run 6. The problem
was caused by radiation damage to 2.5~V voltage regulators.   Prior
to Run 7, all 1.8~V and 2.5~V regulators were replaced on all front-end assemblies.

The large FPGAs installed during the electronics upgrade
(Section~\ref{sec:electronics:dchupgrades}) were subject to
single-event upsets by neutrons. 
These upsets could potentially cause configuration errors. It was found that reconfiguring the FPGA at the start of every run (${\sim} 1$ hour duration) and whenever an
error was detected was sufficient to reduce the problem to negligible
levels. Reconfiguration took 7~s (during which data acquisition was suspended) and was performed automatically.

\subsection{DIRC}

\subsubsection{Overview}
\label{sec:dchover}

The DIRC, a ring-imaging Cherenkov detector of unique design, was the primary detector system designed to separate charged particles of different masses.  Its components, fused-silica radiators, the water-filled 
{\it stand-off box} (SOB), and PMTs were
relatively conventional, and the system proved to be robust and caused very little \babar\ down time.

The DIRC front-end electronics, HV 
distribution, and data transmission were mounted on the back side of the SOB, all enclosed by a large magnetic shield supported by the detector end doors.  Therefore, access to the crates or individual PMTs required the opening of the detector end doors, usually creating several hours of down time.  Thus, the reliability of components inside the shield was of great importance, and crate power and cooling were closely monitored. 

It was crucial to retain as many of the Cherenkov photons as possible for charged tracks traversing the fused silica bars as possible. This required that the bar boxes remained clean and dry, the water in the SOB stayed transparent, and the PMTs and front end electronics retained their full efficiency throughout the ten years of operation. Since the efficiency of the reconstruction of the Cherenkov image was sensitive to the background photon rate, it was essential to control these backgrounds 
as luminosity increased.

In the following, various procedures associated with the long term operation at increasing luminosity are presented, complemented by some lessons learned

\subsubsection{Calibration of PMT Timing}
\label{sec:dirccal}

The  calibration of the PMT time response and delays introduced
by the electronics was performed in two steps: 
the online calibration with a conventional light pulser and 
an offline calibration based on reconstructed tracks
in colliding beam data.

\begin{figure}[htbp!]
\begin{center}
\includegraphics[width=0.46\textwidth]{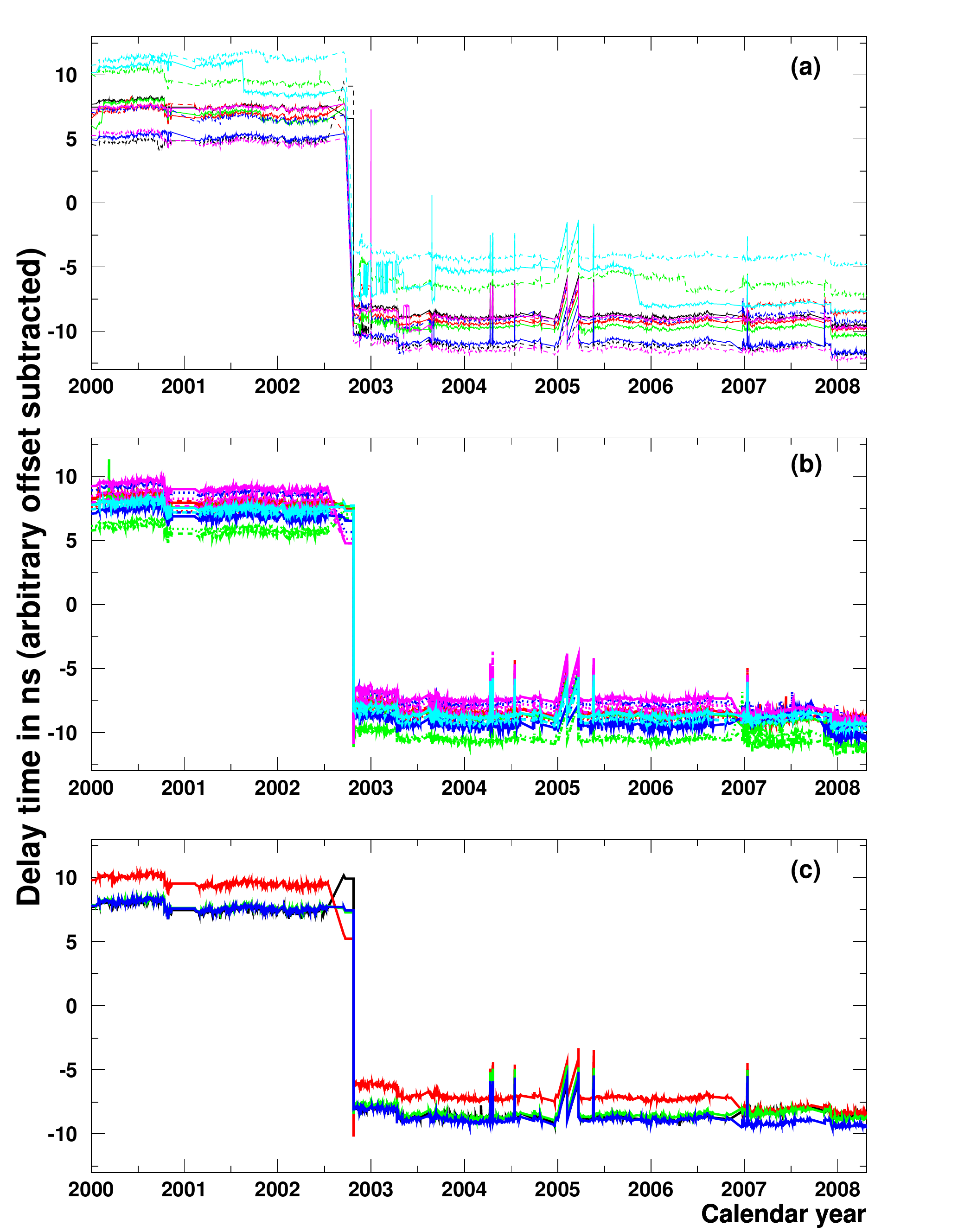}
\end{center}
\vspace{-0.5cm}
\caption{DIRC online calibration:  Time delays of the PMT signals as a function of calendar
time averaged over (a) the 12 DIRC sectors, (b) the 14 DFBs of a single DIRC sector, and
(c) the 4 TDC ICs of a single DFB.}
\label{fig:drc_calib}
\end{figure}

The DIRC online calibration~\cite{Adam:2004fq} was part of the regular global calibration of the detector.
Precisely timed light pulses of $1\ns$ duration were generated by 12 blue LEDs, one per sector.
Figure~\ref{fig:drc_calib} illustrates the stability of the response time during the entire lifetime of the experiment for different elements of the readout.
The chart shows strong correlations between the various elements and good overall stability.  
The steep drop at the end of 2002 corresponds to the TDC upgrade,
which is described in Section~\ref{sec:electronics:drc_fee}.

The offline calibrations used reconstructed tracks from collision data  
to determine the differences between the measured 
and expected photon arrival times.  Initially, these offline measurements were
averaged over all PMTs to determine a global time delay for each run.
From 2003 onwards, the time delays were determined per sector,  
to account for sector-to-sector differences in the online calibration hardware,
and to correct for instabilities that affected the LEDs.
Channel-by-channel data stream calibration was also implemented, 
but never applied routinely because the improvement with 
respect to the existing calibration was insignificant.
The overall resolution of the arrival time of photons was $1.7 \ns$.
It was dominated by the $1.5 \ns$ intrinsic transit time spread of the PMTs.

\subsubsection{Monitoring and Protection System}
\label{sec:dircmon}

Apart from the standard tools to monitor low and high voltages, cooling, fan speed,  and front-end electronics, several special diagnostic tools were available online 
to monitor the key elements of the DIRC PMTs, the SOB water system, the instantaneous background rates, and the occupancy of the PMTs. After several years of improvements and upgrades, the configuration of the monitoring system remained almost unchanged during the last three years of data taking.

\begin{figure}[htbp!]
\begin{center}
\includegraphics[width=0.46\textwidth]{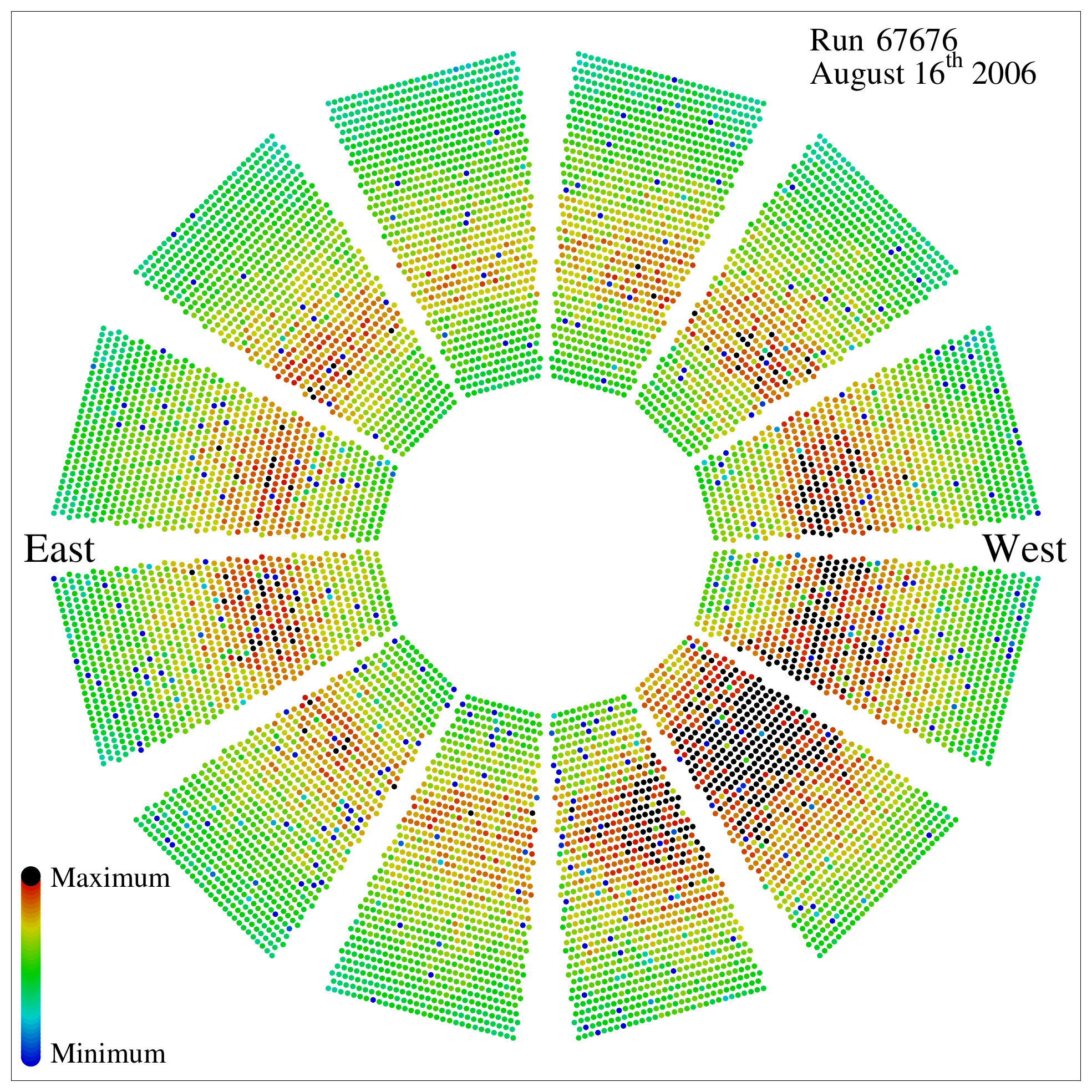}
\end{center}
\vspace{-0.5cm}
\caption{DIRC occupancy map recorded during high-luminosity colliding beams in 2006.  This cross-sectional view shows the 12 sectors viewed from the North side of the SOB. The rate recorded in each of the 10,751 instrumented PMTs is displayed by a color code (see bottom left corner).}
\label{fig:drc_occupancy}
\end{figure}

\paragraph{PMT Occupancy}
One example of DIRC monitoring tools is the occupancy map shown in Figure~\ref{fig:drc_occupancy} for a representative high-luminosity run. In this plot, most hits are due to the background.
This pattern is typical of the running conditions after the installation of the DIRC shielding in 2003. The high rate region appears as a broad ring. The background is dominated by the HER beam and is most prominent in the bottom right sectors (see Section \ref{sec:mdi:background-monitoring}). During this particular run, single PMT rates were stable, ranging from 180-190~\khz on the east side to 250-310~\khz in the lower part on the west side. The average DIRC occupancy of 11\% was stable.

\paragraph{Slow Monitoring}

The DIRC hardware status was displayed in the \babar\ control room.
Most critical were the ${\rm N}_2$ circulation system, designed to keep the bar boxes dry, and the purification system for the closed-circuit water system for the SOB. 

Since water leaking from the SOB to other components of the \babar\ detector was a serious concern, a hardware-based protection system was installed. If an increase in humidity was detected in more than one sensor, an automatic dump would be initiated, emptying the 6,000-liter SOB in a few minutes. Fortunately, all seals remained watertight over the ten years of operation, and such action was not necessary.

In 2007, a water supply line for one of the \pep2\ magnets inside the DIRC tunnel broke, and water leaked into the lowest bar box slots.
Some moisture penetrated into the bar box through a pressure relief valve and the sensors detected an increase in humidity.
The problem was mitigated by increasing the $\rm{N}_2$ flow
through the bar box from 60 to 100 $\cmc/\rm{min}$.

Groups of 16 neighboring PMTs were connected to a single HV channel, and all voltages and currents were continuously monitored. Each group had two nominal voltage settings, one during data taking (ranging between $1000$ and $1400\volt$) and the other during extended periods of beam injection (300\volt\ lower). 
Whenever the current in a given channel exceeded a safety threshold, in 
the range $2 - 2.5$~mA, tuned individually for each channel, the voltage was automatically lowered to 10\volt\ and
the shift operators and DIRC experts were notified to assess the problem.  Persistent trips in the same
channel usually indicated a PMT problem resulting in either a very low or very high rate.  Until the single PMT could be unplugged from the HV distribution (an action which required a maintenance period during which the backward end doors were opened), 
the HV for the 16 PMTs powered by the same channel were kept at 10\volt.
In practice, having 16 out of a total of 10,751 PMTs turned off had
no noticeable effect on the overall DIRC performance.
Only about 30 PMTs (0.3\%) had to be unplugged during the entire data taking period.

The proper operation of the 12 DIRC front-end crates was among the 
requirements for data taking.  
Therefore,  voltages, temperatures, cooling fan speeds and status, and the quality of the optical connections to the DAQ system were closely monitored. 
Given the high number of front-end boards (14 per sector,
168 in total) and the fact that each of them collected data from
a set of PMTs aligned radially in the SOB (a pattern perpendicular
to the Cherenkov rings), a failure of part of a board,
or even a full board, had only minor consequences for the DIRC
performance, and repairs were delayed until the next scheduled maintenance. 
On the other hand, a failure affecting a whole sector or its readout 
required immediate intervention, interrupting the
data taking for several hours. Such problems were very infrequent.

\paragraph{Data Quality}

The online monitoring of the DIRC data included distributions of occupancies and signal timing as well as information on the DAQ status.
A frequent observation was the distortion of the timing
distributions caused by poor trickle injection, a condition 
 which did not impact the quality of the DIRC data.
In addition to the weekly checks by the DQG, a number of detailed studies of the
DIRC detector performance were performed offline. This included the number of
photons per track as a function of time (see Section~\ref{sec:dircage} below) and
the quality of the reconstruction algorithm as a function of the background level.
The alignment of the fused-silica bars, which affect the Cherenkov
angle resolution, was also monitored. Look-up tables updated
on a regular basis were used to correct the angle measured for each
photon, leading to ${\sim} 10$\% improvement in the Cherenkov resolution
per track.

\subsubsection{Impact of Beam Background and Aging of Components}
\label{sec:dircage}

\begin{figure}[htbp!]
\begin{center}
\includegraphics[width=0.46\textwidth]{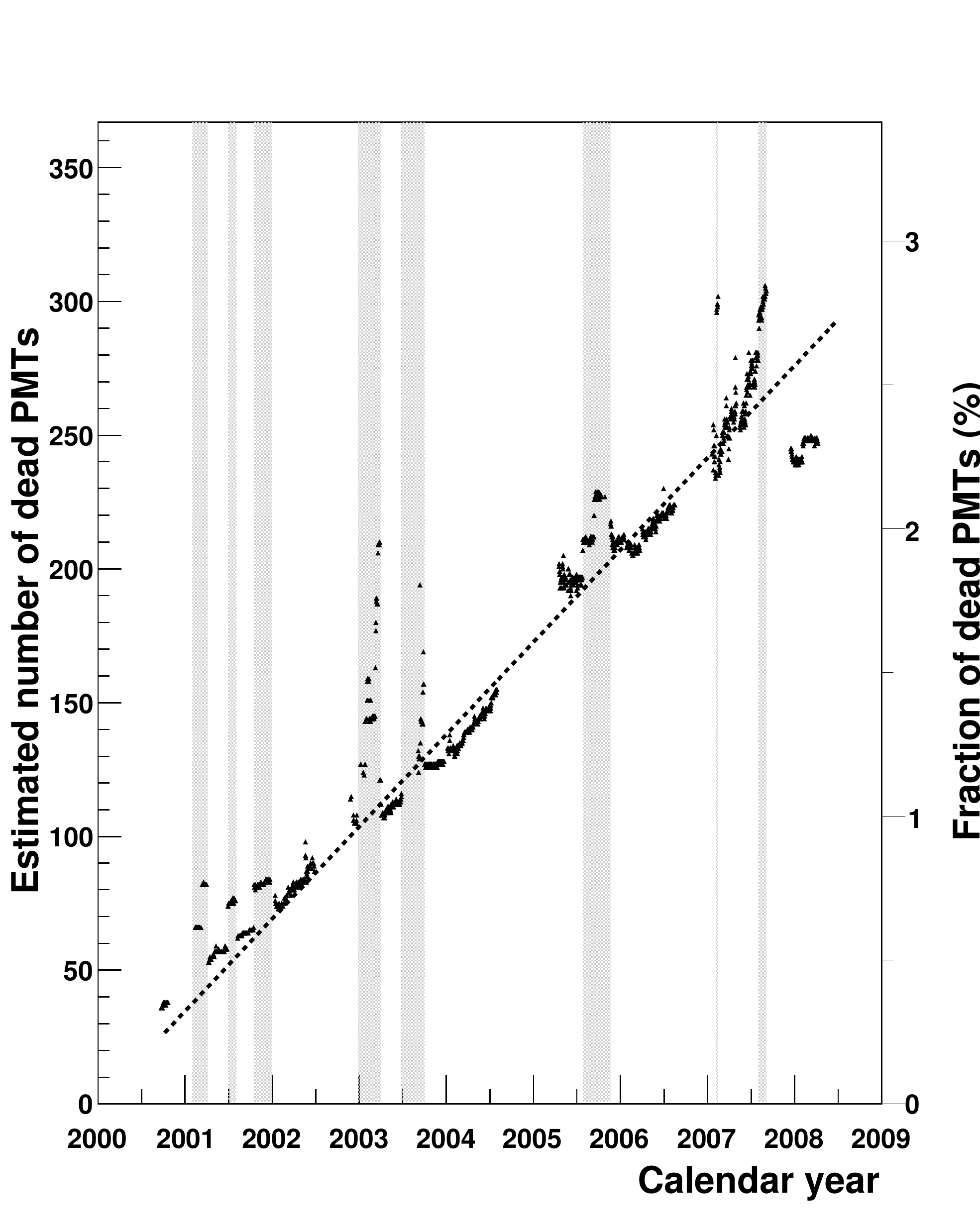}
\end{center}
\vspace{-0.5cm}
\caption{DIRC PMT failures over the lifetime of the experiment.
Left scale: the number of dead PMTs; right scale: the corresponding
fraction, in percent of the total number of PMTs.
The gray bands correspond to periods with known hardware problems.
They are excluded in the linear fit; the result is superimposed on the data.}
\label{fig:drc_dead}
\end{figure}

The DIRC detector performance was continuously monitored for signs of aging components,  
some of which were enhanced by temperature and exposure to ionizing radiation. 
Some aging effects could lead to a gradual reduction in performance, while others limited the lifetimes of components. 

\paragraph{Photomultipliers}

\begin{figure}[htbp!]
\begin{center}
\includegraphics[width=0.46\textwidth]{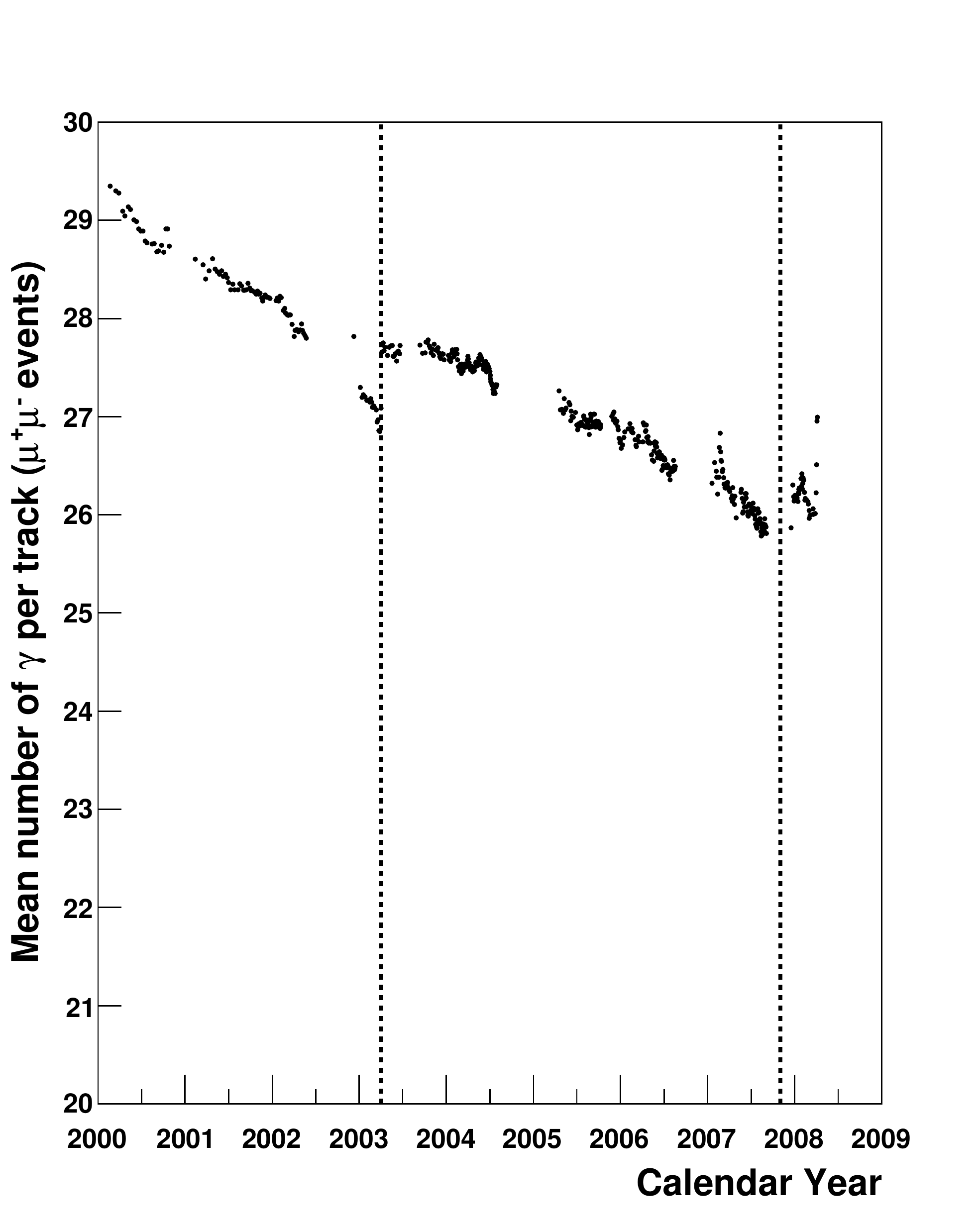}
\end{center}
\vspace{-0.5cm}
\caption{DIRC:  Evolution of the mean number of detected photons per track
in di-muon events.
The two vertical dashed lines mark the times when the PMT HV settings
were raised. Note that the vertical axis is zero-suppressed:
the loss over the whole data taking period is only around 10\%.}
\label{fig:drc_photon}
\end{figure}

Since the PMTs were difficult to replace,
it was important to keep their failure rate very small.
Figure~\ref{fig:drc_dead} shows the evolution of the number of dead 
PMTs over the lifetime of the experiment. 
A dead PMT is defined as a PMT with a counting rate of less than 10\% of 
the average rate of the 64 PMTs connected to the same DIRC front-end board.
Periods with known problems are marked by vertical bands. Excluding these data, the  failure rate, approximately constant as a function of time,
was measured to be 2.8 PMTs per month.  
At the end of \babar\ running, more than 97\% of the PMTs were still 
operational.
The reduction in the number of dead PMTs in the spring of 2003 and the fall of 2007 occurred after
the high voltage settings were increased~-- see below for details.

The likely cause of most of the PMT failures was a gradual breakdown of the vacuum inside the phototubes, causing a glow discharge and light emission. Such discharges may have been triggered by current spikes caused by high instantaneous \pep2\ backgrounds.  These light flashes affected nearby PMTs because their glass enclosures 
were transparent.
It is believed that the reason for the vacuum failure was the unnecessarily thin front windows of the PMTs.  The failure rate further increased due to the corrosion 
of the glass by the ultra-pure water in the SOB.
Since individual PMTs were not easily accessible, the corresponding high voltage channel had to be disconnected until the faulty PMT could be unplugged.  This and other temporary FEE-related problems explain the loss of several PMTs at the same time.

One effect of PMT aging was the reduction of the dynode gain with exposure. 
This effect was mitigated by raising the HV.
Initially, before the PMTs were mounted on the SOB in 1998,
the high voltage for each PMT was set to achieve a single photo-electron signal of $20\mv$.  

Raising the HV was done twice during the decade of \babar\ operation.
In 2003, an ADC calibration revealed that the mean
single photo-electron signal had decreased to $15\mv$, 
implying a 25\% loss of gain in five years.
To restore the signal to $20\mv$, the HV settings were raised 
on average by 40~V in April 2003.
The exact value of the increase was decided individually for each high voltage distribution channel, \ie\ group of 16 PMTs. This procedure was repeated 
in the fall of 2007, when the gain had dropped again by 25\%.  

The effect of the high voltage change on the number of detected  photons per track is clearly visible in Figure~\ref{fig:drc_photon}, showing the result of an offline study of $\mu^+\mu^-$ pairs. The gradual decline in the mean number of
photons per track (less than 2\% per year) had no significant effect on charged particle identification. Its origin remained unknown.

\paragraph{Water System}

The DIRC water system was designed to maintain good transparency at wavelengths as small as 300 nm.  Any significant change in the transmission would have impacted the DIRC performance. 
The DIRC water system is described in detail in an earlier publication~\cite{Adam:2004fq}.  In the following, the focus is on the experience 
with the system over the lifetime of the experiment.

The water purification  was based on filtration with mechanical, charcoal, and resin filters,  reverse osmosis, degassing bubbles by pumping them through a membrane,
and removal of bacteria with UV light of 254\nm\ wavelength.
Water with a resistivity close to $18~M\Omega \cm$
is very aggressive chemically, and therefore most plumbing components were made of stainless steel or polyvinylidene fluoride (Kynar). 
The main concern was the long-term stability of
the water transmission, and, therefore, parameters such as water resistivity,
pH-value, and bacteria levels in the SOB supply and return lines were monitored regularly. Furthermore, the rate of leaching of sodium or boron from the PMT windows
was monitored as a measure of the PMT glass removal rate.

\begin{figure}[htbp!]
\begin{center}
\includegraphics[width=0.46\textwidth]{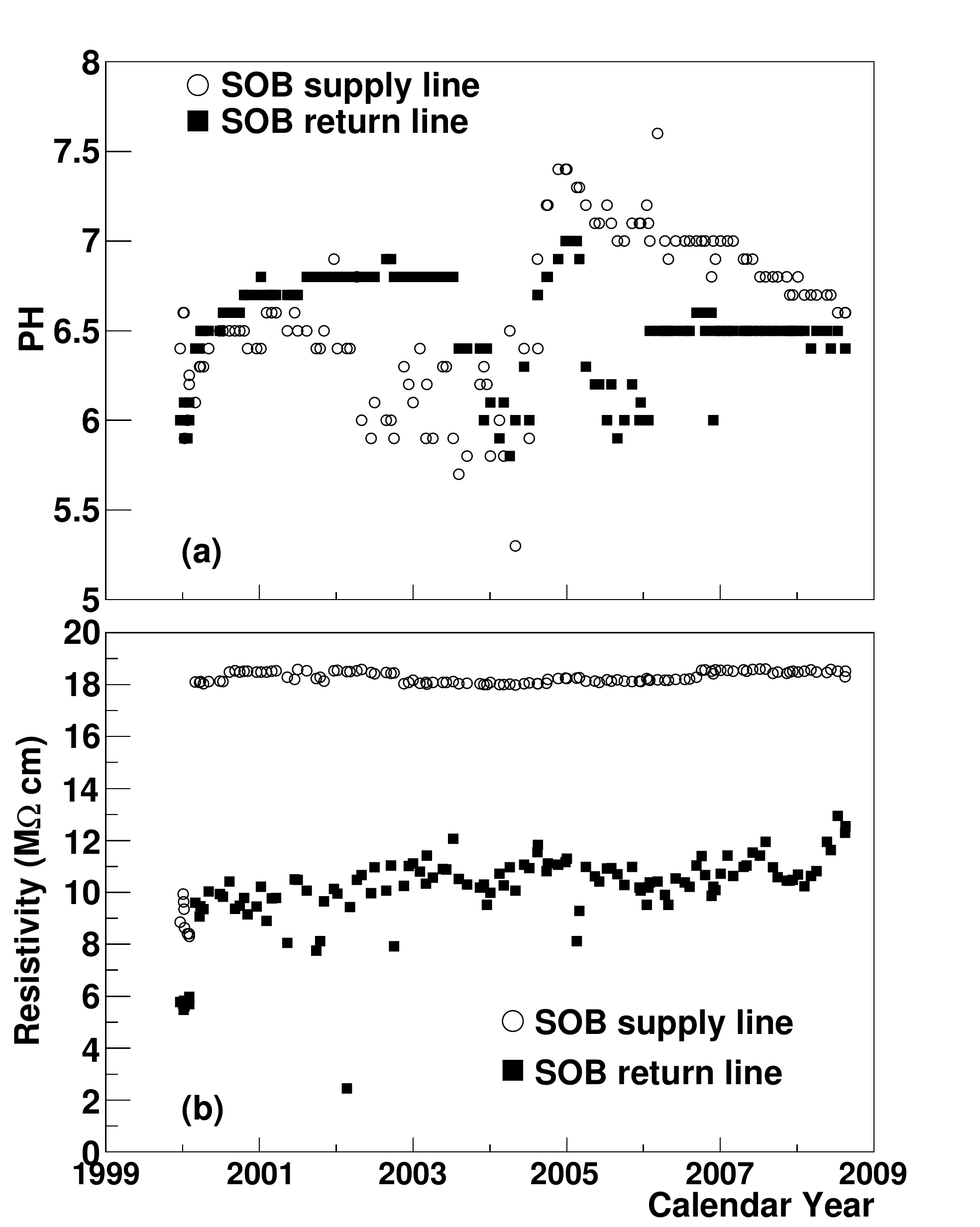}
\end{center}
\vspace{-0.5cm}
\caption{DIRC water system performance over the lifetime of the experiment: 
a) comparison of the pH-value and b) 
resistivity of the water in the SOB supply and return lines.}
\label{fig:drc_water}
\end{figure}

Figure~\ref{fig:drc_water} illustrates the performance of the purification system. 
The pH-value of the supply water was usually close to $6.5$, and 
the difference between pH-values of the supply and return water was kept
to less than 1. The resistivity of the water was typically $18~M\Omega \cm$
in the supply line and $8-10~M\Omega \cm$ in the return line
at a temperature of about $23-26^{\circ}$C.

\begin{figure}[htbp!]
\begin{center} 
\includegraphics[width=0.46\textwidth]{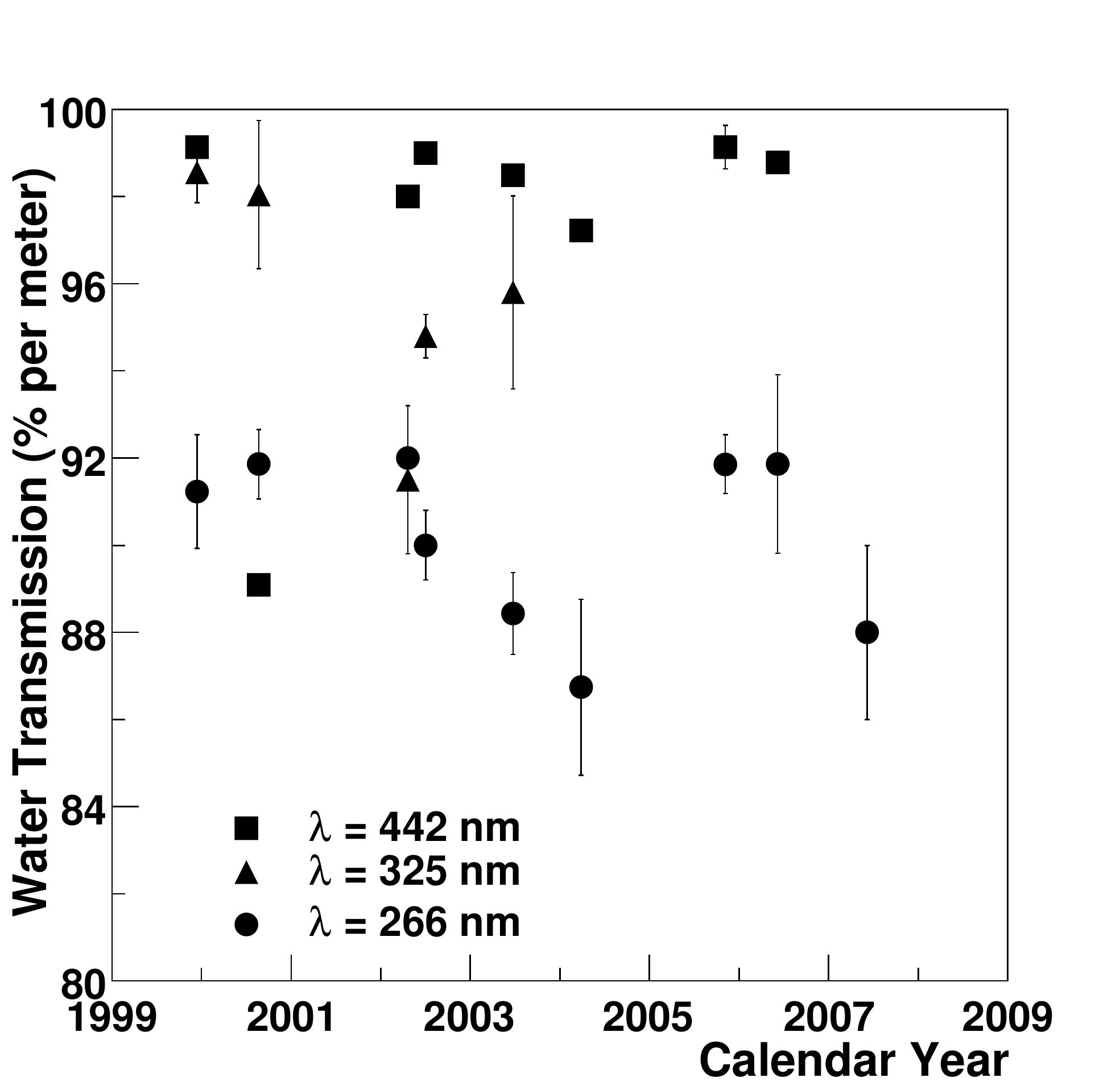}
\end{center}
\vspace{-0.5cm}
\caption{DIRC: Periodic measurements of the transmission of the water in the SOB at three wavelengths.
The vertical axis is zero-suppressed in this graph.
}
\label{fig:drc_trans}
\end{figure}

The water transmission was measured periodically in a one meter-long cell,
which was comparable in length to the optical path in the SOB. As experience was gained,
these measurements were performed less often, typically 
at the beginning and at the end of each running period, because one could rely on automatic measurements
of water resistivity and pH-value during data taking.
Figure~\ref{fig:drc_trans} shows the water transmission at three laser wavelengths.
The measured photon absorption length at $442\nm$ was about 90\m .

\subsubsection{Maintenance and Operational Issues}
\label{sec:dircmaint}

\paragraph{DIRC Maintenance}

Various maintenance and preventive maintenance activities were performed by the DIRC team during shutdowns between periods of data taking. These
were the only times when the DIRC front-end electronics and the PMTs could be easily
accessed, because these components were located inside the \babar\ shield doors. The \babar\ {\it Hot Spares}
policy was applied by the DIRC: spares of electronic boards which had been validated in test stands were installed during shutdowns, when they could be fully tested in situ, replacing boards which were known to function well.  These boards could be trusted and were retained as replacements, in case of a later failure during data taking. Bad PMTs were individually disconnected from HV during these shutdowns. As a result, the 15 other PMTs powered by the same HV channel were recovered after each disconnection.

\paragraph{Hardware Problems}

Failures of hardware components during data taking were rare and random. In almost ten years of data taking,
about 10 crate controller boards (12 were needed to run the detector: one per sector)
and 30 front-end boards (168 were running in parallel in the DIRC) had to be replaced.
Of the six front-end crate power supplies (powering two DIRC sectors each), only one failed. 
A few HV boards failed over the years, but they were readily fixed.
The HV crates had few problems, and one spare crate was sufficient to deal with these isolated problems.

\begin{figure}[htbp!]
\begin{center}
\includegraphics[width=0.46\textwidth]{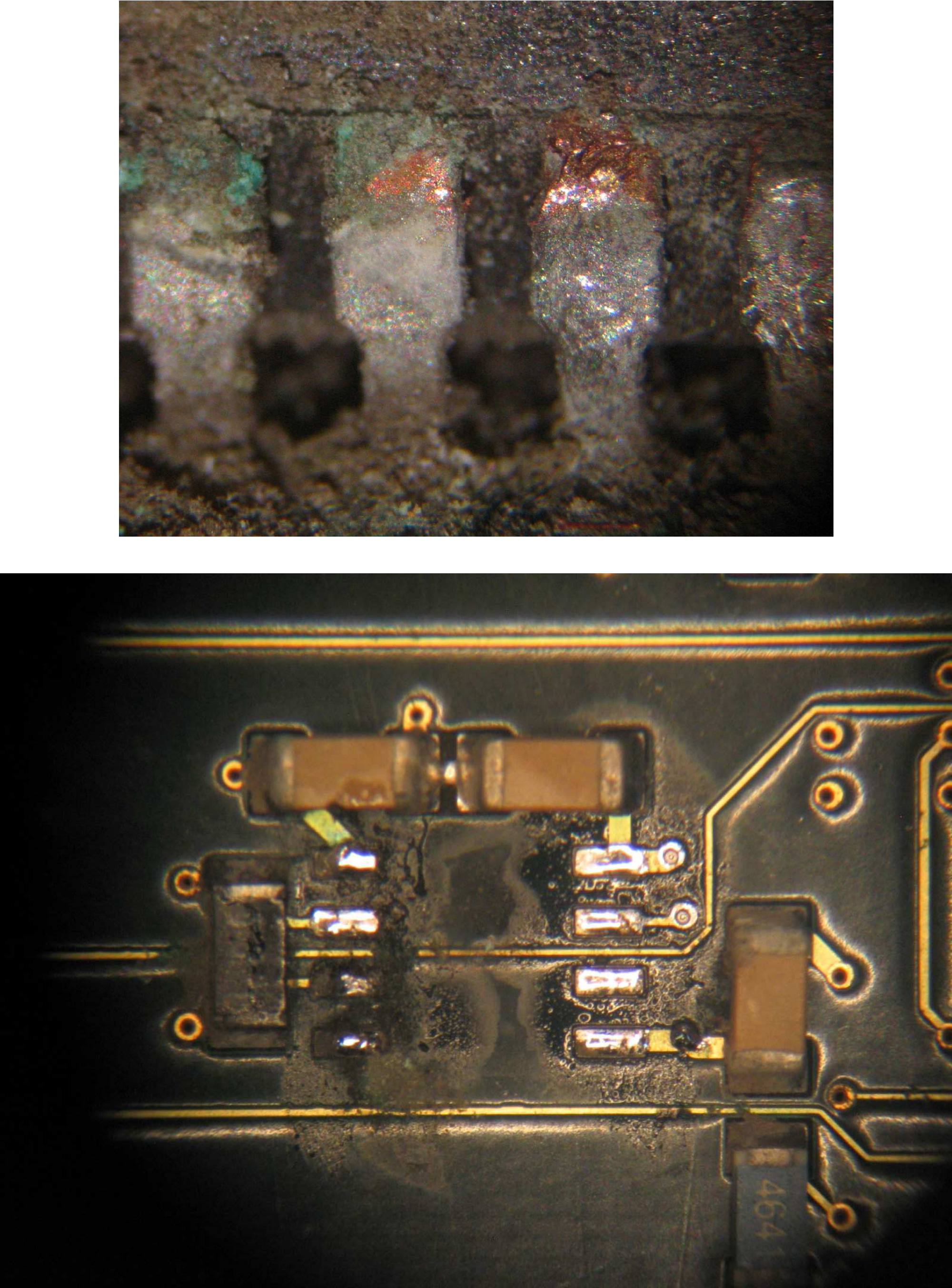}
\caption{ Damage to the DIRC 
front-end boards during the 2006 shutdown, dust, corrosion and open traces.}
\label{fig:drc_dust}
\end{center}
\end{figure}

In 2004, a data quality problem appeared that was correlated with beam background.
The problem was eventually traced to a failure of the reboot of the DIRC ROM
master crate. During the reboot, the TDC ICs received a wrong clock signal
whenever the input rates were unusually  high.
As a result, all times were shifted by $34 \mus$, and only random background hits were recorded.
A change in  the reboot sequence avoided the problem, without introducing
dead-time.

Four fans died in the DIRC front-end crates between the end of 2001 and mid-2005.
Such failures impacted data taking as they caused an automatic
shut-off of the crate. Six of the nine fans per tray could be bypassed through EPICS, whereas the remaining three could not.
If any of these three fans had failed, the \babar\ backward doors would have had to be opened to replace the broken components. 
Studies in 2005 indicated that the fans had nearly reached their lifetime. Therefore, they were all replaced during the fall shutdown.
No new failures occurred during the remaining two and a half years of operation.

Since the \babar\ detector hall was not equipped with air conditioning or filters, dust was always a concern. As a precaution, the fan trays were 
removed from the front-end crates and cleaned during each shutdown.
Fan speed monitoring was added to EPICS and checked daily for fan weakness.

\paragraph{Dust Contamination of Front-end Electronics}

An especially threatening situation arose during the 2006 shutdown, when nearby construction in the detector hall generated substantial dust mixed with vehicle fumes.
During this time, following the usual 
shutdown protocol, the DIRC electronics were fully powered,
both to monitor their status and to allow  online-related tests to proceed
(calibrations, front-end configuration, etc.).
With the backward end doors wide open, dust settled on the DIRC front-end electronics.
The boards, and thereafter crate components, started to fail at an increasing rate.

The failed electronics boards were found covered with dust and showed signs of corrosion,  as illustrated in Figure~\ref{fig:drc_dust}.
Since most of the 180 boards had been exposed to the dust,
they were all removed and cleaned commercially. 
Subsequent tests indicated that about 25\% of the boards failed and had to be repaired. The repairs included preventive maintenance: jumpers were added on all boards to bypass potentially weak traces, and the boards closest to the fans were painted with a conformal coating.  Preventive maintenance was also performed on some components of the front-end crates.
After burn-in, the repaired components
worked well until the end of data taking 16 months later,
with a failure rate similar to that before the shut down. 

\paragraph{Inspection of the SOB}

After the end of data taking, the ultra-pure water was drained and the SOB was kept dry by circulating nitrogen. 
Some months later, the SOB was opened for the first time in many years. 
The goals of this opening were threefold. First, to proceed with the disassembly of the \babar\ detector;
second, to access, extract, and store the DIRC fused silica bars
for future use; and
third, to inspect the PMTs and quartz glass windows.

\begin{figure}[htbp!]
\begin{center}
\includegraphics[height=0.46\textwidth,angle=90]{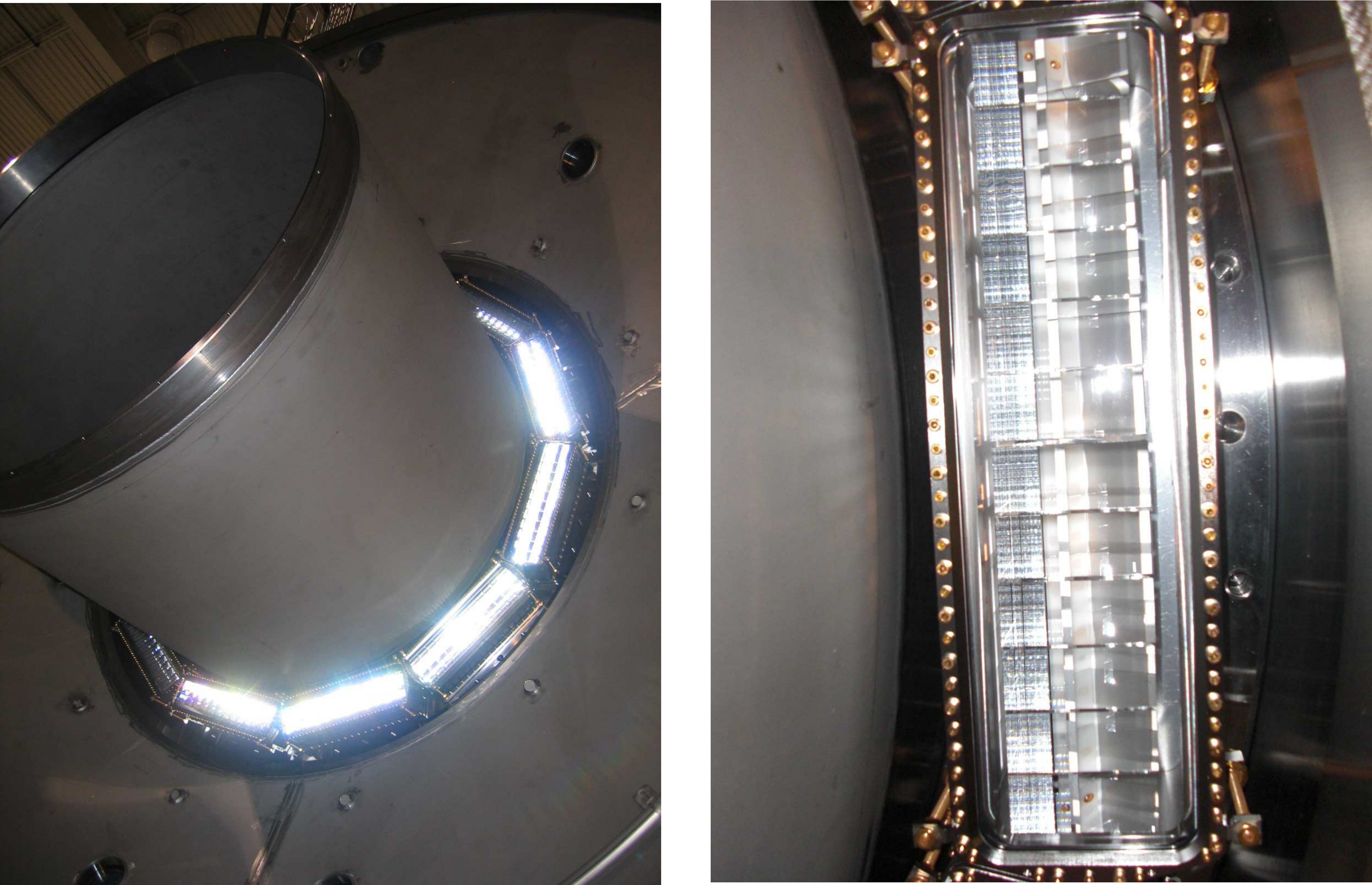}
\caption{DIRC SOB: Inner cylinder and the windows of several of the 12 bar boxes (left), and
the face of a bar box showing the ends of 12 fused-silica bars (right).}
\label{fig:drc_barboxes}
\end{center}
\end{figure}

As shown in Figure~\ref{fig:drc_barboxes}, the ends of the bar boxes 
protruding into the SOB were very clean.  The fact that they are perfectly illuminated
by a flash light is an indication of the quality of the bars,
protected for over ten years by the bar boxes and the ${\rm N}_2$ flow. 
A careful inspection did not reveal any significant aging effects.
The Epotek 301-2~\cite{epotek} epoxy joints connecting the four equal-length fused-silica pieces
that make up the $4.9\m$-long DIRC bars had been of particular concern. The joints did not show any sign of yellowing although
this particular epoxy turned yellow in laboratory tests after
an exposure greater than $10^{19}-10^{20}$ photons per \cma\ at 350~nm
(with a threshold around $5 \times 10^{18}$).
This level is hard to reach in an actual experiment such as \babar.
However, this problem should not be underestimated when preparing an experiment
for which a high light background is anticipated.
Furthermore, data-MC comparisons based on di-muon events also showed
no evidence for degradation of the optical properties of the Epotek joints. 

\begin{figure}[htbp!]
\begin{center}
\includegraphics[width=0.46\textwidth]{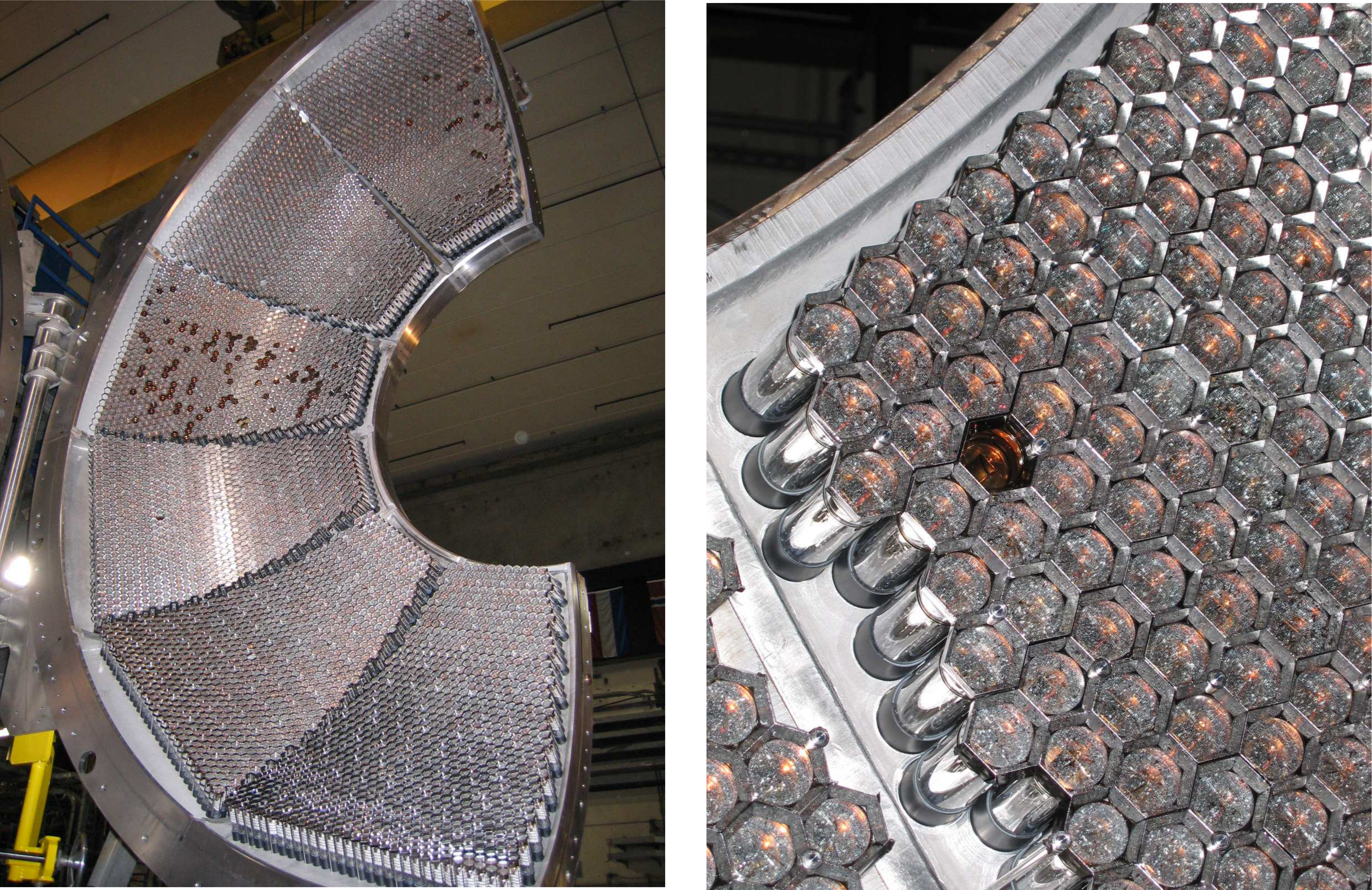}
\caption{DIRC SOB after completion of data taking,
left:   full half of the SOB (six sectors); right: a small section of the walls lined with PMTs, one of them with a glass window still transparent.}
\label{fig:drc_pmts}
\end{center}
\end{figure}

A great surprise on opening the SOB was the appearance of the entrance windows of the PMTs lining its back wall~(see Figure~\ref{fig:drc_pmts}).
Most of the PMT windows appeared white or {\it frosty} due to fine surface cracks, except for about 2\% that remained transparent.
By correlating the location of the few remaining transparent PMTs in the SOB
with their serial numbers, it was found that almost all of them came
from the very first production batch.
Already in 2000, after about a year of operation, some 50 PMTs with these white surfaces had been discovered.  This abnormal appearance was traced to a difference in the chemical composition of the glass
in a single production batch.

While a slow but continuous and uniform decrease of the DIRC photon yield had been observed as a function of time, there was no evidence of a correlation of this effect with the degradation of the PMT front windows.
Some 20 frosty PMTs were removed from the SOB and their photon detection efficiencies in air were measured and found comparable to that of a new tube with a  transparent window.  Therefore, the decrease in photon yield
is more likely due to PMT dynode aging, expected for an estimated accumulated charge of about 150~C in ten years of operation.

\subsection{Electromagnetic Calorimeter}

\subsubsection{Introduction}
\label{sec:emcintro}

The electromagnetic calorimeter consisted of 6580 Thallium doped
Cesium Iodide [CsI(Tl)] crystals. It was divided into a barrel section composed of 48 rings with 120 crystals 
each and a forward endcap composed of 8 rings with 80-120 crystals. The 
crystals, extending from 15.8$^{\circ}$ to 141.8$^{\circ}$ in polar angle, 
were indexed from 1-56 counting from the innermost ring of the endcap, and from 0-119 in
azimuth angle, counting from the horizontal plane.
 
Each crystal was read out by a pair of PIN diodes, connected to a charge sensitive shaping preamplifier. 
These preamplifiers had two integrations of 250~ns and one differentiation of 500~ns.  To avoid pile up due to 
beam background, the shaping times were chosen to be shorter than the scintillation decay time of CsI(Tl) which 
has two components, a fast signal of 680$\,$ns (64$\%$ of the pulse) and a slower signal of 3.34\mus\  duration (36$\%$ of the pulse).
This resulted in a photo-electron yield of $\approx$3900 photo-electrons/MeV as compared to $\approx$7600~photo-electrons/MeV
expected for a shaping time of 2\mus. The choice of shaping time reflects an optimization between photon statistics and beam background effects
on the calorimeter resolution.

The uniformity of the light collection efficiency along the length of the crystals had been adjusted  
by selectively roughing or polishing the crystal surface to reduce or increase reflectivity. 
The adjustment was originally performed with a PMT-based crystal  readout. 
The geometrical difference between photodiodes and PMT readout introduced a small systematic 
non-uniformity, which for photons below 1 \gev worsened the resolution by up to 0.3\%, while for high energy photons the resolution improved slightly~\cite{Hryn'ova:2004sm}.
The light yield uniformity of the crystals changed during the life of the experiment as described in Section \ref{sec:emcage}.

The key requirement for EMC performance was the stability of the energy response
across the full energy range of 1\mev to 10\gev.
To maintain this stability, it was necessary to perform precise calibrations
of crystal light yield and uniformity, because radiation damage was expected 
to change the physical characteristics of the CsI(Tl) crystals.
A study of the effects of high levels of absorbed radiation doses on the 
EMC is presented below.
Instant bursts of radiation during data taking with stable beams, causing high background conditions,
were less of a concern for calorimeter data quality.
The increase in the number of crystal signals per event had a modest effect on
the dead-time of the detector, and beam-background photons overlapping with
showers from physics processes led to a slight increase in the measured
photon energy,  but no significant effect on the energy resolution of neutral particles was observed.

\subsubsection{Routine Operation and Maintenance}
\label{sec:emcops}

During routine operation the EMC required very little maintenance. The Fluorinert chillers for crystal 
cooling and the water chiller that regulated the temperature of the front-end
electronics were continuously monitored and the fluid levels had to be adjusted on a regular basis.
Since temperature variations influence the light yield in CsI(Tl),
the EMC temperature was controlled within $\pm 1\degc$.

For the daily, centrally performed, \babar-wide calibration, a light-pulser system was operated to check each 
channel of the EMC for electronics problems. The light pulser system was also used to monitor the transparency of the crystals
which were affected by radiation damage, as well as the coupling of the 
photodiodes to the crystals.

To ensure precise linearity and accurate pedestal subtraction, a set of 
calibrations of the EMC electronics~\cite{Aubert:2001tu} was performed to determine pedestals, 
gains, and linearity of the readout chain, from the 
PIN diodes to the ADCs, after any hardware intervention.

Electronics repairs were carried out on a regular basis. Front-end failures occurred primarily on the Input-Output-Boards (IOB)
and the ADC-Boards (ADB). Many of the malfunctions were traced back to bad connections between the boards:
since each IOB connected to six ADBs in the barrel and four ADBs in the endcap, poor alignment and expansion with temperature caused bad contacts.
Repairs were necessary about twice per month. 

On a few occasions, water leaks in different cooling systems caused
damage to the EMC front-end electronics, and in some cases also to the bias voltage circuitry for the PIN diodes.
The preamplifiers and PIN diodes, located inside the EMC enclosure, proved to be highly reliable, which was crucial, since they  were inaccessible. Only three of the 6580 crystals could not be 
read out, because both readout chains were permanently damaged soon after the start of the experiment.

\subsubsection{Crystal Light Yield Calibrations}
\label{sec:emclight}

The calibration of the EMC was performed in two steps: 
first, the charge-to-energy conversion for the single crystals (this section)
and second, the photon energy calibration of the EMC clusters (see Section~\ref{sec:emccal}).

To obtain a precise energy calibration for the crystals, two sets of calibrations were performed 
periodically, based on electrons and positrons from Bhabha scattering and photons from a radioactive source.
The radioactive source calibration used photons at 6.13\mev, while the calibration with electrons from Bhabha events covered the opposite end of the energy spectrum
at energies of several GeV. The results of the two procedures were interpolated.
Initially, both sets of calibration constants were determined every 2-4 weeks.
Bhabha events were accumulated during normal data taking using a calibration trigger stream  
to record them at a preselected rate of typically 1~Hz. The energies deposited in the crystals
in Bhabha events were compared to MC predictions to derive a calibration constant for each crystal~\cite{mueller:thesis}.
During the latter part of the experiment, the procedure to determine calibration constants 
from Bhabha events was fully automated.

The radioactive source calibration was based on 6.13\mev photons from decays 
in the neutron-activated fluoro-carbon Fluorinert (FC-77), which was pumped through tubes mounted in front of the crystals.
A detailed description of this system is given elsewhere~\cite{Aubert:2001tu}.
This source calibration required a down time of about two hours.
The data acquisition relied on a cyclic trigger to record the maximum number of decays possible.
The photon spectrum for each crystal was fitted to extract the calibration constants.

\subsubsection{Digital Filter}
\label{sec:emcdf}

The EMC online readout software extracted the time and magnitude of energy deposits from the waveforms recorded for individual crystals.  
A time-domain digital filter was applied to waveforms from the lowest gain amplifier (range maximum equal to 25\mev) to suppress electronics noise and signals from other beam crossings, the combination of which will be referred to simply as noise.
For each raw waveform, a filtered waveform was generated
by convolving the channel raw waveform with a set of filter weights.  The energy deposit was 
estimated from the maximum of a parabolic fit to the highest filtered waveform sample and its
two neighboring samples. 

The filter was constructed to optimize the signal-to-noise ratio, assuming a typical CsI(Tl) signal shape and beam background that is random in time.  The calculation of the filter weights
required the ideal signal shape and noise auto-correlation function for each crystal.
The signal shape was extracted from raw waveforms from cosmic rays recorded 
without beams, while the noise characterization data were recorded
during typical colliding-beam conditions.  
The derivation of the
filter weights $\bf{w}$ from noise waveform data $\bf{n}$ and ideal signal waveform ${\bf s}$ follows from optimal filter design theory:
\begin{eqnarray}
 a(k)   & = & < n_l n_{l+k} > , \nonumber\\
A_{ij}    & = & a( |i-j| )  , \nonumber\\
{\bf w} & = & A^{-1}{\bf s}/{\bf s}^T A^{-1} {\bf s} , 
\label{eqn:filter_def}
\end{eqnarray}
\noindent
Since the convolution of the raw waveform with the predetermined filter weights 
was performed as part of the readout, the processing time impacted the maximum event readout 
rate and could potentially generate data acquisition dead-time.  Consequently, 
the number of filter weights was  limited to eight.  The use of a different set of weights for each calorimeter channel  increased the memory bandwidth, but the channel-to-channel flexibility allowed the highest degree of optimization of the signal-to-noise ratio.  
In the central section of the calorimeter barrel, where the background was relatively light, the filter weights created longer integration times and reduced the contribution from electronics noise. 

On the other hand, the filter weights for
channels in the far forward region of the calorimeter endcap 
created filters with short time constants that reduced the overlap with beam-induced background at the cost of increased signal-to-noise.
With the implementation of the digital filter, the improved single channel energy resolution allowed the energy threshold for the recording of crystal hits  to be lowered from 1\mev\ to 0.75\mev\  without a noticeable increase in the average number of channels read out per event.

\subsubsection{Crystal Aging}
\label{sec:emcage}

\begin{figure}[htbp]
\begin{center}
   \includegraphics[width=0.45\textwidth]{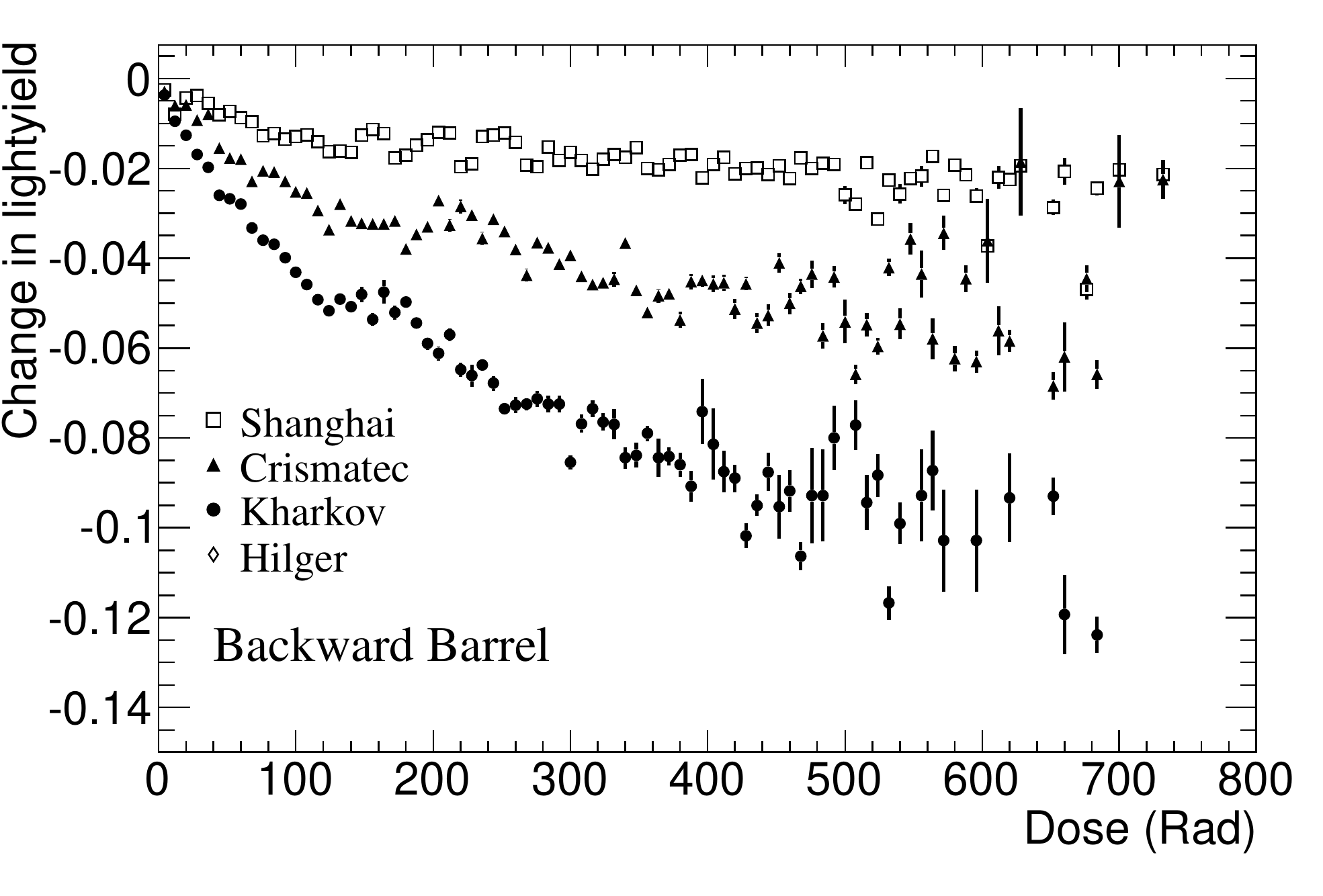}
   \includegraphics[width=0.45\textwidth]{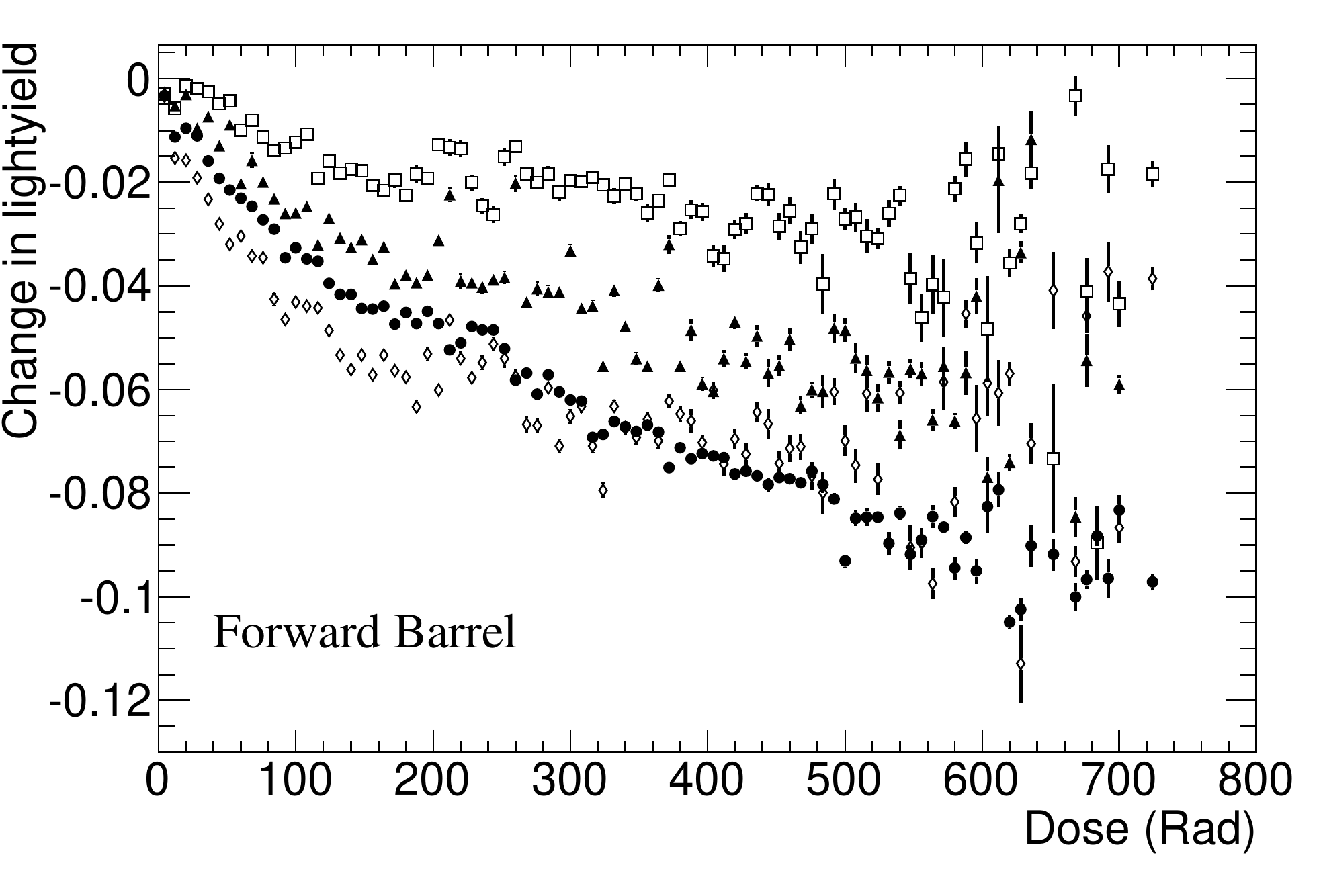}
   \includegraphics[width=0.45\textwidth]{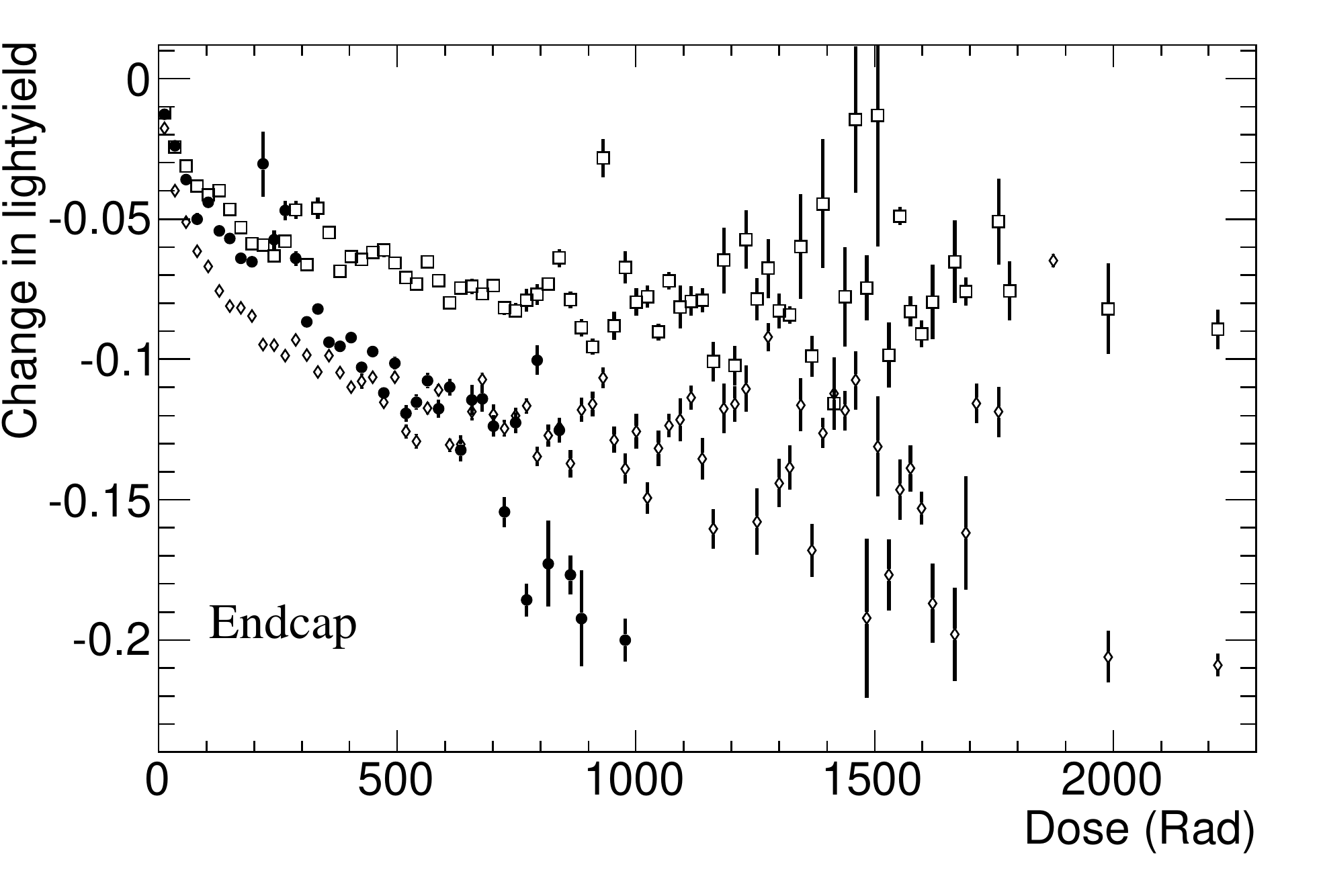}
\caption{
Fractional change in the EMC crystal light yield based on periodic calibrations with a radioactive source;  top: backward barrel, center:  forward barrel, and  bottom: forward endcap, as a function of the accumulated radiation dose averaged over crystals produced 
at different foundries. 
\label{fig:emc_ly_change}}
\end{center}
\end{figure}

During the operation of \babar, the EMC was exposed to intense levels of radiation
which resulted in radiation damage to the crystals.  
Because of the energy asymmetry of the \pep2\ beams and the \babar\ layout, the endcap crystals experienced higher levels of radiation, up to about 2.2~kRad, compared to crystals in the 
backward section of the barrel, typically 500 Rad. The absorbed dose was monitored by 115 RadFETs distributed throughout the EMC (see Section~\ref{sec:mdi:background-monitoring}).

Radiation damage to the crystals had two principal effects: a reduction of the total 
light output of a crystal, and a change in the 
uniformity of the light produced along the length of a crystal, introducing a 
dependence on the depth of the  energy deposition inside a crystal.

The change in the total light output for each crystal as measured during the experiment 
with the radioactive source calibration was -6$\%$ in the backward barrel, -10$\%$ in the forward barrel and 
-14$\%$ in the endcap. Although the backward barrel and
forward barrel had similar radiation exposure, the decrease in light yield
differed. As demonstrated in Figure~\ref{fig:emc_ly_change}, this could be explained by the difference in light yield for crystals supplied by different manufacturers:
Crismatec~\cite{Crismatec}, Hilger~\cite{Hilger}, Kharkov~\cite{Kharkov} and Shanghai~\cite{Shanghai}. 
For the same manufacturer, the light output degradation in the forward barrel
and the backward barrel was similar, although the rates of change for
the barrel and the endcap differed. In general the crystals delivered from Shanghai performed 
best in all parts of the EMC, with Crismatec being second, whereas crystals from 
Kharkov and Hilger had higher sensitivity to radiation. 
It is interesting to point out that while the Kharkov crystals
showed a continuous decrease in light yield, crystals from Shanghai and
Hilger exhibited a slower light-yield decrease, in particular in the endcap,
after the start of the trickle injection in March 2004.
With the initial average light output of crystals of $\approx$3900 photo-electrons/\mev, 
the total decrease in light yield had little impact on the energy resolution. 

The change of the light output uniformity along the length of the crystal was estimated in-situ 
by comparing the results of the light yield changes as measured by source calibration at 6.13 $\mev$ and Bhabha calibration with electrons of up to 9 \gev\ momentum. While the source calibration was only sensitive to light output changes in the front of the crystal, the Bhabha calibration probed the whole length of the crystal. 
As one can see from Table~\ref{tab:emc_source_vs_bhabha} the results of the two measurements 
were significantly different, in particular in the endcap. 
This was an indication that light yield uniformity was affected by irradiation. 
Its variations in various parts of the detector was most likely related to the manufacturer, just as in the case of the total light yield. Note that, in contrast to numbers quoted above,  the table only
shows the degradation of the light yield since 2001. 

\begin{table}[!htb]
\caption{EMC:  Comparison of the total light yield change in various parts of the EMC as measured
by radioactive source calibration and Bhabha calibration from January 2001 until April 2008.}
\label{tab:emc_source_vs_bhabha}
\centering
\begin{tabular}{lcc}
\\
\hline
                   & Source   & Bhabha     \\\hline
Backward Barrel    &  $- 3.7\%$ & $- 3.5\%$  \\
Forward Barrel     &  $- 5.8\%$ & $- 8.0\%$  \\
Endcap             &  $- 6.6\%$ & $- 14.0\%$ \\
\hline
\end{tabular}
\end{table}

Dedicated studies~\cite{Hryn'ova:2004sm} showed that for an exposure of 
2 kRad the light yield uniformity changed by up to $(-3.0 
\pm 0.7)$\%. 
A change of this magnitude would lead to a 
~3$\%$ worsening of the resolution for photons below 500$\,$MeV and an improvement of the resolution 
for photons above that value, due to compensation of the rear shower leakage.

\subsubsection{Photon Energy Calibration}
\label{sec:emccal}

The final step in the energy calibration of EMC clusters relied on photons from radiative processes and $\pi^0$ decays. It followed the calibration of the amplification and readout chain ({see Section~\ref{sec:emcops}), and the charge-to-energy conversion for single crystals (see Section~\ref{sec:emclight}).

Photon energies measured in the calorimeter were lower than their true values because of the following effects:
\begin{enumerate}
\item Energy leakage out the back of the crystals which had a length of 16 - 17.5 radiation lengths;  
\item Material in front of the EMC, mostly the DIRC quartz bars;
\item Material between the crystals consisting of the carbon fiber housing that held the crystals as well
as wrapping material.
\end{enumerate}
In addition to these energy losses, both the data acquisition and the reconstruction software imposed requirements on the minimum crystal energy, optimized for noise rejection.  Because of all of these effects, the measured cluster energy had to be calibrated to be able to
reconstruct the true energy of the incident electrons or photons.

\paragraph{Cluster Energy Calibration}
The cluster calibration required methods which provided a calorimeter-independent determination of the shower energy (mostly from photons).
The predicted 
energies had to be compared to their measured energies in the EMC over the range extending from well below 100\,MeV from the calibration source up to the highest energies from Bhabha scattering.
Since no single process spanned the whole energy range. Photons from 
$\pi^0 \rightarrow \gamma \gamma$ decays  in multi-hadron events were used in the low-energy range
(70\,MeV $< E_{\gamma} <$  2\,GeV), and
the high-energy range (400\,MeV $ <  E_{\gamma} <$ 6\,GeV) was evaluated using photons from radiative muon-pair production, $e^+ e^- \rightarrow \mu^+ \mu^- \gamma$.

The calibration function $C = f(E_{\gamma} , \theta_{\gamma})$
depends on the energy $E_{\gamma}$ and the polar angle $\theta_{\gamma}$ of the incident photon.
The goals of the calibration were twofold,  to correct the measured 
cluster energies to obtain the true photon energies, and  to assure that the detector simulation based on GEANT4\,\cite{Agostinelli:2002hh} 
correctly reproduced the polar angle dependence of these corrections.
The coverage of the full parameter space in  
$E_{\gamma}$ and $\theta_{\gamma}$
was limited by statistics,  so we factorized C and relied on
projections of the calibration function,
\begin{equation}
   C = f(E_{\gamma} , \theta_{\gamma})
     = g(E_{\gamma}) \cdot  h(\theta_{\gamma}) \; \; .
\end{equation}
In a first step, the polar-angle dependent factor
$h(\theta_{\gamma})$ was determined
using radiative muon pairs. After correcting for $h(\theta_{\gamma})$,
 the energy-dependent factor $g(E_{\gamma})$ was extracted over the full energy range by combining results from selected $\pi^0$ decays and  radiative muon pairs.
 
\begin{figure}[htbp!]
\begin{center}
\includegraphics[width=0.45\textwidth]{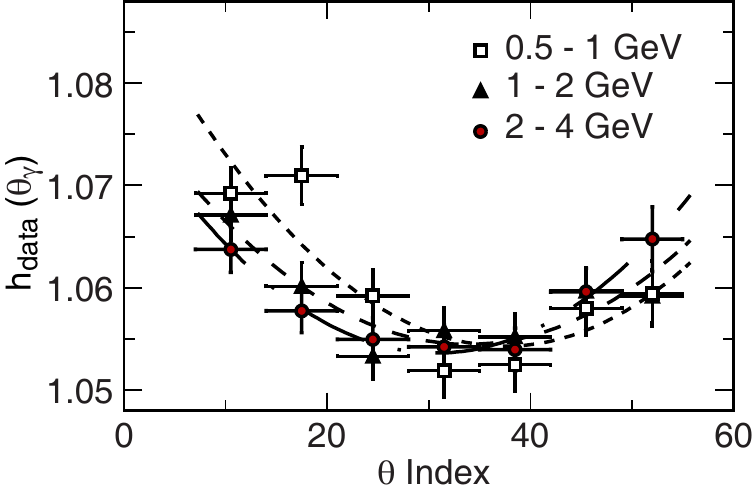} 
\caption{\label{thetacalib}
EMC energy calibration: Polar angle $\theta$ (crystal index) dependence for three photon energy ranges from selected $ \mu \mu \gamma$ events (2006 data). 
The data do not cover the very forward endcap.
The lines represent the parameterization for the three energy ranges.
}
\end{center}
\end{figure}

\paragraph{Calibration With Radiative Muon Pairs}
Since the initial state momenta of the process 
$e^+ e^- \rightarrow \mu^+ \mu^- \gamma$ are known and the muon momenta are
measured in the tracking system, 
the energy and the polar angle of the 
photon can be determined 
by a kinematic fit, independent of any calorimetric measurement. 
The predicted photon energy $E_{\rm fit}$ was compared to $E^{\gamma}_{\rm rec}$, the photon energy measured in the EMC, by considering the
distribution of the ratio $E^{\gamma}_{\rm rec}/E^{\gamma}_{\rm fit}$. The average values  $ \langle E^{\gamma}_{\rm rec}/E^{\gamma}_{\rm fit} \rangle $ were 
quantitatively evaluated by the peak position of a 
fit with a modified log-normal distribution~\cite{fnov}. 
In the MC simulation, the cluster calibration function was extracted using the 
true photon energy $E^{\gamma}_{\rm true}$, instead of the predicted one. 

The asymmetric line shape of the photon response 
distribution  $ E^{\gamma}_{\rm rec}$  
caused systematic shifts of the ratio 
$E^{\gamma}_{\rm rec}/E^{\gamma}_{\rm fit}$ and therefore the data-MC ratio 
had to be used to extract the calibration function.

\begin{figure}[htbp!]
\begin{center}
\includegraphics[width=0.45\textwidth]{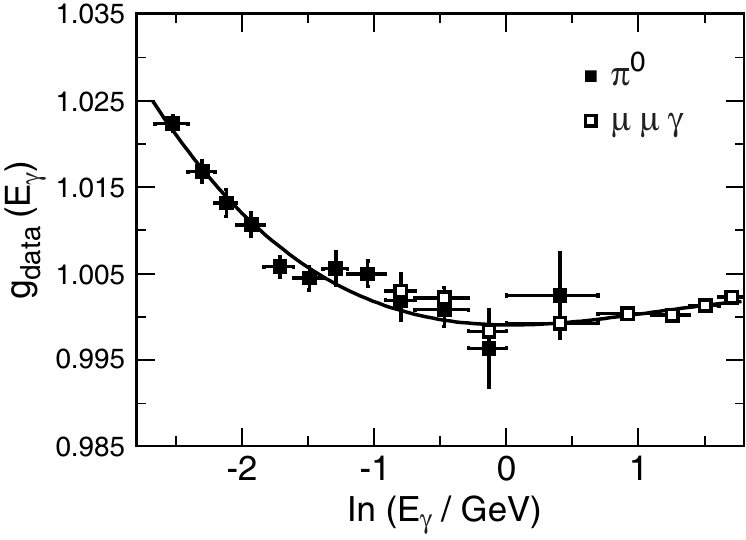}
\caption{\label{energycalib}
EMC:  Photon energy calibration based on $\pi^0$ and  $ \mu \mu \gamma$ data after applying the $\theta$-dependent calibration (2006 data).
The solid line represents the calibration function extracted from a fit to all data points. 
}
\end{center}
\end{figure}

The polar angle dependence $ h(\theta_{\gamma})$ of the calibration function was determined for seven intervals in $\theta^i_{\gamma}$ (averaged over seven crystals each).
The following ratio yielded the $\theta$-dependent calibration function:
\begin{equation}
 h_{\rm data} (\theta^i_{\gamma}) = \frac{  h_{\rm MC} (\theta_{\gamma}) \cdot \langle E^{\gamma}_{\rm rec}/E^{\gamma}_{\rm fit} \rangle_{\rm data} (\theta^i_{\gamma})}  {\langle E^{\gamma}_{\rm rec} /E^{\gamma}_{\rm fit} \rangle_{\rm MC}  (\theta^i_{\gamma})   } \; \; .
\end{equation}
The function was measured for three different ranges of the photon energy as a function of
 the polar angle of the crystals and is shown 
in Figure\,\ref{thetacalib}. 

To derive the energy-dependent calibration function $g(E_{\gamma})$ for the MC simulation, the $\theta$-dependent correction
$ h(\theta_{\gamma})$ was applied and $g_{\rm MC}(E_{\gamma})$ was obtained using the true photon energy. 
The calibration function for data was extracted as, 
\begin{equation}
 g_{\rm data} (E^i_{\gamma}) = \frac{ g_{\rm MC}(E_{\gamma}) \cdot h_{\rm data} (\theta_{\gamma}) \cdot \langle  E^{\gamma}_{\rm rec} /E^{\gamma}_{\rm fit}\rangle_{\rm data} (E^i_{\gamma}) }  {  h_{\rm MC} (\theta_{\gamma}) \cdot \langle   E^{\gamma}_{\rm rec} / E^{\gamma}_{\rm fit} \rangle _{\rm MC}  (E^{i}_{\gamma}) }.
\end{equation}

\paragraph{Calibration with Photons from $\pi^0$ Decays}
To measure $g (E^i_{\gamma})$ in the low-energy regime, $\pi^0$ decays 
with the two photons in the same energy range (symmetric $\pi^0$'s) were selected.
The distribution of the invariant mass $m_{\gamma \gamma}$ of the $\pi^0$ candidates is shown in Figure\,\ref{pi0sig} for photons in the range 70\,MeV $< E_{\gamma_{1}},E_{\gamma_{2}} <$ 90\,MeV.
The combinatorial background contribution was taken from  two photons in two different events. Subtracting this background from the invariant mass distribution yielded a clean $\pi^0$ signal (Figure\,\ref{pi0sig}).

\begin{figure}[htpb]

\begin{center} 
\includegraphics[width=7cm]{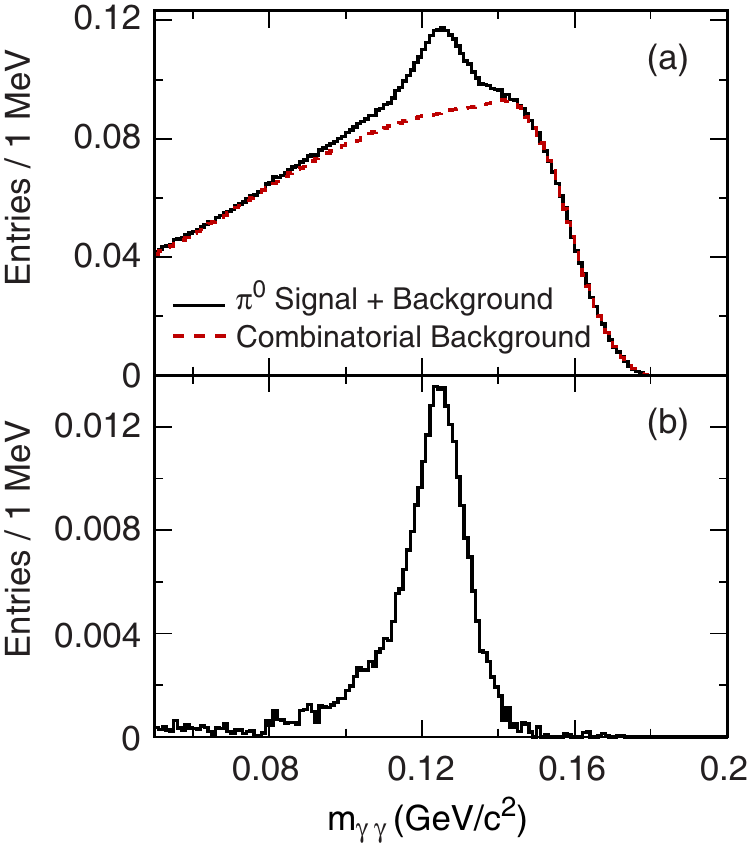} 
\end{center}
\caption{\label{pi0sig}
EMC:  $\gamma \gamma$ invariant mass (a) before and (b) after subtraction of the combinatorial background. The photon energies were restricted, $70\mev < E_{\gamma_{1}},E_{\gamma_{2}} < 90\mev$ (2006 data).}
\end{figure}

The $\pi^0$ mass distribution is a convolution of the two asymmetric photon response functions;
because of the asymmetric shape of the photon response, the invariant mass $m_{\gamma \gamma}$ distribution is shifted relative to 
the nominal $\pi^0$ mass 
, even if the photon response were perfectly calibrated. 
This shift is a function of the photon energy resolution, the tail of the photon response function and 
the photon position resolution. To account for these effects, the calibration function $g_{\rm data}(E_{\gamma})$ in the low-energy regime was determined as the ratio of the average 
invariant mass obtained with calibrated photon energies in the MC simulation and the average invariant mass obtained with uncalibrated photon energies in data,
\begin{equation}
 g_{\rm data} (E^i_{\gamma}) = \frac  { C_{\rm MC}(E_{\gamma},\theta_{\gamma}) \cdot \langle  m_{\gamma \gamma}  \rangle_{\rm MC}(E^i_{\gamma})}
{ h_{\rm data} (\theta_{\gamma}) \cdot \langle  m_{\gamma \gamma} \rangle_{\rm data} (E^i_{\gamma}) } \; \; .
\end{equation}
The average values of the invariant mass distributions $ \langle m_{\gamma \gamma} \rangle $ were
taken as the peak position determined by a 
fit of a modified log-normal distribution\,\cite{fnov}. 
Figure\,\ref{energycalib} shows $g_{\rm data}(E^{i}_{\gamma})$ on a logarithmic $E_{\gamma}$ scale determined from both the $\mu \mu \gamma$ and the $\pi^0$ calibrations.  The results of the two calibrations agree well in the overlap region.

This procedure relied on an excellent description of the data by the MC simulation. Any deviation
would cause systematic shifts in the invariant mass and therefore shifts of the calibration function. 
Based on typical discrepancies between data and MC simulation, the systematic
uncertainties were estimated to be  0.3\,\% for the $\pi^0$ calibration. For the calibration based on radiative muon pairs,
the systematic uncertainties decreased from 0.5\,\% at the lowest energy to 0.1\,\% at the highest.

The invariant mass ratio of data and simulation for the $\pi^0$ sample 
before and after the cluster calibration is shown in Figure\,\ref{pi0datamc}.
Before the cluster calibration, the differences
 between data and MC simulations were as large as 1.5~\% and were
$E_{\gamma}$ and $\theta_{\gamma}$ dependent.   
After the cluster calibration, the data were well reproduced by the MC simulation, 
and they are, within errors, independent of $E_{\gamma}$ and $\theta_{\gamma}$.  

\begin{figure}[htpb]
\begin{center} 
\includegraphics[width=7cm]{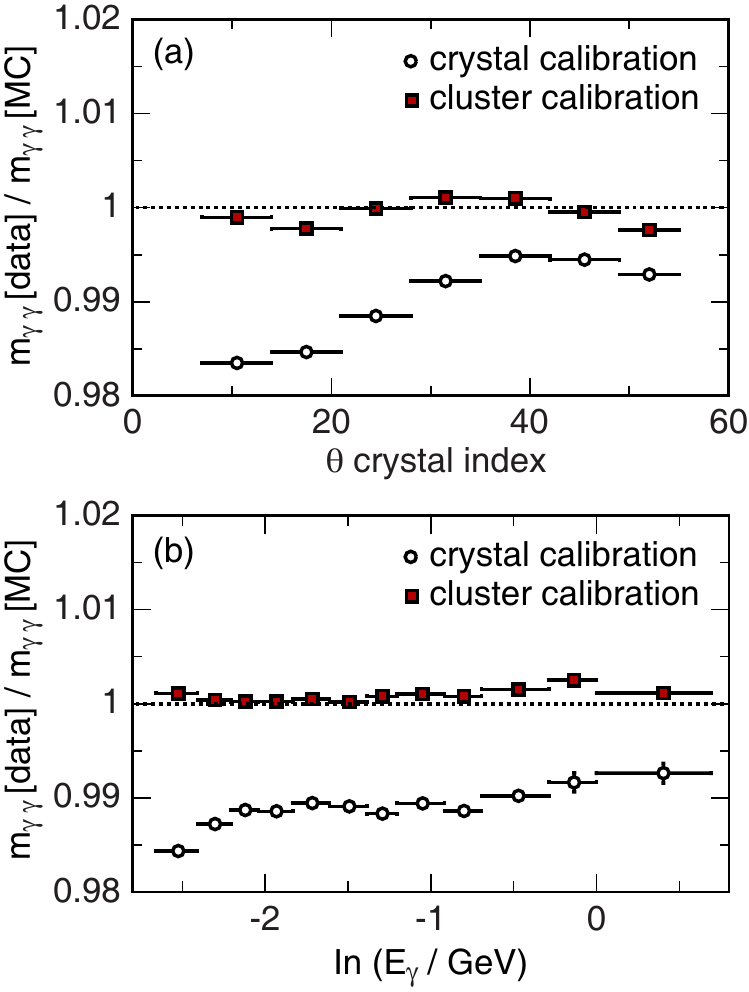}
\end{center}
\caption{\label{pi0datamc}
EMC cluster calibration: Data/MC ratio of $\gamma \gamma$ invariant mass from selected $\pi^0$ decays before and after the cluster calibration as function of (a) the polar angle (crystal index) and (b) the photon energy (2007 data) . 
}
\hspace{1pc}
\begin{center} 
\includegraphics[width=7cm]{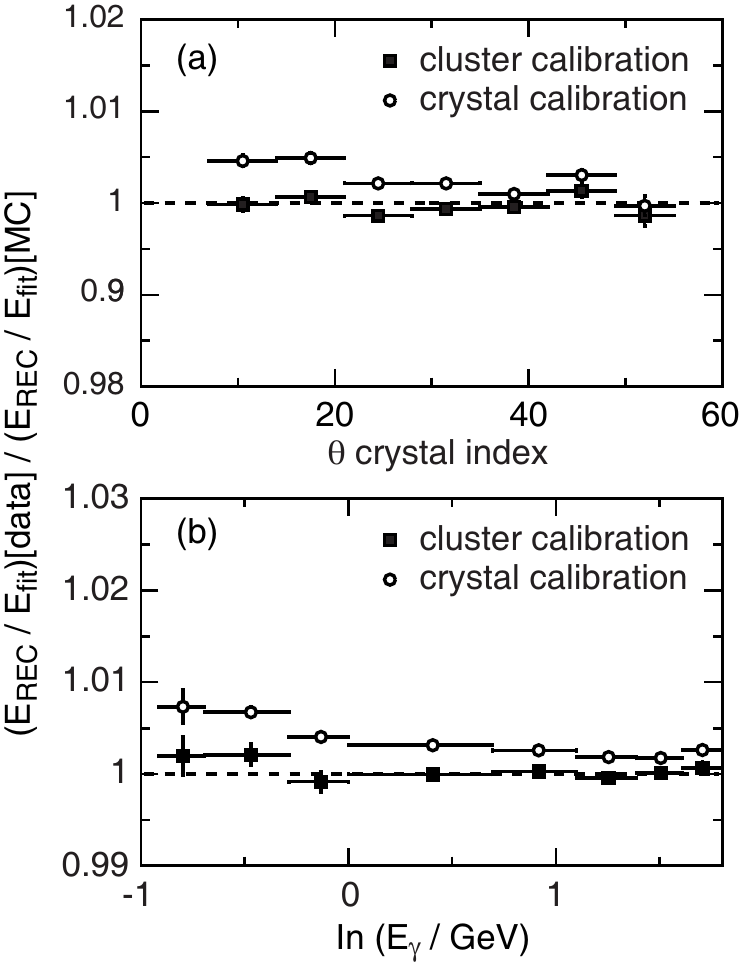} 
\caption{
\label{mmgdatamc} 
EMC:  Comparison of two photon calibration procedures based on selected $\mu^+ \mu^- \gamma$ events:  Data/MC ratio $E^{\gamma}_{\rm rec}/E^{\gamma}_{\rm fit}$ as function of (a) the polar angle (crystal index) and (b) the photon energy (2006 data). }
\end{center}
\end{figure}

The data-MC ratio $E^{\gamma}_{\rm rec}/E^{\gamma}_{\rm fit}$ for selected radiative muon pairs 
is shown in Figure\,\ref{mmgdatamc}. As in the case of the $\pi^0$ sample,
after applying the calibration, the data are well described within 
the estimated uncertainties. 
No remaining dependence on $E_{\gamma}$ and $\theta_{\gamma}$ is
observed. 

The energy resolutions of high-energy photons for Runs 1 and 6 are compared in
Figure~\ref{fig:emc-energy-resolution-data}.   The resolution obtained
for Runs 2--6 is very similar, and significantly better than for Run
1.  Starting from Run 2, the digital filter was employed
to optimize the signal-to-noise ratio for the expected EMC signal
shape (see Section~\ref{sec:emcdf}).  This improved the single-channel
resolution and allowed a lower energy-threshold setting, further improving the
overall energy resolution.
A fit to the Run 6 data with $E_\gamma$ above $3\gev$ results in an
energy resolution which may be parameterized as,
\begin{equation}
\frac{\sigma(E_\gamma)}{E_\gamma} = \frac{0.027}{\sqrt[4]{E_\gamma (\gev)}} \oplus 0.010 . 
\end{equation}

\begin{figure}[htpb]
    \centering
    \includegraphics[width=16pc]{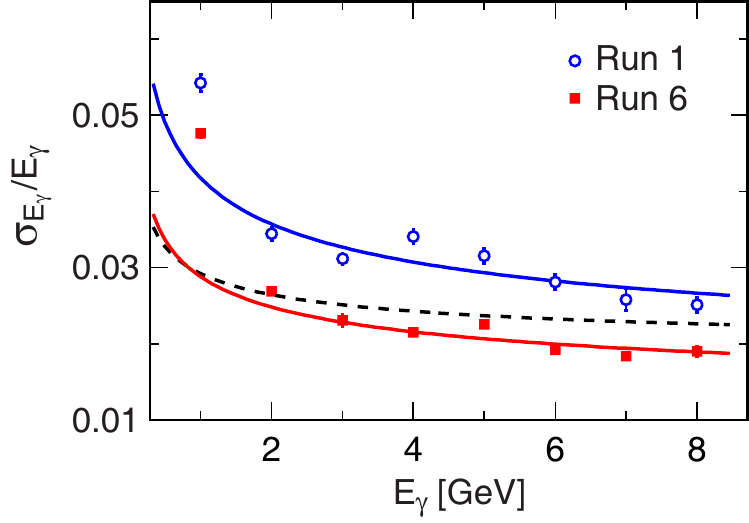}
    \caption{
    EMC energy resolution measured as a function of the photon energy in
      $\mup\mun\gamma$ events, after all calibrations and corrections,
      separately for Run 1 and Run 6, showing the data points and 
      the fits (solid lines) to the data above $3\gev$.  
      For comparison, the dashed curve shows the
      parameterization of the resolution determined in 
      2001~\cite{Aubert:2001tu}, based on photons from $\piz$ decays 
      and electrons and positrons from radiative Bhabha scattering.
     }
    \label{fig:emc-energy-resolution-data}
\end{figure}

\paragraph{Validation of the Cluster Energy Calibration}

In the decay  process $ \Sigma \rightarrow \Lambda \gamma \rightarrow p \pi \gamma $, the photon energy is constrained
and therefore the process can be used to perform independent tests of the calibration in the low photon energy range  ($50 < E_{\gamma} < 400 \mev $). 

The photon energy can be calculated using charged particle tracking and identification, 
\begin{equation}
 E^{\rm calc}_{\gamma} = \frac{M^2_{\Sigma} - M^2_{\Lambda}} {2 (E_{\Lambda} - p_{\Lambda} \cos \theta_{\Lambda \gamma})} \; \; ,
\end{equation}
where $M$ denotes the  mass, $E_X$ the energy, $p_X$ the magnitude of the momentum of the baryons (the subscript $X$ stands for $\Lambda$ or $\Sigma$) and  $\theta_{\Lambda \gamma}$ is the
opening angle between $\Lambda$ and $\gamma$.

The ratio of the predicted to the measured photon energy was measured as a function of $E_{\gamma}$ and $\theta_{\gamma}$ and 
found to be consistent with unity to within the total uncertainties 
of about 1\,\%.

The photon energy scale in the range from 1.5\,GeV to 3.5\,GeV can be probed by the decay process
$B^0 \rightarrow K^{*0} \gamma  \rightarrow K^+ \pi^- \gamma $. 
The difference  between the reconstructed energy $E_{\rm rec}$ of the  $B^0$ and the
beam energy $E^*_{\rm beam}$ (both in the $\FourS$ rest frame)  
\begin{equation}
 \Delta E = E_{\rm rec} - E^*_{\rm beam} 
\end{equation}
contains the cluster energy.
It was determined by using the measured particle momenta and identification, and the photon energy. 
The measurements yielded  $\Delta E_{\rm data} = (-8.7 \pm 2.3)\,\textrm{MeV}$ for the data and 
$\Delta E_{\rm MC} = (-8.5 \pm 0.2)\,\textrm{MeV}$  for the MC simulation. This translates to a deviation of the 
photon energy scale of $(-0.35 \pm 0.09)$\,\%  .

\paragraph{EMC Monitoring with Muons}
The energy loss of minimum ionizing muons from $ e^+ e^- \rightarrow \mu^+ \mu^- (\gamma)$ pair production 
in a single CsI(Tl) crystal was typically 200 \mev .
It was used as a continuous monitor of the stability of the whole calibration procedure.  
The average energy deposition for a muon track passing through a single crystal, normalized to the path length, was recorded as the peak position 
from a fit using a modified log-normal distribution\,\cite{fnov} for each crystal. The time dependence of the average energy loss as a function of the polar angle  is shown in Figure\,\ref{muaverageinphi} for the barrel section of the calorimeter, separately for  Runs 2-7.  For each polar angle, the data were averaged over the 120 crystals in $\phi$.
Since the muon energy increased with
decreasing polar angle $\theta$, the relativistic rise of $dE/dx$ \cite{bethe} led to an increase  of the average energy deposition in the forward direction. 
The variation over the data taking period from 2001 to 2008 was about $\pm$0.5\,\%.

The normalized muon  energy deposition could also be used to determine the light yield calibration factors for each crystal.

\begin{figure}[htpb]
\begin{minipage}{18pc}
\begin{center} \includegraphics[width=16pc]{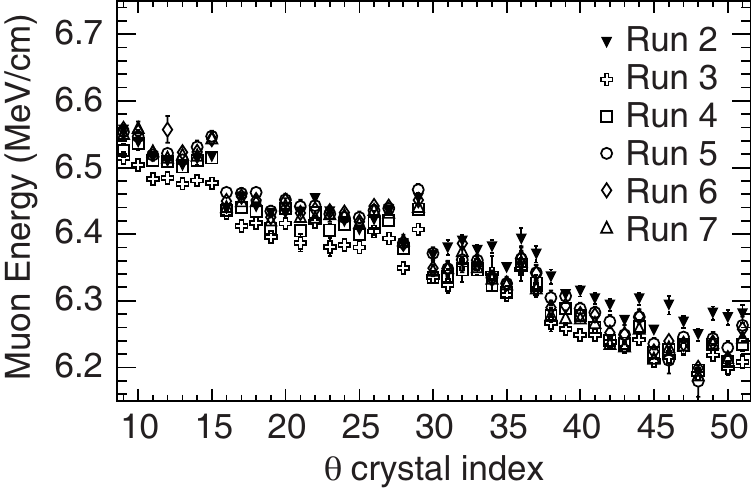} \end{center}
\caption{\label{muaverageinphi} 
EMC stability: Muon energy deposition per track length in the crystal, averaged in $\phi$, as function of the polar angle $\theta$ (crystal index) for different data taking periods.}
\end{minipage}
\end{figure}

\subsubsection{Preshower Photons}
\label{sec:emcpre}
\def\mmg        {\ensuremath{\mu\mu\gamma}\xspace}

As determined from MC simulation, photons produced in \BzBzb events had a probability of 10\% to interact with the material of the inner detector before they reached the EMC.  Such preshowers led to energy losses and resulted in a degradation of the EMC resolution 
(Figure~\ref{fig:emc_energyRes}).
 
Most of the material 
between the IP and the EMC crystals was in the SVT and the DIRC~(\cite{Aubert:2001tu} Figure~3).  
A photon interacting in a DIRC silica bar, at a short radial distance from the EMC, would leave a signal in the DIRC that could in principle be used to identify the preshower and correct for the energy loss. 

\begin{figure}
  \includegraphics[width=\columnwidth]{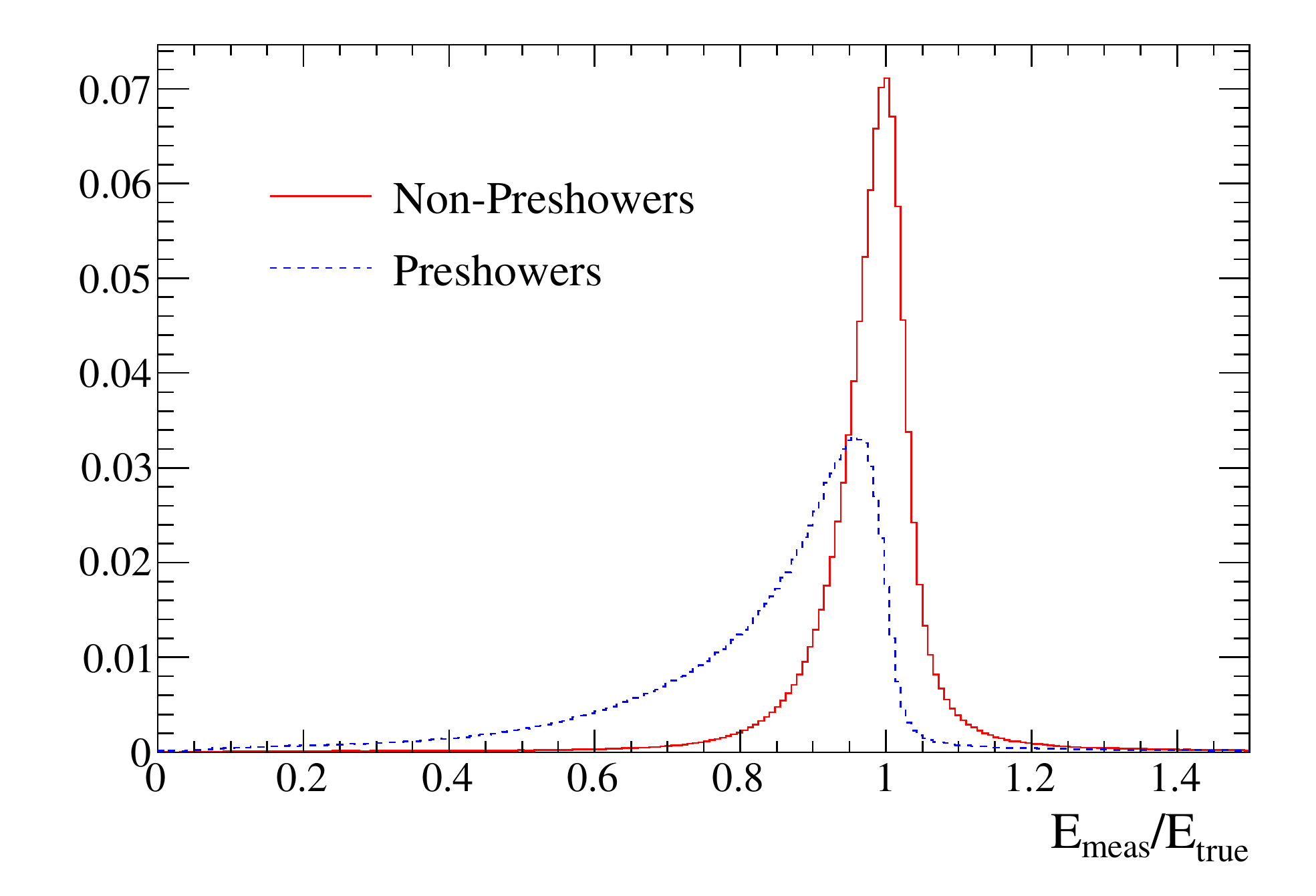}
        \caption{
Ratio of measured EMC energy $E_{\rm meas}$ to true energy $E_{\rm true}$ for photons in simulated \BzBzb events, separately for photons with or without a preshower before reaching the EMC 
}
	  \label{fig:emc_energyRes}
\end{figure}

\paragraph{Identification of Preshowers}
Many of the electrons and positrons originating from a photon conversion pass through the DIRC bars and emit Cherenkov light. It is possible to associate these Cherenkov photons with the neutral EMC cluster produced by the same photon.  

The reconstruction proceeded as follows: an attempt was made to link
the detected Cherenkov photons with charged tracks in the DCH. Cherenkov photons which could not 
be associated with a charged track might be attributed to preshower photons.  To do this, neutral clusters in the EMC were 
used to extrapolate photon trajectories to the DIRC bars, and their 
intersections were used in the same Cherenkov angle reconstruction algorithm that was commonly applied to charged
tracks.  For approximately 80\% of preshowers it was possible to associate DIRC information with the preshower in MC events.

Preshowers were identified using a neural network (NN) based on quantities from the Cherenkov ring fit and a number of EMC
variables  as input to the NN classifier.  
This approach improved the
background rejection while maintaining good efficiency, 
and allowed users to tailor the
efficiency or background rejection to the needs of their analysis.

The NN was trained on simulated \BzBzb events with 45,000
photon candidates each in the preshower and non-preshower samples.
For both samples, $N_{\theta_{C}}$, the number of Cherenkov photons associated with the incident photon, was required to be greater than four.
The DIRC quantities used as inputs
to the NN were:  the Cherenkov angle determined by the ring fit $\theta_{C}$, the error on the Cherenkov angle, the number of Cherenkov
photons $N_{\theta_{C}}$, the ratio of
the number of background Cherenkov photons to signal Cherenkov photons as determined by the ring fit algorithm, and the chi-square probability of the ring fit.  The EMC quantities used as input were:  six shower shape variables
and moments, and the number of crystals in the photon cluster. 
Approximately 4\% of the \mmg events selected by the trigger passed the default NN selection.
Table~\ref{tab:emc_effpur}  gives the efficiency and purity for three
different thresholds on the NN output for preshower candidates in simulated \BzBzb events.

\begin{table}
\begin{center}
\caption{
Identification of photon preshowers in the DIRC: Efficiency and purity of the preshower sample in simulated \BzBzb data for three
thresholds on the NN output.  The default value in the standard analysis  is at 0.5625.}
\label{tab:emc_effpur}
\begin{tabular}{ccc}\hline
NN threshold & efficiency & purity \\ \hline
-0.6   & 0.90       & 0.11   \\
0.5625 & 0.42       & 0.55   \\
0.9    & 0.21       & 0.77   \\ \hline
\end{tabular}
\end{center}
\end{table}

\paragraph{Preshower Correction}
A simple procedure was used to correct for the energy loss $E_{\rm loss}$ due to preshowers  in the DIRC.  A study of simulated preshowers in
\BzBzb events indicated that $E_{\rm loss}$ scaled linearly with the path length in the DIRC bars  and was proportional to the number of signal Cherenkov photons associated with the preshower $N_{{\rm sig},\theta_{C}}$.   The energy correction was calculated
for six evenly spaced intervals in the polar angle $\theta$ ($0.47 \leq \theta
\leq 2.46 \rad$ to cover the acceptance of the DIRC)
and as a function of $N_{{\rm sig},\theta_{C}}$ ($0\! \leq\! N_{{\rm sig},\theta_{C}}\! <\!19$ and $19\!\leq\! N_{{\rm sig},\theta_{C}}$).
The energy correction ranged from approximately $20 \mev$ to $60 \mev$ and did not depend on the initial photon energy.

The performance of the preshower correction was studied with $\epem \to \mumu\gamma$  event samples.  Figure~\ref{fig:emc_mmgEnergyRes} shows the ratio of the measured photon energy $E_{\rm meas}$ to the true photon energy $E_{\rm true}$ which was determined from a kinematic fit.
 The  energy correction shifted the distribution closer to the expected value; the improvement to the overall photon energy resolution was small.

\begin{figure}[htb!]
   \includegraphics[width=\columnwidth]{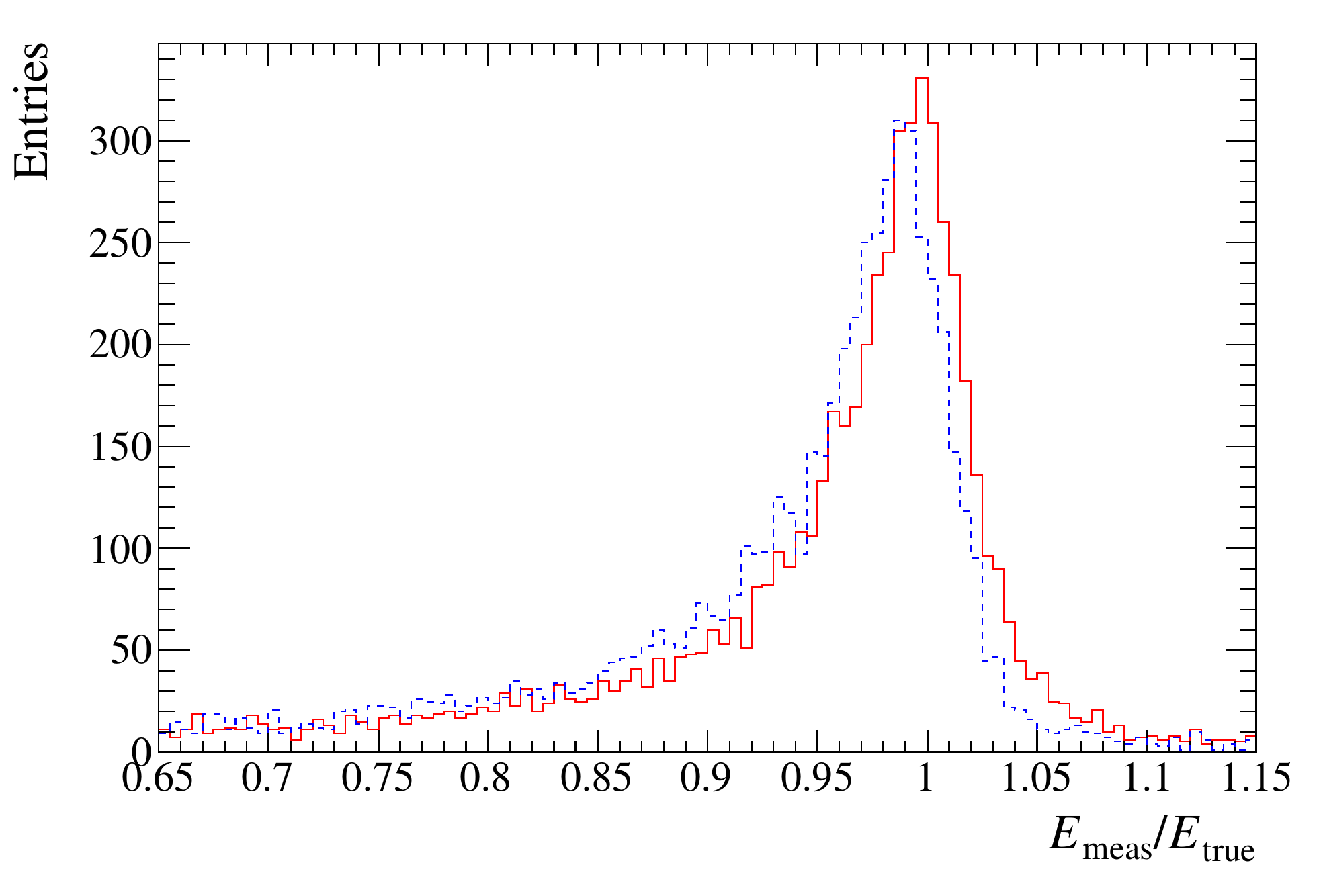}
   \caption{Correction for photon preshowers in the DIRC: Ratio of measured EMC energy to true photon energy with (solid histogram)
   and without (dashed histogram) the energy correction for \mmg events
   data passing the default threshold on the NN output.}
  \label{fig:emc_mmgEnergyRes}
\end{figure}

\subsection{IFR Operation and Performance}
\label{sec:ifrrpc}

The IFR, consisting of a barrel section and two endcaps, was the principal detector system designed to identify muons and 
other penetrating particles.  Due to serious aging problems of the RPCs, the IFR underwent major upgrades. The original chambers were replaced 
in the forward endcap by RPCs of an improved design (see Section~\ref{sec:upgrades-RPC}) and in the barrel by LSTs (see Section~\ref{sec:upgrades-LST}). 
As a consequence, the IFR performance varied strongly with time and was monitored very closely.

The IFR detectors and hardware were controlled and monitored by the EPICS data control software. 
The LST and RPC HV were ramped down (up) before (after) \pep2\ injection in less than a minute. 
Injection voltages were chosen to have very low gas gain to make the chambers insensitive to beam backgrounds. 
Following the adoption of trickle injection, the RPC and LST chambers were constantly kept at 
operational voltage, except for occasional beam aborts.
Chamber currents, voltages, rates, gas flow, beam conditions and temperature were recorded in the conditions 
database.  With the addition of offline efficiency measurements,  the performance of the RPCs and LSTs could be tracked and  changes in efficiencies or currents could be correlated with 
changes in environmental or \pep2\ conditions.

\subsubsection{Resistive Plate Chambers}
\label{sec:ifrmon}
\paragraph{RPC Efficiency}
Average RPC efficiencies per chamber were measured by $\mu$ pairs during normal colliding beam data 
and by cosmic rays during dedicated plateau calibration runs. 
The average single gap RPC efficiencies measured by $\mu$ pair events in the first five years of operation are 
shown in Figure~\ref{RPCeff} for the forward endcap, backward endcap, and central barrel regions.
During the shutdowns between Runs, 
electronics and HV connections were repaired, slightly improving performance.
The degradation of the RPC efficiency with time was dramatic and motivated the replacement of the chambers. 
The forward endcap RPC replacement at the end of 2002
restored the efficiency in this region.
The final layer efficiencies at the end of data taking 
in 2008 were 90.5\%  in the forward endcap and 43.1\% in the backward endcap.
A more detailed history of the barrel RPCs is shown in Figure~\ref{fig:lst:efftime} indicating a similar degradation with time.
\begin{figure}[!htb]
\centering
\includegraphics[width=0.45\textwidth]{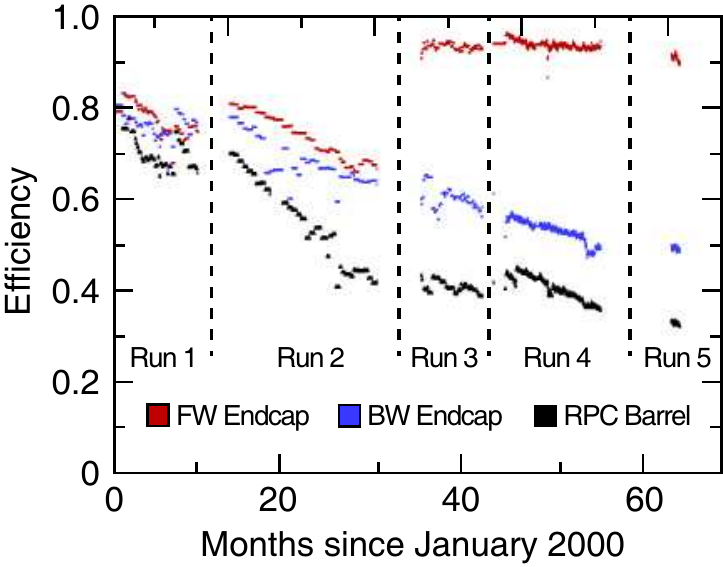}
\caption{Single gap RPC efficiencies determined from dimuon events accumulated during data taking
as a function of time from Jan. 2000 to May 2005. 
The color code differentiates data from the forward, the 
backward, and the barrel RPCs.}
\label{RPCeff}
\end{figure}
\paragraph{RPC Backgrounds}

 The central barrel and backward endcap RPCs were well-shielded from the PEP tunnel and interaction 
region. Noise rates with beam were similar to noise rates without beam. The only significant 
exceptions were associated with failures of the \pep2\ beam abort system.  In 2001 and 2002, there were 
several instances of asynchronous beam aborts in which the high energy beam was not steered onto 
the beam dump but was lost in the beam pipe just before the \babar\ Hall. Large pulses of neutrons 
flooded the entire detector often tripping many RPC HV channels.  Much worse, however, was the 
permanent damage done to the FEC electronic cards as described in Section~\ref{sec:upgrades-RPC}.  Anywhere from 5-40\% 
of the barrel electronics were damaged in each incident. Repair of the broken channels was often 
delayed by several days and took over one week.  
Data taken after the worst of these incidents were unusable for physics analysis 
requiring muon identification.  An interlock was developed to warn shift operators when the beam 
abort system had lost its synchronization. 
RPC HV could then be ramped down as a precaution.
This strategy prevented the reoccurrence of this failure mode.

The RPCs in the forward endcap experienced significant backgrounds in two areas.
Streamer rates produced by backgrounds and signals varied considerably (from 0.1 to 20~Hz/cm$^2$) depending 
on layer and position as shown in Figure~\ref{RPCA05}. 
In the inner layers (1-12) the 
chamber occupancy was highest around the beam line. Signal rates, currents, and occupancy were 
proportional to the \pep2\ luminosity and were typically 30 to 50~kHz per chamber in layer 1 (the innermost one) with 
peak rates of up to 20~Hz/cm$^2$. The rates decreased with increasing layer number, reaching a minimum 
in layer 11 at about 1/4 of the layer 1 rate. The rates in the outer layers (13-16) were sensitive 
to backgrounds from \pep2\ entering the outside of the endcap. The rates in layers 13-14 were typically 
2 Hz/cm$^2$. Rates in layers 15-16 were even higher (sometimes above 10 Hz/cm$^2$) and prevented normal 
operation of these layers until a shielding wall (described in 2.5.3) was installed in the fall of 2004. The twenty 
small RPCs surrounding the endcap had moderate noise rates of $\leq$~3 Hz/cm$^2$.  

\begin{figure}[!htb]
\centering
\includegraphics[width=0.5\textwidth]{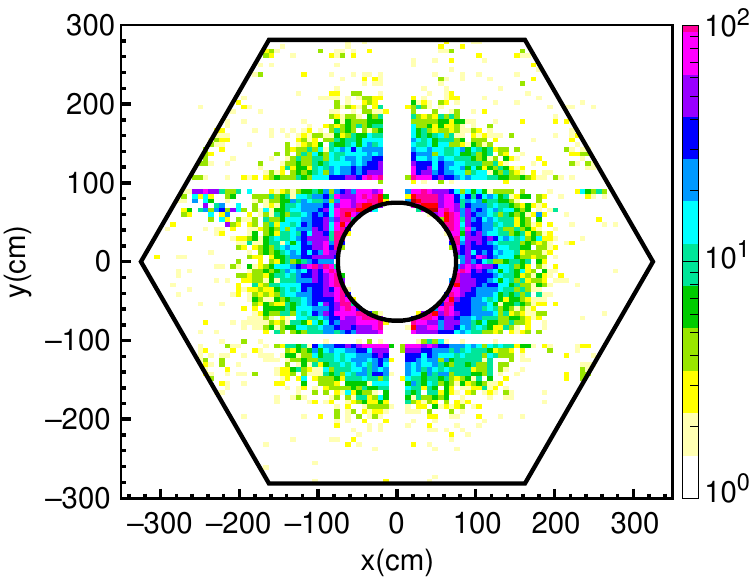}
\caption{Two dimensional occupancy plot during normal data taking for the RPC forward endcap layer 1.
The high rate in the ring around the beam line was proportional to luminosity and exceeded 20 Hz/cm$^2$.
The rates near the outer edges of the layer were consistent with the rate expected from cosmic rays.}
\label{RPCA05}
\end{figure}

\subsubsection{Limited Streamer Tubes}
\label{sec:lstmon}

The LST system performed very well during \babar\ data taking, requiring minimum maintenance and intervention, while dramatically improving the overall  muon identification. To ensure a high operational efficiency,
routine checks and measurements were performed periodically. 
In addition to daily checks of HV currents and trips,
of the gas system, and of online occupancies and multiplicities, 
background conditions were checked once a week to spot anomalous changes in the currents versus rate. Every month, HV plateau  measurements were carried out for all two-cell segments to look for changes that might indicate mechanical or electrical problems, symptoms of aging, or changes in the gas mixture.

\paragraph{Sensitivity to Beam Backgrounds}

The LSTs were designed to cope with steady state beam backgrounds up to luminosities  of a few times  $10^{34}\cms$.
Since the current drawn by a LST is proportional to the counting rates, \ie\ to the exposure to radiation and background,  LSTs in layers close to the beam line were expected to draw higher currents than those at larger radii.  As the \pep2\ luminosity grew, the currents in the inner layers exceeded the expectations by about a factor of three, and this raised concerns about the lifetime of the tubes.  
Beam instabilities resulting in intense bursts of radiation could lead,
in the innermost layers, to
self-sustained discharges in the tubes and high voltage trips which required a power cycle for the entire layer. The current and singles rates during collisions were continuously monitored to track their dependence with increasing luminosity.

Figure \ref{fig:lst:curr} shows  for three different layers the evolution of
the ratio of the average LST current to the \pep2\ luminosity for the period extending from January 2005 to the end of the data taking in April 2008.  After an early period of high currents, induced by
high backgrounds, and some HV problems, the currents dropped by 30\%  and remained stable. Short term variations were caused primarily by beam instabilities.  The currents in the outer layers were about a factor of ten lower than those in the first layer. By comparison, the average currents drawn during periods with no beams showed no significant increase over three years of data taking.  Thus, there was no evidence of aging of the LSTs.
\begin{figure}[!htb]
\centering
\includegraphics[width=0.5\textwidth]{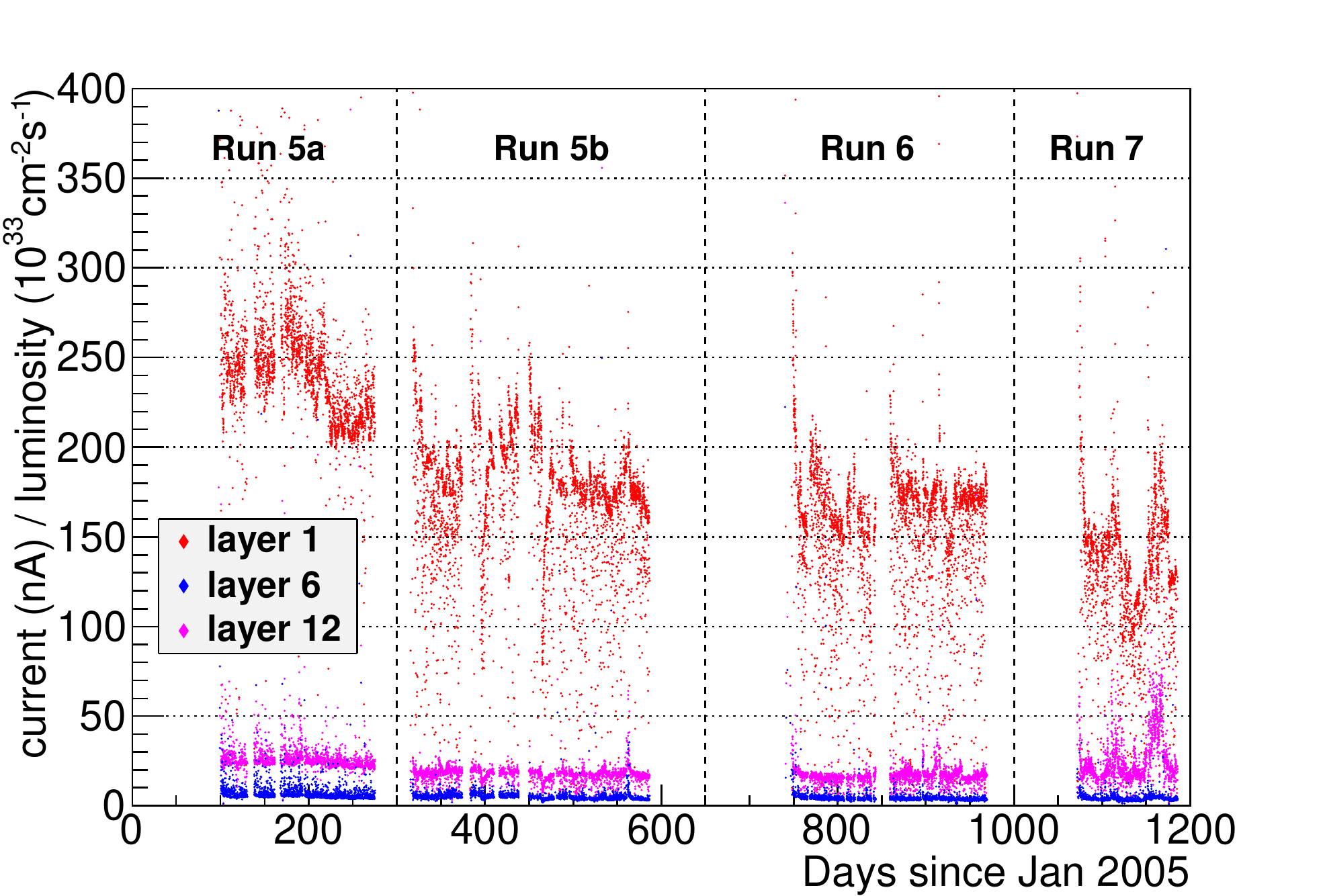}\\
\includegraphics[width=0.5\textwidth]{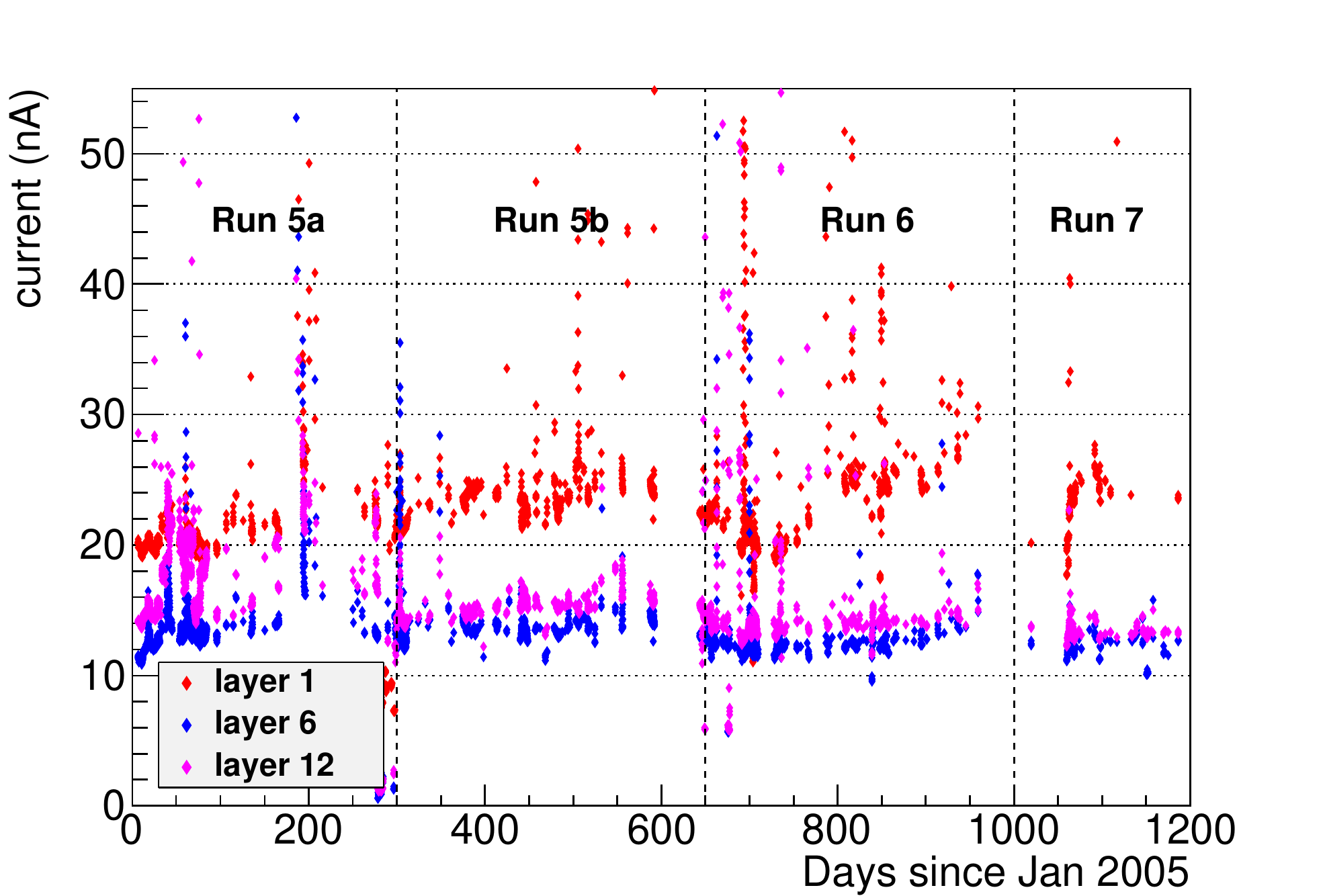}
\caption{Average LST currents as a function of calendar time:
 (top) Ratio of the average currents to the PEP-II luminosity recorded during colliding beams and (bottom) currents drawn for cosmic ray data recorded without beams.
}
\label{fig:lst:curr}
\end{figure}

Due to the higher levels of exposure to beam generated backgrounds, the LSTs in the inner two layers that most frequently caused HV trips were connected to dedicated power supply and kept at lower voltage. A few of them (about 0.5\%) were totally disconnected from the high voltage.
This was the principal source of single layer inefficiency during the data taking.

\paragraph{High Voltage Power Supplies}
In total 18  power supplies (three per sextant) were required to power the LSTs.  These supplies were mounted in three racks. 
In each rack, an additional supply was installed which served as a {\it clinic} to study and fix HV channels that suffered from various problems.  
The high granularity of the custom-designed LST high voltage system  
permitted an efficient redistribution of the connections, such that multiple HV channels could be used for a single tube, minimizing HV trips and avoiding the need to reduce the HV for the whole layer.  
Another important feature was the ability to disconnect individual two-cell HV segments without affecting the remaining segments in the tube.  Often these HV segments were connected to the {\it clinic} supply for diagnostics,  conditioning and repairs.

In three years of operation, the down time caused by the LST HV system was minimal. Most of the down time was caused by self-sustained discharge initiated by high backgrounds or by very high currents due to high humidity. 
Among the few isolated problems were occasional failures of HV regulators and current monitors (5 out of 2000). 

During the early LST operation, it was observed that in conditions of high atmospheric humidity (dew point $> 15\degc$), the HV currents increased by several hundred nA per channel.  Although this did not adversely impact the performance of the LSTs, it caused occasional HV trips and affected the current monitoring.  The likely cause were surface currents from the HV pins of the tubes which were exposed to air. The problem was solved by flowing several liters per minute of dry air into the IFR gaps.

\paragraph{Dead and Noisy Channels}

At the end of the data taking, less than 0.5\% of the signal channels malfunctioned
 (23/4656 wires and 26/5070 z-strips), highlighting the effective quality control during fabrication and installation. No failures of the front-end electronics, such as dead or noisy channels or broken boards or crates, occurred.

\paragraph{Muon Efficiency}
The detection efficiency of the LST detector was measured by using  
pair-produced, high momentum muons. The average efficiency at the end of \babar\ data-taking was about 88\%, slightly below the geometrical acceptance of 92\%.  The main sources of the additional inefficiency were broken connections on strips or wires that had been disconnected or kept at lower voltage.  Figure~\ref{fig:lst:effdist} shows the average efficiency of the 72 LST planes, immediately after the
installation and at the end of the data taking.   The worst results were obtained for the two innermost layers.  However, their efficiency loss did not significantly impact the muon identification.

\begin{figure}[!htb]
\centering
\includegraphics[width=0.5\textwidth]{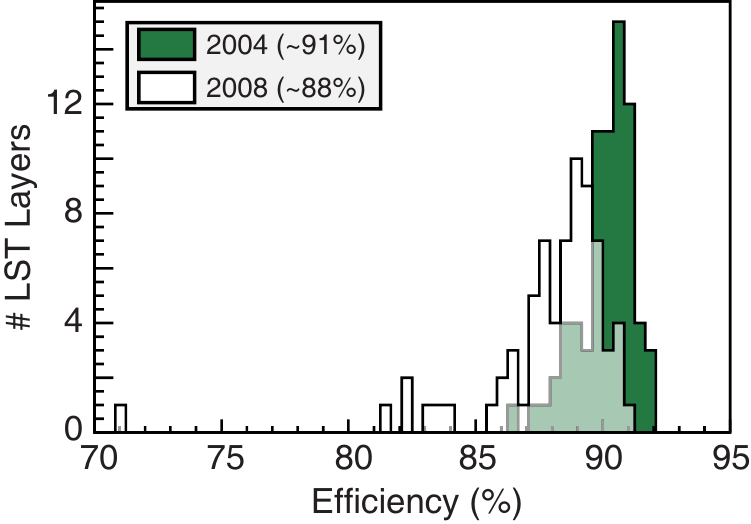}
\caption{Single gap efficiency distribution of all LST layers shortly after the installation in 2004 (shaded) and at the end of the data taking in 2008
(white), with averages of 91\% and 88\%, respectively.
}
\label{fig:lst:effdist}
\end{figure}

Figure~\ref{fig:lst:efftime} shows the evolution of the IFR single layer detection efficiency for single muons for the  barrel sextants 
from Run 1 to Run 7.  The steady decline of the RPC performance is clearly visible, as are the improvements following the installation of the LSTs in 2004 (sextants 1 and 4) and 2006 (sextants 0, 2, 3 and 5).
The efficiencies for all LST sextants were constant until Run 7, when the innermost layers started to be affected by the increasing beam backgrounds. 

\begin{figure}[!htb]
\centering
\includegraphics[width=0.5\textwidth]{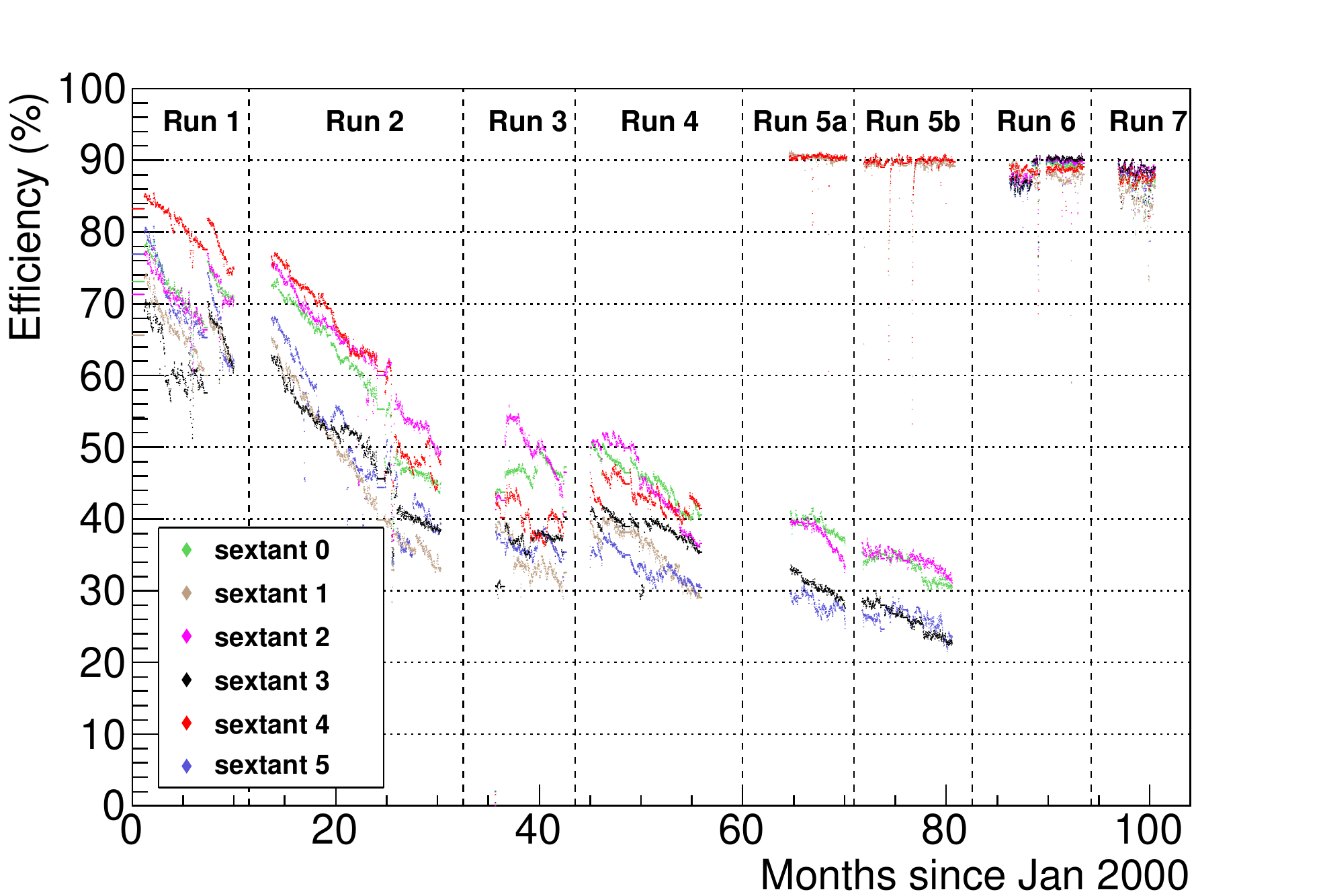}
\caption{Average single gap detection efficiencies for the barrel IFR sextants for Runs 1-7.
}
\label{fig:lst:efftime}
\end{figure}

\subsubsection{Overall IFR Performance} 
\label{sec:ifrperf}
Despite the ongoing effort expended in keeping the original RPCs as efficient as possible, the average 
single gap efficiencies of these RPCs declined steadily until their replacement in the barrel and forward 
endcap.  The backward endcap RPCs which covered $<$ 5\% of the c.m. solid angle were never replaced. The single 
gap efficiencies of the LSTs in the barrel (Figure~\ref{fig:lst:efftime}) and the replacement 
forward endcap RPCs (Figure~\ref{RPCeff}) were stable, except for a small drop in the last months of running due to the 
higher background.

A substantial effort  was also devoted to improve the muon identification algorithms and to reduce their sensitivity to changes in the efficiencies of different IFR layers. 
The algorithm based on bagged decision trees (BDT), described in Section~\ref{section:muonPID}, benefited 
from the large redundancy in the number of detector layers and from additional information from the inner tracking systems and the EMC. The muon identification efficiency and pion misidentification, averaged over the entire detector fiducial volume for tracks
with momenta in the range 1.5  to 2.0 \gevc, are presented in  Figure~\ref{fig:MUIDtime}. 
The muon efficiency depended on the desired level of pion rejection which, for the data presented here, was set at constant levels between 1.2\% and 5\%. Lower pion misidentification rates were achieved at the cost of lower muon efficiency.  For a pion misidentification of 5\%, the muon efficiency exceeded 85\% at all times, whereas for a tighter requirement of 2\% the muon efficiency ranged from 70\% at the end of Run~2, prior to the RPC endcap upgrade, to 80\% after the completion of the LST installation, prior to Run~6. 

\begin{figure}[!htb]
\centering
\includegraphics[width=0.45\textwidth]{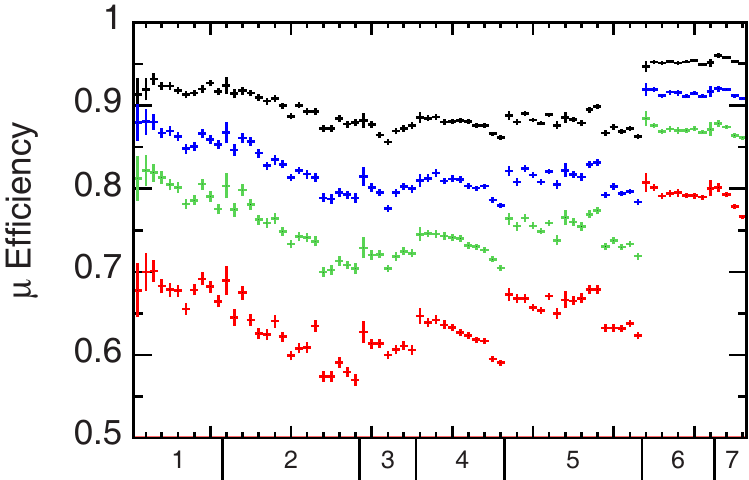}
\includegraphics[width=0.45\textwidth]{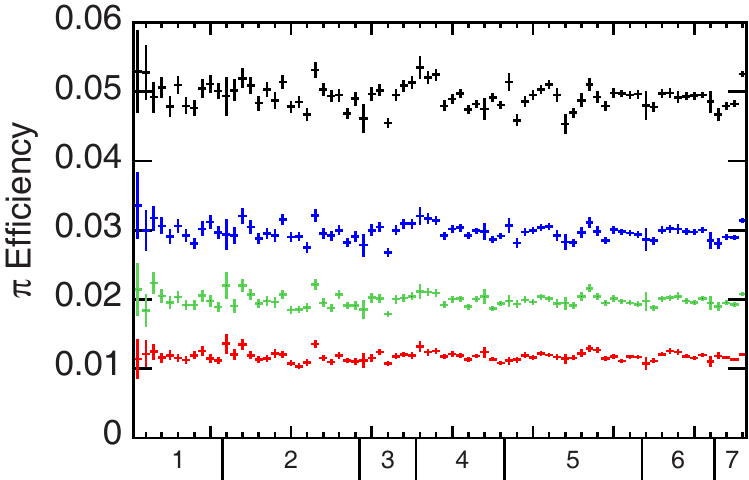}
\caption{Performance of muon identification based on BDT algorithms:
muon identification efficiency (top) and pion misidentification (bottom) for selected charged tracks in the momentum range of 1.5 to 2.0 \gevc\ as a function of calendar time, expressed in terms of Runs 1-7.  The four data series marked in different colors correspond to four constant levels (1.2\%, 2\%, 3\%, and 5\%)
of the average pion misidentification rates and indicate their correlation with the 
muon efficiencies.   
The data are integrated over the full fiducial volume.}
\label{fig:MUIDtime}
\end{figure}

The data recorded after the completion of the LST and forward endcap replacement show close to optimum detector performance.  By comparing the average muon identification of the entire \babar\
dataset to this final period, one can derive a rough measure of the number of muons lost to physics analyses due to RPC hardware problems. This estimate ignores improvements from the additional absorbers installed during the IFR upgrades. The average muon efficiency  for the highest purity (1.2\% pion misidentification) was ~63\% compared to 80\%.  
For a higher pion misidentification rate of 5\%, the muon average was ~89\% compared to 95\% after the upgrades.  This translates to about 6-17\% of  muons lost due to hardware problems.

\subsection{Trigger}
\label{sec:operations-TRG}

\subsubsection{Operation at \FourS}

The L1 trigger configuration was designed for maximum efficiency and
background suppression, and used a mixture of DCT-only, EMT-only,
mixed and pre-scaled triggers.

The entire trigger system was configured from a single source, which
defined the logic of L1 trigger lines and connected them to a variable
number of L3 output lines from which the final logging decision was
formed.
Rather than executing algorithms for specific physics processes, the
 strategy was to employ an {\em open} trigger, based primarily
on one- and two-particle signatures, that used simple topologies
rather than kinematic selections wherever possible.
Such topological triggers became increasingly important as luminosity
and beam-generated backgrounds rose above the design levels.
The replacement of the \pt-discriminator with the $z$-track trigger is
an example for upgrades designed to mitigate higher background levels.
In order to maximize the trigger efficiency, its stability and its
measurability, the general strategy was to employ {\em orthogonal}
triggers, such that physics events were identified by more than one algorithm based on drift chamber and calorimeter information.
The lion share of luminosity-induced events, from Bhabha
scattering, was selected at Level~3 and flagged for calibration purposes, 
and separated from the physics lines by a \emph{veto},
which positively identified these events with very high purity, thus
leaving the efficiency for physics events unaffected.
More than 100 such trigger configurations were deployed over the
lifetime of the experiment, the majority of them to adjust the pre-scale
rates for Bhabha events, while others introduced or refined trigger lines, and
tuned algorithms. 

At the \FourS\ resonance, it was relatively easy to trigger on
generic \BB\ events, nevertheless, special care was taken to 
ensure high efficiencies for rare and low-multiplicity decays.
By virtue of its openness and orthogonality, the trigger system was
able to maintain efficiencies of over 99.5\,\%\ even for rare decays such
as $B \to \mtau \nu$ and $\B \to \ppz$.

Continuous minor modifications to the trigger allowed the dead-time of
the system to be kept near 1\,\%, despite the ever-increasing machine
backgrounds associated with higher luminosity and beam currents.
There were three principal sources of dead-time in the DAQ system:
1)  The system was temporarily {\it busy} while reading out the event buffers with
the minimum command spacing of 2.7\mum between two subsequent
\emph{L1 Accept} transitions (which the FCTS enforced to simplify the
read-out logic).
2) The front-end buffers were {\it full}, usually caused by 
bursts of triggers within a short period of time, or more 
generally when any part of the system was unable to keep up
and exerted {\em backpressure}.
3) An external signal {\it inhibited} an \emph{L1 Accept}, for instance
during trickle injection.
Since the entire L3 processing was in the dead-time path, the
trigger itself could contribute to backpressure.  This, however, never
had a significant impact.
On the other hand, the command spacing resulted in an irreducible,
albeit small, amount of dead-time (0.54\,\% for every 1\,\khz\ rate).
This was the main contribution to the dead-time at high
luminosity, rather than the physical limits of the
system, and the principal reason to keep the trigger rate 
as low as practical.

The maximum L3 output rate was originally set to 120\,kHz, 
to avoid overloading the downstream storage and processing 
capacity of the offline computing system.
This limit was raised as luminosity exceeded the design and more
offline capacity became available.
The average event processing time on a L3 node was about
10\,\ms, dominated by the tracking algorithm.
The overall CPU time grew only slowly over the years, despite the
increasing detector occupancies.  
This was accomplished by diligent and continuous efforts to optimize the CPU
performance and the resource utilization of the L3 software.
A single major upgrade of the online farm was adequate to create
sufficient headroom until the end of data taking. 

Table~\ref{tab:trg-effs} shows the trigger selection efficiencies and
rates for typical running conditions  at the \FourS\ resonance ($1.2
\times 10^{34}\cms$ instant luminosity, 2.7\khz L1 rate).
Triggering on generic \BB\ events was highly efficient, due to high
particle multiplicity and the orthogonality of the trigger lines.
With the volume of data collected, searches for and
measurements of very rare processes could be performed; the rates
for some of these rare processes are also given in
Table~\ref{tab:trg-effs}. 
The efficiencies were assessed based on MC simulations of
events.  The results demonstrate how multiple redundancies built into the
trigger configuration ensured that even very rare decays were
efficiently selected.  For instance, for events with a 
$B^0\to \nu\overline{\nu}$ decay the L1 trigger efficiency
was  estimated to be 99\,\%, \ie\ it worked very well for
a single detectable $B$ decay in the event.

The simulation included details of the 
DCT, EMT, and GLT down to the level of the primitives, and
applied the same L3 software used during data taking.
By employing the original data-taking configurations, this
resulted in a reliable emulation of the L1 decisions and an
effectively \emph{self-simulating} L3 selection.
Being able to thoroughly simulate the trigger had the great advantage
that proposed changes to the trigger configuration could be emulated
and the potential effect on physics processes could be determined 
reliably without any real-time tests interrupting regular data-taking.

\begin{table*}[htb!]
\centering
\caption{Summary of L1 trigger efficiencies (\%) based on MC
  simulations of various production and decay processes at the \FourS\ for actual trigger
  configurations at a luminosity of $11.8 \times 10^{33}\cms$ and a
  2.7\khz\ L1 rate.}
\label{tab:trgeff}
\label{tab:trg-effs}
\begin{tabular}{lcccccccc} \hline
  L1 trigger line &  ${B^+} $ & $B^0$ & $B $ & $B^0$ & $e^+ e^- $ & $e^+ e^- $ & $\tau $ & $\tau $ \\

&  $\to D^0 K^+$ \ & $ \to K^0_s K^0_s$ \ & $ \to \mu \nu$ \ &  $\to \nu \overline{\nu}$ \ &
$\to \tau^+ \tau^-$ \ & $\to \FourS\ \gamma$ \ & $\to e \gamma$ \ & $\to \mu \gamma$\\ \hline
  DCT only & $>$ 99.9 &  98.4  & 93.4 & 43.6  & 70.2 & 93.1 & 68.0 & 69.9 \\
  EMT only &  $>$ 99.9 & 99.8 & 98.4 & 45.9  & 76.1 & 75.0 & 91.0 & 85.8 \\
  Mixed lines & $>$ 99.9 & 99.8  & 99.2 & 62.2 & 88.7 & 93.7 & 91.6 & 91.8 \\
  Total &  $>$ 99.9 & $>$ 99.9 & 99.8 & 99.3  & 91.4 & 95.2 & 95.5 & 94.8 \\ \hline
\end{tabular}
\end{table*}

\paragraph{Masked EMT Channels}

An example of a series of changes that were implemented during
regular operation and significantly improved the trigger operation,
was a response to {\it hot towers} in the EMT. 
The EMT received as input the energy in {\it trigger towers}, \ie\ the 
sums of the energy recorded in groups of $3 \times 8$ crystals.
Erroneous hot towers could be caused by corruption of input data due
to electronic failure, or by failure of the EMT itself, resulting in a
large increase in trigger rates.  These hot towers occurred frequently
during the earlier years of \babar\ operation. In most incidents the
problem could be traced to bad connections or cables and was easily
fixed during maintenance.
When this happened, masks were temporarily introduced in the EMT
selection, and/or the EMC reconstruction in Level 3 as appropriate,
to overcome the high trigger rate. They were subsequently
also applied in the event simulation to assure that the detector
conditions could be accurately reproduced.

A serious system flaw was once introduced when one of the
EMT panels had been erroneously replaced with a spare in which the
firmware had not been updated.
The mismatch in firmware versions caused mismatches in clock phasing
between boards, resulting in bit corruption and hot towers.
This problem led to the adoption of a {\it hot spares}  policy starting in
Run~2:  boards were swapped out during
data-taking to ensure that spares were known to be operating reliably.

A consequence of the varying channel masks during the early years of
operation was that data were recorded under frequently changing
configurations which could affect the analyses.  Thus the actual 
operating conditions had to be logged and implemented in the 
simulation.
This challenge was solved by a technique that overlaid background 
frames from collision data with MC-generated events.
The background sampling was performed by a L3 algorithm that selected
random triggers with a rate proportional to luminosity (using Bhabha
triggers as seeds).
Event time-stamps of the background frames were used to select the
trigger (and other) configurations for overlay of the simulation.
This method guaranteed that configurations, for instance the channel
masks, were applied to the correct data taking periods and in the 
actual proportions to reproduce the recorded data.

\paragraph{Changes to the IFT Timing}

The IFR trigger (IFT) was a standalone trigger for muon diagnostics,
introduced at a late stage.
The IFT was never designed to allow access for configuration, for
the monitoring of processes, or for information on the
trigger registers.
The output was an 8-bit word carrying geometric information which
enabled cosmic ray triggers.
The limitations of this design choice were exposed when LSTs (see Section ~\ref{sec:upgrades-LST}) 
were installed to replace RPCs in the IFR barrel.
The LST signals arrived about 600\,\ns later than the RPC signals, 
and therefore the signals from the remaining RPCs had to be delayed.
The lack of access to information on the board necessitated a
trial-and-error approach in a sequence of time-consuming firmware
adjustments  to diagnose the problem and  test the solution.
This was in contrast to the rest of the trigger system, which was set-up with a
host of monitoring and configuration tools, built-in to allow fast
diagnosis of problems and easy implementation of solutions.

\paragraph{Level~3 Output Rates}

Operationally, the L3 system underwent a significant evolution.
Initially,  the L3 trigger was operating
at roughly 1\,\khz and logging at 120\,\hz.  By the end of Run 7,
the system was routinely running at three times that rate and logging
800\,\hz with ease.  Such logging rates were necessary to efficiently 
record single-photon events at the \ThreeS\ resonance, but were not
anticipated in the system design.
This increase, which enabled important searches for new physics processes, 
was possible because of the spare capacity in the system design.  
As a result, the L3 system never contributed significantly to the
detector dead-time, even at those enhanced rates.

\subsubsection{Run 7 Operation}
\label{trg-run7}

In December 2007, the decision was made
to terminate data taking at the \FourS\ and to
record data at the $\ThreeS$ resonance
during the remaining months of \babar\ operation.  The emphasis was on the
search for new physic processes in decays such as $\ThreeS \to \pi^+ \pi^-
\OneS (\to \mathrm{invisible})$, $\ThreeS \to \pi^0 \pi^0 \OneS (\to
\mathrm{invisible})$, and $\ThreeS \to \gamma X (\to
\mathrm{invisible})$.
The trigger was the only system that was significantly impacted by
this change and had to be adapted quickly.
The flexibility in the design of the trigger system made this possible.

Triggering on events with invisible energy required a shift in trigger 
philosophy and necessitated a trigger capable to detect the 
signatures that accompanied the invisible decays.
The trigger was originally designed to select events with visible
energies of close to 10.58\,\gev, with high track/cluster
multiplicities, or high transverse momentum/calorimeter energy for
low-multiplicity final states.  Events of interest at the
\ThreeS\ typically deposited less than 10\,\%\ of the c.m.~energy in
the detector, had very low track and photon multiplicity,  and
typically only two particles per event.  The visible energy was 
low, because the mass difference between
the \ThreeS and \OneS is 895\,\mevcc.  The challenge was
to capture events containing these particles without allowing the
trigger rates to exceed the data logging capacity.

Three main extensions were made to accommodate this class of decays.
\paragraph{$\ThreeS \to \pi^+ \pi^- \OneS (\to \mathrm{invisible})$}
\label{par:upsthree}

A new $\pi^+ \pi^- + \mathrm{invisible}$ energy trigger was installed, with
four new L1 trigger lines and a new $t_0$-finding algorithm
and L3 track filter.  Both pions in these events have low
energy; fortunately their transverse momenta were close to equal 
and opposite in direction.
Approximately 78\,\%\ of these tracks were emitted into the detector
fiducial volume.  The \FourS\ trigger had an efficiency of 36\,\%\  for this
channel.  A single line in the \FourS\ trigger captured most of the
decays, and the new L1 lines exploited a number of variations on
this line, requiring low-momentum tracks and EMC towers consistent with minimum ionizing charged particles (MIP), and exploring
charge-dependent selections.  After the addition of the new L1 lines, 
the trigger efficiency for this channel increased to 80\,\%.

It turned out that $\ThreeS\ \to \pi^+ \pi^- \OneS (\to
\mathrm{invisible})$ events were rejected by the L3 trigger not
only because of the low-momentum tracks, but because they did not pass
the $t_0$-finding criteria.  This failure was traced to a stringent
requirement on the TSF segments for the track reconstruction.  For
$\BB$ events, this requirement resulted in good $t_0$ resolution at high
efficiency, but it did not function well for two low-energy
tracks in the event.  The L3 $t_0$-finding patterns were adjusted
for the new physics environment.  A second pass was added, to
be tried when the first pass failed.  This had a particular advantage
that no changes to the existing $t_0$ or track fits were necessary.
New events were added which would have previously failed the
L3 requirements.  The other L3 alteration was to lower the
\pt\ threshold for the new single-track filter from 600\,\mev to 200\,\mev.

After implementation of these changes, the trigger rates were observed
to rise moderately, but did not have a significant impact on the dead-time.

\paragraph{$\ThreeS \to \pi^0 \pi^0 \OneS  (\to \mathrm{invisible})$}
\label{par:upsone}

The deployment of a $\ThreeS\ \to \pi^0 \pi^0 \OneS (\to
\mathrm{invisible})$ trigger required two new L1 lines, and two new
L3 filters.  At the \FourS, this decay was
mostly triggered by a single line which required three MIP clusters,
two of which had to be back-to-back.  
At the \ThreeS, this angular requirement would have lost about
30\,\%\ of the $\pi^0 \pi^0 + \mathrm{invisible}$ events, so two new
lines were introduced.  One line required three MIP clusters with one
medium energy cluster, and the other line required only two clusters, 
not associated with tracks.

The effect of this track veto depended strongly on beam-generated
backgrounds.  It was the first time that a veto had been applied at
Level~1, contrary to the trigger design philosophy of
retaining all possible physics events, so that all events could be
studied by analysts.

The L3 efficiency for this kind of events was essentially zero in the
\FourS configuration, so two new L3 filters were developed to deal
with multiple photon final states.  They selected events without 
tracks originating from the IP, but with EMC clusters, 
and computed the invariant mass, visible
energy, and \OneS\ recoil mass.  The recoil mass was required to exceed
9.5\,\gev.  It was deployed in two exclusive instances: a four-photon
version and a three-photon version with a slightly higher visible
energy requirement.  The L3 acceptance rate reached almost 50\,\%\ of
the L1 rate.

After installation of these new lines, trigger rates increased in both L1 and
L3, but remained within operational limits. 

\paragraph{$\ThreeS \to \gamma X (\to \mathrm{invisible})$}
\label{par:upsups}

The second major change was to install a very low-energy single photon
trigger for $\ThreeS\ \to \gamma X (\to \mathrm{invisible})$
decays.  This took the form of two new L3 trigger lines.  The
previously existing single-photon filter had an energy threshold of
2\,\gev, limiting the maximum hadronic mass for the process 
$e^+ e^- \to \gamma X$ to $m_X \approx$ 8.1\,\gevcc.  Two new conditions of the L3 single-photon filter were added, with energy thresholds at
1\,\gev\ and 0.6\,\gev.  In this way, the reach for the hadronic mass
was extended to $m_X \approx$ 9.7\,\gevcc.

The price for this new configuration was a large increase in L3
data logging rates and through the end of operation this single trigger
line became by far the largest contributor to the L3 output.
Single-photon events accounted for one third of the events on tape.  
A side-effect of this was that the limit on the size of the output file
for each run was reached very quickly, and the length of the runs 
had to be shortened administratively.

\paragraph{Performance}

In the last few weeks of operation, the trigger system demonstrated
its full capacity.

The flexibility and spare capacity that was designed into the
\babar\ trigger enabled the trigger efficiency for most physics
channels to remain above 99\,\%\ throughout the lifetime of \babar, while
running on both the \FourS\ and \ThreeS\ resonances, and later above
the \FiveS.  Despite the extraordinary conditions, and the
rapid increase in luminosity, the dead-time was kept very low.


\renewcommand{\secname}{reco_}
\renewcommand{\sectiondir}{Sec05_Reco}
\section{Event Reconstruction}
\label{sec:reco}

\subsection{Overview}

\babar\ event reconstruction methods have evolved significantly since they were first reported in 
2001~\cite{Aubert:2001tu}.  This section focuses on major additions and modifications
to the algorithms for charged particle tracking, the detection of photons and neutral pions,
and the identification of charged particles.  
Some information on muon efficiencies and pion misidentification is presented in Section~\ref{sec:ifrperf}.

\subsection{Charged Particle Reconstruction}
\label{sec:track-rec}

Charged particles were reconstructed as {\em tracks} in the
\dch\ and/or the \svt.
The track reconstruction algorithms used in the 2008 final reprocessing of the \babar\
data were largely the same as described previously~\cite{Aubert:2001tu}.  
The major differences are noted below.

\subsubsection{Track Finding}

Tracks were found using a sequence of {\em pattern recognition} algorithms,
by which individual hits in the \dch\ and \svt\ were associated with and attributed to a single charged particle.
The first stage of pattern recognition was the selection of
tracks found in the \dch\ by the L3 trigger to be retained as offline tracks.
In the original version of this algorithm, all L3
tracks with a reasonable helix fit were selected. 
In the final reprocessing, only tracks with hits at least 25 of the
tracks with a reasonable helix fit were selected. 
In the final reprocessing, only tracks with hits in at least 25 of the
40 possible layers were retained.  This eliminated poor fits with
incomplete or incorrect stereo reconstruction from the offline sample.
The hits from these rejected tracks
were made available as input to the more sophisticated offline pattern recognition
algorithms which followed. 
Overall, this change improved the efficiency for correctly reconstructing tracks
in offline processing, by a few percent.

A new track finding algorithm was introduced in 2003. It allowed the addition of 
\dch\ hits to low transverse momentum tracks found in the \svt.  
This algorithm used the same core software already
developed for adding \svt\ hits to tracks found in the \dch, thus
increasing the number of hits in the Kalman filter fit, provided they were
consistent within errors.   To add \svt\ hits to
tracks found in the \dch, the search for additional hits was performed
sequentially by layer, working radially inwards from the \dch.  
To add \dch\ hits,
the search proceeded radially outwards, from the \svt through the \dch\ layers.
In addition, a search for \dch\ hits in the same layer, but adjacent in azimuth, was
performed, to find the hits generated by curling, low $p_t$ tracks.
Adding \dch\ hits to tracks found only in the \svt\ improved their momentum resolution by more than a factor of two, largely due to the improved
curvature measurement provided by the \dch.  
This procedure improved the resolution on the $D^{*+} - D^0$ mass difference in $D^{*+} \to D^0 \pi^+$ decays, for which the {\em slow} pions tracks were found in the \svt , by a factor of two.

\subsubsection {Track Filtering and Refinement}
\label{sec:track-ref}

Roughly 10\% of tracks found by the  reconstruction
algorithms were duplicates, ${\it i.e.}$,  the majority of their hits
were generated by the same particle.
Furthermore, some of the reconstructed tracks were
created either by interactions of primary particles with the material of
the detector or by the decay of long-lived particles.
In some cases, spurious tracks resulted from the pattern recognition
algorithms, combining hits from unrelated primary particles or
tracks with background hits.
These duplicate and fake tracks impacted physics analyses by increasing
combinatoric backgrounds and errors in event reconstruction.
Prior to 2008, such spurious tracks were filtered at the
analysis stage by imposing restrictions on the reconstructed track 
parameters.  While effective in removing background tracks, 
these quality selection criteria also removed some tracks from primary particles.  
In 2008,
an improved filter for spurious tracks based on the full reconstruction
information was introduced. 
Additionally, a final pass of a refined, hit-level pattern recognition was 
introduced. 
These additional algorithms are described below, in the order 
in which they were applied in the track reconstruction sequence.

\paragraph{Track Subset Selection}

To reduce the computational burden, track filtering and refinements
were performed on a subset of reconstructed tracks which was expected
to benefit from further processing.  Tracks were selected which 
were inconsistent with a good fit, $P(\chi^2)<10^{-5}$, and tracks with
substantially fewer \svt\ hits than expected based on their trajectory.  
In addition, pairs of tracks separated by less than 1~cm anywhere in 
the \dch were added to the selection, when one of them had a distance 
of closest approach (DOCA)
to the primary vertex exceeding 3~cm.
Approximately 20\% of reconstructed tracks were selected as input 
to the filtering and refinement.

\paragraph{$V^0$ Identification}

Tracks originating from the decay of long-lived neutral particles ($K_s^0$ or $\Lambda^0$)
and from photon conversions to $e^+e^-$ pairs in the material of the tracking detectors
often had a large DOCA and missing \svt\ hits.
To avoid selecting and incorrectly reconstructing tracks from these vertices, referred to as
$V^0$s, they were explicitly identified  and removed them from the filtering and refinement subset.

$V^0$s were identified by a geometric vertex algorithm applied to all pairs of tracks of opposite charge in the events, requiring a fitted vertex probability of $P(\chi^2) > 10^{-4}$.
Furthermore, the vertex position along the direction of the momentum vector of the reconstructed $V^0$ had to be significantly displaced from the primary event vertex.  The particle identification and the invariant mass was required to be 
to be consistent with the $V^0$ decay hypothesis. 
The invariant mass distribution of reconstructed
$\Lambda^0 \rightarrow p^+ \pi^-$ candidates is shown in Figure~\ref{fig:v0_mass}.

\begin{figure}[ht]
\center
\includegraphics[width=7.5cm]{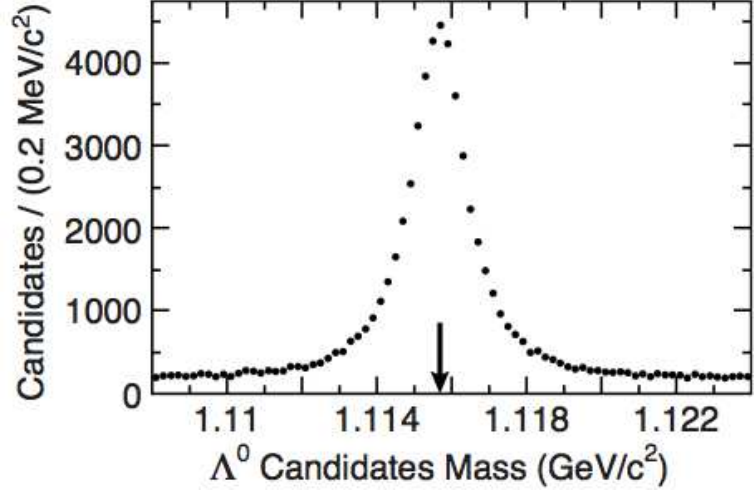}
\caption{Mass of $\Lambda^0 \rightarrow p^+ \pi^-$ candidates reconstructed by
the $V^0$ selection.  The arrow indicates the 
nominal $\Lambda^0$ mass~\cite{Beringer:1900zz}.
}
\label{fig:v0_mass}
\end{figure}

\paragraph{Hit Filtering}

Failure of the pattern recognition algorithms to
discriminate against background hits or hits from other tracks
usually led to poor track fits and incorrect fit parameters.
To overcome this problem, an effort was made to 
identify and remove incorrectly associated hits.

An individual, potentially incorrectly associated hit was selected based on its
contribution to the track fit $\chi^2_\mathrm{s}$, derived from the
unbiased normalized residuals from the Kalman filter fit.
For \svt hits, a term $\chi^2_t$ was added, defined as
the difference squared  between the measured hit arrival time and 
the event time, normalized
to the estimated time resolution.  These total $\chi^2$ was derived as 
$\chi^2_\mathrm{eff} = (\chi^2_\mathrm{s} + w^2 \chi^2_t)/(1 + w^2)$, with
the relative weight $w=2$ obtained from simulation.
For \dch hits, $\chi^2_\mathrm{eff} = f^2 \times( \chi^2_\mathrm{s} + 100\times\Theta(D_\mathrm{wire}-D_\mathrm{max}) )$, where $\Theta$ is the Heaviside step function and  $D_\mathrm{wire}$  refers to the projected drift distance, {\it i.e.,} the distance between the track and the signal wire, and $D_\mathrm{max}= 1.05$~cm is the maximum drift distance.
This term effectively added a penalty whenever the measured drift distance exceeded 
the cell size, as expected for most hits that were incorrectly assigned to the track.
The factor $f=1/3$ accounts for the difference in quality of \svt and \dch hit error estimates; its value was optimized based on simulation.  Figure \ref{fig:doca_cut} compares the distributions of 
\dch\ DOCA for hits associated to tracks in data and MC.
In the MC simulation of
$e^+e^- \rightarrow \mu^+ \mu^-$ events the hits were selected, based on 
MC truth information, to be correctly associated to the track.  The data agree well
with the simulation, although there is some excess above the nominal 
maximum at $1.05\,\mathrm{cm}$.
For the MC simulation of  $e^+e^- \rightarrow \BB$ events, the distribution of wrongly associated hits is presented, demonstrating  that
the selection criterion set at $1.05\,\mathrm{cm}$ has a high efficiency for correctly associated hits, while  rejecting a large fraction of wrongly associated hits.

\begin{figure}[ht]
\center
\includegraphics[width=7.5cm]{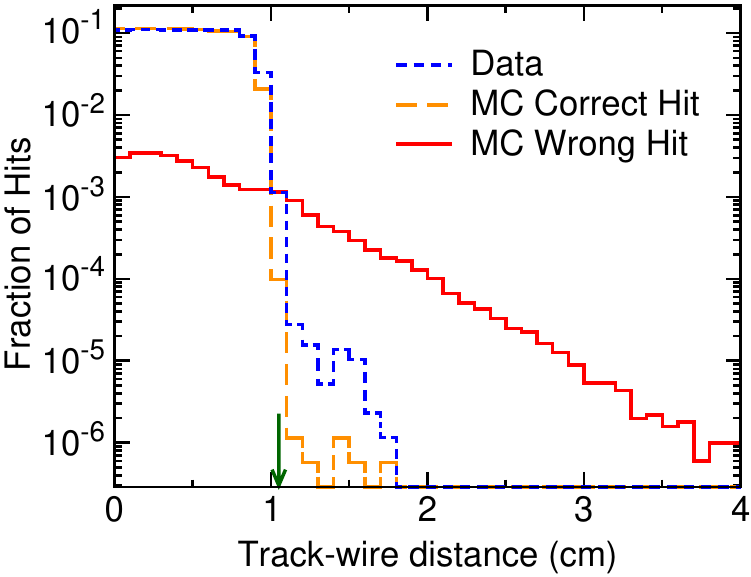}
\caption{DOCA to the wire for \dch hits produced by high momentum tracks in
$e^+e^- \rightarrow \mu^+ \mu^-$  for data and MC (correct hits) events,
and for $e^+e^- \rightarrow B\bar{B}$ for MC (wrong hit) events. The arrow
marks the selection criterion at $1.05\,\mathrm{cm}$.}
\label{fig:doca_cut}
\end{figure}

In this sequential procedure, hits with $\chi^2_\mathrm{eff}>9$ were removed,
provided the $\chi^2$ for the Kalman filter fit improved by at 
least $10$ for each removed hit.
Up to three~hits per track were removed, starting with those
with the largest $\chi^2_\mathrm{eff}$ and making sure that the remaining hits
retained a good over-constrained track fit.
Simulations indicated that the resolution in the track curvature or
the longitudinal impact parameter improved by at least $1 \sigma$ in
$44\%$ of tracks from which hits were removed,
corresponding to roughly $2.5\%$ of all tracks in simulated
\BB\ events.

\paragraph{Addition of More \svt Hits}

As described previously~\cite{Aubert:2001tu}, \svt\ hits were added to tracks found in the
\dch\ as part of the standard track finding procedure.  To prevent \svt\ hits from
background processes or other tracks from being incorrectly assigned to particles which decayed inside or beyond the \svt,
the algorithm required that the pattern of associated \svt\ hits matched expectations for tracks
which originated at the IP. This reduced the efficiency of adding inner-layer \svt\ hits
to tracks, which in turn degraded the impact parameter resolution of some tracks.

To recover inner-layer \svt\ hits excluded by this procedure, a second
pass of adding  \svt\ hits to the selected tracks was implemented.  
As mentioned above, 
tracks associated with $V^0$s had been removed from this sample.
Consequently, we could relax the hit pattern consistency requirement, 
which improved the efficiency of adding inner-layer \svt\ hits.  
As shown in Figure \ref{fig:svthitaddadd}, in simulated \BB\ events
the transverse impact parameter resolution of tracks improved by roughly a factor of three
for the $\sim$1\% of the tracks to which \svt\ hits were added in the
second pass.

\begin{figure}[ht]
\center
\includegraphics[width=7.5cm]{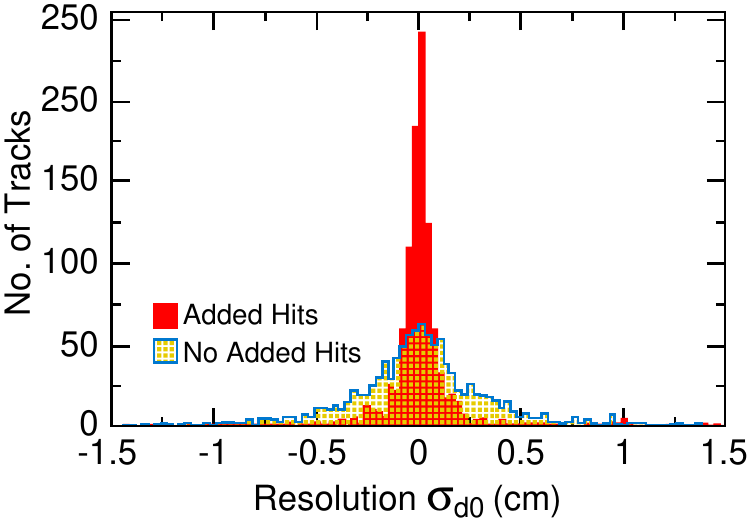}
\caption{Transverse impact parameter resolution of tracks in simulated \BB\ events,
before and after \svt hits were added.}
\label{fig:svthitaddadd}
\end{figure}

\paragraph{Removal of Duplicate Tracks}

Most of the duplicate tracks were due to low-transverse-momentum charged particles which
followed a helical path with multiple loops inside the tracking volume.  Because
the primary \dch\ track finding algorithms worked layer-by-layer, starting at the inner layers,
they were not efficient at adding hits to these tracks since they curved back towards their 
origin at smaller radii. 
Consequently, 
these {\em loopers} were generally found as independent tracks.  This was particularly
the case for particles with glancing incidence on the material just outside the \dch,
which re-entered the \dch with lower transverse momentum.

Loopers were identified as pairs of opposite-sign tracks with azimuthal 
and dip 
angles which differed by $\pi \pm 0.2$. For the
spatial distances between these tracks at locations closest and furthest 
from the $z$ axis it was required that one of those distances
be consistent with zero within the projected track errors.
Pairs which shared a common track were combined recursively
to form  multi-segment loopers.  The track with the smallest longitudinal
impact parameter relative to the event primary vertex were selected as the start of the loop.
All other associated tracks with longitudinal impact parameters that were not consistent within errors with the primary vertex were considered duplicates and removed from the track lists used for physics analysis.

The other major cause of duplicate tracks was the failure of pattern recognition in the \dch, resulting in two separate tracks formed from different subsets of hits generated by the same particle.
In general, the fit for both of these tracks resulted in reasonably accurate parameters in the transverse plane, while one of them had poorly measured longitudinal parameters. 
These duplicates were identified as
pairs of tracks with trajectories in the transverse plane that overlapped within the \dch\ cell size of about 1~cm.  From the two tracks with fewer than $10\%$ of their \dch hits in the same layers, the track with longitudinal impact parameter that was inconsistent with the primary vertex was removed.  Attempts to merge the hits of the selected track with those of the rejected remnant
did not in general not improve the fit. 


All duplicate track removal algorithms combined removed $\sim$50\% of 
the duplicate tracks in simulated \BB\ events.

\paragraph{Identification of Interactions in Material}

A sizable number of tracks originated from interactions of primary particles in the material of the tracking system.
The majority of these interactions resulted in a single track, which could not easily be distinguished from decays of long-lived particles (see below for details).  
Interactions in which the primary particle is itself reconstructed as a track were identified by the charged particle decay identification algorithms and are described below.
In the following, we therefore address only interactions which produce two or more tracks.

 \begin{figure}[ht]   
 \center   
 \includegraphics[width=0.4\textwidth]{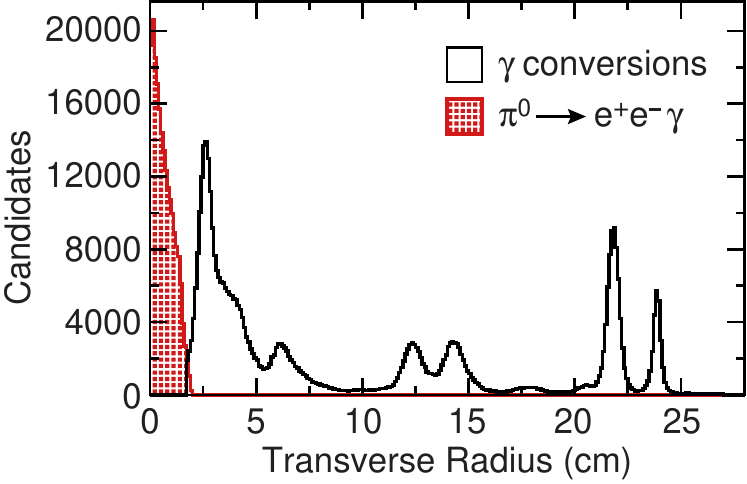}   
 \caption{   
 Transverse radius of the reconstructed vertex for photons converting   
 to $e^+e^-$ pairs, as measured from the center of the beam pipe, and   
 Dalitz decay $e^+e^-$ pairs, relative to the event primary vertex. The   
 principle concentrations of material in the inner detector are clearly   
 visible.   
 }   
 \label{fig:conversion_radius}   
 \end{figure} 

We identified these multi-track interactions by forming a geometric vertex of pairs of tracks, and selecting those with
a radial distance from the beam between 2.5 and 25~cm, where most of the material
was located as shown in Figure~\ref{fig:conversion_radius}.
We also required a consistent vertex fit with $P(\chi^2) > 10^{-3}$,
and a vertex position aligned with the vector momentum sum of the two produced tracks.

We also identified tracks produced by charged particles
that hit the inner surface of the \emc  and either scattered backwards into the tracker,
or interacted and produced secondary tracks which re-entered the tracker.
These {\em albedo} tracks
were found by forming a geometric vertex of pairs of tracks, for which
one of the tracks had large transverse and longitudinal impact parameters, 
while the other was consistent with originating from the primary vertex.
We required that the vertex position was outside the tracker volume and the
distance between the tracks at this vertex was less than 5~cm.

Simulations of \BB\ events predicted that $\sim$25\% of these interactions 
of the primary tracks could be identified by these algorithms.
The identified secondary tracks were removed from the track lists used for physics analysis.
However, because they corresponded to true particles and not detector artifacts,
they were retained for studies of signals in the \emc\ and other  subdetectors.

\paragraph{Identification of Charged Particle Decays}

The decay of $\pi^\pm$ and $K^\pm$ mesons within
the tracking volume can result in either a single track with hits from both the primary particle and the decay product,  
or two distinct tracks.
We have not explicitly dealt with the single-track topology.

Charged particle decays which resulted in a primary and secondary track were identified by the reconstructed decay vertex.
We required that the track separation at the vertex to be at most 1~cm, that the
vertex be inside the tracking fiducial volume, and that the track momenta at the vertex point in roughly the same direction.
We further required that one of the tracks has hits only in the \svt
and inner layers of the \dch, while the other has hits only in non-overlapping outer \dch layers.
This algorithm identified $\sim$50\% of charged particle decays with two reconstructed tracks. The secondary track was removed
from the track lists used for physics analysis, but kept for other detector studies.

\paragraph{Removal of Background Tracks}

A few percent of the tracks were reconstructed predominantly from \svt hits generated by machine backgrounds.
These tracks can have very large  reconstructed momenta, and thus  significantly impact the event reconstruction.
To remove them, we first selected tracks without any \dch hits.  Tracks with hits in only four of the five
\svt layers were removed if their transverse  and longitudinal impact parameter with respect to the interaction vertex were greater than 0.8~cm and 2.0~cm, respectively, or if the average time of the \svt hits differed by more than 32~ns from the event time.
This procedure removed over $90\%$ of background tracks
in simulated \BB\ events, for which machine-related backgrounds had been added by overlaying hits recorded at randomly selected beam crossings to those generated by the simulation.

\paragraph{Impact of Improved Algorithms}

The algorithms described above significantly improved the resolution in $\sim$4\% of tracks,
and eliminated $\sim$30\% of tracks not originating from the decay of the \Bmesons\ or their decay products, as determined from simulated \BB\ events.  They removed fewer than 1\textperthousand\ of the tracks from \Bmesons\ decays.
This set of refined and filtered tracks was provided for physics analysis.
Additionally,
a subset of tracks excluding daughters from well-reconstructed $V^0$ candidates, and restricting the
transverse and longitudinal impact parameters with respect to the primary vertex
to be within $2.0$ cm and $2.5$ cm, respectively, was provided to study particles produced close to the IP.


As an example of the impact of the improved track selection and refinement on physics analysis,
we consider the reconstruction of hadronic \Bmeson\ decays.  The \babar\ {\em Breco} algorithm
combines a reconstructed $D$ or \Dstar\ meson with up to
five charged tracks and four reconstructed neutral particle candidates, selecting
combinations consistent with the most prevalent \Bmeson\ decays to hadrons that do not
exceed the \Bmeson\ mass.
Figure \ref{fig:mes} shows the reconstructed mass of these candidates, computed by constraining
the reconstructed \Bmeson\ energy to the beam energy.  
The signal component due to the \Bmeson\ decays is fit to an
asymmetric Gaussian function~\cite{Skwarnicki:1986xj} and the background to a polynomial threshold function~\cite{Albrecht:1994tb}.
After applying track refinement and filtering,
the reconstruction efficiency improves by $\sim$5\%, and the combinatoric background is reduced by $\sim$15\%.

\begin{figure}[ht]
\center
\includegraphics[width=0.4\textwidth]{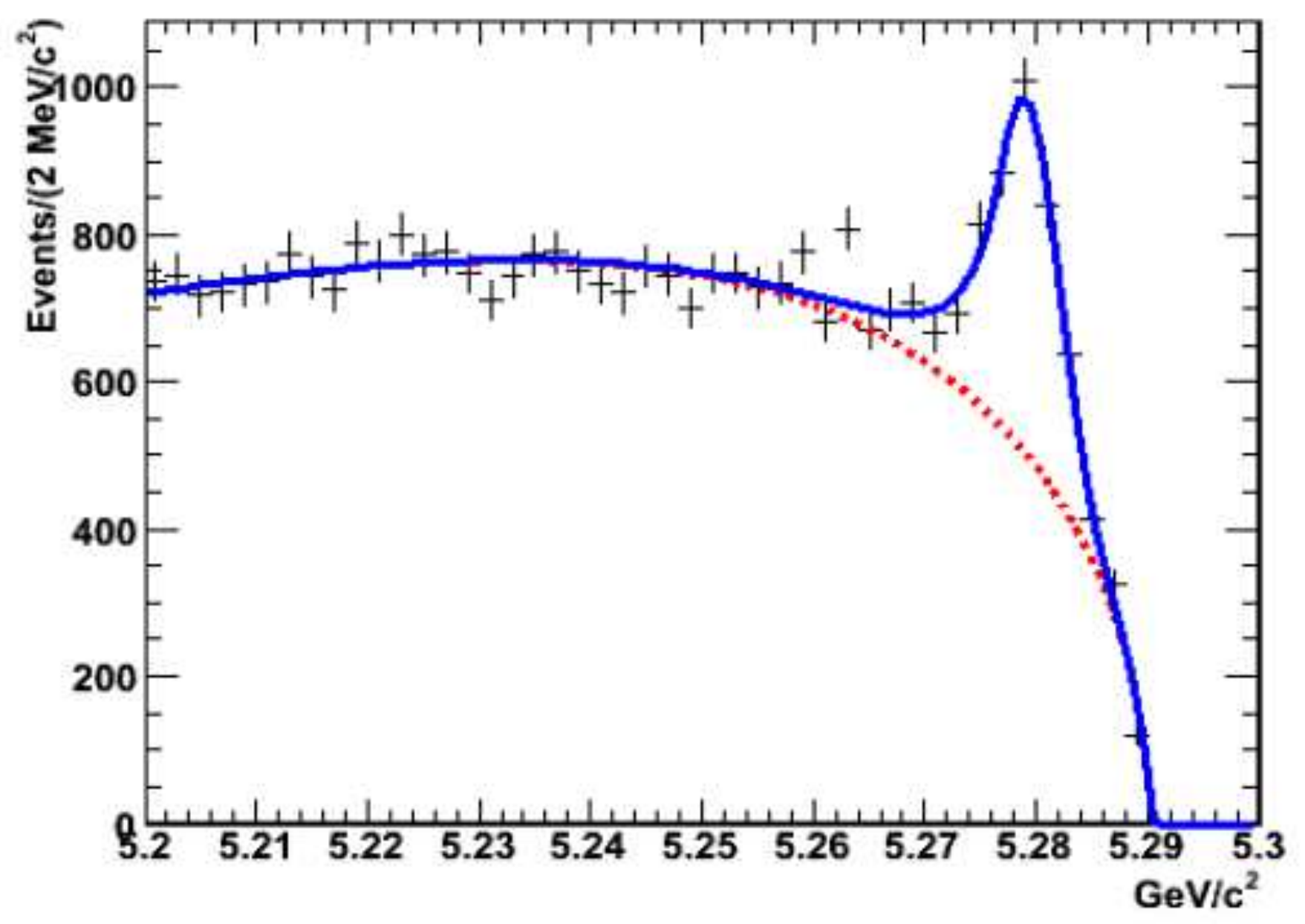}
\includegraphics[width=0.4\textwidth]{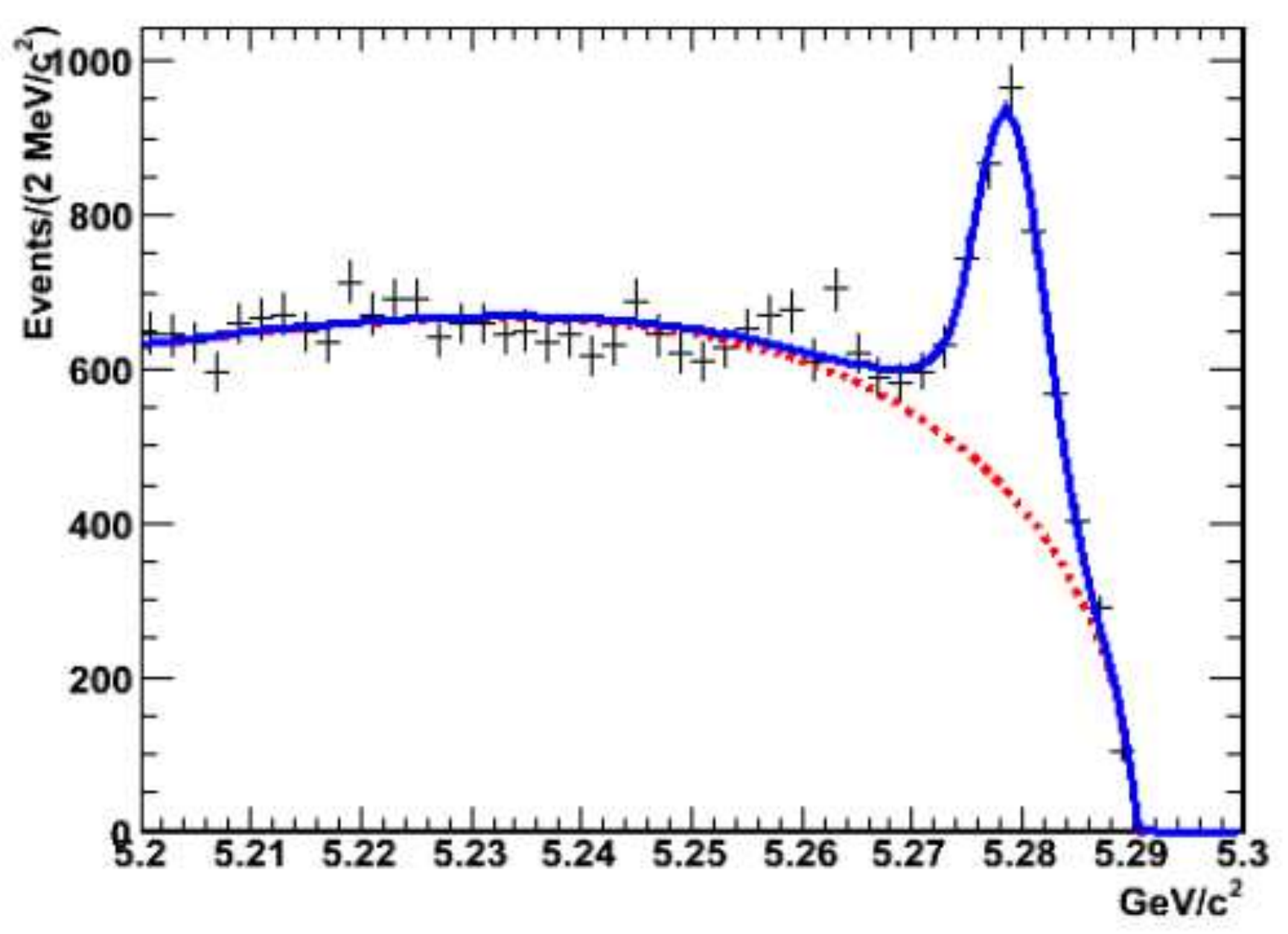}
\caption{Reconstructed mass of hadronic \Bmeson\ candidates, for (a) tracks from the first pass
reconstruction,  and (b) after the implementation of track refinement and selection. The solid line represents the fit to the total distribution, the dashed lines describes the background.}
\label{fig:mes}
\end{figure}

\subsubsection{Track Reconstruction Efficiency}

Many analyses required a precise simulation
of the charged track finding efficiency to determine absolute rates or cross sections.
Different methods were exploited to study efficiency differences between
data and simulation over a wide range of particle momenta and production environments
relevant to most analyses. They are discussed in detail elsewhere
~\cite{Allmendinger:2012ch}.\\

These methods relied on distinct data samples, where additional constraints were applied to 
select certain event topologies. The primary method to study the charged track reconstruction efficiency 
in the data and simulation was based on $\epem \rightarrow\taup\taum$ events, providing 
a sample of 430 million $\tau$ pairs.
Events of interest for the efficiency study involved the leptonic decay of one of the $\tau$ leptons,
$\tau^{\pm}\to\mu^{\pm}\nu_{\mu}\nu_{\tau}$  with $\BR(\tau^{\pm}\to\mu^{\pm}\nu_{\mu}\nu_{\tau}) = (17.36\pm0.05)\%$~\cite{Beringer:1900zz}, and  
the semi-leptonic decay of the other, 
$\tau^{\mp}\to \pimp\pimp\pipm\nu_{\tau}$ with $\BR(\tau^{\mp}\to \pimp\pimp\pipm\nu_{\tau}) = (9.03\pm0.06)\%$~\cite{Beringer:1900zz}, so-called $\tau_{31}$ events.
The $\tau$ pair candidates 
were selected by requiring one isolated muon in combination with two or three charged tracks consistent with being pions.
Through charge conservation, the existence of an additional track was inferred.  The efficiency was computed by
counting the number of times the inferred track was reconstructed.

The variation of efficiency with direction and momentum of the track was among
the largest systematic uncertainties for many physics analyses.  
Due to the presence of multiple neutrinos in the event, 
the momentum vector of the inferred track was not well known.
The systematic error due to this was conservatively derived from the
variation in the agreement between data and simulation as a function 
of the polar angle $\theta$ and transverse momentum $p_{t}$, relying 
on estimator quantities that preferentially 
select $\theta$ or $p_{t}$ regions for the inferred track.
The other main uncertainties resulted from the presence of background events:
radiative Bhabha events with a converted
photon, $\tau$ pair events with a converted photon or a $K_{S}^{0}\rightarrow\pi^{-}\pi^{+}$
decay,  two-photon processes, and $\q\qbar$ events.  Control samples were used to estimate the most significant backgrounds. 
Within the estimated uncertainty of (0.13--0.24)\% (depending on the exact track definition),
this study showed no difference in the track finding efficiency of data and simulation.

The  $\tau_{31}$ method was also used to investigate the stability of track reconstruction over the various run periods. No time-dependent effects in the difference between the data and simulation were observed.

The initial-state radiation (ISR) process $\epem\rightarrow \pipi\pipi\gamma_{{\rm ISR}}$ 
was used to cross-check the $\tau$-based results. Given the absence of neutrinos,  
fits with kinematic constraints to events with at least three detected pions 
could be performed to predict the momentum and angles of the fourth track.
In these events, the high-energy ISR photon was emitted back-to-back to the 
collimated hadronic system. This resulted in a slightly higher probability for track overlap and 
a difference in track reconstruction efficiency for data and simulation of $(0.7\pm0.4)\%$.

Based on the study of low momentum tracks from $D^{*\pm} \rightarrow D^{0} \pi^{\pm}$ 
decays, the systematic uncertainty was estimated to be larger by $1.5\%$ for tracks 
with transverse momenta of $p_t < 180\mevc$. These efficiency studies, 
in combination with the $\taup\taum$ analyses, 
were used to estimate a possible charge asymmetry of the detector
response. This asymmetry was found to be $(0.10\pm0.26)\%$, consistent with zero.

The impact of the displacement of charged tracks from the interaction point 
on their detection efficiencies was
investigated in $\B \to h^+ h^- \KS$ decays (with $h=\pi,K$, and $\KS \rightarrow \pipi$)  
as a function of the \KS\ decay time. 
For the secondary pions, a 
difference in track reconstruction efficiency of $(0.5\pm0.8)\%$ 
between  data and simulation was observed.\\

These studies showed that  the track finding efficiency in data agreed 
within uncertainties with the simulated data~\cite{Allmendinger:2012ch}. 
Thus, the simulated track finding 
efficiencies could be used to estimate signal reconstruction efficiencies
in physics analysis.
However, the analyses must account for the impact of the systematic errors on the tracking efficiency
in the final measurements, depending on the number of tracks 
involved. This method has been adopted for many physics analyses.

\subsubsection{Track-Cluster Matching}
\label{sec:track-cluster-match}

The association of energy clusters with charged tracks entering the \emc\
was critical for the isolation 
of photons (see Section ~\ref{sec:neutrals}), and identification of charged 
particles (see Section ~\ref{sec:cpid}), foremost electrons, 
but also minimum ionizing particles and interacting hadrons. 

As part of the overall event reconstruction, charged tracks were associated with local maxima of
energy deposits in the EMC (so-called `bumps'; see
Section~\ref{subsubsection:ClusterReconstruction}). The algorithm was
updated prior to the 2008 data reprocessing.  
The original algorithm~\cite{Aubert:2001tu}, 
compared the centroid of the bump with the track position projected
to the crystal front face.
However, particles generally deposit their energy throughout the
depth of the crystal.  They also strike the calorimeter face
at non-normal incident angles, due to bending in the magnetic field 
and because the calorimeter crystal axes were intentionally
designed not to project back to the interaction point.  Consequently, this
algorithm introduced biases in the energy and particle-type
dependent track-cluster matching.

The updated algorithm compared the bump centroid with the track position
projected to a depth of
approximately 12.5\,\cm in the calorimeter, which corresponds to the energy-weighted
average shower depth of hadronic showers in the calorimeter.
The track-cluster separation was defined as the
angular difference between the track and the associated bump, computed using the interaction point as origin:
\begin{eqnarray}
\Delta\theta = \theta_{\rm track}-\theta_{\rm cluster},\\ \nonumber
\Delta\phi = \phi_{\rm track}-\phi_{\rm cluster} ,
\end{eqnarray}
where $\theta$ and $\phi$ are the polar and azimuthal angles.

\begin{figure}[ht]
\center
\includegraphics[width=0.4\textwidth]{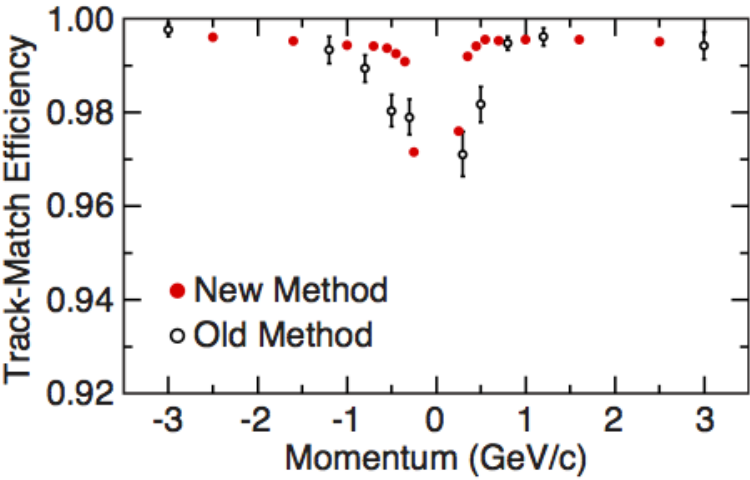}
\caption{
Comparison of the track-cluster matching
efficiency for the two algorithms using pion data.}
\label{fig:trackmatch}
\end{figure}

The track-cluster match consistency, ${\cal C}_{\rm TM}$, was constructed from
the individual consistencies 
${\cal C}_{\Delta\theta}$ and ${\cal C}_{\Delta\phi}$ 
using the following relation~\cite{Janot:1998fz}:
\begin{eqnarray}
{\cal C}_{\rm TM} = {\cal C}_{\Delta\theta}{\cal C}_{\Delta\phi}[1-\ln({\cal C}_{\Delta\theta}{\cal C}_{\Delta\phi})].
\end{eqnarray}
The ${\cal C}_{\Delta\theta}$ and ${\cal C}_{\Delta\theta}$ consistencies were calculated as
one-dimensional two-sided $\chi^2$ integral probabilities,
using a parameterization of the RMS of the $\Delta\theta$ and $\Delta\phi$ distributions as a
function of the track momentum and polar angle, over various fiducial
regions of the detector, as measurement error.  A track-cluster pair was considered to be matched if the corresponding $\chi^2$ consistency exceeded $10^{-6}$, equivalent to a track-cluster separation tolerance of about $\pm 5.5$ standard deviations. The efficiency was 
defined as the fraction of reconstructed track-cluster pairs, generated by the same physical particle, that met the consistency
requirement.  Figure~\ref{fig:trackmatch} compares the track matching
efficiency between the original and new algorithms as a function of the track
momentum, obtained using a data control sample enriched with pions and based on particle
identification that did  not rely on the calorimeter.  We
observe significant improvement, particularly for low momentum tracks.

\subsection{Reconstruction of Neutral Particles}
\label{sec:neutrals}

\subsubsection{Cluster Reconstruction}
\label{subsubsection:ClusterReconstruction}

Typical electromagnetic showers spread across many adjacent crystals
and form \emph{clusters} of energy deposits in the crystals.
Sufficiently close showers can merge into one cluster with more than
one local maximum.  Clusters are reconstructed with a pattern
recognition algorithm which also identifies merged clusters and splits
them into separate shower candidates called \emph{bumps}.  These are
then matched with track candidates as described above in
Section~\ref{sec:track-cluster-match} to determine whether a shower is
generated by a charged or a neutral particle. Shower shape
variables are used to discriminate between electromagnetic and
hadronic showers.

The cluster algorithm uses single crystals with an energy of at least
10 MeV as seeds.  Surrounding crystals are added to the cluster if
their energy is above 1 MeV, or if they are adjacent to a crystal with
at least 3 MeV.   The values for these thresholds are a compromise
between energy resolution and the data volume, given the level of
electronic noise and beam-related background.

Local maxima are determined within a cluster by finding crystals whose
energies exceed that of their immediate neighbors.  Clusters with more
than one local maximum are split into as many \emph{bumps} as there
are local maxima, with the bump energy defined as $E_\mathrm{bump} =
\sum_i w_i E_i$ and the sum running over all crystals in the cluster.
The weights $w_i$ are determined in an iterative procedure from 
\begin{equation}
  w_i = \frac{E_i \exp(-2.5 r_i / r_M)}{\sum_k E_k \exp (-2.5
    r_k/r_M)},
\end{equation}
with
$k$ running over all crystals in the cluster, $r_M = 3.8\,\mathrm{cm}$
being the Moli\`ere radius of CsI(Tl), and $r_i$ the distance between the
crystal $i$ and the centroid of the bump.  
The centroid position is taken as the center-of-gravity with
logarithmic, rather than linear, weights, with $W_i = 4.0 + \ln(E_i /
E_\mathrm{bump})$, and is computed from crystals with positive weight
only.  It is determined using the bump energy 
obtained in the previous iteration, and the procedure iterates until
the bump centroid positions are stable within $1\,\mathrm{mm}$.  This
procedure to calculate the centroid position emphasizes low-energy
crystals but utilizes only the crystals making up the core of the bump.

\subsubsection{Converted Photons}

A dedicated algorithm is used to reconstruct photons that convert into an $e^+e^-$ pair,
starting with track pairs whose particle identification information is consistent with
an electron hypothesis.
A least squares fit is performed by minimizing a $\chi^2$ 
computed from the measured track helix parameters and covariance, comparing those to
the hypothesis that they originated from the conversion of a photon.  Both the position
in space and the momentum vector of the photon are free parameters in the fit.  The fit includes
second-order effects due to the curvature of the tracks,
which is necessary to avoid a singularity in the fit for the conversion
position along the photon flight direction.
The track pair are allowed a small non-zero opening angle at the conversion point,
constrained by an ad-hoc Gaussian added to the $\chi^2$.  We determine the value of the
Gaussian width using a data control sample of photon conversions occurring at the beam pipe,
by statistically subtracting the estimated track angular resolution contribution from the width of
the opening angle distribution measured at the known radius of the beam pipe.
Final photon conversion candidates are selected by requiring their $\chi^2$ consistency and 
vertex position to be consistent with a converted photon.  We further require the
reconstructed photon momentum direction to be consistent with the direction defined by the conversion vertex
position relative to the primary vertex.
Figure~\ref{fig:photon_eff} shows the efficiency for photon conversion and reconstruction in the detector
versus the photon energy in the c.m. frame.  To remove
\piz{} Dalitz decays ($\pi^0\to e^+e^-\gamma$), only conversions
within a radial distance of 1.7 to 27 cm from the beam are included.
\begin{figure}[ht]
\center
\includegraphics[width=0.4\textwidth]{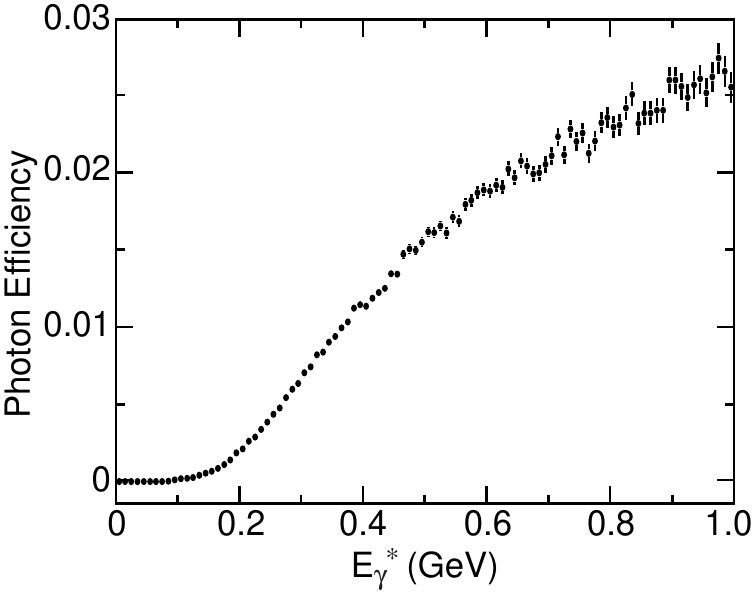}
\caption{
Total efficiency for the conversion and reconstruction of photons versus the
c.m. energy of photons with a radial distance of 1.7 to 27 cm.
}
\label{fig:photon_eff}
\end{figure}

This algorithm is also used to reconstruct pseudo-photons in $\pi^0$
Dalitz decays ($\pi^0\to e^+e^-\gamma$),
using a larger Gaussian width in the opening angle constraint to account for the extra degree of freedom
in that case.  Photon candidates reconstructed from $e^+e^-$ track pairs are included in 
photon candidates reconstructed in the
\emc to form photon candidate lists for physics analysis.
The transverse radius distribution of the reconstructed conversion photons and
Dalitz decay pseudo-photons in the inner
tracking region is shown in Figure~\ref{fig:conversion_radius}.

\subsubsection{Cluster Energy Corrections}

In order to improve the photon energy resolution and the agreement
between data and simulation, we apply two corrections to the
energy of reconstructed clusters:  clusters in simulation and data are
corrected for a lateral non-uniformity, and the energy of simulated
clusters is smeared to better resemble the resolution found in data.

\paragraph{Lateral Non-Uniformity Correction}
Between the CsI(Tl) crystals, dead material cannot be avoided.  As a
result, the reconstructed energy of an electromagnetic shower depends
on the position of the shower maximum projected on the front face of the
crystal. This is observed as a drop in the response 
$ E^{\gamma}_{\rm rec} /E^{\gamma}_{\rm fit} $ at the crystal edges by about 2--3\,\%  
(see Figure~\ref{edgecor} dashed (red) histogram), where 
$E^{\gamma}_{\rm rec}$ is the reconstructed photon energy and
$E^{\gamma}_{\rm fit}$ is the prediction of a kinematic fit.
As in the case of the cluster calibration, photons from the process 
$ e^+ e^- \rightarrow \mu^+ \mu^- \gamma $ are used to obtain
a correction function (called lateral non-uniformity correction) depending 
on the impact position on the crystal surface.   
The correction function is determined in two dimensions and in five energy bins
in the range 400\,\mev $ <  E_{\gamma} <$ 8\,\gev.
Applying the lateral non-uniformity correction, the photon response is found to be independent 
of the impact position as shown in Figure\,\ref{edgecor} as a solid (blue) histogram.  Measured energy resolution
improves by about 10\,\%.
\begin{figure}[htpb]
\begin{center} \includegraphics[width=0.5\textwidth]{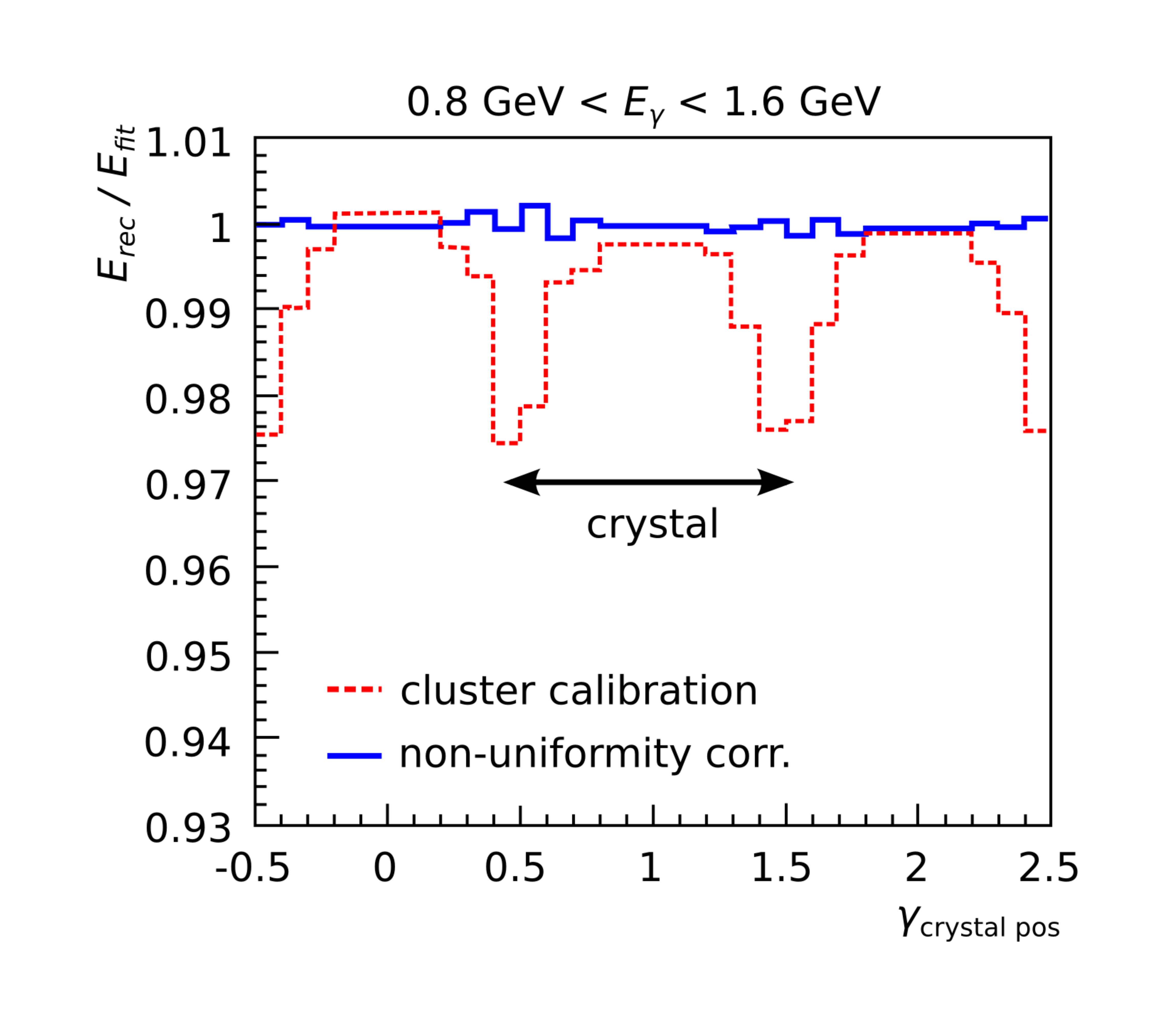} \end{center}
\caption{\label{edgecor}  Normalized response 
$ E^{\gamma}_{\rm rec} /E^{\gamma}_{\rm fit} $ across three CsI(Tl)
crystals before (dashed, red) and after (solid, blue) the lateral non-uniformity correction.}
\end{figure}

\paragraph{Adjustment of Line Shapes of Simulated Photons}
The line shape of the EMC response to photons is not fully
described by the MC simulation. As a consequence,
it is difficult to model selection criteria and efficiencies in physics analyses. The main goal
of this smearing procedure
is the improvement of the MC simulation description by tuning the MC model to data. 
Because of the large amount of MC data already simulated, a simplified approach is to introduce  
an additional term to the simulated photon response.
It reflects the difference between the 
measurement and the MC simulation. Data from the process $ e^+ e^- \rightarrow \mu^+ \mu^- \gamma $ 
allows the determination of a smearing matrix (Students $t$ distribution \cite{student}) $S_{ij}$ which connects the bin i of  
the distribution of the normalized photon response $ E^{\gamma}_{\rm rec} /E^{\gamma}_{\rm fit} $
in MC simulation with the bin j of the data distribution. The smearing matrix depends on the energy 
and the polar angle of the photon candidate. 

After applying the smearing procedure to the MC data, the measured photon line shape in data is very 
well modeled. As an independent test, the $\Delta E$  distribution of   
$B^0 \rightarrow K^{*0} \gamma  \rightarrow K^+ \pi^- \gamma $ processes is compared in data
and MC simulation. The $\Delta E$ line shape in data is very well
described by the smeared MC simulation.

\subsubsection{High-Energy Single-Photon Efficiency}
Precise measurements of hadronic cross sections using the ISR
technique (\eg\ \cite{Lees:2011zi,Lees:2012cr}) need
low systematic uncertainties on the 
single photon efficiency at high photon energies $E_\gamma$.
Simulation can differ from data due to imperfectly simulated shower
shapes or imperfect modeling of the distribution of dead material in
the detector.  In addition, information on crystals with inefficient
or dead readouts was not correctly propagated to the simulation in all
cases. 

To measure the  difference in efficiency for  data and simulation, a clean sample of $\epem \to \mu^+\mu^-\gamma$ events is selected and 
kinematics are used to predict the direction and
energy of the photon from the measured tracks.  In order to avoid
trigger bias, the sample was restricted to events which were 
triggered by the two charged tracks alone.
In total, there were $1.3\times 10^6$
$\epem \to \mu^+\mu^-\gamma$ events with exactly two tracks in which 
the missing momentum pointed
into the fiducial volume of the EMC.  The process $\epem \to \gamma
\pip\pim$ is kinematically very similar, and therefore was included in the sample,
while all other potential backgrounds
were  rejected by the kinematic fit.

The photon reconstruction inefficiency was computed as the fraction of 
events without a detected photon candidate consistent with the kinematic fit.
This inefficiency was evaluated as a function of the
photon polar angle $\theta$ in the lab frame.  Its precise value
depended on the criterion used to match reconstructed photon candidates
with their predicted direction and energy.  

The raw inefficiency was corrected for photons lost due to conversion in the
detector material.  A further correction was applied for events with
internal conversion, in which the radiated photon was virtual and
decayed into an $\epem$ pair at the interaction point.  This type of
event was present in data but was not included in the simulation.  
The simulation overestimated the photon reconstruction
efficiency, with the largest effect in the endcap region of the
EMC.  For high energy photons with $E_\gamma > 3\gev$ and for
typical selection criteria of ISR analyses, an average
difference between the efficiencies in data and simulation of $\Delta
\epsilon = (-1.00 \pm 0.02 (\mathrm{stat}) \pm 0.55 (\mathrm{syst}))\times
10^{-2}$ was observed.

\subsubsection{\KL Selection}
\KL{} can be reconstructed and identified by the hadronic shower
initiated in the EMC or the IFR, although the energy resolution of
these two detectors, when used as hadron calorimeters, is not good enough to
directly use the measured energy in a physics analysis.

The selection in both detectors requires that \KL candidates be well isolated from
tracks.  For the EMC, this translates into a track-cluster association
probability less than $1\%$.  Showers in the IFR are considered isolated from
tracks if they fulfill $\Delta \theta > 350 \mrad$, $q \times \Delta \phi < -350\mrad$ or $> 700
\mrad$, where $q$ is the track charge.  Additionally, \KL candidates
found in the EMC are taken from the list of clusters with energy in the range $0.2
< E_{EMC} < 2 \gev$.  EMC clusters that can form a \piz{} with another
photon of at least $30\mev$, as well as those with two local maxima consistent
with a merged \piz, are rejected.  Showers in the IFR are required to
start before layer 14 and to include hits in at least 2 planar layers.

For some analyses (\eg\ \cite{delAmoSanchez:2010uj}), \KL are
identified by a multivariate classifier based on BDTs~\cite{Yang2005370}.
Input quantities for candidates in the EMC
comprise cluster shape variables, such as the second and the lateral
moments of the energy distribution.  The selector is trained with $\Dz
\to \KL \pip\pim$ events and validated using $\epem\to\gamma\phi(\to
\KS\KL)$ events.

\subsubsection{\piz Efficiency Corrections}
Neutral pions are primarily reconstructed in the decay
$\piz\to\gamma\gamma$.  Differences between simulation and data
in the photon reconstruction efficiency or the overall number and
distribution of extra photon candidates (\eg\ from machine-related
background) influence the \piz{} reconstruction efficiency and need to
be taken into account in measurements of quantities such as branching
fractions.

\paragraph{\piz Efficiency Correction Derived from \mtau Decays}

Differences
between the \piz reconstruction efficiency in data and simulation can
be studied and corrected with \mtau decays.  A \piz efficiency
correction, $\eta_{\piz}$, to be applied for each reconstructed \piz
in simulated events, has been determined from the double ratio
\begin{equation}
  \eta_{\piz} =  \frac{ N^\mathrm{Data}(\mtau \to t^{-} \piz \nut) }
        { N^\mathrm{MC}(\mtau \to t^{-} \piz \nut)} \; \div \;
    \frac{ N^\mathrm{Data}(\mtau \to t^{-} \nut) }{ N^\mathrm{MC}(\mtau \to t^{-} \nut)}.
\end{equation}
Here, $t$ denotes a charged particle that fails the identification criteria for
an electron.  
$N^\mathrm{Data}$ and $N^\mathrm{MC}$ are the numbers of selected $e^{+}e^{-} \to
\taum \taup$ events in data and simulation where one of two \mtau's
decayed into a $t^{-} \piz \nut$ or $t^{-} \nut$ final state.  
The second \mtau is reconstructed in the $\taum \to
e^{-} \nueb \nut$ channel; the $\taum \to \mu^- \numb\nut$ channel
suffers from poorly understood systematic effects and is disregarded for
this study.  This method relies on the precise
knowledge of the \mtau branching fractions used to derive the
correction, and assumes that the efficiencies to reconstruct $\piz$ and
the rest of the event are independent. 

The \piz candidates are reconstructed in their dominant decay into two
photons. The photons are reconstructed in the
barrel region of the EMC ($0.473 < \theta_{\gamma}<2.360\rad$) only and are
required to have an energy $E>75\mevcc$ and a non-vanishing lateral
moment.  This last requirement removes photon candidates consisting of
single crystals; most of these are background hits which are not
well simulated.  Events that contain photons not associated with a \piz
candidate are rejected from the selected samples.

The $\taum \to t^{-} \piz \nut$ and $\taum \to t^{-} \nut$ event
samples are selected with a purity of about $94\%$ and about $95\%$, 
respectively. The branching fractions of the decays where $t=
\pi, \mu, K$ are known to sufficient precision or are small
\cite{Beringer:1900zz}. 
Most systematic effects not related to
the \piz reconstruction cancel to a large extent in the double ratio.
Showers in the EMC initiated by hadrons can contain neutral particles,
such as neutrons or $\Kz$, which can travel larger distances in the
crystals before possibly producing a new shower.  These so-called
{\it split-offs} may subsequently be reconstructed as extra photon
candidates near, but well separated from, a hadron track.  The
simulation underestimates the number of split-offs found in data, and
correction for this effect is applied.

Table \ref{tab:pi0Correction} lists $\eta_{\piz}$ determined
separately for each running period. The agreement between the Runs is good. The
momentum-averaged efficiency correction is $0.960 \pm 0.001 (\mathrm{stat}) \pm
0.008 (\mathrm{syst})$. The main sources of systematic uncertainty are the
\mtau branching fractions and the split-off correction described
above, neither of which cancel in the double ratio.

For use in other analyses, where the momentum spectrum of the \piz
is different than in \mtau events, the \piz correction is also
determined as a function of the \piz momentum. The momentum-dependent
weights are shown as the red triangles in
Figure~\ref{fig:pi0Correction}.

\begin{table}[htb!]
  \caption{\piz efficiency correction for each Run from the
    $\tau$-based study. The errors are statistical only.}
  \label{tab:pi0Correction}
  \begin{center}
    \begin{tabular}{ccc}
 	  	\hline
   	  	Run  &  $\eta_{\piz}$	 \\
 	 	\hline
   	1 &  $0.956 \pm 0.003$ \\
   	2 &  $0.966 \pm 0.002$ \\
   	3 &  $0.965 \pm 0.002$ \\
   	4 &  $0.959 \pm 0.001$ \\
   	5 &  $0.959 \pm 0.001$ \\
   	6 &  $0.959 \pm 0.001$ \\
   	\hline
   	Total &  $0.960 \pm 0.001$\\
  	\hline
    \end{tabular}
  \end{center}
\end{table}

\begin{figure}[htb!]
	\includegraphics[width=1.0\columnwidth]{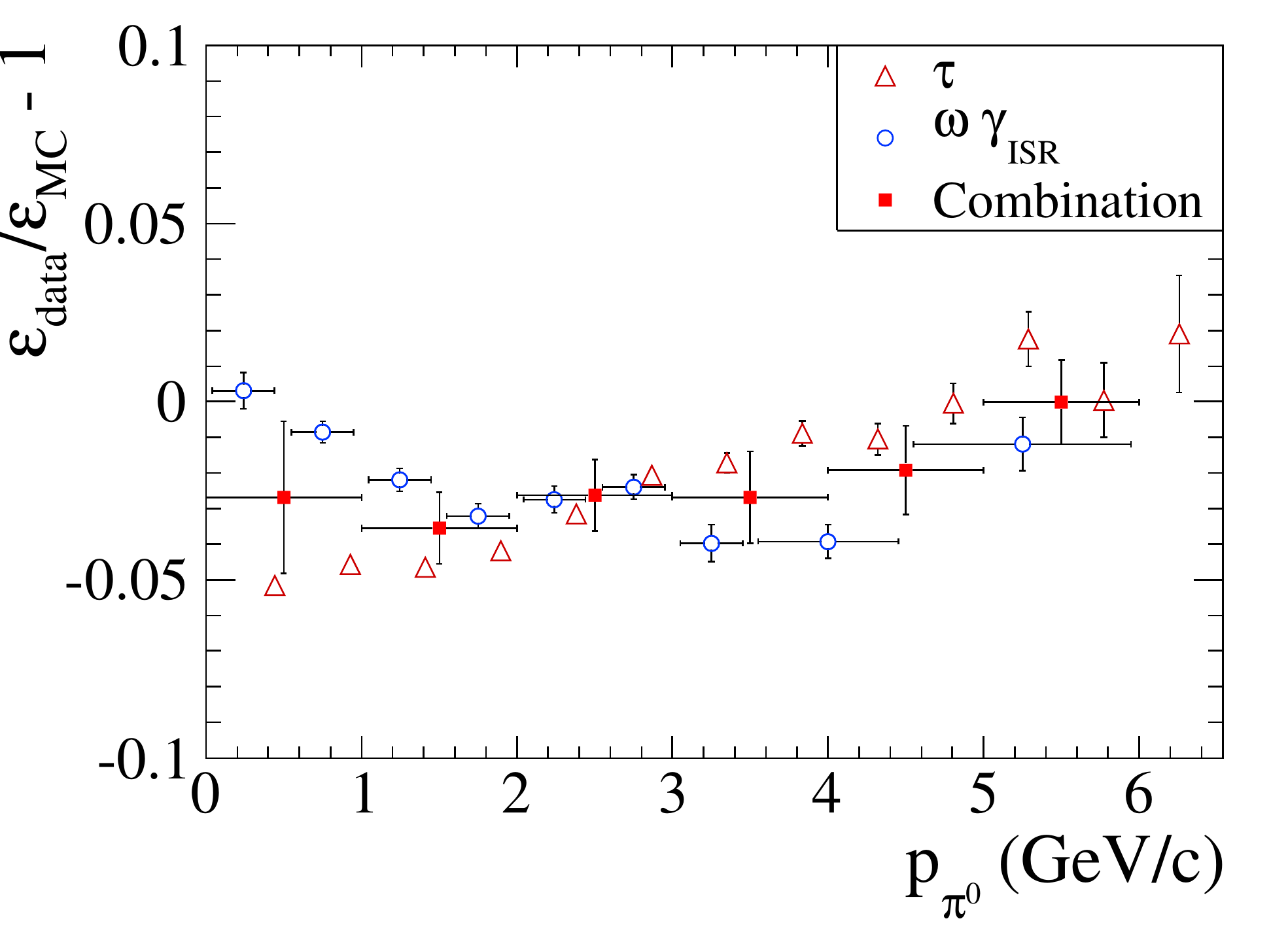}
	\caption{\piz efficiency correction as a function of the \piz
          momentum.  The (red) triangles show the efficiency
          correction
          $\varepsilon_\mathrm{data}/\varepsilon_\mathrm{MC} -1$
          derived from $\tau$ decays, while the (blue) circles show
          the correction obtained from ISR production of $\omega$.
          Uncertainties shown are statistical only.
          The filled squares show the averaged $\piz$ efficiency
          correction with its total uncertainty.}
	\label{fig:pi0Correction}
\end{figure}

The correction is validated with $\mtau \to \pim \piz \piz \nut$
decays, which have a well-known branching fraction. The correction is applied
twice in such decays, \ie\ the effect of a wrong correction would be
doubled, resulting in a measured branching fraction which would deviate from
the expectation. The obtained $\mtau \to \pim \piz \piz \nut$
branching fraction is consistent with the expected value.

\paragraph{\piz Efficiency Correction Based on $\omega$ Decays}
An alternative approach to measuring the reconstruction efficiency of
\piz uses the production of $\omega$ mesons in initial state radiation events
$(\epem \to \omega \gamma_{{\rm ISR}})$
and the subsequent decay $\omega\to\pip\pim\piz$.  This approach
exploits the closed and well-measured kinematics of these types of
events and the narrow line shape of the $\omega$.

Using the known kinematics of $e^+$ and $e^-$ in the initial state and
the measured energies and momenta of the final state particles
$\gamma_{ISR}$, $\pip$ and $\pim$, the four-vector of the
\piz\ can be predicted from a kinematic fit.  Selecting on the $\chi^2$ of
the kinematic fit results in a sample of $\omega\to\pip\pim\piz$
candidates with a small and well-controlled background from other ISR
channels.  The event is then searched for $\piz$ candidates and 
second kinematic fit is performed to determine whether a candidate is compatible
with the kinematics of the rest of the event. 
The resulting efficiency correction as a function of the predicted \piz
momentum is shown as blue circles in Figure~\ref{fig:pi0Correction}.

\paragraph{Global \piz Efficiency Correction}
As can be seen in Figure~\ref{fig:pi0Correction}, the correction
factors derived from the two approaches differ significantly at low
$\piz$ momenta.  

We take as the correction factor the weighted average of the two
correction factors, derived from the $\mtau$- and the
$\omega$-ISR-based analyses, and assign half the difference or $1\%$,
whichever is larger, as overall uncertainty of the correction.  The
overall correction factor is also shown in
Figure~\ref{fig:pi0Correction}.  Averaged over typical \piz momentum
spectra, the \piz reconstruction efficiency is about $3\%$ lower in
data than in simulation, with an overall systematic uncertainty of
$1.5\%$. Since this efficiency correction is determined from very clean
events in which the detector shows little other activity, the applicability 
of the correction to higher multiplicity events is validated
by using the ratio of the decays $\Dz \to \Km\pip\piz$ to
$\Dz\to\Km\pip$.  The $\piz$ efficiency correction derived from this
decay is systematically limited by the knowledge of the branching
fractions and compatible within uncertainties with the more precise
determination of the correction factor.

\subsection{Charged Particle Identification}
\label{sec:cpid}

Discrimination between charged particle types over a large kinematic
range is an essential requirement for many precision measurements and
searches for rare processes in the decays of charm and bottom mesons
produced at the \pep2\, \BF. In addition, efficient and high-purity
tagging of the flavor of $\Bz$ mesons requires well-identified
leptons and kaons. The \babar\, detector was designed
to meet these goals, and algorithms of increasing sophistication have
been implemented over the lifetime of the experiment, 
which use detector information more effectively and 
provide improved PID.

The five types of long-lived charged particles of interest
are $\Kp$, $\pip$, $\ep$, $\mup$ and protons.
Charge conjugation is
implied throughout, and the charge sign ``+'' is implied in any
reference to these five particle types throughout this section, unless explicitly
noted otherwise.

Hadrons are distinguished from each other, 
and from leptons, mainly with DIRC and
$\dedx$ measurements in the DCH and SVT.
The SVT provides $\dedx$ information for low-momentum tracks
that do not enter the DCH; in  fact it is the only PID information available for such tracks,
and so this is of great importance for the separation of pions from electrons in this energy range,
in particular for pions from $D^*$ decays. 
Electrons are identified using energy depositions in the
EMC and \dedx\ measurements in the tracking systems.
Muons are discriminated from hadrons through the
pattern of energy depositions of a track traversing
the layers of the IFR.

\subsubsection{PID Control Samples}

To characterize the  detector response with respect to
particle identification, it is first necessary to isolate very
high-purity charged track control samples covering the full range
of laboratory momenta and polar angles.
Particle types for simulated events are known from the event generation.
However, data control samples must be selected without PID information through
purely kinematic reconstruction of particular decay modes;
reconstruction of these same modes in simulated events allows
tuning of the \geant\cite{Agostinelli:2002hh} detector simulation.
Additional \emph{a posteriori} corrections, discussed below,
are applied to PID track selections in simulated events, so that
they mirror PID performance in actual data.

Very high-purity samples of \Kp and \pip are obtained from the decay
$\Dstarp \to \pip_{{\rm slow}} \Dz (\to \Km \pip_{{\rm fast}})$ using narrow windows of the
$\Km \pip_{{\rm fast}}$ invariant mass and the $\Dstarp-\Dz$ mass difference.
The charge of the low-momentum slow pion $\pi_{{\rm slow}}$ uniquely identifies the pion and kaon from the $D$ meson decay. 
Additional selections on vertex probability and momentum of the $\Km \pip_{{\rm fast}}$ system are
made to suppress residual backgrounds. This decay is observed both in cascade
decays from $B$ and in non-$\BB$ fragmentation events, and provides $K$ and $\pi$
samples over a large momentum range. This sample is the sole source for $K$ calibration tracks.
The $K$ and $\pi_{{\rm fast}}$ purity is $\gae 99.8\%$ for tracks with momentum $p_{\rm lab} >0.5 \gevc$.

Two other decays provide calibration pion samples: $\tautau$ pairs decaying with a 3-1
topology and $\KS \to \pip \pim$.
Here, $\taup\taum$ pairs with a 3-1 topology are events in which
one tau decay has three charged tracks and the other a single track.
The single track is required to be identified as a lepton,
which also defines the expected charge of the parent of the three-daughter system.
The three-daughter decay is preferential to a non-resonant
three-pion system; however, there is a momentum-dependent probability of ${\sim} 2-3\%$
that one of the three charged daughters is actually a charged $K$. This sample is therefore
unsuitable as a source of calibration tracks to discriminate between hadrons, but it
does provide pions at higher momenta than are available from the $\Dstarp$ sample so that they can be
used to characterize muon--pion discrimination. The decay $\KS \to \pip \pim$ provides
a very high-statistics sample of pions at momenta lower than the {\it fast}
pions from $\Dstarp$ decays~-- see definition above. A narrow $\pip \pim$ invariant mass window,
vertex probability and displaced vertex selections are imposed such that the resulting
sample of pions is essentially background-free.

Proton calibration tracks are obtained from the decay $\Lambda^{0} \to p \pim$. As with
pions from $\KS$ decays, the imposition of invariant mass, vertex probability and
displaced vertex criteria provides a sample of protons that is almost pure.

Muon tracks are obtained from $\epem \to \mumu \gamma$ events.
These QED events are dominated by decays with a low-energy photon
and therefore high-momentum muons. However,
these events are so copiously produced that large samples of muons, covering a very wide
momentum range down to approximately $0.5\gevc$, can be obtained by requiring a minimum
photon energy $E_{\gamma} > 1.6\gev$. An almost pure sample of muons can be obtained
by requiring that the {\it other} muon in an event passes a very loose muon particle identification selection.
This selection on the other muon induces essentially no bias with respect to PID discriminants for
the control sample muon. In an analogous manner, an electron control sample
with purity $\gae99.9\%$ and good momentum coverage is obtained from
$\epem \to \epem \gamma$ events through selections on the
photon energy and the {\it other} electron in an event.

\begin{figure}[t!]
\begin{center}
\subfigure{\includegraphics[width=0.49\textwidth]{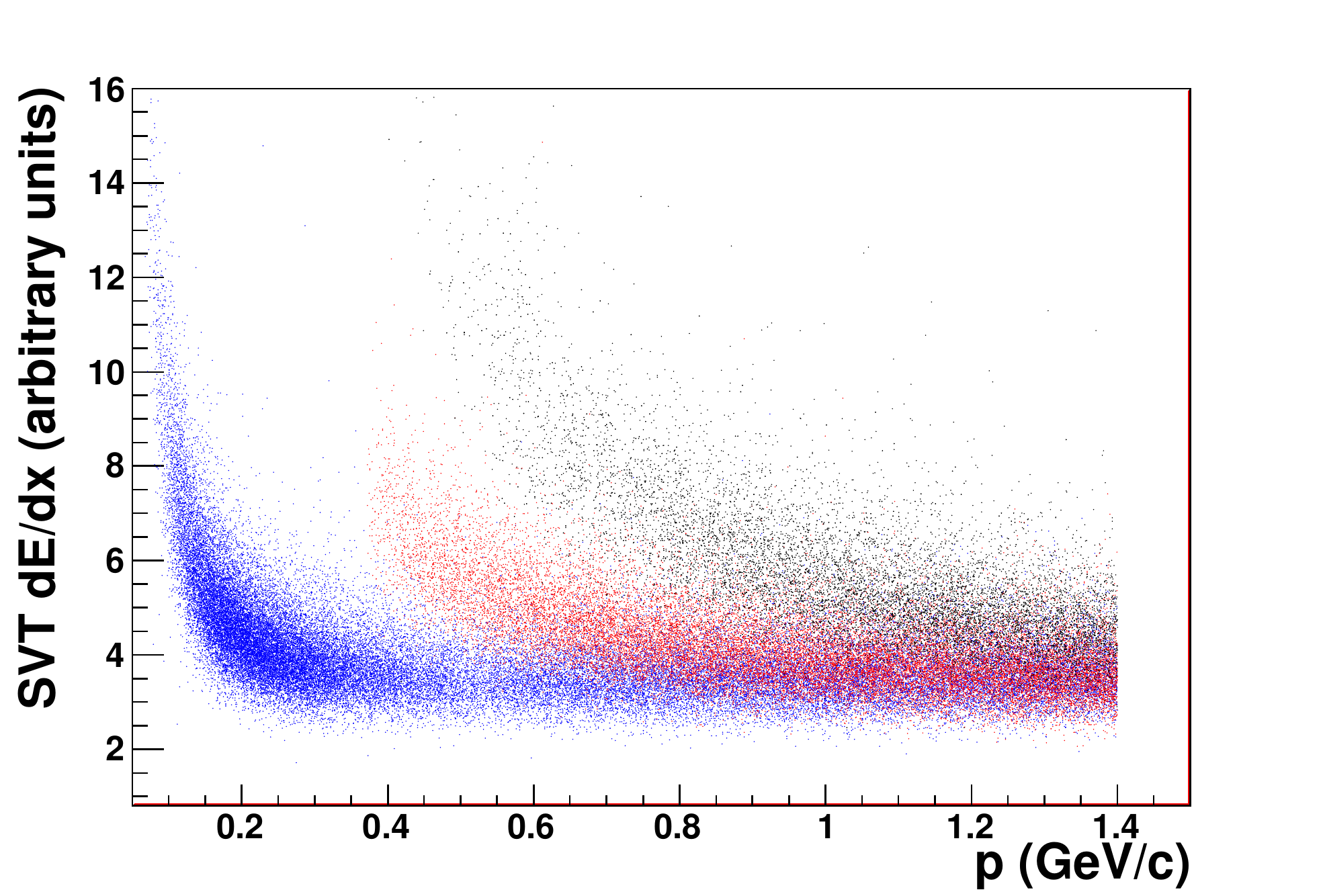}}
\subfigure{\includegraphics[width=0.49\textwidth]{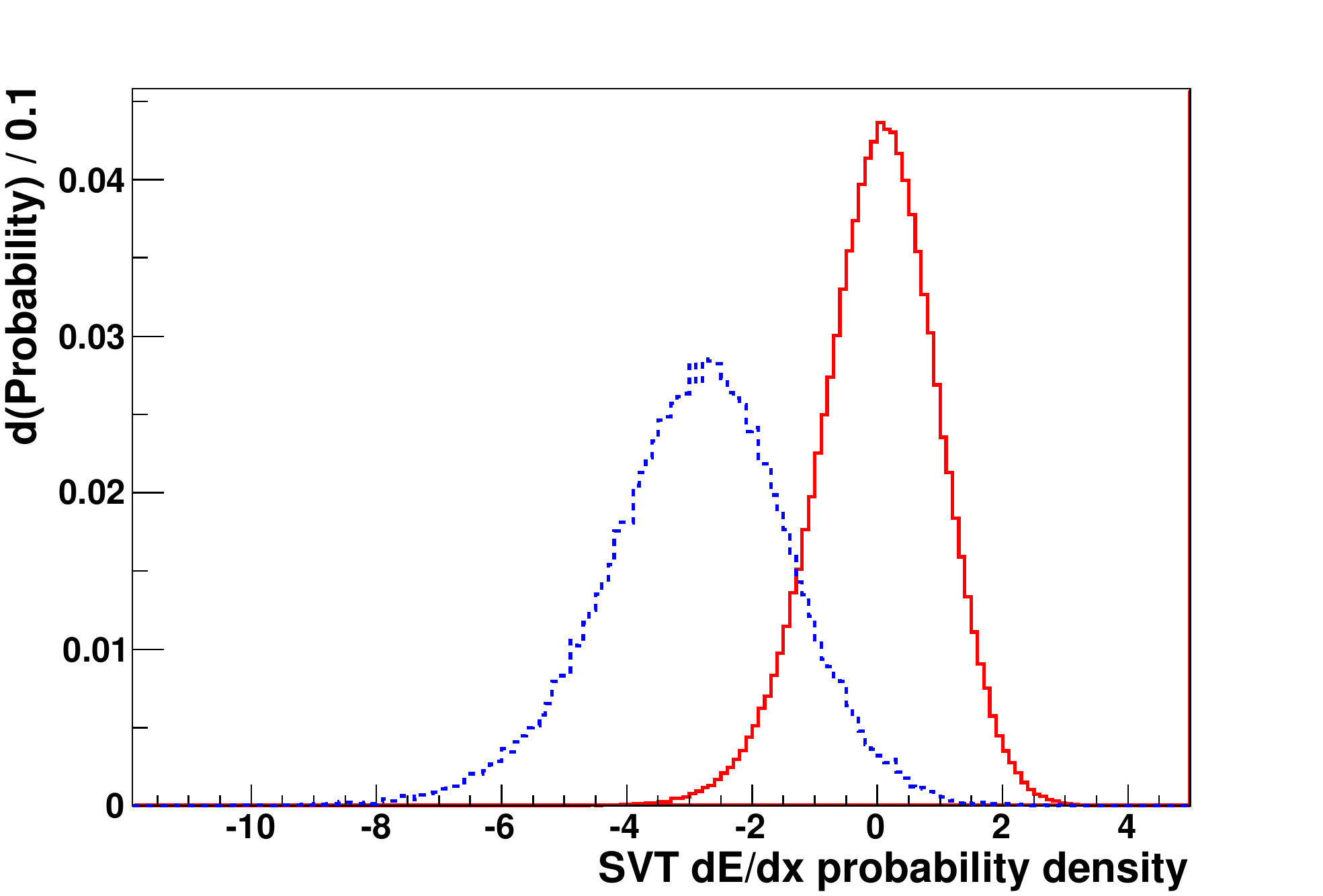}}
\caption{
(top) SVT $\dedx$ vs $p_{\rm lab}$ for $\pi$ (blue, lower band), $K$ (red, middle band), protons (black, upper band);
(bottom) SVT $\dedx$ probability density function of the $K$ hypothesis for true $K$ (solid red) and $\pi$ (dashed blue) particles with $0.5<p_{\rm lab}<0.6 \gevc$.}
\label{fig:svtpid}
\end{center}
\end{figure}

\begin{figure}[t!]
\begin{center}
\subfigure{\includegraphics[width=0.49\textwidth]{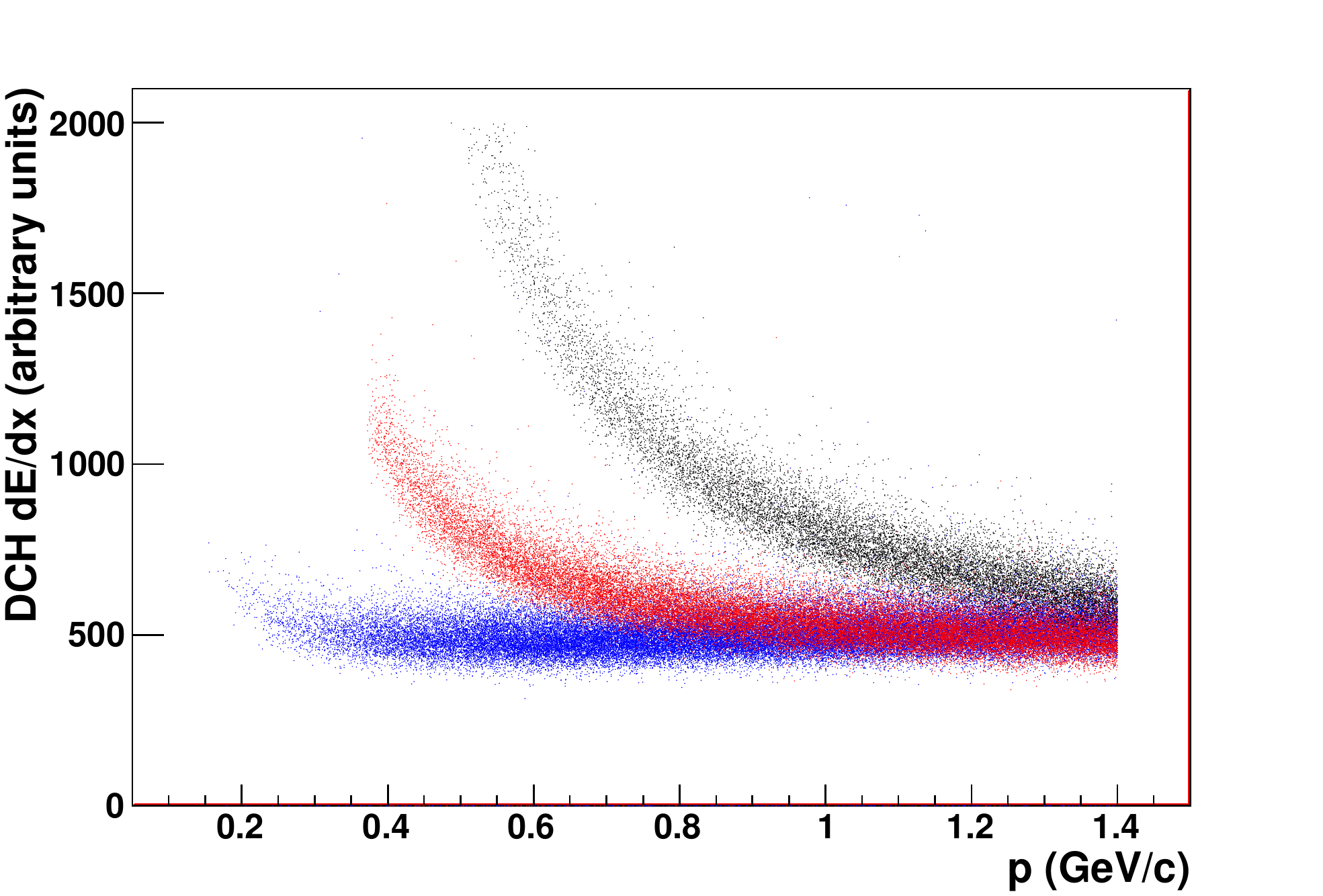}}
\subfigure{\includegraphics[width=0.49\textwidth]{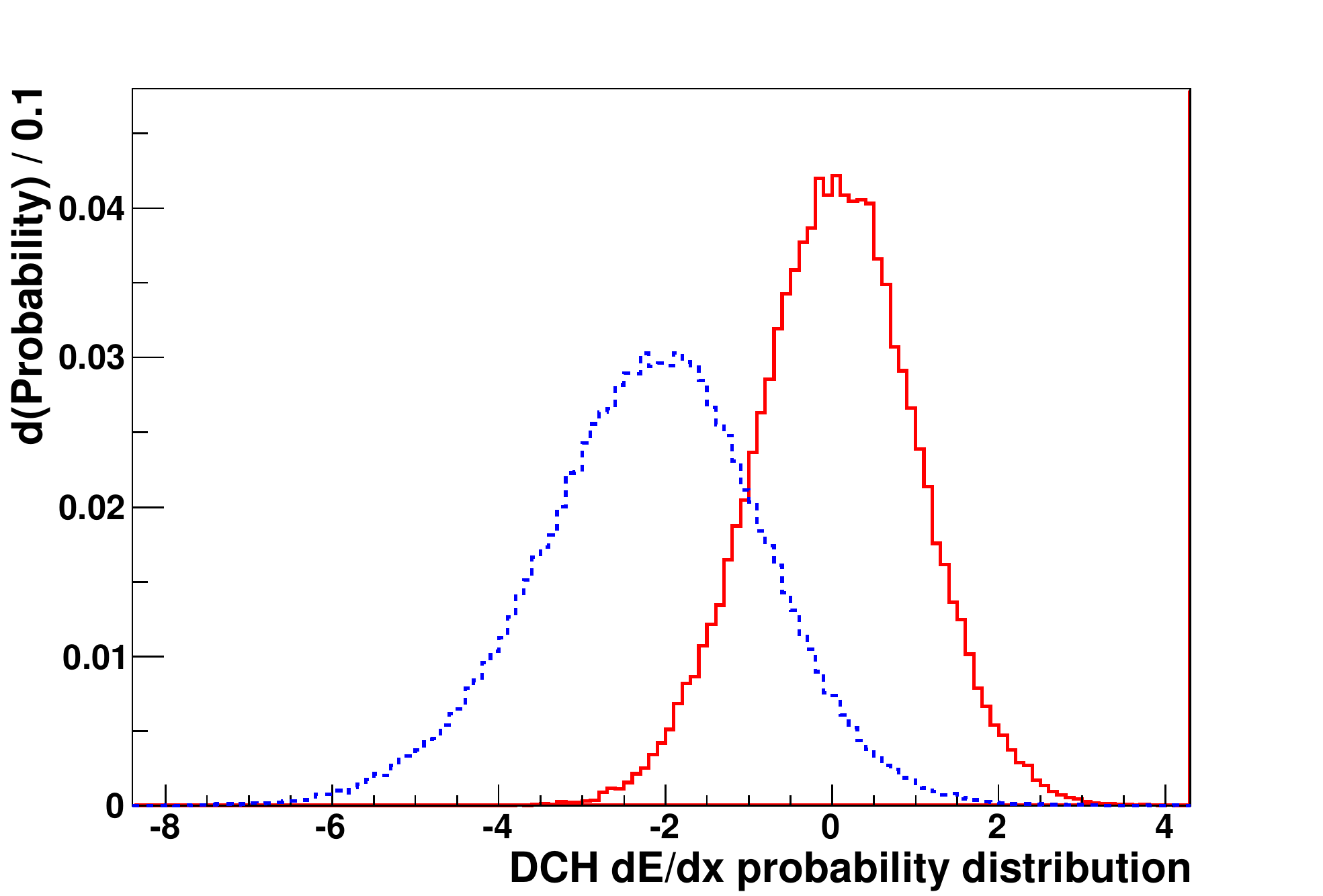}}
\caption{(top) DCH $\dedx$ vs $p_{\rm lab}$ for $\pi$ (blue, lower band), $K$ (red, middle band), protons (black, upper band);
(bottom) DCH $\dedx$ probability density function of the $K$ hypothesis for true $K$ (solid red)
and $\pi$ (dashed blue) particles with $0.9<p_{\rm lab}<1.0 \gevc$.}
\label{fig:dchpid}
\end{center}
\end{figure}

\subsubsection{PID Information from Subdetectors}

The five-layer SVT provides up to ten measurements of $\dedx$ per
charged track. For every track with at least four
measurements, a truncated mean from the lowest 60\% of $\dedx$ measurements
is calculated. For MIPs, the $\dedx$ resolution is ${\sim} 14\%$.

The DCH provides $\dedx$ measurements derived from the total charge deposited in each
of its 40 layers, and the specific energy loss is computed as a
truncated mean from the lowest 80\% of the individual $\dedx$
measurements. Several corrections are applied to remove
sources of bias that can degrade the accuracy of the measurement:
\begin{itemize}
\item changes in gas pressure and temperature,
leading to $\pm 9\%$ variation in $\dedx$, are corrected
using a single overall multiplicative correction;
\item differences in cell geometry and charge collection
($\pm 8\%$ variation) are corrected by a set
of multiplicative corrections for each wire;
\item signal saturation due to space charge
buildup ($\pm 11\%$ variation) is corrected
as a function of the dip angle $\lambda$,
defined as the opening angle between the
track momentum and the $r \phi$-plane;
\item non-linearities in the most probable energy
loss at large dip angles ($\pm 2.5\%$ variation)
are corrected with a fourth-order Chebychev polynomial
as a function of $\lambda$; and
\item variations in cell charge collection as a function of
entrance angle to the DCH volume ($\pm 2.5\%$ variation)
are corrected using a sixth-order Chebychev polynomial
in the entrance angle $\phi$.
\end{itemize}
\noindent The overall gas gain is updated continuously based on rolling
calibrations derived from prompt reconstruction
of colliding beam data; the remaining corrections are
determined once for a given high-voltage setting and gas mixture.

Figures~\ref{fig:svtpid} and~\ref{fig:dchpid} show
SVT and DCH $\dedx$ for $\pi$, $K$ and $p$
tracks as a function of the laboratory momentum $p_{\rm lab}$. There is good
$K / \pi$ separation in the SVT for $p_{\rm lab} \lae 0.5 \gevc$ and
in the DCH for $p_{\rm lab} \lae 0.8 \gevc$. However, above these momenta, the $\dedx$
bands for $K$ and $\pi$ begin to significantly overlap.
Protons are clearly distinguishable from both $K$ and $\pi$ up to higher momenta,
$p_{\rm lab} \lae 0.7$ in the SVT and $p_{\rm lab} \lae 1.2$ in the DCH.
Control sample $\dedx$ distributions binned by polar and
azimuthal angles, laboratory momentum and charge, are used in
the construction of probability densities which are subsequently
incorporated into the PID algorithms which rely
on $\dedx$ information. Figures~\ref{fig:svtpid} and~\ref{fig:dchpid}
also show the $K$ and $\pi$ response for the SVT and DCH $\dedx$
probability densities of the $K$ particle hypothesis,
demonstrating the $K/\pi$ separation achievable with SVT $\dedx$ for
$0.5 < p_{\rm lab} < 0.6$ tracks, and with DCH $\dedx$ for $0.7 < p_{\rm lab} < 0.8$ tracks.

Since the inner tracking system is able to provide reasonable $\pi/K$ separation
only up to about 700~\mevc, the \dirc\ was designed to provide particle
identification for tracks with higher momenta.
Data from the \dirc\ is reconstructed using two different methods.
The first algorithm consists of a maximum likelihood fit for the value of the opening
angle of a Cherenkov ring with respect to the track direction, referred to as the Cherenkov angle $\theta_C$,
for every track above Cherenkov threshold with at least one associated photon.
The relatively large number of photons per track make it possible to determine not only $\theta_C$
but also simultaneously the number of signal and background photons.
Figure~\ref{fig:dircpid} shows the good discrimination among hadron
particle types obtained by using the ring fit results for both $\theta_C$ 
and the number of signal photons. For the latter plot, the different behaviors at
high momentum come from the correlation between the track momentum and
angle in the laboratory frame, plus the fact that the DIRC has a better
acceptance with tracks that are more forward.

The second reconstruction method uses an unbinned maximum likelihood
incorporating all \dirc\ information for an entire
event to calculate a likelihood for each track. These
likelihoods are calculated in an iterative process, at the level of
individual DIRC photomultiplier tube hits, by maximizing the global event likelihood value
while testing different hypotheses for each track. For a given track, the likelihoods $L_i$ for the five particles species are calculated while holding all
other tracks at their current best hypothesis.
Using only the global likelihood, a $K$ efficiency $\gae 90\%$ is generally possible, with
pion misidentification rates increasing as the $K$ and $\pi$
Cherenkov angle bands begin to converge at higher momenta.

The EMC was designed
to detect electromagnetic showers with good
energy and angular resolution over the energy
range from 20\mev to 4\gev. Although primarily designed for reconstruction
of photons, the EMC provides excellent electron/hadron
separation using the ratio of shower energy to track
momentum $E/p$, along with the shower geometry.
Pattern recognition algorithms are used to form
clusters of crystals containing the shower energy~-- see
Section~\ref{subsubsection:ClusterReconstruction}.
A shower is associated with a
track by projecting the track to the reconstructed position
of the maximum of the energy deposition within the EMC volume.
The measured energy resolution is a few percent.

\begin{figure}[t!]
\begin{center}
\subfigure{\includegraphics[width=0.49\textwidth]{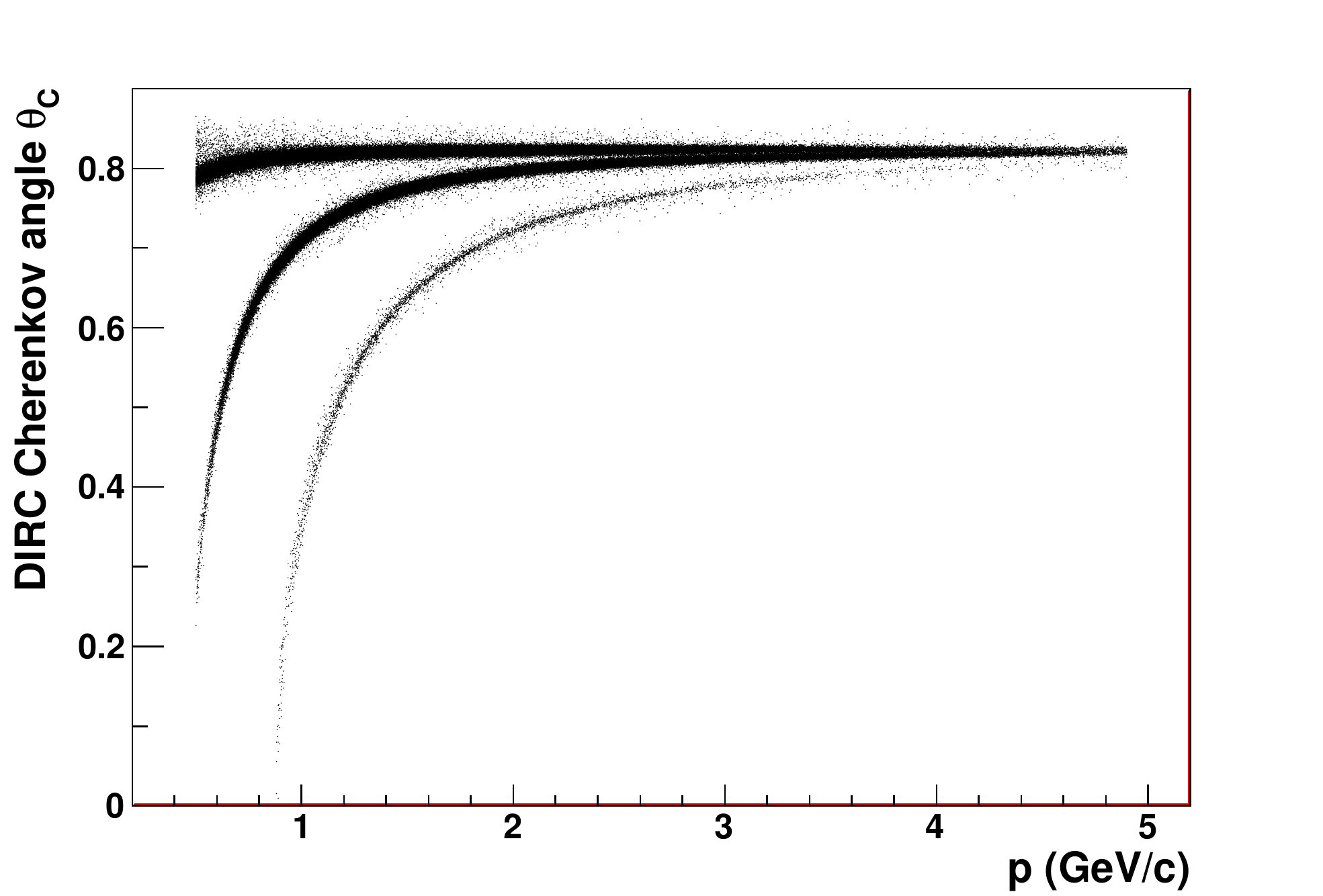}}
\subfigure{\includegraphics[width=0.49\textwidth]{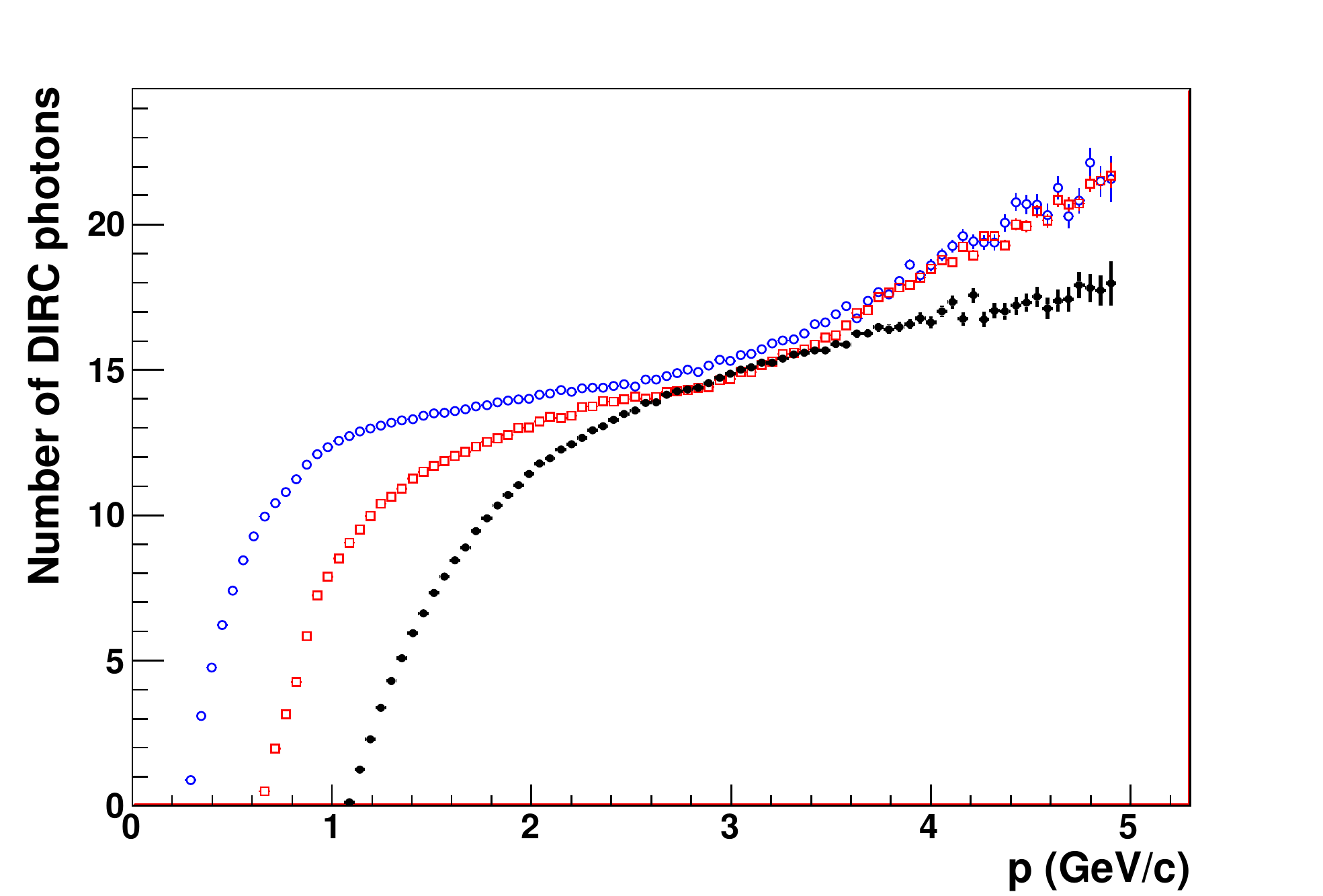}}
\subfigure{\includegraphics[width=0.49\textwidth]{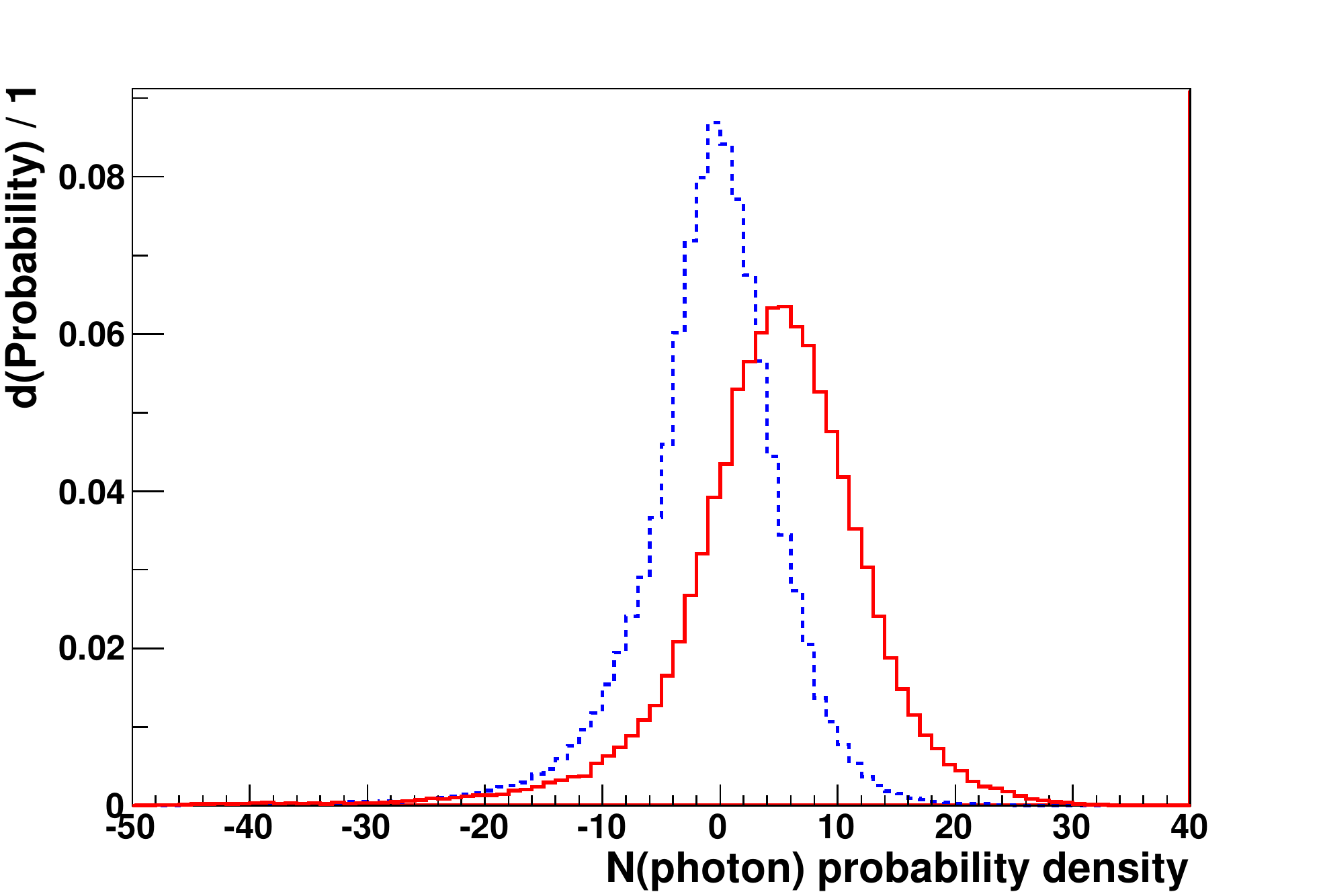}}
\caption{(top) DIRC ring fit $\theta_{C}$ vs $p_{\rm lab}$ for $\pi$ (upper band), $K$ (middle band), protons (lower band);
(middle) Mean number of DIRC signal photons vs $p_{\rm lab}$ for $K$ (red), $\pi$ (blue) and protons (black);
(bottom) Probability density function of the difference between the measured and expected numbers of DIRC signal photons.
The plot is drawn for true $K$ (solid red) and $\pi$ (dashed blue) particles, with $0.9<p<1.0 \gevc$ and assuming the $K$ hypothesis.}
\label{fig:dircpid}
\end{center}
\end{figure}

The IFR was designed to identify muons with high efficiency
and good purity.
Charged tracks are extrapolated to the IFR taking into account
the non-uniform magnetic field, multiple scattering, and
average track energy loss. The projected intersections with the
RPC/LST planes are computed, and for each readout plane all
signals detected within a predefined distance from the predicted
intersection are associated with the track.
Several PID variables are defined for each IFR cluster:
\begin{itemize}
\item the total number of interaction lengths;
\item the difference between the measured number of interaction
lengths and the number of interaction lengths predicted for a
muon of the same momentum and angle;
\item the average number and the RMS of the distribution
of hits per layer;
\item the $\chisq$ for the geometric match between
the projected track and the centroids of
clusters in different layers; and
\item the $\chisq$ of a polynomial fit to the
two-dimensional IFR clusters.
\end{itemize}

\subsubsection{Charged Hadron Identification}

\begin{figure*}[]
\begin{center}
\subfigure{\includegraphics[width=0.49\textwidth]{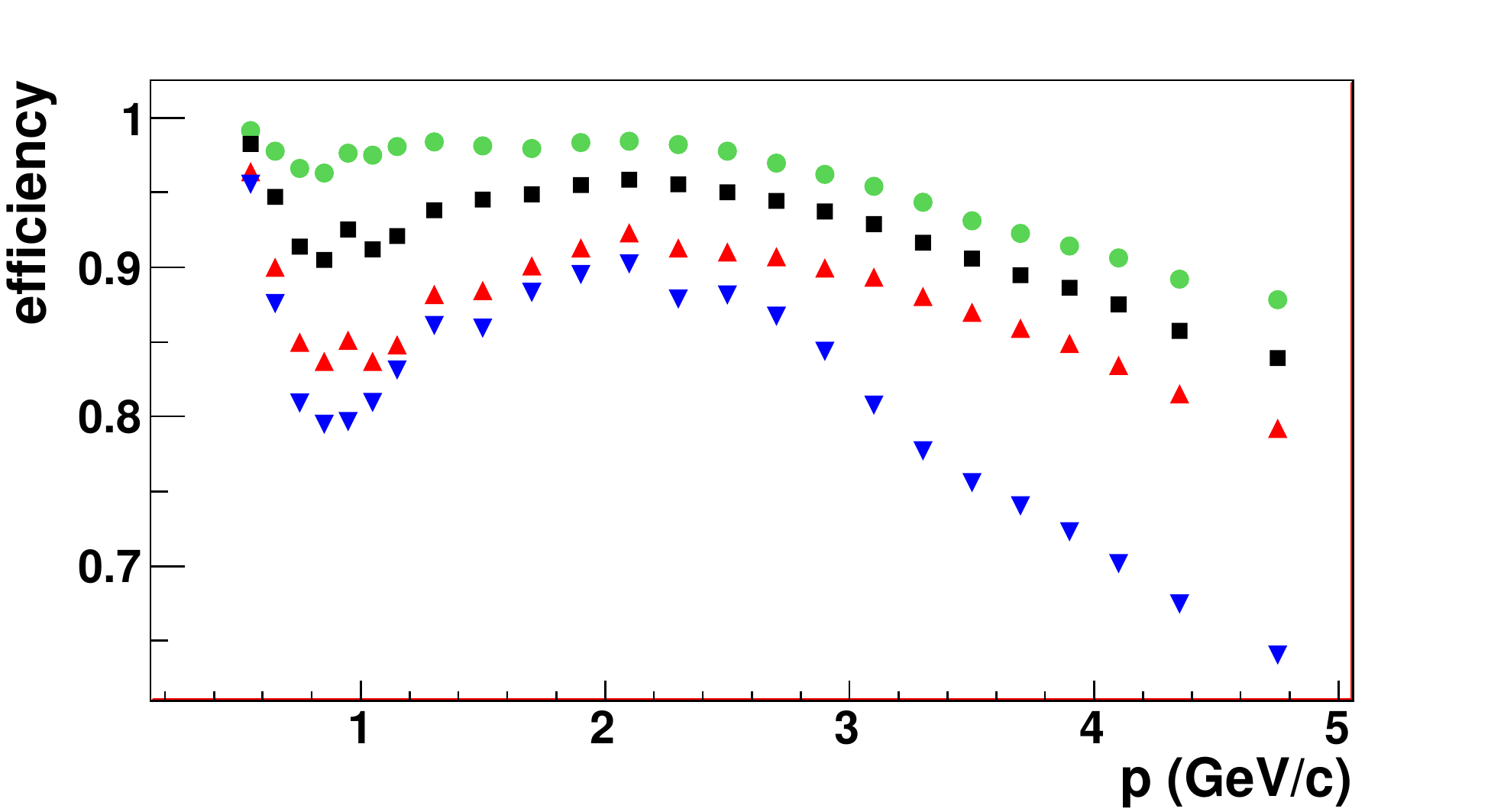}}
\subfigure{\includegraphics[width=0.49\textwidth]{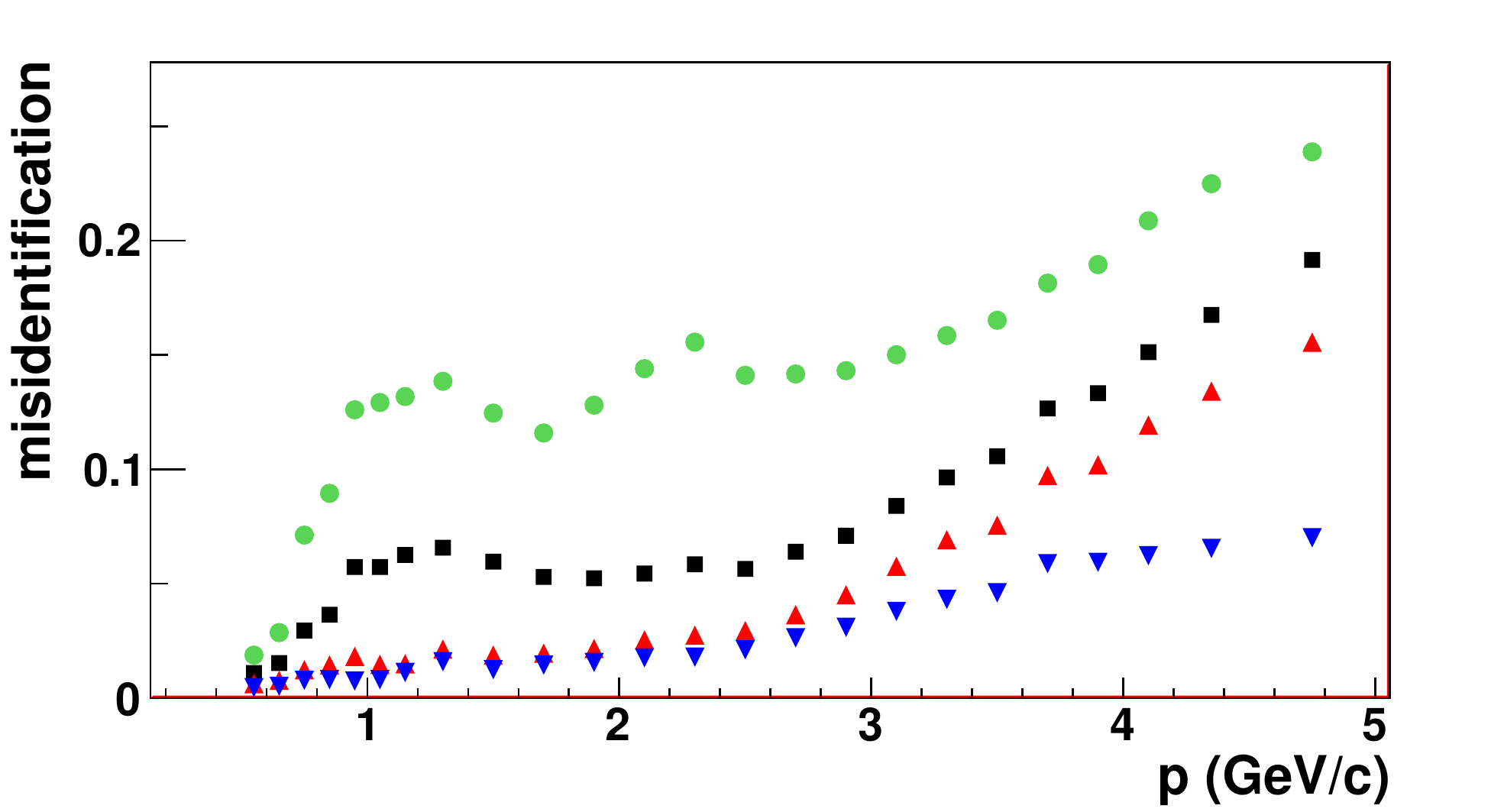}} \\
\subfigure{\includegraphics[width=0.49\textwidth]{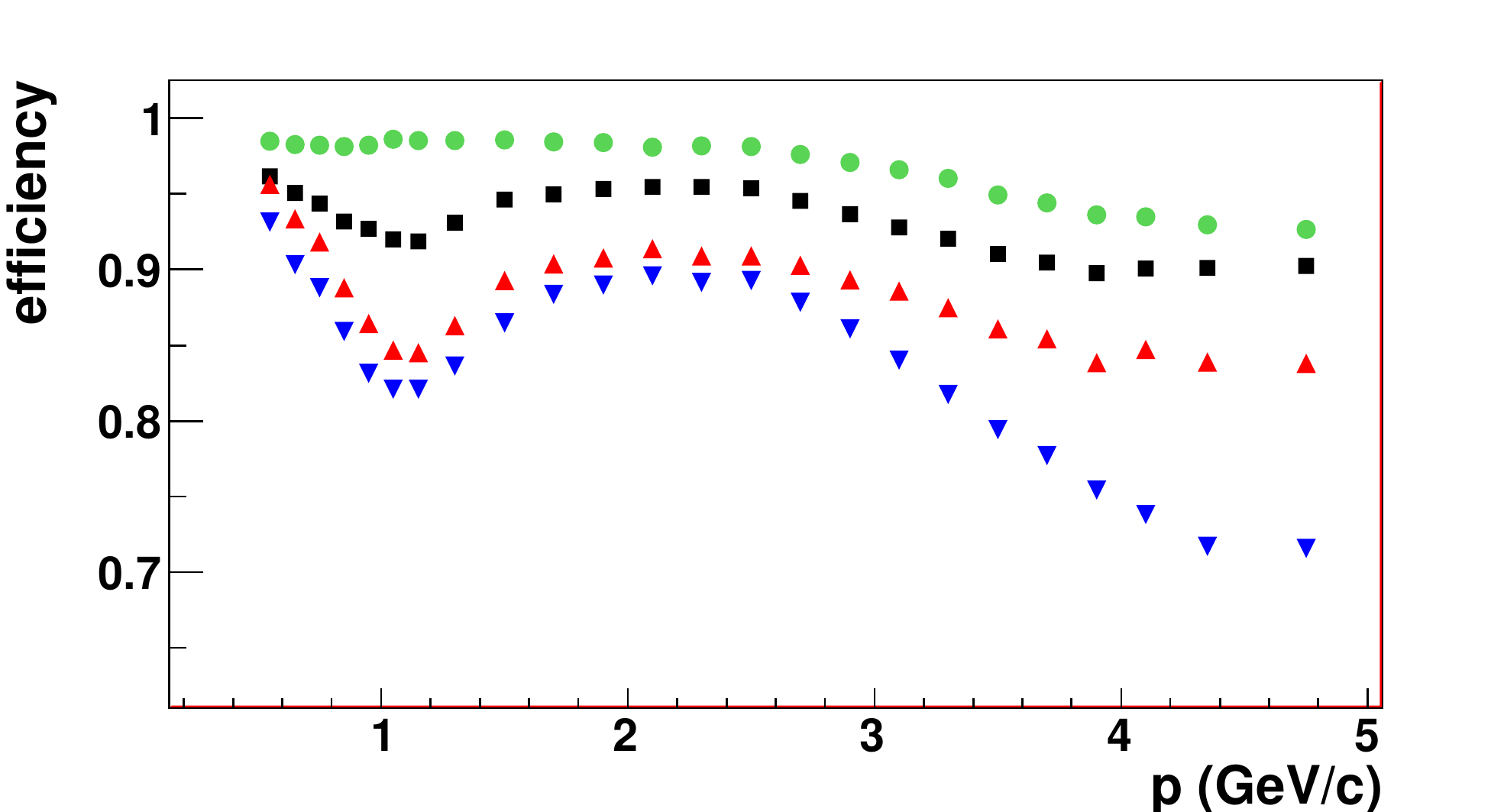}}
\subfigure{\includegraphics[width=0.49\textwidth]{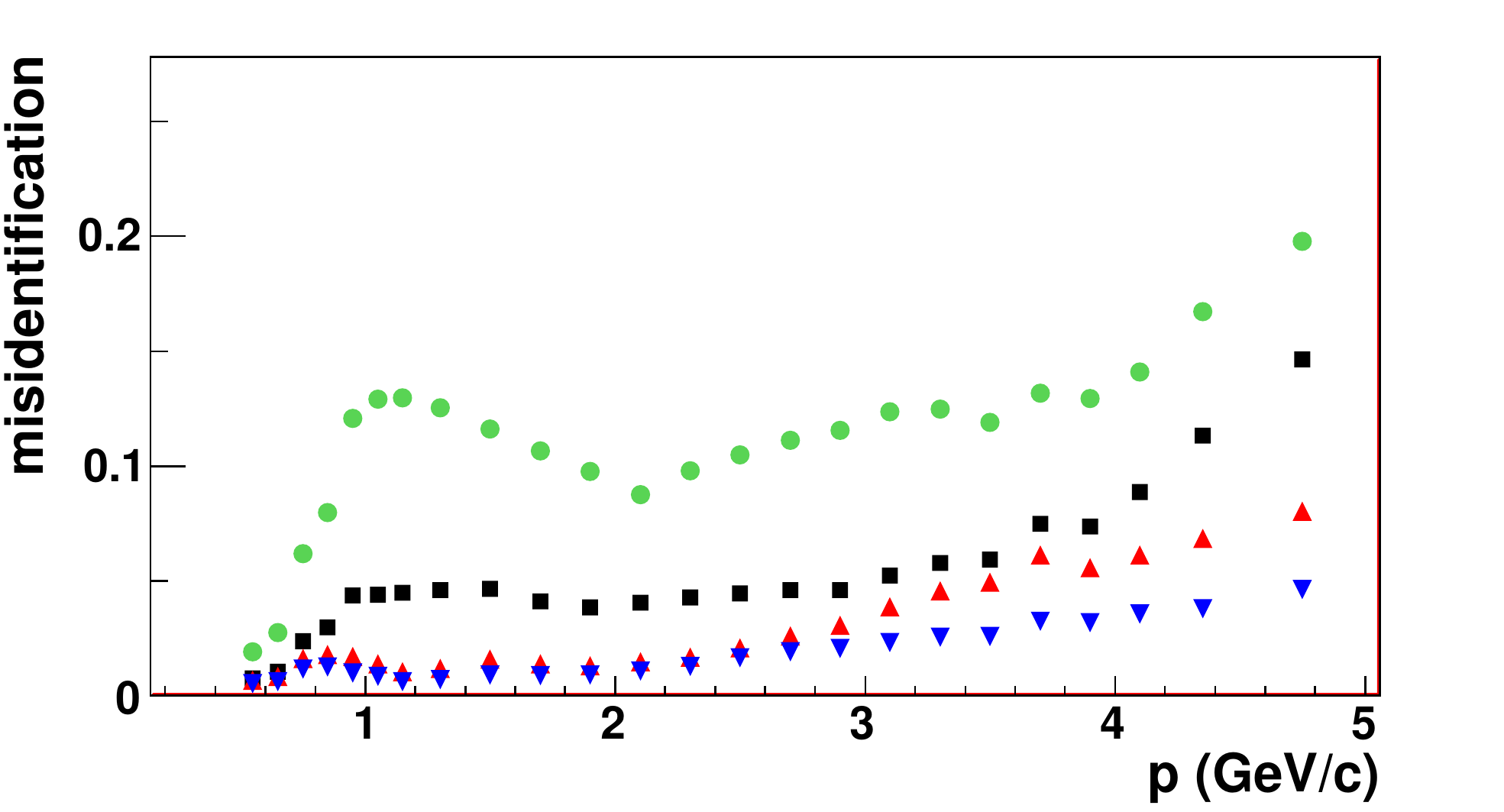}} \\
\subfigure{\includegraphics[width=0.49\textwidth]{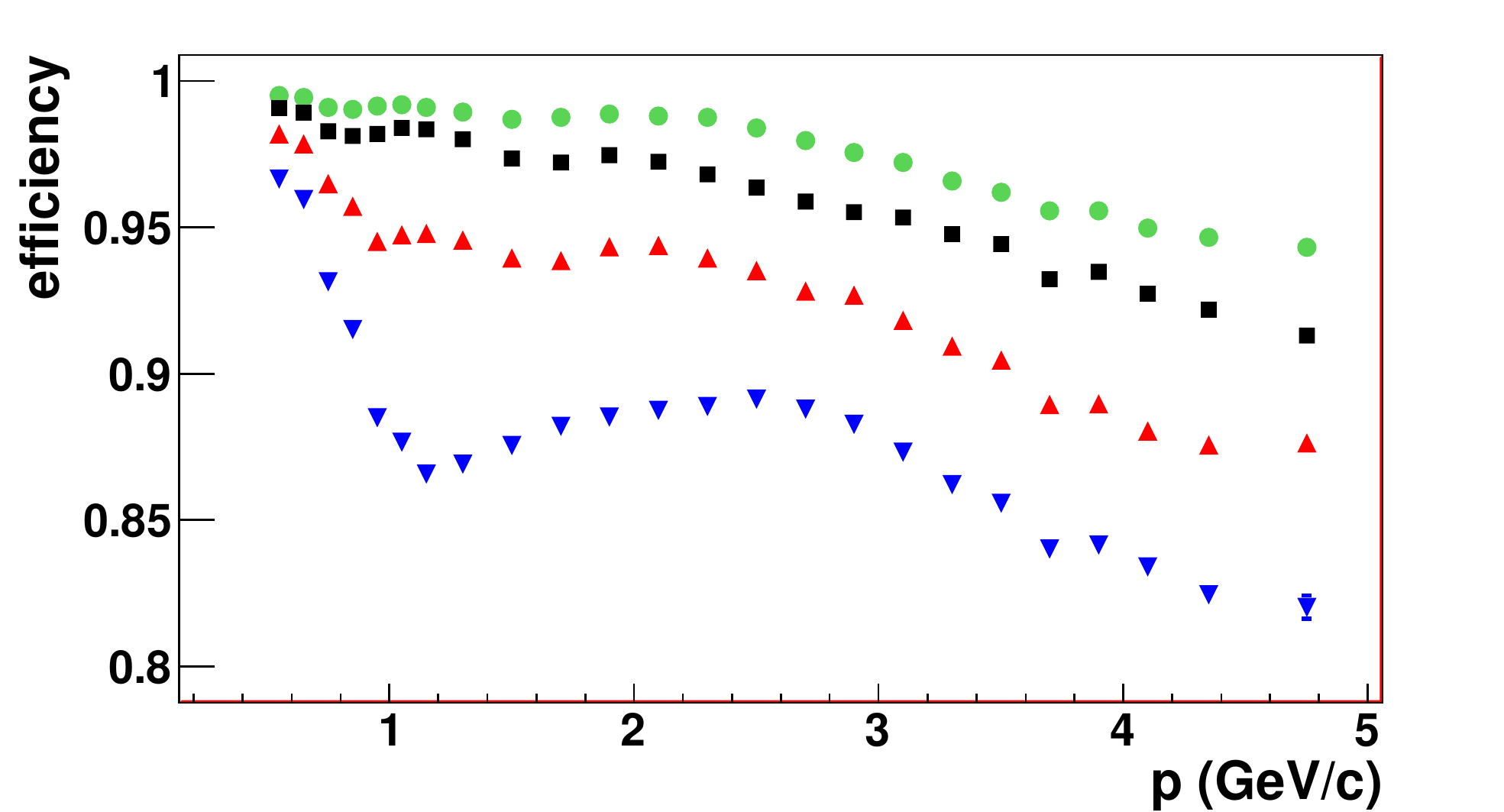}}
\subfigure{\includegraphics[width=0.49\textwidth]{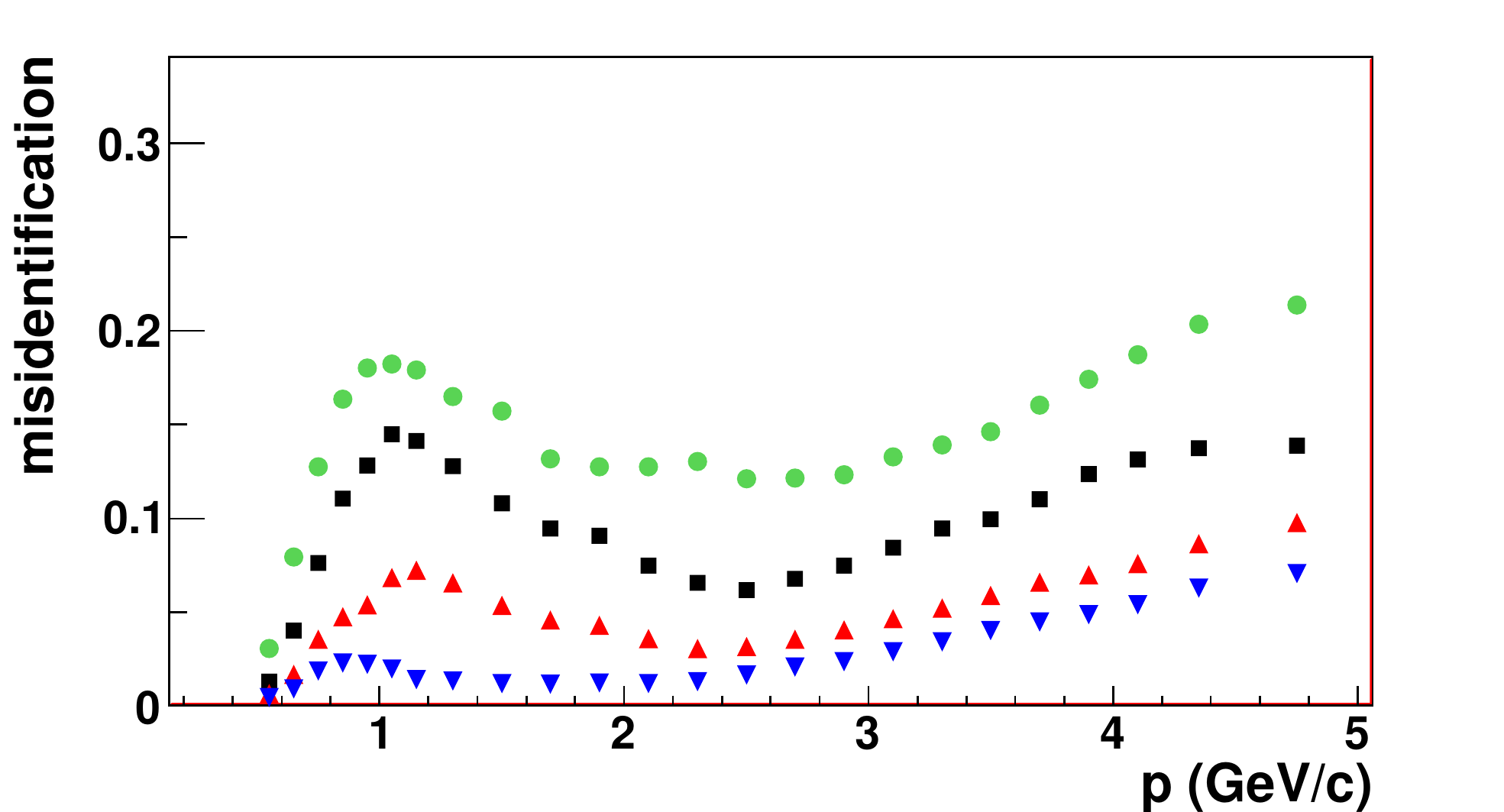}} \\
\subfigure{\includegraphics[width=0.49\textwidth]{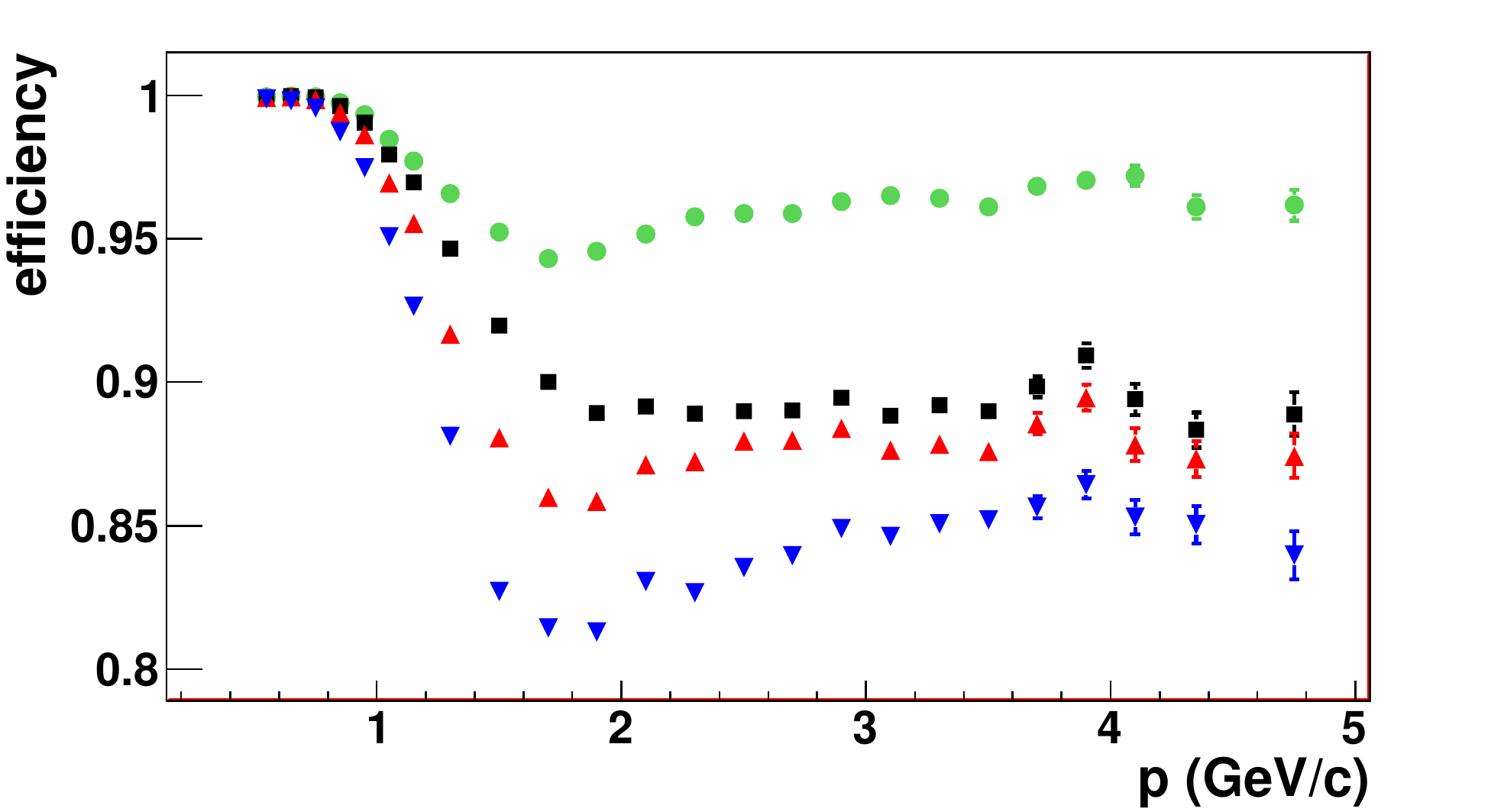}}
\subfigure{\includegraphics[width=0.49\textwidth]{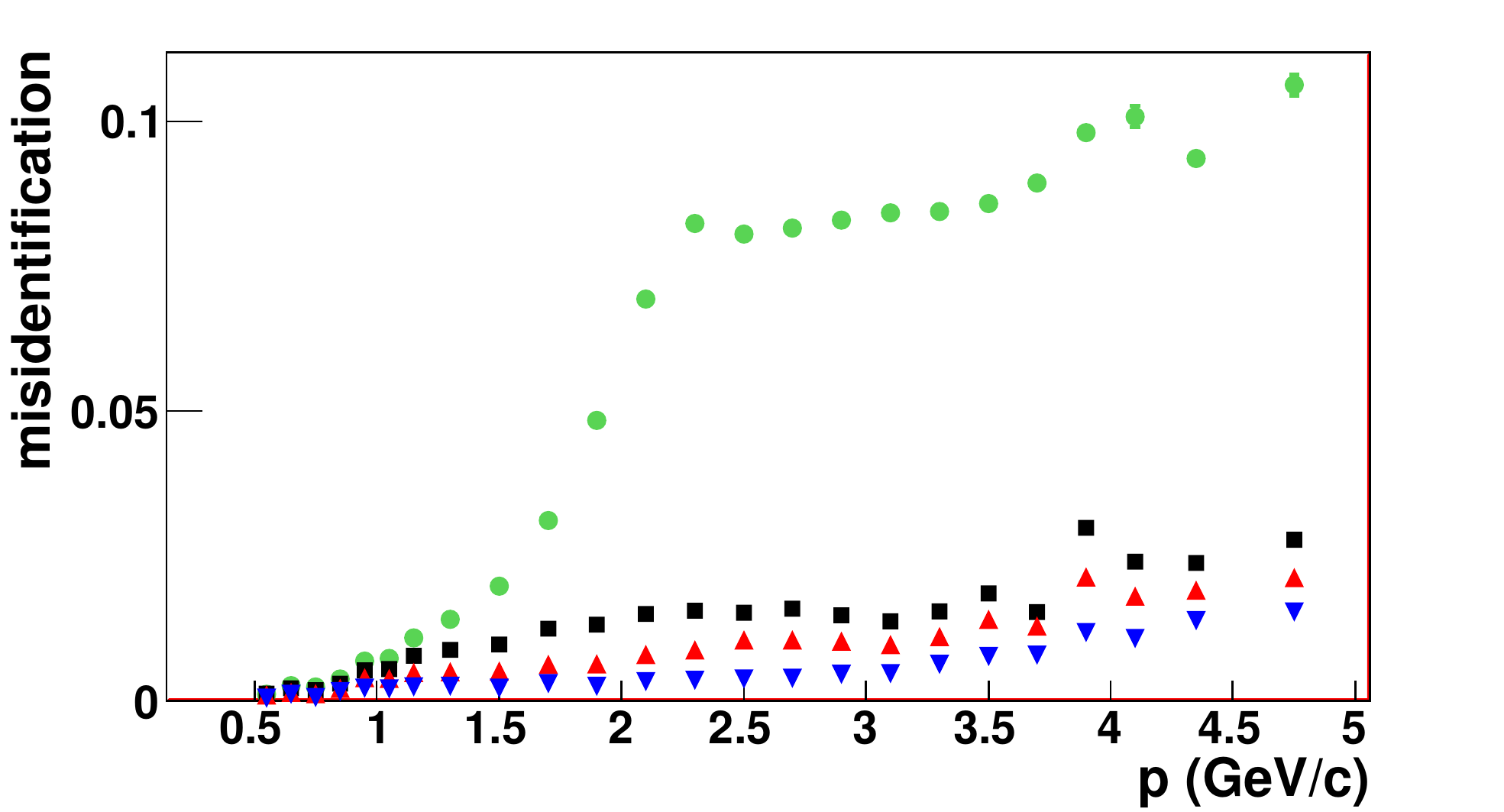}}
\caption{Hadron PID efficiency and misidentification as a function of laboratory momentum:
(all plots) {\it VeryLoose} (green circles), {\it Loose} (black squares), {\it Tight} (red up-pointing triangles)
and {\it VeryTight} (blue down-pointing triangles) PID criteria;
(top row) (left) $K$ LH efficiency, (right) $K$ LH $\pi$ misidentification rate;
(second row) (left) $K$ ECOC efficiency, (right) $K$ ECOC $\pi$ misidentification rate;
(third row) (left) $\pi$ ECOC efficiency, (right) $\pi$ ECOC $K$ misidentification rate;
(bottom row) Proton ECOC efficiency, (right) Proton ECOC $K$ misidentification rate.}
\label{fig:hadpid}
\end{center}
\end{figure*}

A number of algorithms have been developed in \babar\ data analyses 
to discriminate hadrons. The earliest hadron particle identification scheme was based on a
combined likelihood (LH) ratio test, with individual SVT and DCH likelihoods based on $\dedx$, and
DIRC likelihoods based on Cherenkov angle, number of photons, and track quality,
with likelihoods computed for each particle type.
This PID method is compared below to a later, more sophisticated algorithm
based on decision trees and error-correcting output code (ECOC)~\cite{ECOC95}, which incorporates
additional detector information. Several other hadron PID algorithms have also been
developed, most notably a neural network-based hadronic PID algorithm used in
the tagging of the decay flavor of $\Bz$ mesons in many early analyses.
However, the LH ratio and the ECOC-based hadronic PID algorithms
have been used by the great majority of all analyses, so these two hadronic PID
schemes are discussed in detail below.

The strategy underlying hadronic PID by using likelihoods is to determine
a combined likelihood for each particle hypothesis $i$:
\begin{equation}
L_{i} = L_{i}^{{\mathrm DIRC}} \times L_{i}^{{\mathrm DCH}} \times L_{i}^{{\mathrm SVT}},
\end{equation}
\noindent As shown in Figures~\ref{fig:svtpid} and~\ref{fig:dchpid} above,
the SVT and DCH likelihoods are quasi-Gaussian in form, and are parameterized
as the difference of expected and measured $\dedx$ normalized to the experimental
uncertainty on the measured $\dedx$. The DIRC likelihood cannot be expressed
in as straightforward a fashion since there are significant tails on the
fitted Cherenkov angle and number of photons. Therefore, a three-dimensional likelihood
look-up table binned in laboratory momentum (100\mevc bins), Cherenkov angle (three bins
corresponding to the $\pi$, $K$ and proton bands) and number of photons (four
bins) is created from simulated tracks. This binned DIRC likelihood is designed to
adequately characterize the tails in the DIRC response, particularly near particle thresholds. At
momenta $p\gae 1.5\gevc$, well above the DIRC threshold, this binned
likelihood does not separate the DIRC angle bands well, so it is multiplied by a
purely Gaussian Cherenkov angle likelihood.

Tracks of a given particle type are chosen using likelihood
ratios which allow selection of the desired particle type at a
given efficiency either for the target particle or for one or more
background types. The likelihood ratio [LR] is defined as
\begin{equation}
LR_{i} = \frac
{L_{i}^{{\mathrm DIRC}} \times L_{i}^{{\mathrm DCH}} \times L_{i}^{{\mathrm SVT}}}
{\sum \alpha_{h} (L_{h}^{{\mathrm DIRC}} \times L_{h}^{{\mathrm DCH}} \times L_{h}^{{\mathrm SVT}})}
\end{equation}
\noindent where $i=K$ or $\pi$ or $p$ is the target particle type,
$h$ runs over all signal and background particle types, and
$\alpha_{h}$ is a scaling factor that characterizes the differences in
expected particle multiplicities.

Error-correcting output code is a technique for combining multiple binary classifiers
to form a multi-class classifier.
The ECOC-based PID uses 36 input parameters including particle momentum, charge,
polar and azimuthal angles, the LH ratios described above, along with the inputs
used to calculate them, the DIRC likelihoods for $\pi$/$K$/proton, the number of signal and
background DIRC photons, the last DCH layer hit, the number of SVT layers hit, and the EMC
energy along with ten other EMC-derived parameters.
These parameters are used to construct seven separate bootstrap aggregate decision
trees ({\it bagged} decision trees, BDTs) trained to collectively separate $K$, $\pi$, $e$ and protons.
The seven bagged trees $t_i$ are trained using the same 36 inputs but with differing groups
of particle types defined as ``signal'' and ``background'', as shown in Table~\ref{tab:ecoc_matrix}.
The $4 \times 7$ matrix covers all combinations of signal and background available given the four particle types here.

To classify a given track, each classifier gives an
output between -1 and 1, according to its particular
definition of signal and background. The output of
all seven binary classifiers can be represented as a
sequence of real numbers of length seven. To classify
a given track, this sequence is used to find the
sum of the squared difference of the outputs from the
corresponding row of Table~\ref{tab:ecoc_matrix}.
This is analogous to computing a hamming distance $H$ \cite{Hamming}
in this information space, which results in four numbers $H_{i}$,
each of which corresponds to a different particle
hypothesis. The hypothesis with the minimum hamming
distance was adopted the most probable particle type.

\begin{table}[htb!]
\caption[ECOC PID training samples matrix.]
{ECOC PID training samples matrix. Each entry indicates whether a given particle
type should be considered as signal (1) or background (-1) for each of the
seven trees $t_i$.}
\label{tab:ecoc_matrix}
\begin{center}
\begin{tabular}{cccccccc}
\\ \hline
Class  & $t_1$ & $t_2$ & $t_3$ & $t_4$ & $t_5$ & $t_6$ & $t_7$ \\ \hline
$K$    &   1   &  1    &  1    &  1    &  1    & 1     &  1    \\
$\pi$  &   -1  &  1    &  -1   &  1    &  -1   & 1     &  -1   \\
proton &   1   &  -1   &  -1   &  1    &  1    & -1    &  -1   \\
$e$    &   1   &  1    &  1    &  -1   &  -1   & -1    &  -1   \\
\hline
\end{tabular}
\end{center}
\end{table}

Although it is possible in principle to arrive at an optimized PID selection for
any particular analysis through direct use of either LH ratios or ECOC outputs,
in practice standardized track lists with several selected target
particle efficiencies, typically characterized as {\it VeryLoose},
{\it Loose}, {\it Tight} and {\it VeryTight}, ordered by decreasing
target particle efficiency and misidentification rates, have been created for each
particle type and PID algorithm.
Figure~\ref{fig:hadpid} shows LH and ECOC $K$ efficiencies and $\pi$ misidentification rates
as a function of momentum for these four different target particle efficiencies.
As expected, the ECOC-based PID shows superior pion rejection, at all momenta for
any given $K$ efficiency, relative to the LH-based algorithm.
The figure also shows the converse PID performance for pion selection using the ECOC algorithm.
The ECOC-based PID is generally superior to other algorithms for hadronic PID
because it optimally incorporates all information used in the other
algorithms, and then supplements this with a relatively large number of additional
weaker discriminants.

\begin{figure}[t!]
\begin{center}
\subfigure{\includegraphics[width=0.49\textwidth]{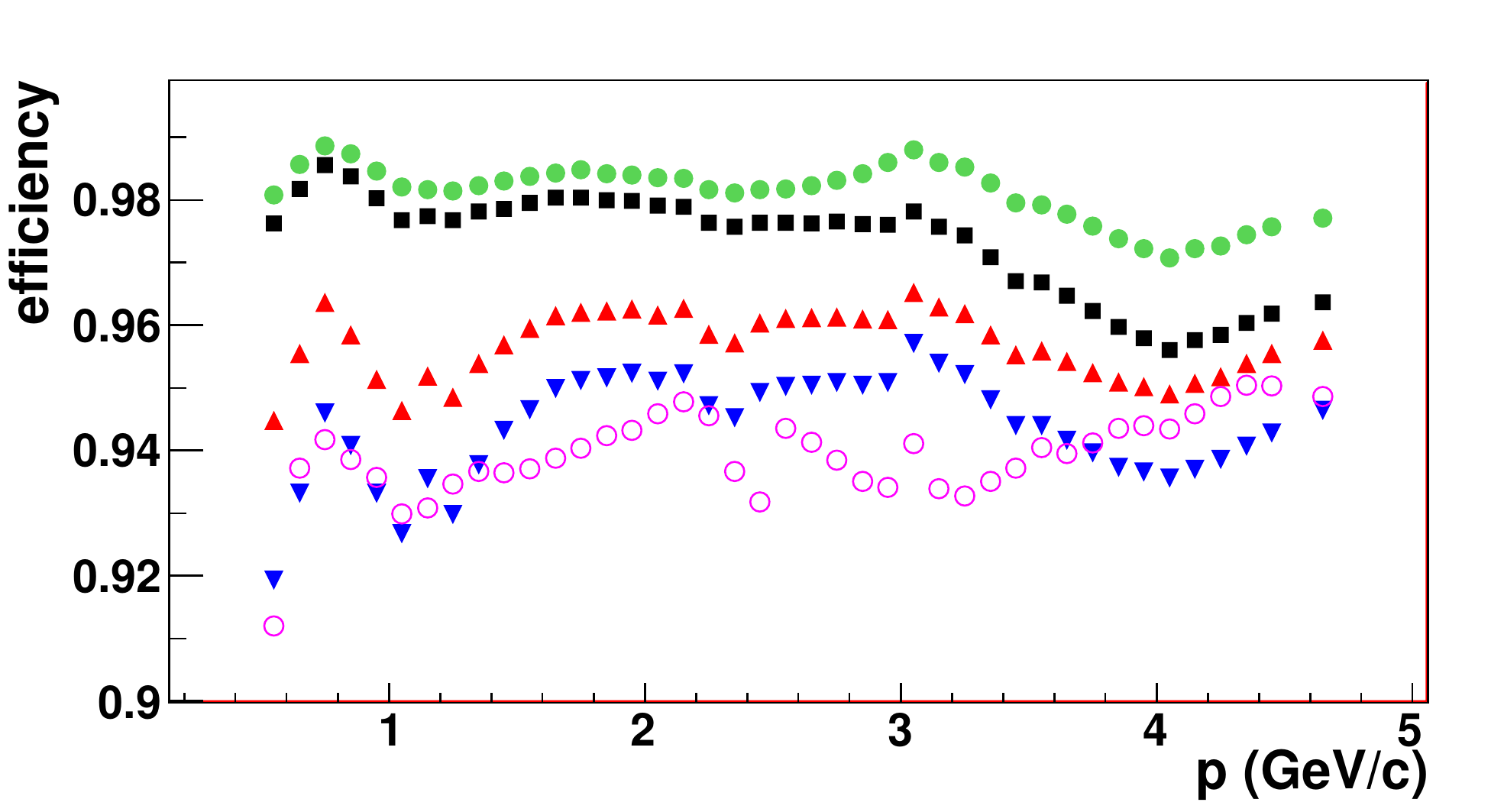}}
\subfigure{\includegraphics[width=0.49\textwidth]{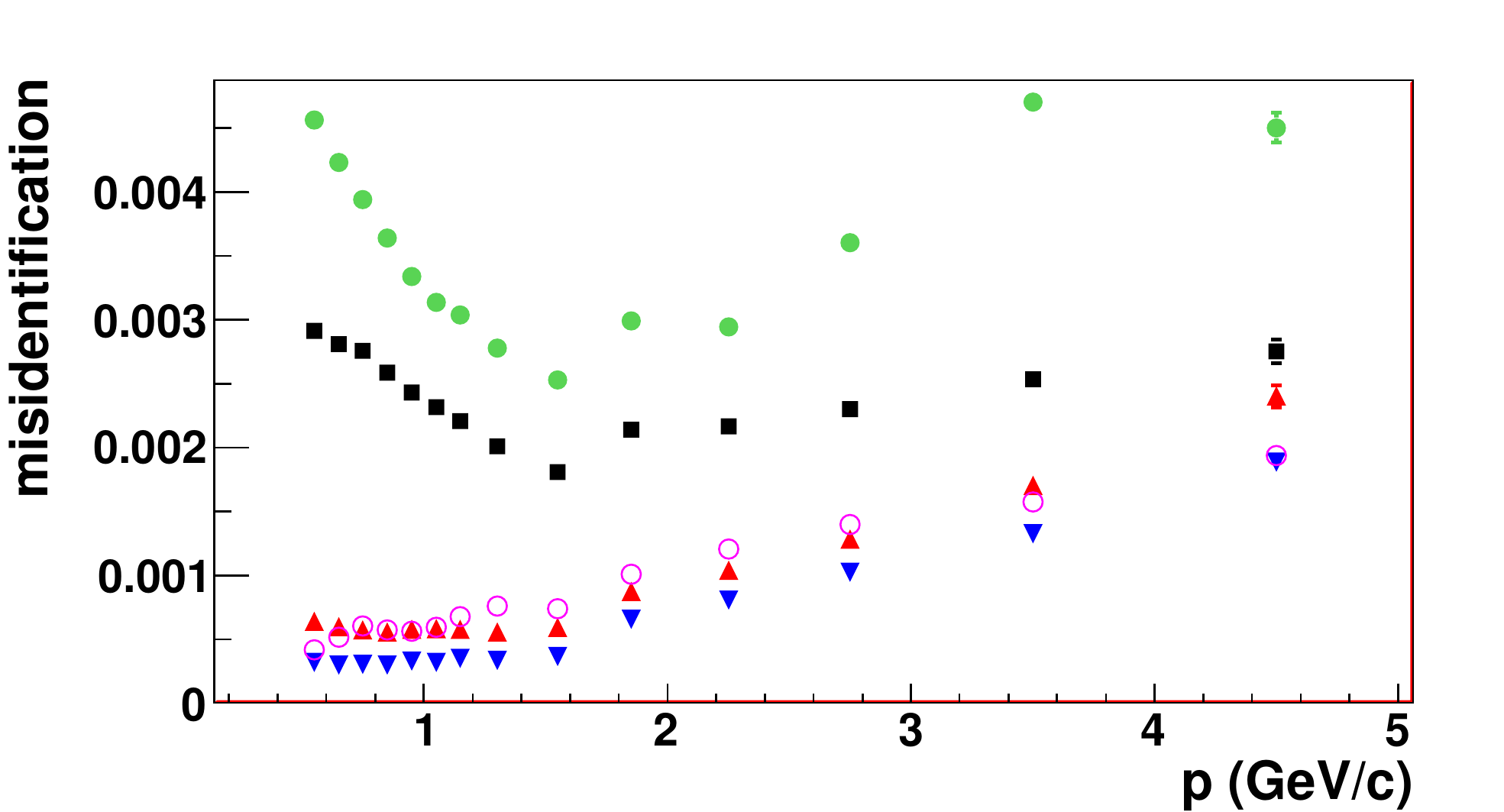}}
\caption{Electron efficiency and pion misidentification as a function of laboratory momentum:
(both plots) ECOC {\it VeryLoose} (green circles), ECOC {\it Loose}
(black squares), ECOC {\it Tight} (red up-pointing triangles), ECOC {\it VeryTight} (blue down-pointing triangles) and
LH-based (cyan open circles) PID criteria.}\label{fig:epid}
\end{center}
\end{figure}

To select tracks at the above four target levels of particle efficiency,
selection criteria are placed on $LR_{h}$ for the LH PID, or on the hamming distances and
ratios of hamming distances shown in Table~\ref{tab:pidham}. The actual thresholds
vary widely, \eg\ $L_K/(L_K+L_{\pi}) > 0.5 \, (0.9)$ for {\it VeryLoose} ({\it VeryTight}) kaons,
and $L_K/(L_K+L_{\pi}) < 0.98, (0.2)$ for {\it VeryLoose} ({\it VeryTight}) pions.
The selections on $H$ and ratios of $H$ for the four levels of ECOC efficiencies
are similar to the likelihood ratio thresholds.
Muons were not included as a part of the ECOC PID since the IFR discriminants provide
essentially no discriminating power between hadron species, and electron-initiated
EM showers are typically completely contained in the EMC and no IFR information is
available.

\begin{table}[htb!]
\caption{ECOC PID hamming distance and hamming distance ratio criteria used to select each particle type.}
\label{tab:pidham}
\begin{center}
\begin{tabular}{cc}
\\ \hline
Particle type  & Selection criteria \\ \hline
$K$    & $H_{K}$,   $H_{\pi} / H_{K}$, $H_{p}   / H_{K}$, $H_{e}   / H_{K}$     \\
$\pi$  & $H_{\pi}$, $H_{K}   / H_{\pi}$, $H_{p} / H_{\pi}$, $H_{e} / H_{\pi}$ \\
proton & $H_{p}$,   $H_{K}   / H_{p}$, $H_{\pi} / H_{p}$, $H_{e}   / H_{p}$     \\
$e$    & $H_{e}$,   $H_{K}   / H_{e}$, $H_{\pi} / H_{e}$, $H_{p}   / H_{e}$     \\
\hline
\end{tabular}
\end{center}
\end{table}

\subsubsection{Electron Identification}

As with hadron particle identification, likelihood-based and ECOC algorithms
are the principal means by which electrons are discriminated from other
charged particle species. Electron PID using likelihoods has been used by essentially all
analyses requiring electron discrimination until they were superseded by ECOC-based PID
in the more recent years. The LH-based electron algorithm uses information similar to
LH-based hadronic PID, supplemented by likelihoods describing the ratio of energy
deposited in the EMC to track momentum, and several EMC parameters such as
longitudinal and lateral shower shapes. All of these likelihood
distributions are binned in polar and azimuthal angle, momentum and charge for
each charged particle species except muons. Similar to LH-based hadron PID,
a selection is made on the ratio of the total combined electron likelihood to the
sum of the total combined likelihoods for all particle types. As discussed in the previous section, electrons are also
one of the target particle species used in ECOC-based PID, and the performance of
the ECOC and LH algorithms for electrons is shown in Figure~\ref{fig:epid}.

\begin{figure*}[]
\begin{center}
\subfigure{\includegraphics[width=0.49\textwidth]{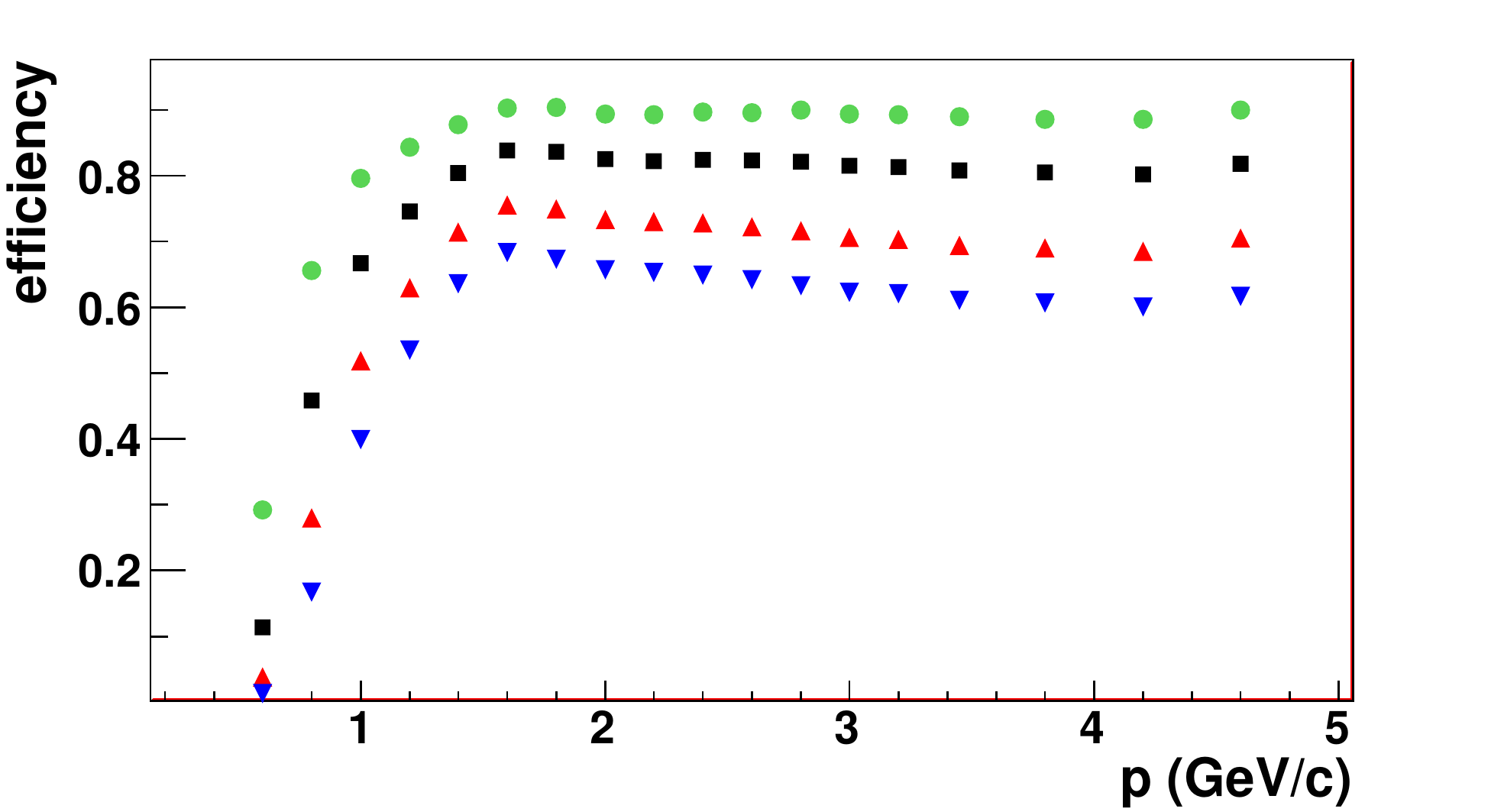}}
\subfigure{\includegraphics[width=0.49\textwidth]{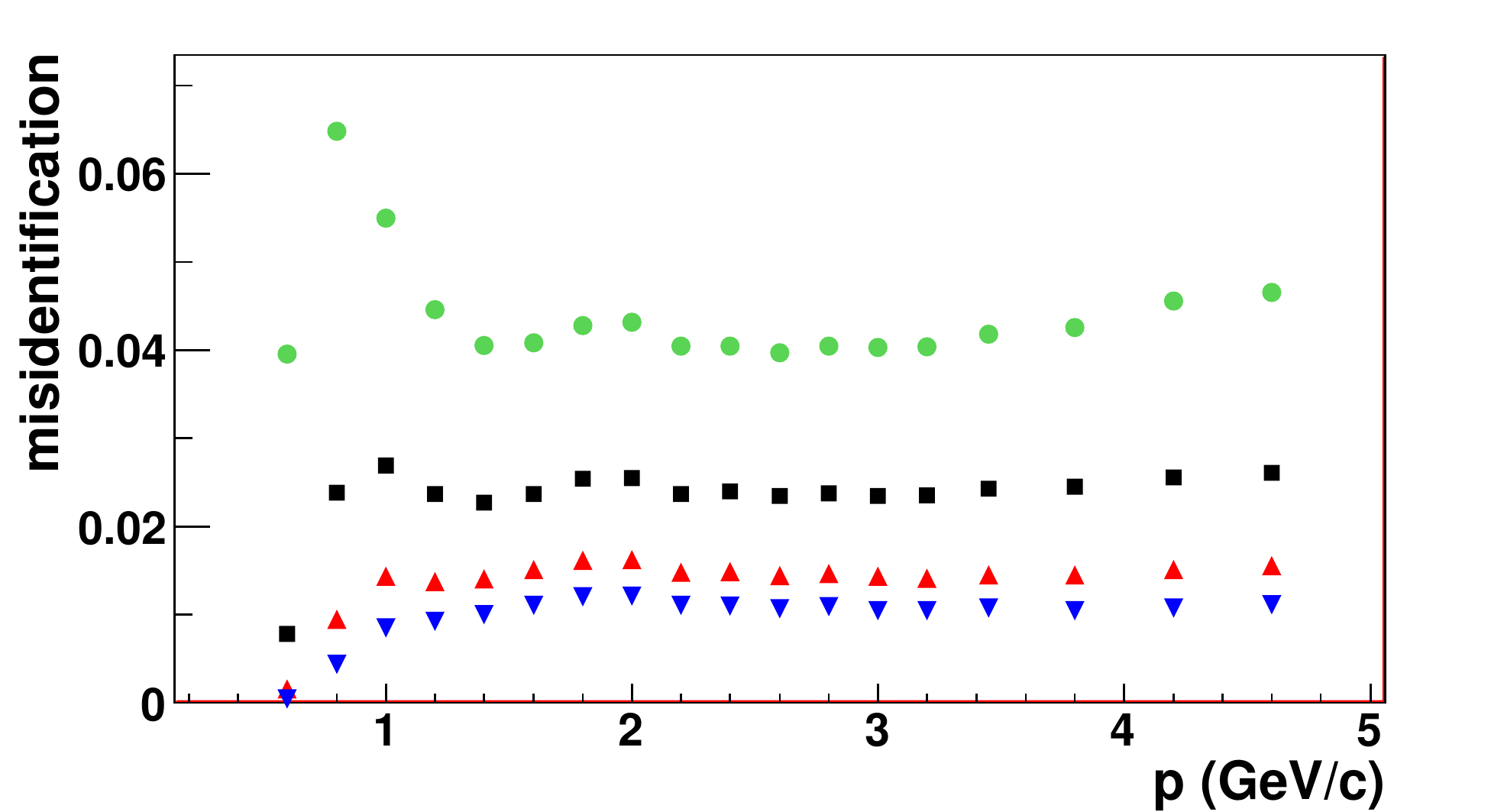}}\\
\subfigure{\includegraphics[width=0.49\textwidth]{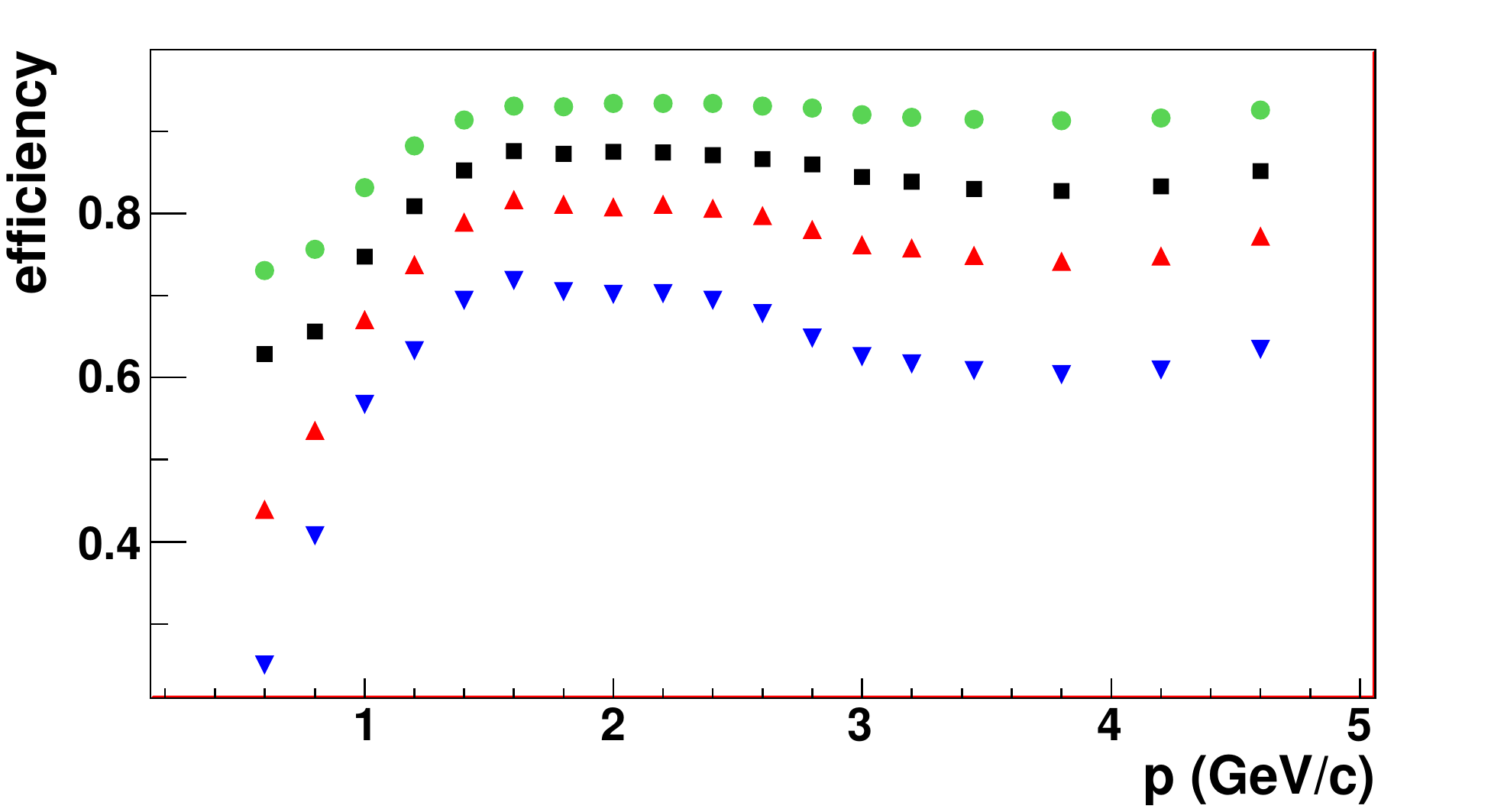}}
\subfigure{\includegraphics[width=0.49\textwidth]{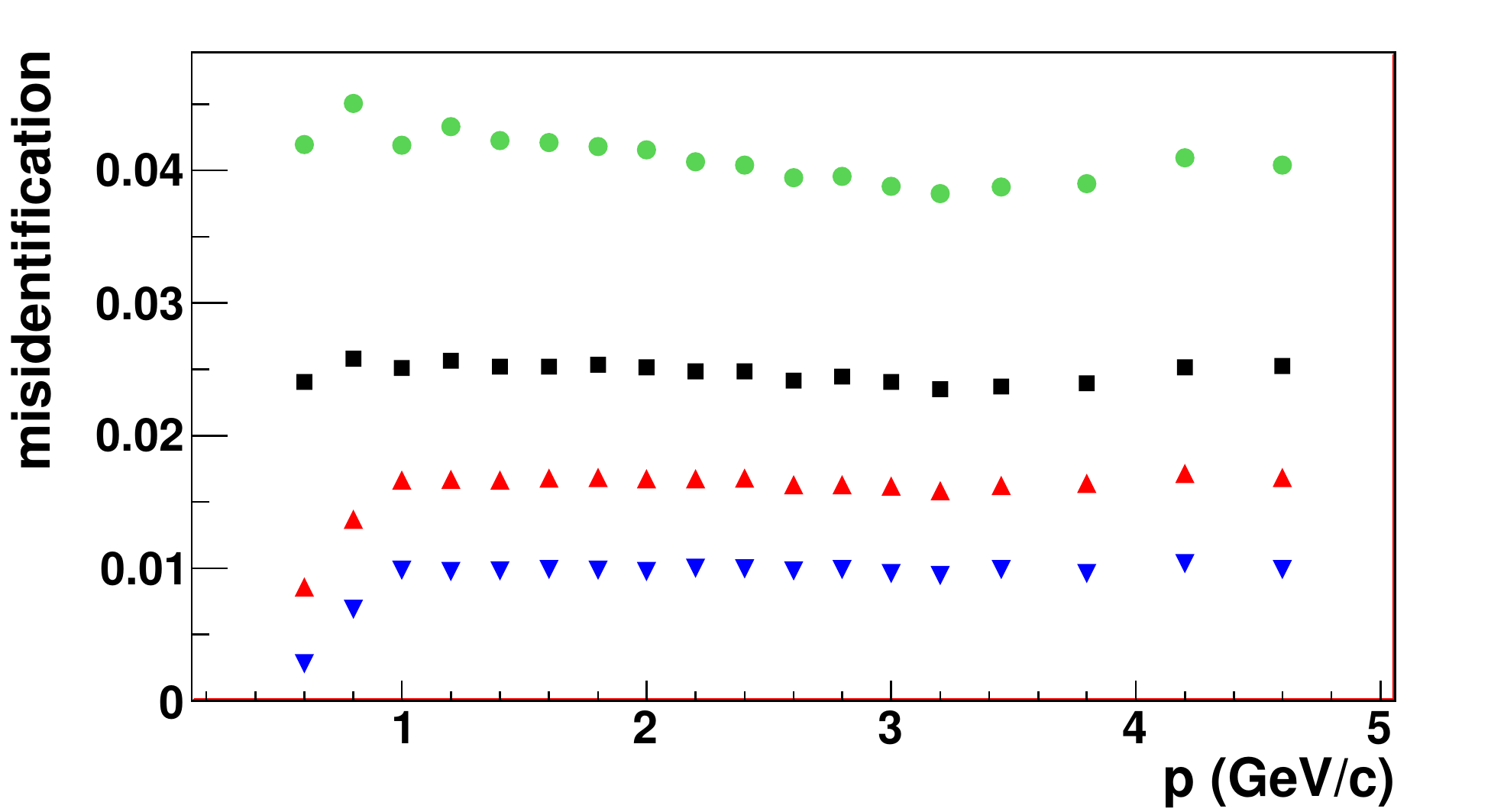}}
\caption{Muon efficiency and pion misidentification as a function of laboratory momentum:
(all plots) {\it VeryLoose} (green circles), {\it Loose}
(black squares), {\it Tight} (red up-pointing triangles), {\it VeryTight} (blue down-pointing triangles) PID criteria;
(top row) (left) $\mu$ NN efficiency, (right) $\mu$ NN $\pi$ misidentification;
(bottom row) (left) $\mu$ BDT efficiency, (right) $\mu$ BDT $\pi$ misidentification.}
\label{fig:mupid}
\end{center}
\end{figure*}

\subsubsection{Muon Identification}
\label{section:muonPID}

Unlike the particle identification performance described above,
which was essentially stable during the \babar\ detector operation,
muon PID was time-dependent, initially because of RPC
hardware problems and subsequently due to upgrades to major portions of the IFR.
The original muon PID algorithms employed rectangular selection criteria on several
important IFR discriminants. In 2004, neural networks 
utilizing the same input variables were
developed. Most recently, the use of BDTs,
in which information from the inner
detectors can be effectively combined with the IFR
discriminants, has resulted in substantial performance
improvements over the previous muon PID algorithms, especially for
data taken during periods of poor IFR performance.
The additional SVT, DCH, DIRC and EMC discriminants
are essentially identical to those used for the ECOC-based
PID. The IFR discriminants include the number of
interaction lengths traversed by a track, the
difference of the actual and expected number of
interaction lengths traversed by a muon with kinematic
properties similar to the candidate track, the first
and last IFR layer hit as well as the total number
of hit layers, and the $\chisq$ per degree of freedom of the IFR hit
strips with respect to the track extrapolated to the IFR.
Figure~\ref{fig:MUIDtime} from Section~\ref{sec:ifrperf} 
shows the time-dependent variation
of muon PID efficiency at four fixed levels of pion misidentification.

Figure~\ref{fig:mupid} compares the performance
averaged over the lifetime of the \babar\ detector for
the NN and BDT muon PID, with both algorithms
tuned to give similar pion misidentification rates.

\subsubsection{Systematic Uncertainties}

A measure of the PID systematic uncertainties can be provided by evaluating
PID efficiencies using several different data control samples,
ideally sampling events with widely varying detector occupancies, kinematic
properties, background levels, etc., and searching for statistically
significant differences among measured PID efficiencies.
If no such significant differences are found, then the purely statistical uncertainties
associated with the efficiency measurements are reasonably likely to cover any PID systematic.
If statistically significant differences between PID efficiencies are observed,
then the spread of efficiencies is also reasonably likely to cover possible systematic effects.
There are, however, few such multiple data control samples for which charged particles
of \emph{a priori} known type can be used to reconstruct decays with and without PID
selections applied.

The most useful alternative PID control samples are from $B$ meson decays
to final states with a $\jpsi \to \ellell$ daughter. Although not as
copiously produced as the PID control samples
described above, final states such as $\jpsi K$, $\jpsi \pi$ and $\jpsi K^{*}$
have very high signal-to-background ratios and occur
in a detector environment more similar to that of most analyses than the
lepton control samples obtained from QED processes.
Systematic uncertainties in PID efficiencies for $e$, $\mu$, $K$ and $\pi$
can be studied by comparing signal yields from $B \to \jpsi K^{(*)}/\pi$ events
obtained with and without PID selection(s) in data and simulated events
in which PID efficiencies have been calibrated using the nominal PID control samples.
With this method, PID efficiency systematic uncertainties of ${\sim} 0.7\%$ per electron,
${\sim} 1\%$ per muon, ${\sim} 1.1\%$ per kaon, and ${\sim} 0.2\%$ per pion
are found. Similar results are observed when using a $B \to D^{*}\pi$
control sample.

The above systematic uncertainties are adopted by
most analyses of $B$ and $D$ decays, particularly those with
statistics-limited results where any PID uncertainty is likely a
small fraction of the total uncertainty. However, for systematics-limited
measurements, it is often possible to refine these estimates by
examining variations in event selection, which typically lead
to differences in backgrounds and allows studies of possible biases
due to background subtraction techniques, along with the effects of
decays in flight, material interactions, event multiplicity, and
a host of other possible sources of systematic bias.


\renewcommand{\secname}{offline_}
\renewcommand{\sectiondir}{Sec06_Offline}
\section{Offline Computing}
\label{sec:offline}

\subsection{Processing Chain}

The offline computing system supports the following key processing
steps:

\begin{itemize}

\item Calibration and reconstruction of raw detector data;

\item Simulation of physics processes, the detector, and its response,
  taking into account the recorded conditions of data-taking; and
  reconstruction of the resulting data;

\item Skimming --- that is, the creation of a variety of overlapping
  subsets of the real and simulated data matching signatures for
  various physics processes --- created to improve the efficiency of
  analyses; and

\item Distribution of the resulting datasets to a variety of sites,
  making them available to users.
\end{itemize}

In this section, we describe their implementation in the releases used
to prepare the final consistent version of the full dataset.  In the
following section we discuss some of the evolution of the system that
led to that final version.

\begin{figure*}[htbp!]
\begin{center}
  \includegraphics[width=15cm]{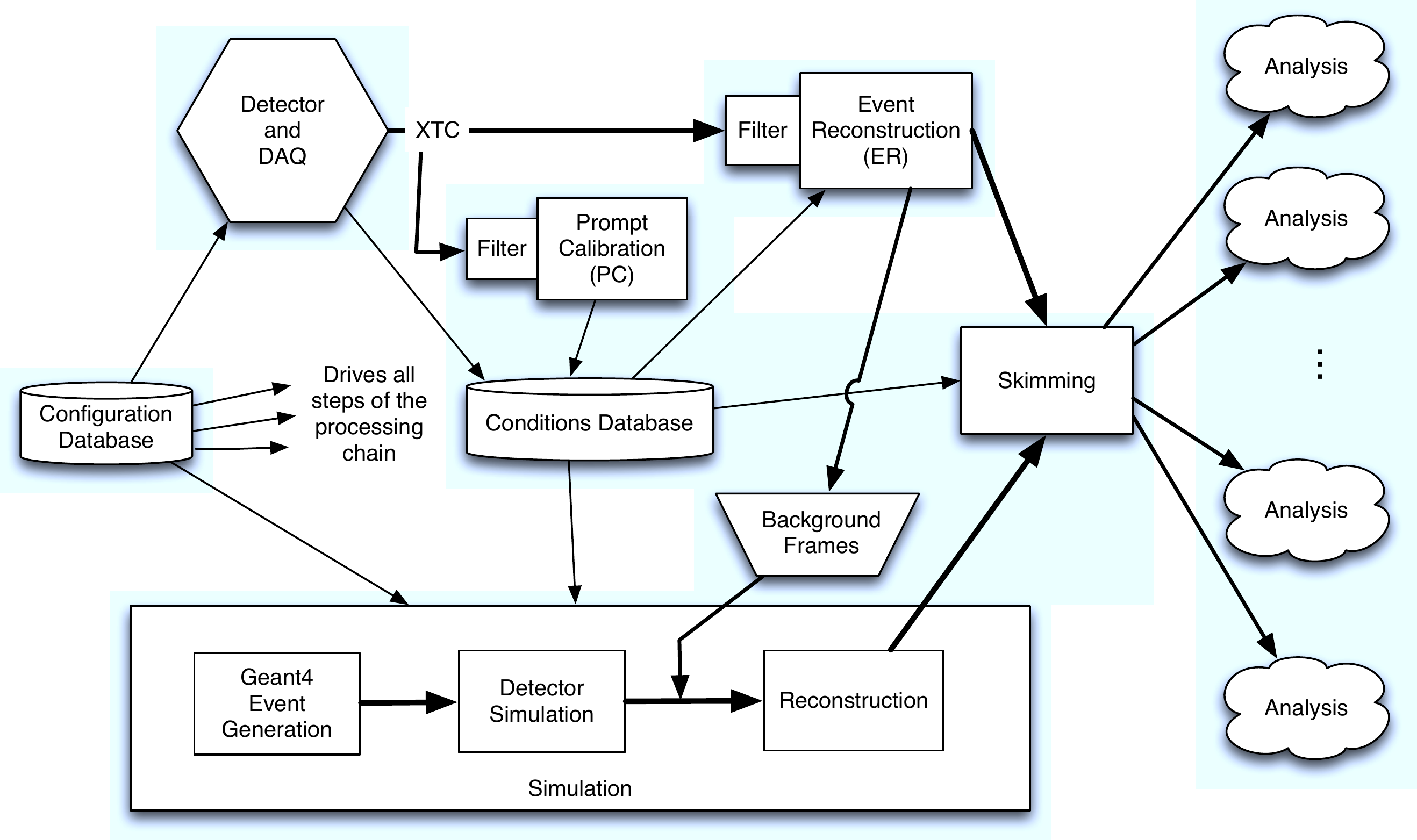}
\caption{Schematic diagram of the dataflow in event reconstruction and analysis.}
\label{fig:offline_reco}
\end{center}
\end{figure*}

\subsubsection{Raw Data Handling, Filtering, and Calibration}
\label{sec:offline-calib}
Raw data from the detector were acquired in units of runs, defined as
periods of operation under stable conditions, bounded to limit the
resulting file sizes. Typical runs lasted for 30--60 minutes and
contained $10^5$--$10^6$ events accepted by the L3 trigger.

The detector systems required a variety of calibrations to support the
event reconstruction algorithms.  Derived calibration constants were
stored in the conditions database in a cross-subsystem framework
defined in the original offline software.  The acquisition and
processing of calibration data fell into two classes: dedicated-data
{\it online} calibrations, described above in
Section~\ref{sec:onl-runcon}, and {\it offline} calibrations based on
regular colliding-beam data.  Of the latter, most were carried out
in an extensive automated framework. Other calibrations required very long
periods of data accumulation and/or extensive human intervention, and were
carried out manually.

Each automated offline calibration relied on a narrowly defined class
of colliding-beam events, typically low-multiplicity events with
high-momentum tracks, such as di-lepton final states.  The event
selections were defined at the original design luminosity of \pep2\ to
produce sufficient statistics to measure the calibrations at least as
rapidly as the expected rate of change of the detector behavior. As
the luminosity increased, it was not necessary to retain the same
effective cross-section for selected events, allowing a nearly
constant-rate subset of events to be selected.

Taking advantage of this, data reconstruction was divided into two
passes, Prompt Calibration (PC), which performed the automated offline
calibrations and operated only on the calibration event subset, and
Event Reconstruction (ER), which performed the reconstruction of the
full physics data stream.

In the PC processing for each run, a full reconstruction of the
selected events was performed using the calibration constants from the
previous run.  New constants were generated as permitted by the
accumulation of sufficient statistics. This processing had to be
performed in the order of data-taking, in order to follow the
evolution of detector behavior, and accumulate statistics across
multiple runs. The new constants from PC were written to the
conditions database.  Once this was done, the ER pass was able to
process each run in an order-independent way.

Among the automated offline calibrations, the duration of
colliding-beam data-taking required for a statistically significant
calibration varied from minutes to days. Supporting this range of data
requirements was a key design goal for the offline calibration system.
The calibration processing was separated into two steps: the
accumulation of statistics, and the fitting of calibration constants
to the resulting data.  In the first phase, data were reconstructed,
with the calibration constants from the previous run, and, for each
type of calibration, a specific set of reduced data (typically some
form of per channel statistics such as a histogram) was accumulated
per run.  For each type, once sufficient data for a valid calibration
had been accumulated, the data were combined and the second phase of
calibration --- constants fitting --- was performed.  The resulting
constants were automatically validated against defined standards and
checked for unexpectedly large changes from the previous set. They were
then stored in the Conditions database for application to any further
processing of the present and subsequent runs.

The subset of events used in PC was selected in two stages.  A simple
filter was first applied to the raw data, selecting calibration event
candidates based on the output of the L3 trigger algorithms.
This selection was performed within the framework of the online system
and produced a parallel set of smaller {\it calibration XTC} files in 
addition to the full raw data XTC files.
An additional stage of filtering was applied within the calibration
processing application in PC, further reducing the final calibration event
stream with respect to the L3 rate.  Because the scheme for
integrating calibrations over multiple runs required data to be
processed in order, keeping up with data acquisition required the
processing time of PC for any single run to be kept comparable to the
length of the run.  
Limiting PC to the constant-rate stream of clean calibration events greatly
reduced its CPU and I/O requirements and helped in limiting its latency.

In the final years of \babar, keeping up with the rate of data acquisition
required the parallelization of PC processing across 40--60 CPU cores.
Results from the multiple processes were accumulated using a version
of the same mechanism used for accumulation over multiple runs. 
The PC processing was performed at SLAC.

The systems for accumulation of data across multiple parallel jobs,
and across successive runs, were known as the {\it spatial} and {\it
  temporal} databases, respectively.  An early implementation of these
databases, in Objectivity, is described in Ref.~\cite{Bartelt:1998dd}.  In
the final version of the offline system these databases were
re-implemented in the ROOT framework.

\subsubsection{Reconstruction and Reprocessing}

With calibration constants in hand from PC, the full reconstruction of
the data could proceed in ER.  Here, too, an initial stage of
filtering, based on L3 trigger results and some additional
analysis, eliminated some beam background and low invariant-mass final
states.

Using the calibration constants from the PC pass, runs could now be
processed in any order during the ER pass. This provided useful
flexibility in the scheduling of processing, allowed multiple runs to
be processed in parallel, and relieved restrictions on the maximum
time required to fully process a single run.

The full reconstruction was generally performed at the Padova
computing site, following network transfer of the XTC files and
associated calibration constants from the PC processing at SLAC.
The operation of PC and ER was orchestrated by a \babar-developed
{\it Reconstruction Control System} which is described in detail in
\cite{Ceseracciu:2005jm}.

The production and data bookkeeping systems supported periodic full
reprocessings from raw data, including both the PC and ER passes.
Users were provided the ability to select current or previous
processings.

\subsubsection{Simulation}

A full MC simulation of the detector and interaction region
was developed.  The simulation of the basic \epem\ collision physics
was performed using a variety of event generators.  These were
primarily based on the EvtGen~\cite{Lange:2001uf} toolkit developed
for CLEO and \babar.  The \geant\ toolkit~\cite{Agostinelli:2002hh}
was then used to simulate the propagation of particles through the
detector geometry, and produce a record of their interactions with the
detector elements. \babar-specific software was then used to convert
this record into simulated digitized raw data, which was then
subjected to full reconstruction. The L1 and L3 trigger and
offline filter selection algorithms were evaluated as part of the
reconstruction of all simulated data, facilitating the understanding
of their effects on detection efficiency for physics processes.  The
entire process was combined into a single application, which greatly
simplified the production process by eliminating the use of
intermediate data files.

The production of simulated events was driven by an automated
production request system, which handled the production of both very
large collections of {\it generic} event samples, covering all relevant
\epem annihilation processes, and collections of events with specific
\B\ or $\tau$ decay modes or other physics processes.

Simulation production~\cite{Smith:2004ay} was distributed across the
national laboratory computing centers and a variety of institution
clusters, with about 25 sites producing a total of over 50 billion
events to date. Simulation production requests, consisting of event
generator selections, random number seeds, and detector conditions
setups were sent to the production sites. The resulting reconstructed
event collections were returned to SLAC.  The final simulated data
set, corresponding to the version of the offline software described
here, contains approximately seven billion events for specifically
requested physics processes and 20 billion generic events. The numbers
of generic events generated were chosen to correspond to specific
multiples of the numbers of events produced at the \FourS\ resonance.
The sizes of the generated samples, as multiples of the production
cross sections for specific final states, are shown in
Table~\ref{tab:generic-sim} for the final simulation production.
Earlier production cycles had multiples in the range of one to three.

\begin{table}[h]
\caption{\label{tab:generic-sim} Size of generic simulation production samples
relative to the production cross sections for specific final states 
at the \FourS\ resonance. 
}
\centering
\begin{tabular}{lcc}
\hline
Final State & Cross Section (nb)& MC Sample \cr
\hline
\BpBm, \BzBzb           &   1.05             &   $\times$ 10.8    \cr
\ccbar                  &   1.30             &   $\times$ 10.3    \cr
\uubar, \ddbar, \ssbar  &   2.09             &   $\times$ 4.1     \cr
\tautau                 &   0.94             &   $\times$ 4.1     \cr
\mumu ($\gamma$)        &   1.16             &   $\times$ 1.3     \cr
\hline
\end{tabular}
\end{table}

\paragraph{Handling of Detector Conditions and Trigger Evolution} The
simulation system was designed to track the evolution of the detector
and accelerator conditions.  For each calendar month of data-taking, a
representative set of detector configurations (including the trigger
configuration) and conditions (including the constants produced by PC)
were selected by detector system experts.  In simulation production,
event generation requests were divided into increments for each
data-taking month, in proportion to the recorded integrated luminosity
for the month, and produced using the corresponding configurations and
conditions.  This produced a luminosity-weighted simulated dataset
that reflected the actual history of data-taking.  Trigger and filter
algorithms corresponding to the monthly configurations were applied to
the simulated data as part of their generation and reconstruction.
This had the interesting software-maintenance consequence that the
final simulation application had to be able to apply all the software
trigger algorithms used in the history of the experiment, including
ones long since retired from use in data acquisition.

\paragraph{Background Mixing} Non-colliding-beam backgrounds from the
accelerator were not simulated. Instead, these backgrounds were
sampled by acquiring events from quasi-random triggers,
created in the FCTS at periodic intervals, unsynchronized to the
accelerator RF timing structure. The L3 trigger selected a
luminosity-weighted subset of the random triggers by accepting them in
proportion to the number of well-identified Bhabha scattering events
in the data stream.  The resulting background events for an entire
month were collected, randomized, and distributed to the simulation
production sites, and mixed into generated simulated events.

\subsubsection{Skimming}

The final dataset contains more than nine billion events passing the
pre-reconstruction filter described above, primarily taken on the
\FourS\ resonance.  This dataset has been inconveniently large for
frequent access by the several hundred collaboration physicists.  In
order to decrease the number of events needing to be processed as
input for analyses and decrease contention from many users accessing
the same data, one further step of central processing was applied to
the data, the {\it skimming} process, which performs two actions:
selecting subsets of events deemed to be interesting for various
groups of analyses, and pre-processing those events to compute
commonly-needed quantities and perform multi-particle combinatorics to
reconstruct short-lived intermediate states, such as neutral kaons and
$D$ mesons.  Skim data streams correspond to individual analyses, or
small subsets of analyses using common information, with each stream
representing a different event selection algorithm. Skimming produced
up to two hundred parallel streams of output; individual events can
appear in more than one stream.

The fraction of input events selected varies from stream to stream,
with some selecting as much as 15\% of the input data, (\eg\ a
skim searching for a $\tau$ lepton in the event, or looking for a \Dz\
decay with an $\eta^{(')}$ meson in the final state), and others as
little as one in $10^5$ events (\eg\ $\B \to \ccbar \Kstar$ or
$\B \to \etac K$).

Skimming was applied centrally, both to the colliding-beam data and to
the large samples of generic simulated events.  Skim
production~\cite{offline:tm2} was distributed across several computing
sites. All skim outputs were returned to SLAC and then redistributed
to the computing sites for user analysis.

In addition to automatically evaluating the trigger and filter
selections, the software framework facilitated the application of the
skimming event selections by users to individual samples of simulated
events from specific decay chains, allowing the estimation of the
efficiency of each of these stages of selection.

\subsubsection{Analysis Environment and Framework}

The offline processing framework, shared by production and user
analysis activities, as well as the L3 trigger filter processes
within the online system, provides services for the location of and
access to event data, association of these data with its corresponding
configuration and conditions data, and the delivery of these data in a
C++ object model to user and production code.

Much of this software framework was jointly developed with the CDF
collaboration~\cite{jacobsen:1998,frank:1997}.  In the framework,
processing flows are modeled as sequences of modules, sharing data
through the \emph{event data structure}, a C++-type-safe key-value
dictionary~\cite{frank:1997}.  The framework supports a limited
model of branching and selection on event flows, a capability that
was of particular use in the construction of the L3 trigger 
and offline filters.

Persistence of the event data is handled by a translation layer which
converts the native C++ \babar\ object model to an external form based
on the object-persistency framework provided by the ROOT
toolkit~\cite{Brun:1997pa}.  The transient-persistent translation
insulates the reconstruction and analysis code from the choice of
persistency mechanism. This was crucial to the ability to evolve the
persistency model, as described below.

The persisted data is in the form of ordinary POSIX~\cite{posix} files containing
data in the ROOT format, together with a meta-data catalog maintained
in a conventional relational database.  The catalog is supported by a
set of tools that permit the versioning and evolution of datasets over
time and facilitate production and user analysis tasks running over
the large number of files comprising the data~\cite{Smith:2005zza}.
These tools provide provenance tracking for datasets and ensure the
integrity of datasets (completeness and avoidance of inclusion of
multiple processings of the same run or unit of simulated data).

The \babar\ analysis framework supports reading and writing data
either via POSIX APIs, covering a variety of file system types
(including NFS~\cite{IETF:NFSv3},
GPFS~\cite{Schmuck:2002:GSF:1083323.1083349}, as well as locally
attached disk), and via the \babar-developed \xrootd\
protocol~\cite{Hanushevsky:2012:SSC:2357488.2357759,SLAC:xrootd}.  At
the production computing facilities, almost all data access uses
\xrootd.  The \xrootd\ system provides a simple POSIX-like
random file access API, optimized for large transfers, with support
for fault-tolerance, load-based replication, and a site-wide
server-independent namespace.

At times when hundreds of analyses were in progress, the provision of
sufficient compute capacity required contributions from all the major
national computing sites. Each topical analysis working group, and the
skim output streams it required, were assigned to a specific computing
site. In this manner, widely used skims could be made available at
multiple sites, reducing the contention for data access.

\subsection{Computing Model Evolution}
\label{sec:offline:evolution}

The offline computing model described above is that of the final year
of \pep2\ operation and the subsequent period of data analysis, until
the transition to the archival data analysis system described in
Section~\ref{section:LTDA} below. It differs in many respects from the
model in place at the start of data-taking.  That model displayed a
variety of problems with performance, scaling to higher luminosities,
and the support of many simultaneous physics analyses.  In the ensuing
years, the computing model was extensively modified to address these
problems, with changes to specific technology choices as well as a
complete redesign of the data processing architecture, while
preserving the user analysis environment and the bulk of the software
developed for reconstruction and simulation.

\subsubsection{Data Persistence}

The transformation of \babar\ offline computing incorporated a
substantial change to the data persistence model.

High energy physics events naturally lend themselves to being
represented as structured objects with complex inter-object
relationships.  Since this data model could not be supported directly
in the dominant programming language in the field in the 1980s and
1990s, FORTRAN-77, virtually every HEP experiment of that era had
devised its own approach to data organization.  In the years
immediately preceding the start of \babar\ software development, many
experiments had converged on a few complex systems supporting the
creation and persistence of structured data in FORTRAN-77, notably the
ZEBRA~\cite{cernlib:zebra} libraries developed at CERN.  Largely
motivated by the fragility and unnatural coding of these systems,
\babar\ early on decided to develop its software in the C++ language,
allowing a native implementation of the event data model.

The use of C++ still left the data persistence problem unresolved.
However, just as the experiment was being designed, a new
{\it object-oriented database} technology was being adopted in the
commercial software market, driven by the rapid acceptance of
object-oriented languages in professional software engineering.  This
technology offered the promise of directly persisting complex objects
and their relationships, freeing developers from implementing custom
object and reference serialization solutions.  It also provided a
framework for the organization and cataloging of datasets, another
long-standing challenge for HEP experiments. The use of a true
database also offered hope that the complex data filtering and
selection frameworks independently developed by many experiments could
be replaced by the use of structured database queries.

Motivated by the very large datasets expected for the LHC, various
research projects in the mid-1990s had explored the use of both
relational and object-oriented database technologies for the
persistence of HEP data, as an alternative to the use of ZEBRA-like
systems and serial tape storage~\cite{Malon:1997hq}. The results of
these explorations suggested that object-oriented database systems did
have attractive features for HEP data.  In particular, the CERN
RD-45~\cite{Binko:1995ud,cern:rd45} study, aimed at selecting a
technology for LHC use, ultimately recommended the use of a specific
commercial object-oriented database, Objectivity/DB~\cite{objectivity}.

Based on these studies and additional performance and scaling tests,
\babar\ adopted this recommendation, becoming the first major
HEP experiment to put this technology into production for event data
storage.

By the start of data-taking in 1999, significant problems with this
technology choice had become apparent.  While the object-persistence
features of an object-oriented database had proved useful, other aspects of the database
architecture were emerging as ill-suited to HEP data production and
analysis, or were a poor fit to other design choices made by \babar.
Moreover, specific performance problems with the vendor implementation
were encountered.

The transaction model supported by Objectivity proved inappropriate to
the write-once, append-only, read-many access pattern typical for HEP
event data.  Writing to the database generally involved the creation
of locks which obstructed the ability to write simultaneously from
many sources, in direct conflict with the natural parallelism of HEP
data processing.  While improvements in the vendor implementation and
in the use of locking by \babar\ allowed significant increases in
scalability, ultimately the need to lock meta-data objects in the
database could not be avoided. This limitation was partially
alleviated by maintaining multiple copies of the database for
production and user analysis. However, maintaining the consistency of
these replicas introduced considerable fragility and complexity to the
system.

Objectivity provided mechanisms to control of the placement of data,
but they lacked the flexibility needed to support optimization of the
I/O patterns and to avoid the performance penalties of random-access
loads on disk drives.  In addition, the Objectivity bulk data and lock
server processes presented significant scaling and fault-tolerance
problems. 

Intensive work with the vendor did result in substantial performance
improvements, sufficient to support the initial years of \babar\
operation. However, the software engineering effort by the experiment
to support this work was found to be unsustainable, especially because
of the lack of access to the proprietary Objectivity source code
inhibited the detailed understanding of the performance of the
database. The complexity of the system developed to overcome the
limitations of the database limited the experiment's ability to
deliver newly-acquired data to analysts in a timely manner, producing
considerable dissatisfaction in the collaboration, especially in view
of the rapid growth and high quality of the dataset.

Further, the original expectation that analyses would be performed on
a single event store, with preselections performed via database
queries or through the use of lists of pointers to previously
identified events of interest, proved impractical to implement, as it
would have generated intolerable random-access loads on disk-based
storage. The model actually deployed required the production of up to
200 parallel {\it streams} of events, with significant levels of
duplication of data, in order to relieve the random-access bottleneck
and other scaling problems. However, this approach then ran into
limitations on the total numbers of object collections in the internal
Objectivity data model, producing serious operational constraints.

The complexity of both the \babar\ and vendor database implementations 
also became a significant obstacle to distributed computing.
The difficulty of installing and supporting the database made it 
infeasible to perform data analysis at small sites and on 
personal computers, especially in the absence of network connectivity.
\babar\ relied on a group of 10-20 institutions to perform a large
fraction of simulation production, and the support of the Objectivity
database at each site represented the majority of the system administration
effort involved.
The movement of data among the established installations was error-prone
and required complex operational procedures.

Finally, because of the decision of the LHC experiments to abandon the use of
Objectivity for event data, the expected development of Objectivity
expertise in the wider HEP community failed to materialize.

\subsubsection{Offline Calibration System}

The performance problems of Objectivity exacerbated a poor choice in
the original offline system design: the computation of calibration
constants was performed in the same application as the reconstruction
of the full event data stream. Since the averaging process needed to
compute calibration constants spanning multiple runs required all runs
to be processed in order, this design required the full processing of
a single run's data to be completed in no more than the time required
to acquire it. 

The reconstruction of multiple runs could therefore not be performed
in parallel; all parallelization had to be at the event level. This
required each run to be processed on a large cluster of CPUs: in the
early data-taking, 100-200 hosts working in parallel were needed to
complete a single run in time. The scaling problems with Objectivity
then affected both the writing of reconstructed event output and the
computation of calibrations, where the use of the Objectivity database
had been incorporated into the mechanisms for combining calibration
results across multiple CPUs and multiple runs.

\subsubsection{Solutions}

As early as the initial months of data-taking, the difficulties with
supporting user analysis in Objectivity led to the creation of an
experimental alternate event data model limited to analysis objects
(the so-called {\it micro DST} containing only the final track momenta,
energies, and particle identification assessments).  This system,
based on the use of the CERN ROOT toolkit for object persistence, was
in place by 2000~\cite{Adye:2002qv}.

Its introduction was facilitated by the existence of the
transient-persistent translation layer mentioned above, already a
feature of the Objectivity-based design, motivated by the risks
inherent in using a novel and commercially supplied technology.  The
existence of the translation layer permitted the replacement of the
persistence technology for event data with virtually no changes to the
reconstruction, simulation, or analysis code.  It did, however,
require the development of a new infrastructure for cataloging the
ROOT-format event data files.  This was developed using standard
relational database technology.  The micro-DST data corresponding to
the first three years of \babar\ data acquisition yielded
approximately $3\times 10^5$ files comprising a total of 4.2\tbyte\.
These files were all kept on disk, stored on a set of NFS file
servers.  The largest file sizes in this scheme were a few hundreds of
megabytes, to be compared to file sizes of tens of gigabytes for
Objectivity, driven by the need to keep the total number of files
below a vendor-imposed limit.  The ROOT file format naturally
supported compression, with a column-oriented data model that led to
significantly improved compression statistics compared to the data
compression later offered by Objectivity.

The Objectivity-based production systems for reconstruction and
simulation were not changed, but their output was converted to the new
micro-DST format in an additional processing step.

The experience with this system was very positive.  It greatly
simplified data cataloging and distribution to remote sites, which
could now maintain copies of the event data even with limited
technical expertise.  The reduced event size led to reduced I/O loads
and facilitated the scaling up of analysis resources at SLAC and other
sites.

The success of this initiative led to a decision to re-implement the
entire event store along these lines, during the period 2003--2005.
This required supporting features of the event store not needed in the
experimental analysis-only system, most notably the ability to persist
pointer relationships between analysis and reconstruction objects,
tracing the derivation of analysis quantities from lower-level
reconstructed data.  This feature was particularly crucial to support
the new \emph{event summary data} event component that had been
deployed following the start of \babar\
production~\cite{Brown:2003qd}.

A full reimplementation of event persistence also required the use of
a much more robust file serving system to support the scale and
fault-tolerance required for the combination of production and
analysis needs, and the expected growth of the \babar\ dataset with
increasing luminosity.  The \xrootd\ system described above was
developed to meet this need.

The non-event databases that had been developed using Objectivity ---
conditions, configuration, and ambient --- did not have comparably
serious performance and scaling problems, and were left intact in the
new computing model.  In the final years of operation, however, all
remaining uses of Objectivity were replaced with combinations of ROOT
for bulk data storage and a conventional relational database for
catalog meta-data.  This was motivated primarily by the desire to
reduce the operational and support costs associated with commercial
software.  The re-implementation of the conditions database
persistence was more difficult and more disruptive than it had been
for the event store, primarily because its original design lacked a
transient-persistent translation layer.  Significant sections of
calibration, reconstruction, and simulation code thus depended
directly on the persistent objects defined in Objectivity and required
modifications when the new ROOT persistence technology was
introduced. Additional details of the migration of the non-event
databases are described in Section~\ref{sec:onl-dbimprove} above.

Other than the introduction of the non-Objectivity event store, the
most significant architectural change to offline computing in the new
computing model was the separation of the generation of calibration
constants from the reconstruction of the full event data stream.  The
final design, described above in Section~\ref{sec:offline-calib},
provided a nearly luminosity-independent number of events per unit
time to the calibration processes, far smaller than the full event
rate.  This made the calibration processing virtually independent of
accelerator upgrades or changes in background, and able to execute on
many fewer CPUs.

With calibration performed as a separate pass, the subsequent
reconstruction of the full event data stream benefited because a)
calibration constants derived from a given run could be used in its
full reconstruction, b) the latency to complete the processing of a
single run was no longer constrained by the time to acquire the run,
and c) it could be completed out of order.  This greatly improved
performance, fault-tolerance, and operational flexibility.

With the benefits of these and other lesser improvements,
and the extensive use of distributed computing,
the redesigned and upgraded offline computing system was able to 
continue to scale to increased luminosity and total data sample size
throughout the life of the experiment.  
Notably, it also proved able to handle the much higher hadronic event rates at
the \TwoS\ and \ThreeS\ resonances explored in the final months of data-taking.


\renewcommand{\secname}{datapreservation_}
\renewcommand{\sectiondir}{Sec07_Data_Preservation}
\section{Long Term Data Access}
\label{section:LTDA}

\subsection{Overview}

As the end of \babar\ data taking approached, it became evident that
physics analyses would continue for several years.  The large
data sample offered unique opportunities for future studies, extending
over the succeeding decade. Furthermore, in response to new
theoretical insights, improved experimental techniques, the necessity
for cross checks mandated by new results elsewhere, or the quest for
higher sensitivity of combined analysis, could motivate extended or
often highly sophisticated new analyses of the data.

The final \babar\ dataset produced with the most recent software
release, referred to as R24, consists of over $9\times 10^9$
reconstructed physics events and $27\times 10^9$ simulated events
that, including the skimmed data, are stored in $1.6 \times 10^6$
files.  The total data amount to about 2\pbytes, specifically
700\tbytes\ of raw data, and 985\tbytes\ and 653\tbytes\ of
reconstructed real and simulated data for the two most recent software
releases, R24 and the earlier R22, both being actively used in ongoing
analyses. These data are referred to as {\it Legacy data}.

The \babar\ collaboration decided that the best way to preserve this
most valuable heritage of the experiment was to create a computing
system that would sustain the full analysis capability, including the
preservation of the data and the reconstruction and simulation
software, accompanied by detailed documentation for future users. The
primary challenge was to design and implement such a system at a time
when resources available to the collaboration were already declining
and experts for data analysis and computing were transitioning to
other experiments. 

To assess the scope of such a data preservation
project the collaboration determined the number of actual and
projected publications per year. Not accounting for new initiatives,
the future research activity is expected extend to at least 2018 and
result in more than 60 publications, as can be seen in
Figure~\ref{PublicationTrend}.

\begin{figure}[hbpt!]
\begin{center}
\includegraphics[width=6.5cm]{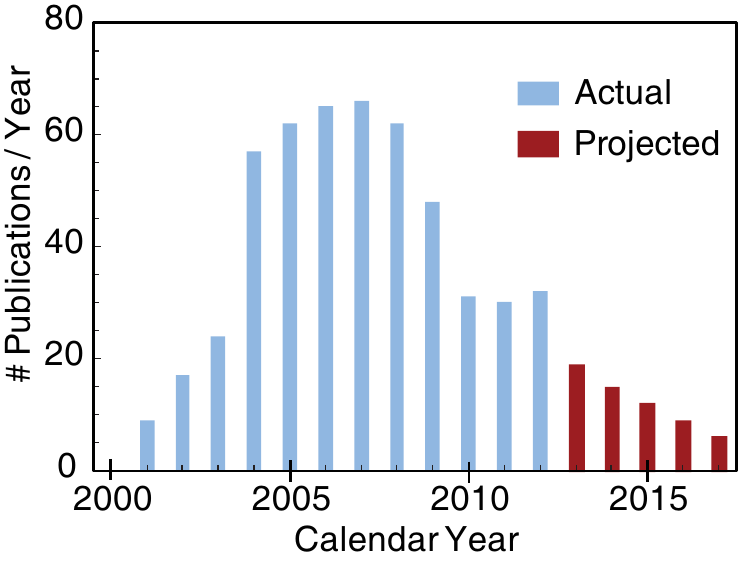}
\caption{
The number of \babar\ publications per calendar year,
actual up to 2012 and projected for the years beyond.}
\label{PublicationTrend}
\end{center}
\end{figure}

This results of this study led to the inception of a formal \babar\
data preservation project in 2008, referred to as the {\it Long Term
  Data Access} (LTDA) project.  After intense R\&D, consultation and planning,
the LTDA hardware acquisition and setup was completed by the end of
2011. A period of software installation, configuration, testing, and
benchmarking followed until the LTDA system was officially ready at
SLAC in March 2012. By this time, several \babar\ collaboration
members were already using the system; since then the LTDA use has
substantially increased. Conceptually, the LTDA system operates
as a separate logical {\it site}, similar
to the remote analysis and production sites associated with \babar.

In the following, the LTDA requirements and goals are presented, and
the implementation, benchmark testing, and operational experience are
described.

\begin{figure*}[hbpt!]
  \begin{center}
    \includegraphics[width=0.8\textwidth]{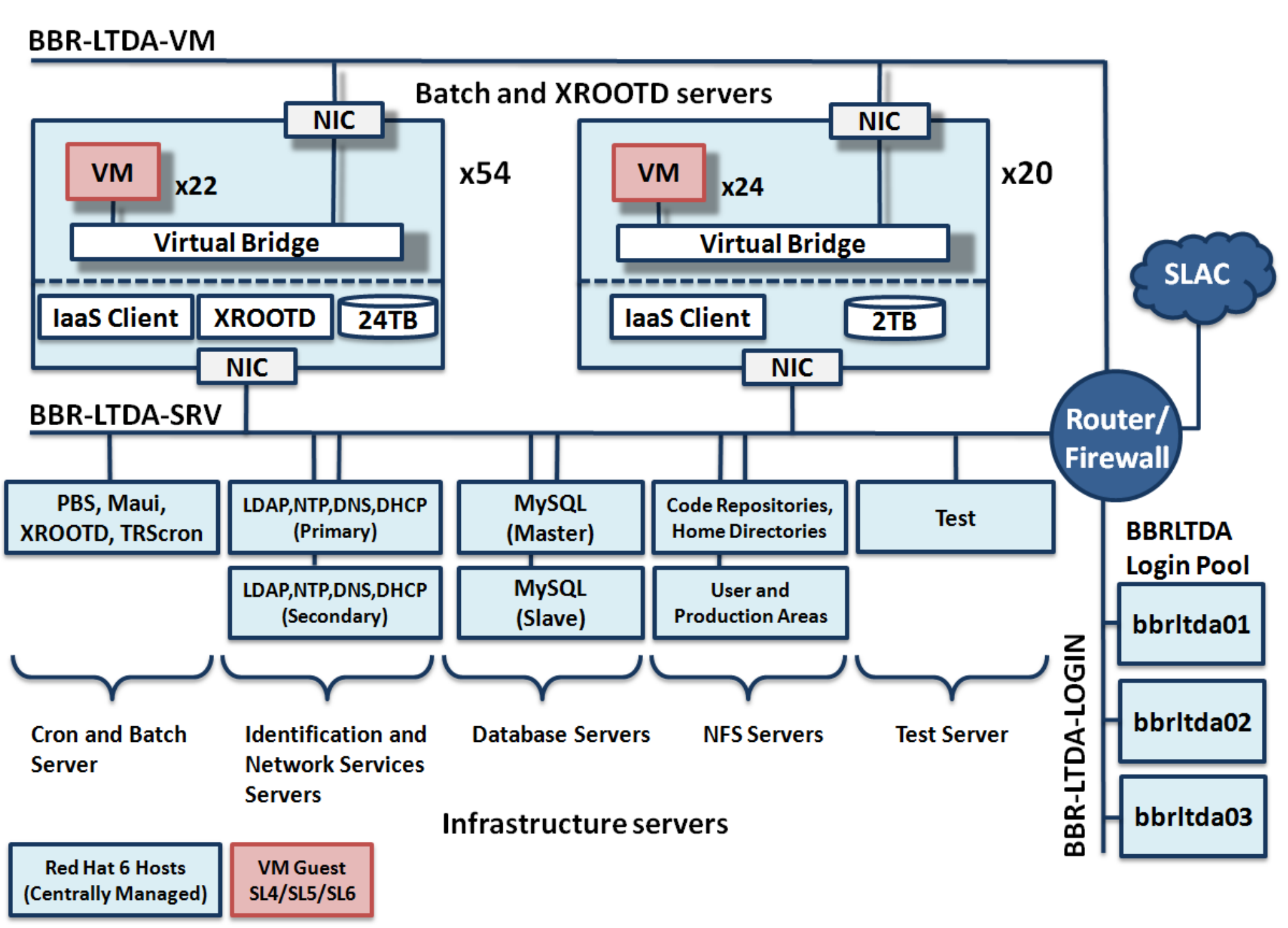}
    \caption {Schematic layout of the LTDA cluster showing the 74
      batch servers hosting the virtual machines, the infrastructure
      servers, the login machines, and the networks to which the
      servers are attached, with connections to the router enforcing
      the firewall rules.  }
    \label{FIG:LTDAnet}
  \end{center}
\end{figure*}

\subsection{Goals and Requirements}

The primary LTDA design goal was to preserve the full ability to
analyze the \babar\ data at least until the end of 2018. Future
analyses may require revised event selections, improvements to
calibrations and event reconstruction algorithms, simulations of new
signatures, and new studies of efficiencies and corrections to the
simulation. Therefore, in addition to being able to analyze the R24
and R22 data sets, the LTDA system also needs to preserve full access
to raw data, simulated data and control samples for calibrations;
allow performance of event reconstruction, skimming, analysis, and fitting
procedures; and provide the complete support of code repositories,
database and tools, accompanied by accurate documentation for
current and future users. 

With the expected decline of the availability of computing professionals 
after the cessation of data taking, and the continuing
rapid changes in storage technology, computing, and software systems,
maintenance and the migration of
the data analysis infrastructure to newer systems posed a major challenge.
Thus a principal goal of the LTDA design had to be the minimization of
maintenance. In particular, means had to be found to deal with
frequent updates to the operating system (OS) and the need to validate
the \babar\ code after substantial changes to the hardware or software
environment.

Over time, the LTDA system will become the primary \babar\ facility for data
analysis.

\subsection{Design}

The design approach chosen to address the above requirements was that
of a {\it contained computing system}, allowing the creation of a
controlled environment that simplifies the support of the \babar\
analysis framework within a frozen operating system, shielded from
upgrades. This also simplifies software maintenance, documentation and user
support and requires less personnel to maintain the system. Users are
able to access the frozen platform interactively for code development,
debugging, and via batch queues for bulk data processing. All
frequently accessed data are made available on disk to provide good
performance in a standalone environment, and to limit the need for
access to the SLAC tape system which serves primarily as a data backup
system.

The estimate of the required CPU power was based on the projected analysis activity, as
reported in Figure~\ref{PublicationTrend}.
The amount of disk storage was determined to be 1.3\pbytes based on
the size of the R22 and R24 legacy data sets; raw data are not kept on
disk since they are not directly usable for physics analysis.
The entire legacy data set was migrated to
new tape media at SLAC in 2010 and, in preparation for the long term
data analysis period, backed up at the
IN2P3 computing center in Lyon (France).

\subsection{Implementation}

The central element of the \babar\ LTDA system is an integrated computing and
storage cluster that uses virtualization to permit the preservation of
past, stable and validated operating system platforms ({\it
  back-versioned OSs}) on current and future hardware and software
platforms; for scalability it relies on distributed computing and
storage. Because of the need for running a back-versioned and
potentially insecure OS, particular attention had to be paid to
computer security.

In order to reduce the security risk to an acceptable level a
risk-based analysis and design approach was used. It was assumed that
any computer that could be easily compromised would be compromised at a
privileged level.  Based on this the LTDA vulnerabilities were
assessed and controls put in place to prevent or mitigate unauthorized
access, accidental modification or destruction of data, persistency of
compromised elements, and any unplanned events. It is important to
note that virtualization itself does not guarantee security. Without
appropriate mitigations, a virtual machine (VM) running an old,
exploitable or compromised OS is not very different from a physical
machine running the same system.

The primary line of defense is a network firewall that isolates the
VMs with deprecated versions of software components from the SLAC
network. The VMs are prevented from connecting to the SLAC network and
only well-defined services are permitted between the VMs and the other
LTDA servers. Moreover, the VMs are only allowed to write to very
specific areas of the LTDA system; for example, they have no write access to
the user home directories.

A schematic outline of the LTDA cluster is shown in
Figure~\ref{FIG:LTDAnet}. The cluster is connected to the main SLAC
network though a router which also implements the network isolation
rules. Within the LTDA system there are three separate networks to which
different elements of the cluster are attached. All back-versioned
components, {\it i.e.,} the VMs running back-versioned OSs, are
connected to the BBR-LTDA-VM network. Network firewall rules define
which connections are allowed to the service network (BBR-LTDA-SRV)
and the login network (BBR-LTDA-LOGIN). The physical machines that
host the VMs, providing batch services and \xrootd~\cite{xrootd}
disk space, and other infrastructure servers, are connected to the
BBR-LTDA-SRV network. The infrastructure servers provide user
identification and network services. They support the databases and
run the batch system daemon. Code repositories, user home
directories, work and production areas are stored on two NFS servers
(Sun X4540), each providing 32\tbytes\ of disk space protected with
raidz2 redundancy. The login servers, grouped in a load-balanced
pool, are attached to the BBR-LTDA-LOGIN network.

The infrastructure servers and the physical storage and compute
servers do not require special security measures because they only use
managed platforms (currently Red Hat Enterprise Linux 6) which are
updated regularly by SLAC computing services, and thus pose no
specific security threats other than those common to the central SLAC
computing system.

\begin{figure*}[hbpt!]
\begin{center}
\includegraphics[width=14cm]{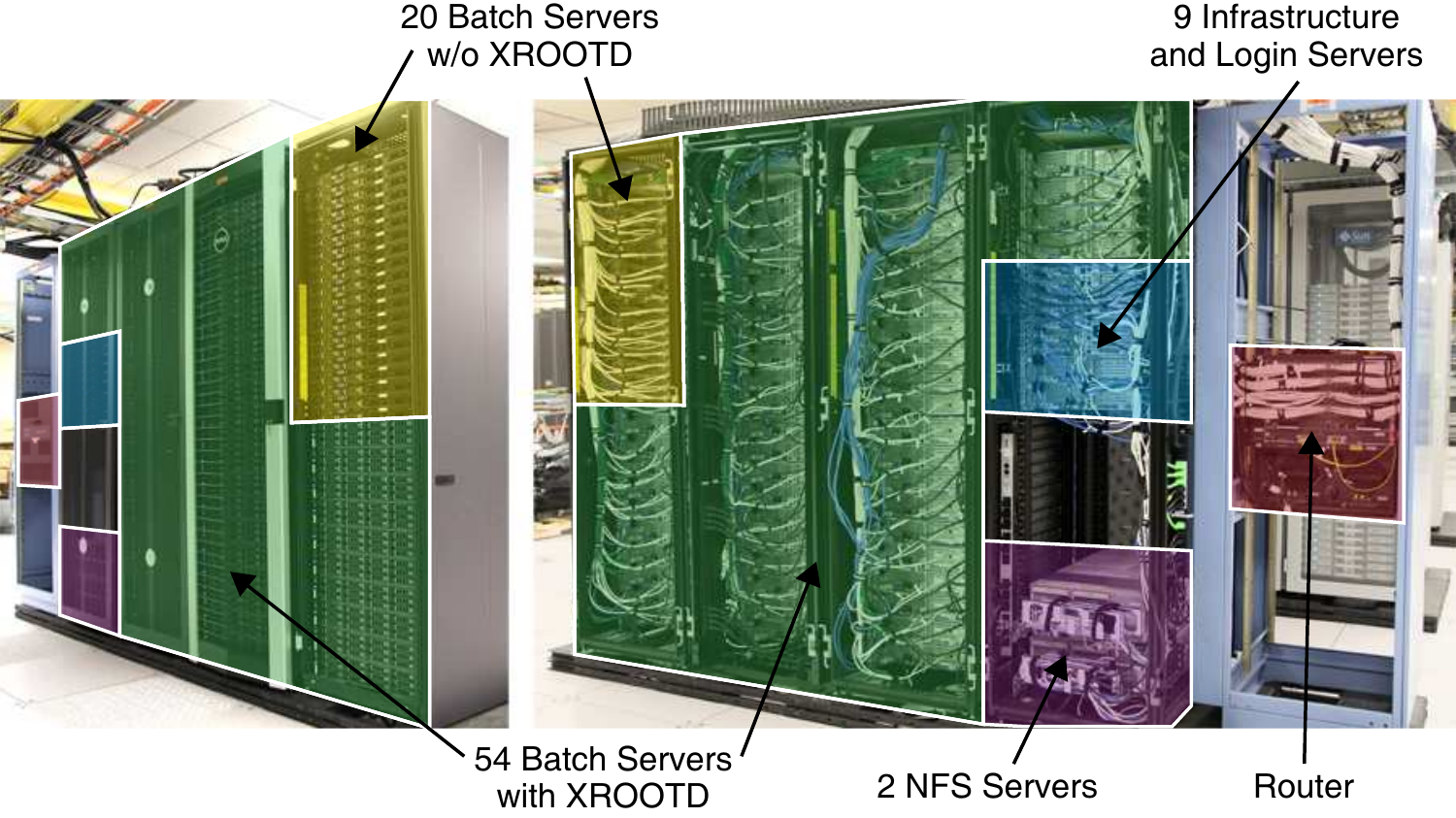}
\caption{LTDA hardware: five racks, front (left) and back (right),
  containing the batch and \xrootd\ files servers, the infrastructure
  and login servers, the NFS servers, and the router implementing the
  firewall to isolate the virtual machines.}
\label{ltdaphot}
\end{center}
\end{figure*}

The LTDA hardware fits into five standard racks that are installed in
the SLAC computing center (see Figure~\ref{ltdaphot}).

The compute and storage cluster consists of 74 servers, each with two
six-core CPUs (3.06\ghz\ Intel Xeon 5674) and 48\gbytes\ of RAM.
Fifty-four of those are Dell R510, configured with 24\tbytes\ of disk
for system, scratch and distributed data storage; the remaining 20
servers (Dell R410) have only 2$\times$2\tbytes\ in a mirrored
configuration for system and scratch space.  The distributed data
storage is managed by \xrootd, effectively turning the cluster into an
efficient low-cost high-bandwidth storage array that holds the data to
be analyzed. These files are easily recovered from the SLAC tape
system, obviating the need for RAID protection and maximizing the
amount of disk space.

Each server has 12 physical CPU cores---taking advantage of
hyper-threading~\cite{hyperthreading}, each core can operate as two
independent logical processors, resulting in a considerable gain of
CPU power under full load. On the 54 compute and storage servers, 2 of
the 24 logical cores are reserved for \xrootd\ and system I/O, leaving
22 logical cores to run VMs. On the 20 compute-only servers all 24
logical cores can be utilized by VMs. 

The LTDA system uses the TORQUE Resource Manager~\cite{TORQUE} with the Maui
Scheduler~\cite{MauiScheduler} as a batch system; virtualization is
done by QEMU (Quick EMUlator)~\cite{qemu} using KVM (Kernel-based
Virtual Machine)~\cite{kvm} for hardware virtualization. Scripts were
developed that allow the batch system to create VMs on
demand, run a job within the VM, and destroy the VM after completion
of the job. 
The guest OS instances are managed by providing \babar-specific system
images that are made available to the VM hosts through NFS. These
images contain the base operating system; all the additional
\babar\ software is provided via NFS. The LTDA system supports the most
current major release of Scientific Linux OS (SL6), as well as the two
previous major OS releases. The base images are read-only and the
QEMU/KVM hypervisors use copy-on-write to a local scratch disk to
permit modifications of a guest OSs' system disks while the VM is
executing.  Any modifications to a guest's system disk are lost when
its VM is deleted at the end of the job, a very important aspect of the
security design that limits such problems to the lifetime of a VM.

\subsection{Performance}

To benchmark the data processing capabilities of the LTDA system and compare
them with those of the SLAC central computer system, the entire
\babar\ dataset was analyzed on both systems using a sample physics
analysis. The physics results were identical. The difference in the
total CPU time required to process the 1500 jobs of the sample
analysis was negligible.

Before deciding on the final configuration of the batch servers, a
series of extensive tests were performed. Specifically, to understand
the impact of hyper-threading, different configurations were tested by
repeatedly processing the sample physics analysis. It was found that
by enabling hyper-threading and doubling the number of VMs per server,
the total processing time for an analysis could be reduced by about
$18\%$.

Tests of the transfer speed for data files provided by \xrootd\ on all
batch servers to the VMs showed that the \xrootd\ cluster is capable
of delivering up to 12\gbytesps without significant impact on data
processing performed on the remaining cores of the servers.

The application of KSM~\cite{ksm} combines identical memory pages
among different processes into a single one which results in a
reduction of memory usage of about $45\%$, a very significant gain for
the parallel operation of many VMs on a single batch machine.  The
unused memory provides additional cache on the batch machine that can
lead to gains in performance under heavy load. 

Based on these and other tests, the standard LTDA system was configured to
use KSM and hyper-threading with up to 22 VMs per compute and storage
server and up to 24 VMs per compute-only server. Overall, no
performance decrease for data processing with virtualization compared
to the conventional processing environment has been observed.

In summary, the LTDA system has been successfully implemented and is
gradually taking the place of the original \babar\ computing system.
It is currently used for tens of physics analyses by users, most of
them working from remote home institutions.  Through the use of
virtualization and a unique combination of processing and storage
systems, this novel system meets the \babar\ analysis needs, while
maintaining the familiar analysis environment. The use of \xrootd,
interfaced with the SLAC tape library, mitigates the impact of
occasional disk failures.  Virtualization through KVM and QEMU images
with KSM memory management allows easy control of multiple stable
platforms that are quickly bootable on demand with little overhead.


\renewcommand{\secname}{summary_}
\renewcommand{\sectiondir}{Sec08_Summary}
\section{Summary}

The diverse \babar\ physics program was based to a very large degree on technological achievements, the delivery of stable beams of ever increasing luminosity by the \pep2\ storage rings, and the reliable operation and excellent performance of the \babar\ detector and its associated software, both online and offline.  

\pep2\ exceeded its design luminosity within its first full year of operation, 
and its performance continued to improve in the following years. 
By the end of data taking in 2008, the instantaneous luminosity had reached a peak of $12 \times 10^{33} \cms$, and up to 20\invfb were delivered per month. 
This was achieved through changes to the mode of operation and upgrades, the most prominent were the increase in beam currents, and the implementation of trickle injection.  The beam orbits, tunes, and other critical parameters were optimized frequently.  Beyond these operational changes, there were substantial upgrades to the machine hardware components and instrumentation, above all, the addition of klystrons to sustain the increasing beam currents. 

To handle the increase in both signal and background rates, the detector underwent a series of upgrades and changes in operation.  The surprisingly high beam-related backgrounds required the identification and strength of the dominant sources.  This led to the insertion of absorbers to shield the effected detector components, and the installation of a variety of monitors to record the background rates and instrumentation to measure radiation damage.  

Initially, very high levels of radiation in the SVT and unusually high currents in the DCH and the RPCs could initiate  beam aborts to protect the detector components from damage.
It was realized that most incidences of high background were of short duration, and that frequent beam aborts followed by beam injection contributed significantly to the absorbed radiation dose. Consequently, more emphasis was placed on detailed monitoring of \babar\ background conditions, shared with 
\pep2\ operators, and the signal for beam aborts was only activated under severe conditions extending over longer periods of time. 

The efficient \babar\ operation was enabled by the Online System, which spanned from Detector and Run Control, Online Databases, Dataflow, to the online computer farm, servers, data switches, and networks. With the increasing demands of higher data rates and the rapid increase in speed and capacity, the hardware and software, including the operating platforms, had to be replaced, most of them more than once.  Over the years, the main goal was to improve the execution of the many critical tasks and to adjust to changing conditions. Special emphasis was placed on automation, simplification of online control and monitoring, as well as on error diagnosis and recovery.  
 Driven by constant surveillance and evaluation of lessons learned by dedicated teams of experts, \babar\ was able to maintain a better than 96\%\ operational efficiency.

Given the flexibility and spare capacity of the trigger system, the only major upgrade to its hardware was the addition of stereo layer information to the charged track trigger which resulted in a more effective suppression of background tracks originating from outside the luminous region of the beams.
Trigger conditions could be readily adjusted to keep dead-times at a minimum.
Trigger efficiencies for most physics processes of interest were maintained at over 99\%, and triggers for new physics signatures were easily added.

A central feature of the \babar\ detector was the
common architecture of the electronics.  It not only reduced the number of different components, but it also unified calibration and monitoring procedures, 
and thereby simplified the diagnosis of problems.  It also
simplified many routine maintenance tasks, upgrades, and the replacement of 
components, thereby reducing detector down times. 

To cope with higher background rates and maintain the nearly dead-time-free 
operation, upgrades to the front-end electronics and data acquisition were successfully implemented.  For the DIRC, faster TDCs were installed and a much faster feature extraction was developed, while for the SVT, the sensitive time window was narrowed. For the DCH, the feature extraction was transferred from the ROM to a new Xilinx FPGA on the interface card that was mounted on the detector endplate.  All of these changes were fully simulated prior to implementation. 

The only detector system which required major upgrades and a rebuild was the IFR.  In the barrel section, the original RPCs were replaced by layers of LSTs, and at the same time, the absorber thickness was increased to improve the pion rejection.
Given the high background rates in the forward endcap, the RPC were replaced with new chambers of similar design; more stringent quality and test procedures during the assembly eliminated most of the reliability problems of the past.   
However, after several years of operation, rate dependent inefficiencies were observed that were attributed to the large streamer discharge.  This problem was resolved by converting the chambers to operate in avalanche mode.  

The scale of the offline computing and data storage requirements were unprecedented in HEP and presented enormous challenges.  The implementation and continuous upgrades of the system's hardware and software required innovation in many areas and the resources of the whole collaboration.  It led to the pioneering development of distributed computing supported by seven large computing centers. An elaborate system of checks assured the quality of the data and integrity of the software used by hundreds of scientists. Substantial redesigns and hardware and software adaptation continuously enhanced the capabilities of the system and were accomplished without interfering with the physics analyses.
The diverse needs of the hundreds of scientists analyzing the data inspired the design of a flexible and configurable mechanism to access the results of event reconstruction and calibrations, both for events recorded by \babar\ and simulated by Monte Carlo techniques.

Over the years, event reconstruction and simulation underwent substantial refinements.  They included innovative techniques for precision alignment of the SVT and DCH, benefiting  not only  the overall efficiency and precision of the track reconstruction, but also the matching of tracks with signals in the DIRC and EMC.  The application of multivariate analysis methods 
combining all available information pertinent to the identification of charged and neutral particles resulted in much improved background suppression and efficiencies.  Background suppression in the EMC and the reconstruction of charged tracks to lower momenta significantly enhanced the overall event reconstruction efficiencies.

In summary, the \babar\ experiment supported a research program that far exceeded the design goals, largely because of the higher than expected luminosity, the stable performance of the \babar\ detector, and the advanced analysis methods developed by the scientists of the collaboration.  The discovery of \CP-violation in neutral $B$ mesons in 2001 was followed by comprehensive studies in many different $B$ meson decays to determine the parameters of the CKM matrix and to test predictions for \CP-violating effects. A few of these tests and other precision measurements have revealed differences between measured quantities and Standard Model expectations at the level of $2-3 $ standard deviations.  The high rate of direct \CP-violation in charged $B$ mesons and the discovery of mixing in neutral $D$ mesons were unexpected,
and recently the violation of time reversal in the mixing and decay of neutral $B$ mesons was demonstrated, independent of \CP violation. 
The large sample of $\tau$ pairs led to precision measurements of many decays of $\tau$ lepton and sensitive tests of lepton universality.
Beyond that, searches for rare processes, among them many not expected in the framework of the Standard Model, were pursued and this pursuit continues. On the basis of data recorded at higher c.m. energies and at the lower mass \TwoS\ and  \ThreeS\ resonances, limits for the production and decays of axions and other dark matter candidates were established.   

Recently, a system referred to as LTDA was installed that preserves access to the \babar\ data and standard analysis and simulation software to support future analyses of the large \babar\ data sample.  This novel system will also be valuable for the preparation of experiments at future storage rings operating at or near the $\FourS$ resonance.  Beyond that, the LTDA system is recognized as a flagship for data preservation for current and future high energy physics experiments. 


\section*{Glossary}

\BaBarAcronym{ADB}{Analog-to-Digital converter Board}{converting raw EMC signals to digital format}
\BaBarAcronymShort{ADC}{Analog-to-Digital Converter}
\BaBarAcronym{API}{Application Programming Interface}{a communication interface between software components}
\BaBarAcronym{AToM}{A Time-over-threshold Machine}{a radiation-hard ASIC to determine SVT pulse height in front-end electronics}

\BaBarAcronymShort{B1,B2}{\pep2\ dipole magnets in the interaction region}
\BaBarAcronym{BCM}{Bunch Current Monitor}{a device monitoring the charge of each bunch}
\BaBarAcronym{BDT}{Bagged Decision Tree}{a multi-variate algorithm for statistical classification of data samples} 
\BaBarAcronym{BLT}{Binary Link Tracker}{a component of the DCT}
\BaBarAcronym{BPM}{Beam Position Monitor}{determining the transverse positions of the beams}
\BaBarAcronym{BW}{Backward}{the \babar\ endcap facing the incoming HER beam}

\BaBarAcronym{CCE}{Charge Collection Efficiency}{a quantity relevant for SVT sensors}
\BaBarAcronym{cron}{}{time-based job scheduler in Unix-like computer operating systems, enabling users to schedule jobs to run periodically at certain times}

\BaBarAcronym{DAQ}{Data AcQuisition system}{it selects and records events}
\BaBarAcronym{DCCT}{DC Current Transformer}{used by \pep2\ to monitor accurately the current in each beam}
\BaBarAcronym{DCH}{Drift CHamber}{the outer tracking system used to measure charged particle momenta and \dedx}
\BaBarAcronym{DCT}{Drift Chamber Trigger}{one of the two main components (with the EMT) of the first level (hardware-based) trigger}
\BaBarAcronym{DCZ}{Drift Chamber Z-trigger}{upgrade to the DCT which uses longitudinal track information}
\BaBarAcronym{DFB}{DIRC Front-end Board}{receives data from 64 PMTs in the DIRC}
\BaBarAcronym{DIRC}{Detector of Internally Reflected Cherenkov light}{the \babar\ ring-imaging Cherenkov detector for charged particle identification}
\BaBarAcronym{DMA}{Direct Memory Access}{a computing functionality used in the online system}
\BaBarAcronym{DQG}{Data Quality Group}{team of experts (one per system) responsible for validating the data reconstruction, from the raw data to the physics analysis input}
\BaBarAcronym{DOCA}{Distance Of Closest Approach} {a variable used in track and vertex reconstruction to measure the distance of a charged track relative to the IP or the primary or decay vertex}

\BaBarAcronym{ECOC}{Error-Correcting Output Code}{a tool for multi-class categorization of problems used in the most powerful PID algorithms}
\BaBarAcronym{EMC}{ElectroMagnetic Calorimeter}{it measures photon and electron energies in the barrel and forward endcap sections}
\BaBarAcronym{EMT}{ElectroMagnetic calorimeter Trigger}{one of the two main components (with the DCT/DCZ) of the first level (hardware-based) trigger}
\BaBarAcronym{EPICS}{Experimental Physics and Industrial Control System}{a toolkit for building control systems}
\BaBarAcronym{ER}{Event Reconstruction}{the second step (following the prompt calibration) of the data reconstruction} 

\BaBarAcronym{FADC}{Flash Analog-to-Digital Converter}{ADC comparing input voltage to a fast sequence of reference voltages}
\BaBarAcronym{FCTS}{Fast Control and Timing System}{provides the accelerator time reference to the online system}
\BaBarAcronym{FEC}{Front-end Electronics Card}{part of the IFR FEE}
\BaBarAcronym{FEE}{Front-end Electronics}{the generic name for electronic circuits receiving and processing the raw signals from the detector systems}
\BaBarAcronym{FEX}{Feature EXtraction}{the online processing of raw data in the ROMs}
\BaBarAcronym{FM}{Fast Monitoring}{the framework used for live monitoring of the recorded data}
\BaBarAcronym{FFC}{Flexible Flat Cable}{used in the LST system}
\BaBarAcronym{FPGA}{Field-Programmable Gate Array}{a configurable integrated circuit}
\BaBarAcronym{FR4}{Flame Retardant Grade 4}{a widely accepted international grade designation for fiberglass reinforced epoxy laminates that are flame retardant}
\BaBarAcronym{FW}{Forward}{the \babar\ endcap facing the incoming LER beam}

\BaBarAcronym{GLT}{Global Level~1 Trigger}{the system deciding to accept (for further processing) or reject an event}

\BaBarAcronym{HEP}{High Energy Physics}{}
\BaBarAcronym{HER}{High Energy Ring}{the \pep2\ storage ring for 9.0\gev\ electrons}
\BaBarAcronym{HOM}{High-Order Mode}{non-fundamental transverse modes of RF cavities or waveguides}
\BaBarAcronym{HPSS}{High Performance Storage System}{the main data storage system on tape used in \babar}

\BaBarAcronym{IAD}{Integration-Amplification Daughter}{a component of the LST FEE}
\BaBarAcronym{IC}{Integrated Circuit}{a micron-scale electronic circuit on semiconductor substrate}
\BaBarAcronym{IFB}{IFR FIFO Board}{a component of the IFR FEE}
\BaBarAcronym{IFR}{Instrumented Flux Return}{the \babar\ system to identify muons and particles penetrating the segmented steel flux return}
\BaBarAcronym{IFT}{Instrumented Flux return Trigger}{the trigger for muon efficiency studies in the IFR}
\BaBarAcronym{IOB}{Input-Output Board}{a component of the FEE}
\BaBarAcronym{IOC}{Input/Output Converter}{circuits used by EPICS}
\BaBarAcronym{IP}{Interaction Point}{the center of the luminous region of the colliding \pep2\ bunches}
\BaBarAcronym{IR2}{Interaction Region 2}{the location of the \babar\ detector in the \pep2 tunnel}
\BaBarAcronym{ISR}{Initial State Radiation}{emission of photons by the colliding \ep\ and \en}

\BaBarAcronym{KSM}{Kernel Samepage Merging}{a software to allow sharing of identical memory pages amongst different processes or virtualized guests}
\BaBarAcronym{KVM}{Kernel based Virtual Machine}{a virtualization infrastructure for the Linux kernel}

\BaBarAcronym{L1}{Level 1 trigger}{the first level (hardware) trigger}
\BaBarAcronym{L3}{Level 3 trigger}{the second level (software) trigger}
\BaBarAcronym{LED}{Light-Emitting Diode}{used to generate controlled pulses of light, for instance for the DIRC calibration}
\BaBarAcronym{LER}{Low Energy Ring}{the storage ring for 3.1\gev\ positrons}
\BaBarAcronym{LH}{LikeliHood}{a mathematical tool commonly used in \babar\ analysis to estimate the parameters of a statistical model}
\BaBarAcronym{LM}{Logging Manager}{a component of the online system transferring data to disks}
\BaBarAcronym{LST}{Limited Streamer Tube}{the final technology used to instrument the barrel IFR, replacing the original \babar\ RPCs}
\BaBarAcronym{LTDA}{Long Term Data Access}{the computing system to permit access to \babar\ data and software for future analyses}

\BaBarAcronym{MC}{Monte-Carlo}{the generic method used to simulate events}
\BaBarAcronym{MIP}{Minimum-Ionizing Particle}{a particle with an energy loss rate (\dedx) close to the minimum}

\BaBarAcronym{NEG}{Non-Evaporable Getter}{a type of vacuum pumps used by \pep2}
\BaBarAcronym{NIEL}{Non Ionizing Energy Loss}{energy causing displacements in Silicon crystals}
\BaBarAcronym{NIOS}{Altera Trademark}{a versatile embedded processor widely used in FPGAs}
\BaBarAcronym{NN}{Neural Network}{a multi-variate algorithm for classification of data samples}

\BaBarAcronym{ODB}{Online Databases}{store detector-related information (configurations, conditions and calibration constants)}
\BaBarAcronym{ODC}{Online Detector Control}{controls and monitors the environmental detector  conditions}
\BaBarAcronym{ODF}{Online DataFlow}{component of the online system to control the extraction of information from the FEEs for event building}
\BaBarAcronym{OEP}{Online Event Processing}{the processing of complete \babar\ events}
\BaBarAcronym{ORC}{Online Run Control}{the top-level control for the detector operation and data taking}

\BaBarAcronym{pCVD}{Polycrystaline Chemical Vapor Deposition}{the material of the two SVTRAD diamond sensors}
\BaBarAcronym{PC}{Prompt Calibration}{the first step of the offline data reconstruction}
\BaBarAcronym{PCB}{Printer Circuit Board}{a component of the LST modules}
\BaBarAcronym{PID}{charged Particle IDentification}{algorithms to identify charged tracks crossing the detector (e, $\mu$, $\pi$, K or p)}
\BaBarAcronym{PBS}{Portable Batch System}{software to perform job scheduling in UNIX clusters}
\BaBarAcronym{PMC}{PCI Mezzanine Card}{small-form-factor card for the PCI bus which can be stacked on top of a host card, typically a single board computer}
\BaBarAcronym{PMT}{PhotoMulTiplier}{an extremely sensitive light detector}
\BaBarAcronym{PTD}{transverse momentum (\pt) Discriminator}{component of the DCT}

\BaBarAcronymShort{Q1-Q4}{\pep2\ quadrupole magnets in the interaction region}
\BaBarAcronymShort{QC}{Quality Control}
\BaBarAcronym{QCOW2}{Qemu Copy-On-Write version 2}{is the Copy-on-Write feature
of QEMU. Copy-On-Write is an optimization technique based on the fact that as long as multiple programs running on a host need read only acess to a given data structure (a memory allocation, data on a disk, ...), a pointer to the same data can be given to each of these programs. If one of these programs needs to execute a write, only then a private copy is created and made available to the program}
\BaBarAcronym{QEMU}{Quick EMUlator}{a generic open source emulator and virtualizer for the KVM kernel module in Linux achieving near native performances by executing the guest code directly on the host CPU}

\BaBarAcronym{RADFET}{RADiation sensitive Field-Effect Transistor}{sensors monitoring the absorbed dose in the EMC}
\BaBarAcronym{RAID}{Redundant Array of Independent Disks}{a storage technology that combines multiple disk drive components into a logical unit}
\BaBarAcronymShort{RF}{Radio-Frequency}
\BaBarAcronymShort{RH}{Relative Humidity}
\BaBarAcronym{ROM}{ReadOut Module}{it processes raw data from the FEE prior to transfer to the DAQ}
\BaBarAcronym{RPC}{Resistive Plate Chamber}{the initial technology used to instrument the IFR. The original \babar\ RPCs were replaced by the LSTs in the barrel and upgraded in the forward endcap}
\BaBarAcronym{RTEMS}{Real Time Executive for Multiprocessor Systems}{an operating system}

\BaBarAcronym{SLM}{Synchrotron Light Monitor}{measuring the bunch profiles}
\BaBarAcronym{SOB}{StandOff Box}{the DIRC photon camera, a large tank filled with ultra-pure water, located at the backward end of \babar}
\BaBarAcronym{STP}{Signal Transfer Plane}{a component of the LST modules}
\BaBarAcronym{SVT}{Silicon Vertex Tracker}{the inner tracking system}
\BaBarAcronym{SVTRAD}{Silicon Vertex Tracker RADiation protection system}{mounted on the SVT to sense high radiation levels and abort the beams whenever a preset limit is exceeded}

\BaBarAcronymShort{TDC}{Time-to-Digital Converter}
\BaBarAcronym{TORQUE}{Terascale Open-Source Resource and QUEue Manager}{a distributed resource manager controling batch jobs}
\BaBarAcronym{TRG}{Trigger}{the two-stage \babar\ event selection system}
\BaBarAcronym{TSF}{Track Segment Finder}{a component of the DCT}
\BaBarAcronym{TSP}{Titanium Sublimation Pump}{a type of vacuum pumps used by \pep2}

\BaBarAcronym{UPILEX}{Upilex}{a heat and radiation resistant polyimide film formed from biphenyl tetracarboxylic dianhydride}
\BaBarAcronym{UDP}{User Datagram Protocol}{a protocol in the Internet Protocol (IP) suite that provides connectionless ("datagram") communications}

\BaBarAcronym{VM}{Virtual Machine}{a software that emulates a computer environmenton which an operating system can run. A VM, also called guest, behaves like it were a separate computer, isolated from the physical computer, the host, on which the emulation software is running}
\BaBarAcronym{VME}{short for VMEbus}{a bus standard widely used by data acquisition systems}

\BaBarAcronymShort{\xrootd}{a generic software framework for fast, low latency and scalable data access}
\BaBarAcronym{XTC}{eXtended Tagged Container}{the format of the \babar\ raw event data}

\BaBarAcronym{ZFS}{ZFS File System}{a combined file system and logical volume manager}
\BaBarAcronym{ZPD}{$z$-\pt\ Discriminator}{a component of the DCZ trigger}
\BaBarAcronym{ZTB}{$z$-strip transition board}{an element of the LST electronics}

\section*{Acknowledgments}
The authors are grateful for the tremendous support they have received from their home institutions
over the almost ten years of \babar\ and \pep2\ operation.
The excellent performance of \pep2, the data taking, the data processing and the data analysis would
not have been possible without the extraordinary expertise, dedication and collaboration of the
supporting staff at SLAC and other \babar\ institutions, as well as at participating computing
centers. The collaborating institutions wish to thank SLAC for the kind hospitality extended to them.
Members of the International Finance Committee should be commended for their continued assistance
in securing resources for the operation and upgrades of the experiment.

This work has been supported by the US Department of Energy and National Science Foundation,
the Natural Sciences and Engineering Research Council (Canada), the Commissariat \`a l'Energie
Atomique and Institut National de Physique Nucl\'eaire et de Physique des Particules (France),
the Bundesministerium f\"ur Bildung und Forschung and Deutsche Forschungsgemeinschaft (Germany),
the Istituto Nazionale di Fisica Nucleare (Italy), the Foundation for Fundamental Research on
Matter (The Netherlands), the Research Council of Norway, the Ministry of Education and
Science of the Russian Federation, Ministerio de Econom\'{\i}a y Competitividad (Spain),
and the Science and Technology Facilities Council (United Kingdom).
Individuals have received support from
the Marie-Curie IEF program (European Union) and the A.~P.~Sloan Foundation (USA).


\def\bibname{References}
\bibliographystyle{NIMU}
\bibliography{NIMU}

\begin{thebibliography}{143}
\bibitem[{Onl(????)}]{Online:EmbeddedLinux}
\bibinfo{note}{The Embedded Linux distribution and its documentation are no
  longer available.}
\end{thebibliography}

\end{document}